%% file: master.tex
\documentclass[a4paper,12pt,twoside]{book}
\usepackage{psfig}
\usepackage{epsfig}
\usepackage{amssymb,wasysym,bbm,rotating,subfigure,here,citesort}
\usepackage{graphicx,color}
\usepackage{letterspace}
\input{aliases.tex} 
\begin{document}
%
%
%
\pagestyle{empty}
%
\input{Title_Abstract} 
%
%
\newpage
\pagenumbering{roman}
\pagestyle{plain}
\tableofcontents
\addtocontents{toc}{\protect\enlargethispage{-0.5cm}}
%
%
\newpage
\pagenumbering{arabic}
\pagestyle{plain}
%
\makeatletter
\@addtoreset{equation}{section}
\makeatother
\renewcommand{\theequation}{\thesection.\arabic{equation}}
%
\input{SectionIntroduction}
%
%
\input{High_Energy_Scattering}
\input{string_decomposition}

\input{Phenomenology}

\input{SectionConclusion}
%
%
\begin{appendix}
%
\input{Appendix}

\end{appendix}
%
%
  
%

\newpage
\pagestyle{empty}
\input{Acknoledgements}

%
\end{document}

%% file: aliases.tex
\font\myfonta=cmbx12 scaled 1330

\newcommand{\alphaS}{\alpha_s}
\newcommand{\alphaEM}{\alpha}
\newcommand{\SVM}{SVM}
\newcommand{\pert}{P}
\newcommand{\nprt}{N\!P}
\newcommand{\soft}{\mbox{\scriptsize soft}}
\newcommand{\hard}{\mbox{\scriptsize hard}}
\newcommand{\pt}{\mbox{\scriptsize pt}}
\newcommand{\WW}{Wegner-Wilson}

\newcommand{\pbar}{{\bar{p}}}
\newcommand{\qbar}{{\bar{q}}}

\newcommand{\Reggeon}{I\!\!R}
\newcommand{\Pomeron}{I\!\!P}
\newcommand{\be}{\begin{equation}}
\newcommand{\ee}{\end{equation}}
\newcommand{\bea}{\begin{eqnarray}}
\newcommand{\eea}{\end{eqnarray}}
\newcommand{\benn}{\begin{displaymath}}
\newcommand{\eenn}{\end{displaymath}}
\newcommand{\beann}{\begin{eqnarray*}}
\newcommand{\eeann}{\end{eqnarray*}}
\newcommand{\barray}{\begin{array}}
\newcommand{\earray}{\end{array}}
\newcommand{\inv}{\frac{1}}
\newcommand{\gtsim}{\gtrsim}
\newcommand{\ltsim}{\lesssim}
\newcommand{\fm}{\mbox{fm}}
\newcommand{\mb}{\mbox{mb}}
\newcommand{\nb}{\mbox{nb}}
\newcommand{\MeV}{\mbox{MeV}}
\newcommand{\GeV}{\mbox{GeV}}
\newcommand{\TeV}{\mbox{TeV}}
\newcommand{\G}{{\cal G}}       
\newcommand{\GG}{\hat{\cal{G}}} 
\newcommand{\Identity}{{1\!\rm l}}

\newcommand{\impactT}{{\cal T}}
\newcommand{\Pc}{{\cal P}}      
\newcommand{\Ps}{{\cal P}_S}    
\newcommand{\Tr}{\mbox{Tr}}             
\newcommand{\rTr}{\Tr_{r}}        
\newcommand{\ronexrtwoTr}{\Tr_{r_1 \otimes r_2}}  
\newcommand{\nrTr}{\tilde{\Tr}_{r}}             
\newcommand{\nroneTr}{\tilde{\Tr}_{r_1}}              
\newcommand{\nrtwoTr}{\tilde{\Tr}_{r_2}}              
\newcommand{\nronexrtwoTr}{\tilde{\Tr}_{r_1 \otimes r_2}}
\newcommand{\roneIdentity}{\Identity_{r_1}}
\newcommand{\rtwoIdentity}{\Identity_{r_2}}
\newcommand{\ronexrtwoIdentity}{\Identity_{r_1 \otimes r_2}}
\newcommand{\im}{\mbox{Im}}    
\newcommand{\Projector}{\mbox{P}} 
\newcommand{\tensor}{t_{r_1 \otimes r_2}} 
\newcommand{\fundamental}{\mbox{\scriptsize N}_c}
\newcommand{\adjoint}{\mbox{\scriptsize N}_c^2\!-\!1}
\newcommand{\Fundamental}{\mbox{N}_c}
\newcommand{\Adjoint}{\mbox{N}_c^2\!-\!1}
\newcommand{\befig}{\begin{figure}}
\newcommand{\efig}{\end{figure}}
\newcommand{\betab}{\begin{table}}
\newcommand{\etab}{\end{table}}


%% file: Title_Abstract.tex
%
\begin{titlepage}
\pagestyle{empty}
%
%
%
\begin{center}
  \Huge \bf \sc
  Anatomy of QCD Strings and\\ 
  Saturation Effects in High-Energy Scattering
\end{center}
\vfill
\begin{center}
  {
  \Large Dissertation\\}
  \vspace{0.3cm}
  {\large
  submitted to the\\
  Combined Faculties for the Natural Sciences and for Mathematics\\
  of the Ruperto--Carola University of Heidelberg, Germany\\
  for the degree of\\ 
  Doctor of Natural Sciences}
\end{center}
\vfill
\begin{center}
\begin{large}
  presented by\\
  \vspace{0.5cm}
  {\LARGE\sc Arif Shoshi}\\
  \vspace{0.5cm}
    born in Sche{\ss}litz, Germany\\
\end{large}
\end{center}
\vfill
\begin{center}
\begin{large}
\begin{tabular}{ll}
Referees: & Prof.~Dr.~Hans-J\"urgen Pirner\\
& Prof.~Dr.~J\"org H\"ufner\\
\end{tabular}

Oral examination: \, February 13, 2003 \, \hphantom{.}

\end{large}
\end{center}
\newpage
%
%
\thispagestyle{empty}
\vspace{8ex}
\begin{center} 
{\bf Zusammenfassung}
\end{center}

Wir erstellen ein Modell, um Hochenergiereaktionen von Hadronen und
Photonen zu berechnen. Der Gluon-Austausch beschreibt die perturbative
und das Modell des Stochastischen Vakuums die nichtperturbative
Wechselwirkung. Letzteres f{\"u}hrt zum Quark-Confinement in Dipolen
via eines Strings von Farbfeldern.  Wir erforschen die QCD-Struktur
der Dipol-Dipol Wechselwirkung im Impulsraum, vor allem die
Wechselwirkung zwischen den Strings. Wir stellen den String als eine
Ansammlung von stringlosen Dipoles dar, zeigen Confinement-Effekte in
der Hochenergiestreuung und berechnen die unintegrierte
Gluonverteilung von Hadronen und Photonen. Im Sto{\ss}parameterraum des
Streuprozesses untersuchen wir das Unitarit{\"a}tslimit der
$S$-matrix. Wir berechnen die Sto{\ss}parameter-Profile der Hadron-Hadron
und Photon-Hadron Streuung, bestimmen die Energiewerte bei welchen die
Profile das Black-Disc-Limit erreichen, sch{\"a}tzen die
sto{\ss}parameterabh{\"a}ngige Gluonverteilung des Protons ab und
diskutieren die Gluons{\"a}ttigung. Wir vergleichen die Resultate des
Modells f{\"u}r $pp$, $\pi p$, $Kp$, $\gamma^* p$ und $\gamma\gamma$
Reaktionen mit experimentellen Daten und identifizieren
S{\"a}ttigungseffekte in experimentellen Messgr{\"o}{\ss}en.
\begin{center}
\newcommand{\ganzelaenge}{30}
\newcommand{\halbelaenge}{15}
\newcommand{\ganzehoehe}{50}
\setlength{\unitlength}{1pt}
\begin{picture}(\ganzelaenge,\ganzehoehe)(0,0)
\end{picture}
\end{center}

\begin{center}{\bf \normalsize Abstract }\end{center}

We develop a model to compute high-energy reactions of hadrons and
photons. The perturbative interaction is described by gluon exchange
and the non-perturbative interaction by the stochastic vacuum model
which leads to quark-confinement in dipoles via a string of color
fields. We study the QCD structure of the dipole-dipole scattering in
momentum space focussing especially on interactions between strings. We
represent the string as a collection of stringless dipoles, show
confinement effects in high-energy scattering and calculate
unintegrated gluon distributions of hadrons and photons.  In the
impact parameter space of the scattering process we investigate the
unitarity limit of the $S$-matrix. We calculate the impact parameter
profiles for proton-proton and photon-proton scattering, determine the
energy values at which the profiles saturate at the black disc limit,
estimate the impact parameter dependent gluon distribution of the
proton and discuss gluon saturation. We compare the model results for
$pp$, $\pi p$, $Kp$, $\gamma^* p$ and $\gamma\gamma$ reactions with
experimental data and identify saturation effects in experimental
observables.

\normalsize \pagestyle{empty}
\cleardoublepage
%
%
%
\end{titlepage}
%

%% file: SectionIntroduction.tex
%
\chapter{Introduction}
\label{Sec_Introduction}

%
%

On the basis of the numerous tests in experiments, one can confidently
say that Quantum Chromodynamics (QCD) is the correct theory of strong
interactions. The asymptotic freedom of QCD has allowed us to
understand the interaction of particles at high momentum transfers.
The non-Abelian nature of QCD leading to color confinement prevents
first-principle calculations of hadronic interactions in the
non-perturbative region.  In fact, it is a key issue in high-energy
physics to understand and describe the non-perturbative structure of
hadronic scattering processes.  The unravelling of confinement effects
in such interactions would be especially important.

%
%

So far Lattice Gauge Theory~\cite{Wilson:1974sk} constitutes the only
access to non-perturbative QCD physics from first principles.
Numerical simulations of QCD on Euclidean lattices~\cite{Rothe:1997kp} give
strong evidence not only for color confinement but also for chiral
symmetry breaking and dynamical mass generation from the QCD
Lagrangian.  Unfortunately, lattice QCD cannot be applied in
Minkowskian space-time to simulate high-energy reactions since it is
limited to the Euclidean formulation of QCD.

%
%

Experiments show a rise of high-energy total cross sections with
increasing center-of-mass (c.m.) energy. The rise is slow for
large-size particles (protons, pions, kaons, or real
photons~\cite{Groom:2000in}) and becomes steep for small-size
particles (highly virtual photons~\cite{Adloff:1997mf+X,Adloff:2001qk}
or charmonia~\cite{Breitweg:1997rg+X}) involved in the scattering
process. It is highly unsatisfactory that we do not have a genuine
understanding of the growing cross sections on the basis of the QCD
Lagrangian. The most rigorous result we have is the Froissart
bound~\cite{Froissart:1961ux+X} stating that hadronic cross sections
cannot grow faster than $\ln^2(s/s_0)$ for asymptotic c.m. energies
$\sqrt{s}$. It is derived on the basis of very general principles such
as unitarity and analyticity of the scattering matrix. Its derivation,
however, does not provide any physical mechanism realizing the squared
logarithmic increase.  At present no quantum field theoretical
understanding of the rising hadron extension at high energies is
available.

%
%

It is hopelessly difficult to solve any of the mentioned problems
of high-energy scattering from first principles alone. Consequently,
models that approximate quantum chromodynamics are required. To be
convincing, however, the proposed models should include the basic
features of QCD, reproduce many experimental high-energy data and
agree with lattice QCD simulations.

%
%
In this work we develop a model to describe the high-energy scattering
of hadrons and photons in the eikonal
approximation~\cite{Shoshi:2002in}. Its central element is the
gauge-invariant light-like {\em Wegner-Wilson loop} $W_{r}[C]$ with
the boundary $C$ and the representation $r$ of
$SU(N_c)$~\cite{Wegner:1971qt,Wilson:1974sk}.  The scattering
amplitude factorizes into the vacuum expectation value of two
correlated Wegner-Wilson loops $\langle W_{r_1}[C_1] W_{r_2}[C_2]
\rangle$ and reaction-specific wave
functions~\cite{Nachtmann:1991ua,Nachtmann:ed.kt,Kramer:1990tr,Dosch:1994ym}.
The Wegner-Wilson loops describe the path of color-dipoles and the
{\em loop-loop correlation function} $\langle W_{r_1}[C_1]
W_{r_2}[C_2] \rangle$ the dipole-dipole scattering. The most
interesting are Wegner-Wilson loops in the fundamental and adjoint
representation of $SU(3)$ which represent color-singlet
quark-antiquark dipoles and glueballs (adjoint dipoles), respectively.
In our framework color-dipoles are given by the quark and antiquark in
mesons or photons and in a simplified picture by a quark and diquark
in baryons. The size and orientation of the color-dipoles in the
hadrons and photons are determined by appropriate light-cone wave
functions.  In this sense the loop-loop correlation function
constitutes the basis for a unified description of hadrons and
photons.


We evaluate the loop-loop correlation funtion $\langle W_{r_1}[C_1]
W_{r_2}[C_2] \rangle$ in the approach of Berger and
Nachtmann~\cite{Berger:1999gu} which resumes gluonic exchanges in the
scattering process using a matrix cumulant expansion and the Gaussian
approximation of the functional integrals in the gluon field
strengths. The resumation is crucial to guarantee the $S$-matrix
unitarity and to investigate saturation effects in high-energy
reactions. In additon to the interaction of two dipoles in the
fundamental representation of $SU(3)$ considered by Berger and
Nachtmann~\cite{Berger:1999gu}, we compute also the interaction of a
fundamental with an adjoint dipole in the representation of $SU(N_c)$.


We express the loop-loop correlation function $\langle W_{r_1}[C_1]
W_{r_2}[C_2] \rangle$ in terms of the bilocal gluon field strength
correlators integrated over two connected surfaces. The surfaces enter
by using the non-Abelian Stokes'
theorem~\cite{Bralic:1980ra+X,Nachtmann:ed.kt} to transform the line
integrals over the gluon potentials in the Wegner-Wilson loops of the
functional integral
approach~\cite{Nachtmann:1991ua,Nachtmann:ed.kt,Kramer:1990tr,Dosch:1994ym}
into surface integrals over gluon field strengths. We use for the
first time explicitly {\em minimal surfaces}, i.e., planar surfaces
bounded by the Wegner-Wilson loops. In Euclidean space-time, this
surface choice is usually used to obtain Wilson's area
law~\cite{Dosch:1987sk+X}.
We have recently shown that
minimal surfaces are actually required to ensure the consistency of
our results for $\langle W_{r}[C] \rangle$ and $\langle W_{r_1}[C_1]
W_{r_2}[C_2] \rangle$ by using low-energy theorems~\cite{Shoshi:2002rd}. The simplicity of
the minimal surfaces is appealing: It allows us to show for the first
time the analytic structure of non-perturbative interactions in
momentum space. The previous pyramid mantle choice for the
surfaces~\cite{Dosch:1994ym,Berger:1999gu,Rueter:1996yb+X,Dosch:1997ss,Dosch:1998nw,Rueter:1998up,D'Alesio:1999sf,Dosch:2001jg}
did not allow such an analysis.


We use in the gluon field strength correlator perturbative and
non-perturbative correlations as needed to describe interactions of
small and large-size particles. The perturbative correlator results
from QCD while the non-perturbative correlator is modelled by
the stochastic vacuum model (\SVM)~\cite{Dosch:1987sk+X}. This
combination allows us to describe short and long distance correlations
in agreement with lattice calculations of the gluon field strength
correlator~\cite{DiGiacomo:1992df+X,Meggiolaro:1999yn}.  Moreover,
this two component ansatz leads to the static quark-antiquark
potential with color-Coulomb behavior for small and confining linear
rise for large quark-antiquark separations as
expected~\cite{Bali:2000gf}.

We use in the non-perturbative correlator the {\em exponential
  correlation function} adviced by lattice QCD investigations of
long-distance correlations~\cite{Meggiolaro:1999yn}. This correlation
function stays positive for all Euclidean distances and, thus, is
compatible with a spectral representation of the correlation
function~\cite{Dosch:1998th}. This represents a conceptual improvement
since the correlation function that has been used in earlier
applications of the \SVM\ becomes negative at large
distances~\cite{Dosch:1994ym,Rueter:1996yb+X,Dosch:1998nw,Rueter:1998up,D'Alesio:1999sf,Berger:1999gu,Dosch:2001jg}.

%
%

In spite of the numerous improvements our approach is still incomplete
as it leads to energy-independent cross sections in contradiction to
the experimental observation. This is because of the missed gluon
radiation in the present model. In this work, however, the energy
dependence is introduced phenomenologically. In line with the
experimental observation mentioned above, we ascribe a strong and weak
energy dependence, respectively, to the perturbative and
non-perturbative component.  Assuming the same mechanism for the
energy dependence of hadron and photon interactions (motivated by the
successful two-pomeron fit of Donnachie and
Landshoff~\cite{Donnachie:1998gm+X,Donnachie:1992ny}) we implement a
{\em powerlike energy dependence} in the loop-loop correlation
function $\langle W_{r_1}[C_1] W_{r_2}[C_2] \rangle$. In this way a
unified description of the energy behaviour of hadron-hadron,
photon-hadron, and photon-photon reactions is obtained. The powerlike
ansatz in combination with the {\em multiple gluonic interactions}
resulting from the resumation method of Berger and Nachtmann is
crucial to guarantee the Froissart bound~\cite{Froissart:1961ux+X}.
The phenomenological energy dependence, of course, can only be an
intermediate step. For a fundamental understanding of the energy
dependence of cross sections quantum evolution has to be incorporated
in our model.

%
%
We adjust the model parameters to reproduce a wealth of high-energy
scattering data, i.e.\ total, differential and elastic cross
sections, structure functions and slope parameters for many different
reactions over a large range of c.m.\ energies. In this way we have
confidence in our model predictions for future experiments (LHC,
THERA) and for energies beyond the experimentally accessible range.



The non-perturbative component of our model gives
confinement~\cite{Dosch:1987sk+X} due to flux-tube formation of
color-electric fields between the color-sources in the
dipole~\cite{DelDebbio:1994zn,Rueter:1995cn,Shoshi:2002rd}.
The thickness of flux-tubes or confining QCD strings saturates for
large dipole-sizes at the value of about one
Fermi~\cite{Shoshi:2002rd}.  The general representation
$r$ of $SU(N_c)$ kept for all computations leads to an exact Casimir
scaling for static color-dipole potentials and QCD
strings~\cite{Shoshi:2002rd} in agreement with lattice
QCD simulations~\cite{Deldar:1999vi,Bali:2000un}. In this work we show
the intrinsic composition of QCD strings, the structure of string
interactions, and manifestations of strings in high-energy scattering.

%

We show for the first time the QCD structure of the perturbative and
non-perturbative dipole-dipole interaction in momentum space within our
model. We reproduce the known results for perturbative interactions
between the dipoles, the two-gluon
exchange~\cite{Low:1975sv+X,Gunion:iy}, and give new insights into the
non-perturbative dipole-dipole scattering process.  This comes out as
a sum of two parts: The first part describes the non-perturbative
interaction between the quarks and antiquarks of the two dipoles and
exhibits the same structure as the perturbative contribution. The
second part represents the interaction between the strings of the two
dipoles.  The latter shows a new structure different from the
perturbative two-gluon exchange.


%
%
We find an extremely nice feature for the QCD string that confines the
quark and antiquark in the dipole: The QCD string of length
$|\vec{r}_{\!\mbox{\tiny\it D}}|$ can be exactly represented as an
integral over stringless dipoles of sizes
$\xi|\vec{r}_{\!\mbox{\tiny\it D}}|$ with $0 \leq \xi \leq 1$ and
dipole number density $n(\xi) = 1/\xi^2$. This outstanding result is
very similar to the perturbatively computed wave function of a $q{\bar
  q}$ onium state in the large\,-\,$N_c$ limit where the numerous
emitted gluons inside the onium state are considered as
dipoles~\cite{Mueller:1994rr,Mueller:1994jq}.

%
%
The decomposition of the QCD string into stringless dipoles allows us
for the first time to extract the microscopic structure of the
unintegrated gluon distribution of hadrons and photons ${\cal
  F}_{\!h}(x,k_{\!\perp}^2)$ from our dipole-hadron and dipole-photon
cross section via $|\vec{k}_{\!\perp}|$\,-\,factorization. We compare
the unintegrated gluon distribution of the proton with those
obtained in other approaches.  

%
%
The unintegrated gluon distribution of hadrons and photons ${\cal
  F}_{h}(x,k_{\!\perp}^2)$ is a basic, universal
quantity convenient for the computation of many scattering observables
at small $x$. It is crucial to describe processes in which transverse
momenta are explicitly exposed such as dijet~\cite{Nikolaev:1994cd+X}
or vector meson~\cite{Nemchik:1997xb} production at HERA. Its explicit
$|\vec{k}_{\!\perp}|$\,-\,dependence is particularly suited to study the
interplay between soft and hard physics. Moreover, the unintegrated
gluon distribution is the central object in the
BFKL~\cite{Kuraev:fs+X} and CCFM~\cite{Ciafaloni:1987ur+X} evolution
equations. Upon integration over the transverse gluon momentum
$|\vec{k}_{\!\perp}|$ it leads to the conventional gluon distribution
$xG_{h}(x,Q^2)$ used in the DGLAP evolution
equation~\cite{Gribov:ri+X}. 

%

We show the manifestations of the QCD string explicitly in
dipole-hadron cross sections at large dipole sizes and in unintegrated
gluon distributions at small transverse momenta. We find further that
the $|\vec{k}_{\!\perp}|$-factorization which is known in perturbative
physics can be extended also to non-perturbative dipole-hadron interactions
within our model.
 
%

To study saturation effects that manifest the $S$-matrix unitarity, we
investigate the scattering amplitude in impact parameter space where
the $S$-matrix unitarity imposes the rigid black disc limit on the height
of impact parameter profiles. We show explicitly that our profiles
respect the black disc limit.  Furthermore, the width of the impact
parameter profiles is shown to increase logarithmically at asymptotic
energies as needed to guarantee the Froissart
bound~\cite{Froissart:1961ux+X}.

%
%
We compute the impact parameter profiles for hadron-hadron and
longitudinal photon-proton scattering for different c.m. energies. We
determine the energy values at which the impact parameter profiles
saturate at the black disc limit for small impact parameters. The
impact parameter profiles provide an intuitive geometrical picture for
the energy dependence of the scattering process as they illustrate the
evolution of the size and opacity of the interacting particles with
increasing energy.

%
%
We estimate the impact parameter dependent gluon distribution of the
proton $xG(x,Q^2,|\vec{b}_{\perp}|)$ using its relation to the impact
parameter profile for longitudinal photon-proton scattering. We find a
low-$x$ saturation of $xG(x,Q^2,|\vec{b}_{\perp}|)$ as a manifestation
of the $S$-matrix unitarity. Consequently,  the $x$-dependence of
the integrated gluon distribution $xG(x,Q^2)$ slows down from a
powerlike to a squared logarithmic rise in agreement
with complementary investigations of gluon saturation at low $x$.

%
%

We compare the model results with experimental data and provide
predictions for future experiments. We compute total cross sections
$\sigma^{tot}$, the structure function of the proton $F_2$, slope
parameters $B$, differential elastic cross sections $d\sigma^{el}/dt$,
elastic cross sections $\sigma^{el}$, and the ratios
$\sigma^{el}/\sigma^{tot}$ and $\sigma^{tot}/B$ for proton-proton,
pion-proton, kaon-proton, photon-proton, and photon-photon reactions
involving real and virtual photons as well.  The successful unified
description of all these reactions indicates indeed the assumed
universal pomeron contribution to the above reactions.

%
%

Making use of the explicit saturation of the impact parameter profiles
at the black disc limit, the energy values at which it is reached, and
the logarithmic rise of the black disc radius, we show explicitly
manifestations of $S$-matrix unitarity limits in experimental
observables. A squared logarithmic behavior of total cross sections of
hadrons and photons is found to set in for c.m. energies $\sqrt{s}
\geq 10^6\,\GeV$. The ratios $\sigma^{el}/\sigma^{tot}$ and
$\sigma^{tot}/B$ become universal for such energies, i.e., independent
of the hadron species considered. For asymptotic energies, also the
total hadronic cross sections become universal and increase in
agreement with the Froissart bound~\cite{Froissart:1961ux+X}.

%
%
The outline of the work is as follows: In chapter~\ref{Sec_The_Model}
we present the model and give the model parameters. In transverse
momentum space considered in chapter~\ref{Chapt_string_decomposition}
we show the structure of QCD string interactions, decompose the string
into stringless dipoles, elaborate the total dipole-hadron cross
section, and extract the unintegrated gluon distribution. We compare the latter
with ones obtained from other approaches and use it also to
compute the integrated gluon distribution of the proton. In impact
parameter space investigated in chapter~\ref{$S$-Matrix Unitarity and
  Gluon Saturation} we study the $S$-matrix unitarity limits,
calculate the impact parameter profiles for proton-proton and
photon-proton scattering, determine the energy values at which the
profiles saturate at the black disc limit, estimate the impact
parameter dependent gluon distribution of the proton, and discuss
gluon saturation. Finally, in chapter~\ref{Sec_Comparison_Data} we
compare the model results with experimental data and identify
saturation effects in experimental observables. The appendices contain
conventions, hadron and photon wave functions, the analytic
continuation of the non-perturbative correlation functions from
Euclidean to Minkowski space-time, and the non-forward scattering
amplitude.


%% file: High_Energy_Scattering.tex
\chapter{The Loop-Loop Correlation Model}
\label{Sec_The_Model}
%
In this chapter we present the loop-loop correlation model. We
describe briefly the functional integral approach to high-energy
scattering in the eikonal approximation and compute in detail the
expectation value of one {\WW} loop and the correlation of two {\WW}
loops within a Gaussian approximation in the gluon field strengths.
Perturbative dipole-dipole interactions (long-distance correlations)
are described by perturbative gluon exchange and non-perturbative
dipole-dipole interactions (short-distance correlations) are modelled
by the stochastic vacuum model (\SVM)~\cite{Dosch:1987sk+X}. We use
the minimal surfaces to calculate the $S$-matrix element. Finally, we
introduce the energy dependence in a phenomenological way and specify
the parameter values.

\vspace*{0.3cm}
\section[Functional Integral Approach to High-Energy Scattering]
{\hspace{-0.4cm}\myfonta{\letterspace to .88\naturalwidth{Functional Integral Approach to High-Energy Scattering}}}
\label{Sec_Functional_Integral_Approach}
%
\vspace*{0.3cm}
The $T$-matrix is the central quantity in scattering processes. It
enters every observable we intend to look at and is obtained from the
$S$-matrix by subtracting the trivial case in which the final state
equals the initial state,
\be
        S_{fi} = \delta_{fi} + i (2 \pi)^4 \delta^4(P_f - P_i) T_{fi} 
        \ ,
\label{Eq_T_matrix_element}
\ee
where $P_i$ and $P_f$ represent the sum of incoming and outgoing
momenta, respectively. Based on the functional integral approach to
high-energy loop-loop scattering in the eikonal
approximation~\cite{Nachtmann:1991ua,Nachtmann:ed.kt,Kramer:1990tr,Dosch:1994ym},
the $T$-matrix element for the reaction $ab \rightarrow cd$ at
transverse momentum transfer ${\vec q}_{\!\perp}$ ($t = -{\vec
  q}_{\!\perp}^{\,\,2}$) and c.m.\ energy squared~$s$ reads
\bea
        \!\!\!\!\!\!\!\!\!
        &&\hspace{-1cm}
        T_{ab \rightarrow cd}(s,t) =
        2is \int \!\!d^2b_{\!\perp} 
        e^{i {\vec q}_{\!\perp} {\vec b}_{\!\perp}}
        \int \!\!dz_1 d^2r_1 \!\int \!\!dz_2 d^2r_2      
        \hphantom{\hspace*{5.cm}}   
        \nonumber \\ 
        \!\!\!\!\!\!\!& \!\!\times \!\! &  
        \psi_c^*(z_1,{\vec r}_1)\,\psi_d^*(z_2,{\vec r}_2)
        \left[1-S_{r_1r_2}(s,{\vec b}_{\!\perp},z_1,{\vec r}_1,z_2,{\vec r}_2)\right]
        \psi_a(z_1,{\vec r}_1)\,\psi_b(z_2,{\vec r}_2) 
        \ , 
\label{Eq_model_T_amplitude}
\eea
with the $S_{r_1r_2}$ matrix element 
\be
        S_{r_1r_2}(s,{\vec b}_{\!\perp},z_1,{\vec r}_1,z_2,{\vec r}_2)
        = \frac{\Big\langle W_{r_1}[C_1] W_{r_2}[C_2] \Big\rangle_G}
               {\Big\langle W_{r_1}[C_1]\Big\rangle_G
               \Big\langle W_{r_2}[C_2] \Big\rangle_G} \ .
\label{Eq_loop_loop_correlation_function}
\ee
The crucial quantity in~(\ref{Eq_loop_loop_correlation_function}) is
the light-like QCD Wegner-Wilson loop~\cite{Wegner:1971qt,Wilson:1974sk}
\be
        W_r[C] = 
        \nrTr\,\Pc
        \exp\!\left[
        -i g \oint_{\scriptsize C} dz^{\mu}\,\G_{\mu}^a(z)\,t_r^a 
        \right]      
        \ ,
\label{Eq_WW_loop}        
\ee
where the subscript $r$ indicates a representation of $SU(N_c)$, $N_c$
is the number of colors, $\nrTr = \rTr(\cdots)/\Tr \Identity_r$ is the
normalized trace in the corresponding color-space with unit element
$\Identity_r$, $g$ is the strong coupling, and $\G_{\mu}(z) =
\G_{\mu}^a(z) t_r^a$ represents the gluon field with the $SU(N_c)$
group generators in the corresponding representation, $t_r^a$, that
demand the path ordering indicated by $\Pc$ on the closed path $C$ in
space-time. A distinguishing theoretical feature of the {\WW} loop is
its invariance under local gauge transformations in color-space.
Therefore, it is the basic object in lattice gauge
theories~\cite{Wegner:1971qt,Wilson:1974sk,Rothe:1997kp} and has been
considered as the fundamental building block for a gauge theory in
terms of gauge invariant variables~\cite{Migdal:1983gj}. Physically,
the {\WW} loop represents the phase factor acquired by a color-charge
in the $SU(N_c)$ representation $r$ along the light-like trajectory
$C$ in the background field. In other words, the {\WW} loop describes a
color-singlet dipole in the representation $r$ of $SU(N_c)$.

{\WW} loops in the fundamental representation of
$SU(N_c = 3)$ are especially important since they 
represent color-singlet quark-antiquark dipoles.
In this representation, the $S_{r_1r_2}$ matrix
element~(\ref{Eq_loop_loop_correlation_function}) describes the
elastic scattering of two quark-antiquark dipoles with transverse size
and orientation ${\vec r}_i$.  The longitudinal momentum fraction of
color-dipole $i$ carried by the quark is $z_i$. The impact parameter between
the dipoles is~\cite{Dosch:1997ss}
\be
        {\vec b}_{\!\perp} 
        \,=\, {\vec r}_{1q} + (1-z_1) {\vec r}_{1} 
            - {\vec r}_{2q} - (1-z_2) {\vec r}_{2} 
        \,=\, {\vec r}_{1\,cm} - {\vec r}_{2\,cm} 
        \ ,
\label{Eq_impact_vector}
\ee
where ${\vec r}_{iq}$ (${\vec r}_{i\qbar}$) is the transverse position
of the quark (antiquark), ${\vec r}_{i} = {\vec r}_{i\qbar} - {\vec
  r}_{iq}$, and ${\vec r}_{i\,cm} = z_i {\vec r}_{iq} + (1-z_i){\vec
  r}_{i\qbar}$ is the center of light-cone momenta. In the eikonal
approximation to high-energy scattering the $q$ and ${\bar q}$ paths
form straight light-like trajectories.
Figure~\ref{Fig_loop_loop_scattering_surfaces} illustrates the (a)
space-time and (b) transverse arrangement of the loops. 

The ${\vec r}_i$ and $z_i$ distribution of the color-dipoles is given
by the {\em wave functions} $\psi_{a,b}$ and $\psi_{c,d}$ that
characterize the interacting particles.  In this framework, the
color-dipoles are given by the quark and antiquark in the meson or
photon and in a simplified picture by a quark and diquark in the
baryon. Concentrating in this work on reactions with $a = c$ and $b =
d$, only squared wave functions
$|\psi_1(z_1,\vec{r}_1)|^2:=\psi_c^*(z_1,{\vec r}_1)\,\psi_a(z_1,{\vec
  r}_1)$ and $|\psi_2(z_2,\vec{r}_2)|^2:=\psi_d^*(z_2,{\vec
  r}_2)\,\psi_b(z_2,{\vec r}_2)$ are needed. We use for hadrons the
phenomenological Gaussian wave
function~\cite{Dosch:2001jg,Wirbel:1985ji} and for photons the
perturbatively derived wave functions with running quark masses
$m_f(Q^2)$ to account for non-perturbative effects at low photon
virtuality $Q^2$~\cite{Dosch:1998nw} as discussed explicitly in
Appendix~\ref{Sec_Wave_Functions}.

The QCD vacuum expectation value $\langle \ldots \rangle_G$ in the
$S_{r_1r_2}$ matrix element~(\ref{Eq_loop_loop_correlation_function})
represents functional integrals~\cite{Nachtmann:ed.kt} in which the
functional integration over the fermion fields has already been
carried out as indicated by the subscript $G$. The model we use for
the QCD vacuum works in the {\em quenched approximation} that does not
allow string breaking through dynamical quark-antiquark production in
color-dipoles in the fundamental representation of $SU(3)$. The linear
rise of static color-dipole
potentials~\cite{Dosch:1987sk+X,Shoshi:2002rd}
and dipole-hadron cross-sections at large dipole sizes as well as the
$1/|\vec{k}_{\!\perp}|$ behavior of unintegrated gluon distributions
at small transverse momenta $|\vec{k}_{\!\perp}|$ display explicitly
the quenched approximation as shown in this work.

In this work we perform computations with Wegner-Wilson loops in a
general representation $r$ of $SU(N_c)$. This allows several
investigations: (a) The {\em large\,-\,$N_c$ limit} of our scattering
amplitudes can be examined which may give additional insights into the
scattering process. (b) The high-energy scattering of color-dipoles in
different representations describing different physical objects can be
studied. (c) In Euclidean space-time, different representations of the
Wegner-Wilson loop are appropriate to study the Casimir scaling
hypothesis of static potentials or chromo-field distributions of
color-dipoles~\cite{Ambjorn:1984dp,Shoshi:2002rd}. (d)
Since our model works in the quenched approximation, the investigation
of the string breaking behavior of the static potential for
color-dipoles in the fundamental and adjoint (glueballs)
representation of $SU(N_c)$ is possible: {\em string breaking} cannot
occur in fundamental dipoles as dynamical quark-antiquark production
is excluded but should be present for adjoint dipoles because of
gluonic vacuum
polarization~\cite{Shoshi:2002rd,Kallio:2000jc}.


%
\befig[h!]
  \begin{center}
        \epsfig{file=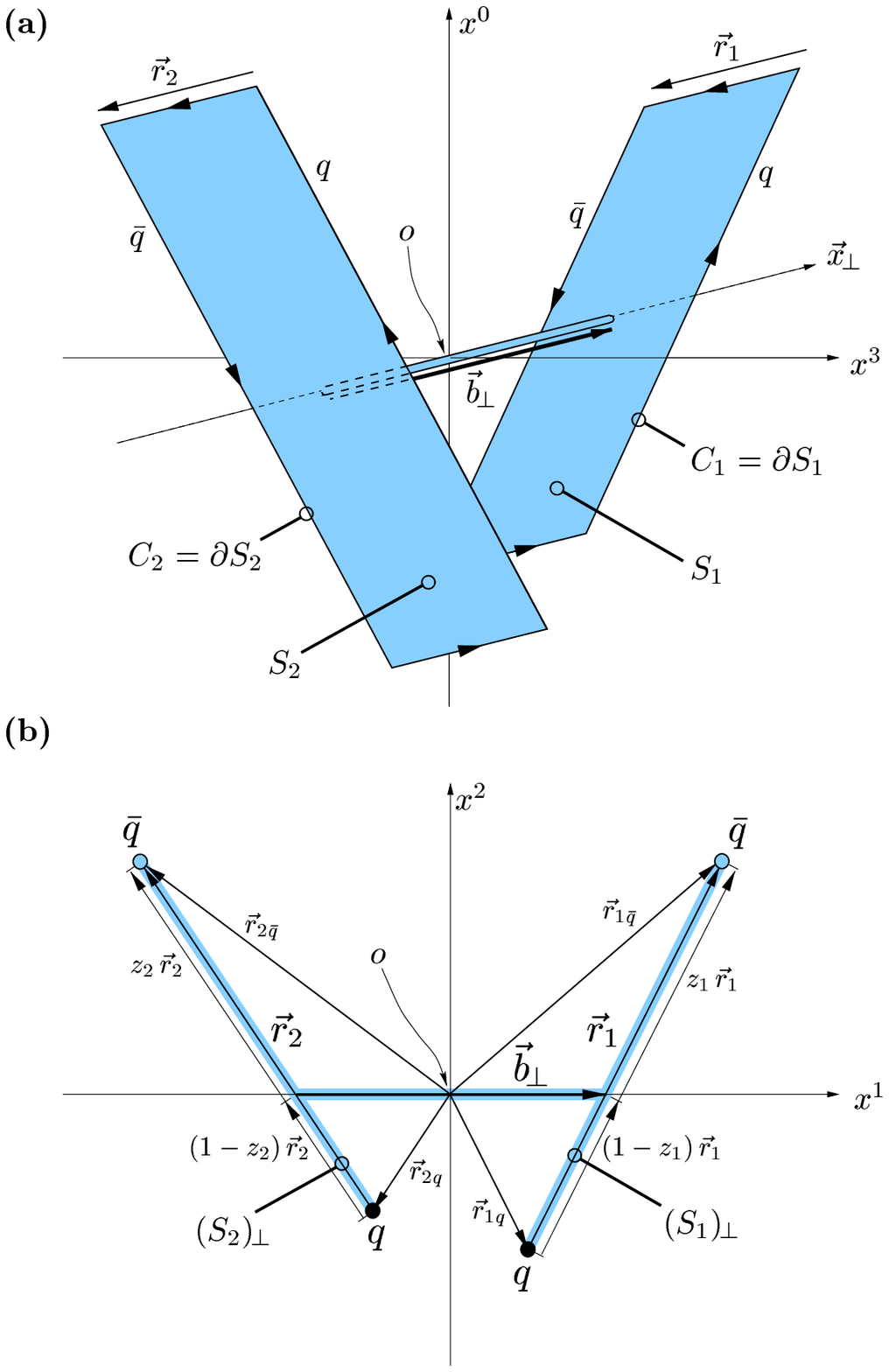,width=10.5cm}
  \end{center}
\caption{\small High-energy dipole-dipole scattering in the eikonal
  approximation represented by Wegner-Wilson loops in the fundamental
  representation of $SU(3)$: (a) space-time and
  (b) transverse arrangement of the Wegner-Wilson loops. The shaded
  areas represent the strings extending from the quark to the
  antiquark path in each color dipole.  The thin tube allows to
  compare the field strengths in surface $S_1$ with the field
  strengths in surface $S_2$. The impact parameter $\vec{b}_{\perp}$
  connects the centers of light-cone momenta of the dipoles.}
\label{Fig_loop_loop_scattering_surfaces}
\efig
%

\section[Vacuum Expectation Value of one Wegner-Wilson Loop]{\myfonta \hspace{-0.4cm}\letterspace to .89\naturalwidth{Vacuum Expectation Value of one Wegner-Wilson Loop}}
\label{Sec_VEV_Of_One_Wegner-Wilson_Loop}
To compute the expectation value of a {\WW} loop~(\ref{Eq_WW_loop}) in the QCD vacuum 
\be
        \Big\langle W_r[C] \Big\rangle_G
        = \Big\langle
        \nrTr\,\Pc
        \exp\!\left[-i\,g \oint_{\scriptsize C} dz^{\mu}\,\G_{\mu}^a(z)\,t_r^a \right]      
        \Big\rangle_G
        \ ,
\label{Eq_<W[C]>}
\ee
we transform the line integral over the loop $C$ into an integral over the surface $S$ with $\partial S = C$ by applying the {\em non-Abelian Stokes' theorem}~\cite{Bralic:1980ra+X,Nachtmann:ed.kt}
\be
        \Big\langle W_r[C] \Big\rangle_G
        = \Big\langle
        \nrTr\,\Ps
          \exp \left[-i\,\frac{g}{2} 
                \int_{S} \! d\sigma^{\mu\nu}(z) 
                \G^a_{\mu\nu}(o,z;C_{zo})\,t_r^a 
          \right] 
        \Big\rangle_G
        \ ,
\label{Eq_Non-Abelian_Stokes_<W[C]>}
\ee
where $\Ps$ indicates surface ordering and $o$ is an arbitrary reference point on the surface $S$. In Eq.~(\ref{Eq_Non-Abelian_Stokes_<W[C]>}), the gluon field strength tensor, $\G_{\mu\nu}(z) = \G_{\mu\nu}^{a}(z)\,t^{a}$, is parallel transported to the reference point $o$ along the path $C_{zo}$
\be
        \G_{\mu\nu}(o,z;C_{zo}) 
        = \Phi(o,z;C_{zo})^{-1} \G_{\mu\nu}(z) \Phi(o,z;C_{zo})
\label{Eq_gluon_field_strength_tensor}
\ee
with the QCD Schwinger string
\be
        \Phi(o,z;C_{zo}) 
        = \Pc \exp 
        \left[-i\,g \int_{C_{zo}} \!\!dz^{\mu}\G^a_{\mu}(z)\,t_r^a \right] 
        \ .
\label{Eq_parallel_transport}
\ee
Due to the linearity of the functional integral, $\langle \nrTr \ldots
\rangle = \nrTr \langle \ldots \rangle$,
Eq.~(\ref{Eq_Non-Abelian_Stokes_<W[C]>}) reduces to
\be
        \Big\langle W_r[C] \Big\rangle_G
        = \nrTr 
        \Big\langle
        \Ps \exp \left[-i\,\frac{g}{2} 
                \int_{S} \! d\sigma^{\mu\nu}(z) 
                \G^a_{\mu\nu}(o,z;C_{zo})\,t_r^a 
          \right] 
        \Big\rangle_G
        \ .
\label{Eq_Tr_<W[C]>}
\ee
For the evaluation of~(\ref{Eq_Tr_<W[C]>}), a {\em matrix cumulant expansion} is used as explained in~\cite{Nachtmann:ed.kt} (cf.\ also~\cite{VAN_KAMPEN_1974_1976+X})
\bea
        && \Big\langle 
                \Ps \, \exp 
                \left[-i\,\frac{g}{2} 
                \int_{S} \! d\sigma(z) \G(o,z;C_{zo})
                \right] 
           \Big\rangle_G \nonumber \\
        && 
        = \exp[\,\,\sum_{n=1}^{\infty}\frac{1}{n !}(-i\,\frac{g}{2})^n
        \int d\sigma(x_1)\cdots d\sigma(x_n)\,K_n(x_1,\cdots,x_n)]
        \ ,
\label{Eq_matrix_cumulant_expansion}
\eea
where space-time indices are suppressed to lighten notation. The cumulants $K_n$ consist of expectation values of {\em ordered} products of the non-commuting matrices $\G(o,z;C_{zo})$. The leading matrix cumulants are
\bea
        K_1(x)       
        & = & \langle \G(o,x;C_x) \rangle_G, 
\label{Eq_K_1_matrix_cumulant}\\
        K_2(x_1,x_2) 
        & = & \langle \Ps
        [\G(o,x_1;C_{x_1})\G(o,x_2;C_{x_2})]\rangle_G 
        \nonumber\\
        &   & - \frac{1}{2}
        \left(\langle \G(o,x_1;C_{x_1})\rangle_G 
        \langle \G(o,x_2;C_{x_2})\rangle_G 
        + (1 \leftrightarrow 2)\,\right) \ .
\label{Eq_K_2_matrix_cumulant}
\eea
Since the vacuum does not prefer a specific color direction, $K_1$ vanishes and $K_2$ becomes
\be
        K_2(x_1,x_2) 
        = \langle\Ps [\G(o,x_1;C_{x_1})\G(o,x_2;C_{x_2})]\rangle_G
        \ .
\label{Eq_K_2_matrix_cumulant<-no_color_direction_preferred}
\ee
Now, we approximate the functional integral associated with the
expectation values $\langle \ldots \rangle_G$ as a {\em Gaussian
  integral} in the gluon field strength. Consequently, the cumulants
factorize into two-point field correlators such that all higher
cumulants, $K_n$ with $n>2$, vanish\footnote{We are going to use the
  cumulant expansion in the Gaussian approximation also for
  perturbative gluon exchange. Here certainly the higher cumulants are
  non-zero.} and $\langle W_r[C] \rangle_G$ can be expressed in terms
of $K_2$
\bea
&& \!\!\!\!\!\!\!\!\!\!\!
        \Big\langle W_r[C] \Big\rangle_G = 
        \nrTr 
        \exp\!\left[-\frac{g^2}{8} \!
          \int_{S} \! d\sigma^{\mu\nu}(x_1) \!
          \int_{S} \! d\sigma^{\rho\sigma}(x_2) 
        \right.
        \nonumber \\
&& \!\!\!\!\!\!\!\!\!\!\!
        \hphantom{\Big\langle W_r[C] \Big\rangle_G = \nrTr \exp}
        \left.
          \Big\langle \Ps\, 
          [\G^a_{\mu\nu}(o,x_1;C_{x_1 o})\,t_r^a\,\,
          \G^b_{\rho\sigma}(o,x_2;C_{x_2 o})\,t_r^b] 
          \Big\rangle_G 
        \right]
\label{Eq_matrix_cumulant_expansion_<W[C]>}
\eea
Due to the color-neutrality of the vacuum, the gauge-invariant bilocal gluon field strength correlator contains a $\delta$-function in color-space,
\be
        \Big\langle
        \frac{g^2}{4\pi^2}
        \left[\G^a_{\mu\nu}(o,x_1;C_{x_1 o})
        \G^b_{\rho\sigma}(o,x_2;C_{x_2 o})\right]
        \Big\rangle_G
        =: \inv{4}\delta^{ab} 
        F_{\mu\nu\rho\sigma}(x_1,x_2,o;C_{x_1 o},C_{x_2 o}) 
\label{Eq_Ansatz}
\ee
which makes the surface ordering $\Ps$ in~(\ref{Eq_matrix_cumulant_expansion_<W[C]>}) irrelevant. The quantity $F_{\mu\nu\rho\sigma}$ will be specified in Sec.~\ref{Sec_Non-pert_Pert_Cont}.  With~(\ref{Eq_Ansatz}) and the quadratic Casimir operator $C_2(r)$, 
\be
        t_r^a\,t_r^a = t_r^2 = C_2(r)\,\Identity_r
        \ ,
\label{Eq_quadratic_Casimir_operator}
\ee
Eq.~(\ref{Eq_matrix_cumulant_expansion_<W[C]>}) reads
\be
        \Big\langle W_r[C] \Big\rangle_G
        = \nrTr 
        \exp\left[
        - C_2(r)\,\chi_{SS}\,\Identity_r
        \right] 
        = \exp \left[-i\,\frac{C_2(r)}{2}\,\chi_{SS}\right] 
        \ ,
\label{Eq_final_result_<W[C]>}
\ee
where 
\be
        \chi_{SS}
        := - i\,\frac{\pi^2}{4} 
        \int_{S} \! d\sigma^{\mu\nu}(x_1) 
        \int_{S} \! d\sigma^{\rho\sigma}(x_2)
        F_{\mu\nu\rho\sigma}(x_1,x_2,o;C_{x_1 o},C_{x_2 o}) 
        \ .
\label{Eq_chi_SS}        
\ee
Our ansatz for the tensor structure of $F_{\mu\nu\rho\sigma}$ in
Minkowski space-time --
see~(\ref{Eq_F_decomposition}), (\ref{Eq_MSV_Ansatz_F}),
and~(\ref{Eq_PGE_Ansatz_F}) -- leads to $\chi_{S S} = 0$ for
light-like loops, as explained in Sec.~\ref{Sec_Chi_Computation}.
Consequently, the vacuum expectation value of the Wegner-Wilson loop
becomes
\be
        \Big\langle W_r[C] \Big\rangle_G
        = 1 \ .
\label{VEV_WWL_1}
\ee
The deviation of $\Big\langle W_r[C] \Big\rangle_G$ from one is just a
measure of the strength of parton splitting processes, e.g., $q \to q
+ G$. Our result~(\ref{VEV_WWL_1}) is consistent with the quenched
approximation and remains valid for arbitrary surface choices.

In Euclidean space-time one obtaines $\chi^E_{S S} \neq 0$ since the
loops are not light-like and $F^E_{\mu\nu\rho\sigma}$ is different
from the Minkowskian one. This allows one to compute the potential of
static color-dipoles~\cite{Dosch:1987sk+X,Shoshi:2002rd} from
the expectation value of one Wegner-Wilson loop in Euclidean
space-time as discussed briefly at the end of this chapter.

\section{The Loop-Loop Correlation Function}
\label{Sec_loop_loop_func}

The computation of the {\em loop-loop correlation function} $\langle
W_{r_1}[C_{1}] W_{r_2}[C_{2}] \rangle_G$ starts also with the
application of the the {\em
  non-Abelian Stokes' theorem}~\cite{Bralic:1980ra+X,Nachtmann:ed.kt}
that is used to transform the line integrals over the loops $C_{1,2}$ into integrals over surfaces $S_{1,2}$ with $\partial S_{1,2} =
C_{1,2}$ as in the previous section
\bea
        &&
        \Big\langle W_{r_1}[C_{1}] W_{r_2}[C_{2}] \Big\rangle_G 
        = \Big\langle 
        \nroneTr\,\Ps
          \exp \left[-i\,\frac{g}{2} 
                \int_{S_1} \! d\sigma^{\mu\nu}(x_1) 
                \G^a_{\mu\nu}(o_1,x_1;C_{x_1 o_1})\,t_{r_1}^a 
          \right] 
        \nonumber \\
        &&
        \quad\quad\quad\quad\quad
        \times\,\nrtwoTr\,\Ps
          \exp \left[-i\,\frac{g}{2} 
                \int_{S_2} \! d\sigma^{\rho\sigma}(x_2) 
                \G^b_{\rho\sigma}(o_2,x_2;C_{x_2 o_2})\,t_{r_2}^b 
          \right] 
        \Big\rangle_G
\label{Eq_Non-Abelian_Stokes_<W[C1]W[C2]>}
\eea
where $o_{1}$ and $o_{2}$ are the reference points on the surfaces
$S_{1}$ and $S_{2}$, respectively, that enter through the non-Abelian
Stokes' theorem. In order to ensure gauge invariance in our model, the
gluon field strengths associated with the loops must be compared at
{\em one} reference point $o$. Therefore, we require the surfaces
$S_{1}$ and $S_{2}$ to touch at a common reference point $o_{1} =
o_{2} = o$.

Following the Berger-Nachtmann approach~\cite{Berger:1999gu}, the product of the two traces, $\nroneTr(\cdots)\,\nrtwoTr(\cdots)$, over $SU(N_c)$ matrices in the $r_1$ and $r_2$ representation, respectively, is interpreted as one trace $\nronexrtwoTr(\cdots):=\ronexrtwoTr(\cdots)/\ronexrtwoTr(\ronexrtwoIdentity)$ that acts in the tensor product space built from the $r_1$ and $r_2$ representations
\bea
        \Big\langle W_{\!r_1}[C_{1}] W_{\!r_2}[C_{2}] \Big\rangle_G 
        \!\!\!\! & \!\!\! = \!\!\! &\!
        \Big\langle 
        \nronexrtwoTr
          \!\left\{\!\!
            \Big[\Ps \exp\!\big[\!-\!i\frac{g}{2} \!
                \int_{S_{1}} \!\!\!\! d\sigma^{\mu\nu}(x_{1}) 
                \G^a_{\mu\nu}(o,x_{1};C_{x_{1} o})\,t_{r_1}^a \big] 
                \,\otimes\,\Identity_{r_2}\Big]
          \right.
        \nonumber \\
        &&\!\!\!\!\!\!\!\!\!\!\!\!\!
        \times
          \left. 
          \Big[\Identity_{r_1}\,\otimes\,
          \Ps \exp\! \big[\!-\!i\frac{g}{2} \!
                \int_{S_{2}} \!\!\!\! d\sigma^{\rho\sigma}(x_{2}) 
                \G^b_{\rho\sigma}(o,x_{2};C_{x_{2} o})\,t_{r_2}^b \big]
          \Big]
          \!\!\right\}\!
        \Big\rangle_G 
\label{Eq_trace_trick_<W[C1]W[C2]>}
\eea
Using the identities
\bea
        \exp\left(\,t_{r_1}^a\,\right) \,\otimes\, \Identity_{r_2} 
        & = & \exp\left(\,t_{r_1}^a \,\otimes\, \Identity_{r_2}\,\right) \\
        \Identity_{r_1} \,\otimes\, \exp\left(\,t_{r_2}^a\,\right) 
        & = & \exp\left(\,\Identity_{r_1} \,\otimes\, t_{r_2}^a\,\right)
\label{Eq_exp(t^a)_times_1_identities}
\eea
the tensor products can be shifted into the exponents. With matrix multiplication in the tensor product space
\bea
        \big( t_{r_1}^a \,\otimes\, \Identity_{r_2} \big)
        \big( t_{r_1}^b \,\otimes\, \Identity_{r_2} \big) 
        & = & t_{r_1}^a t_{r_1}^b \,\otimes\, \Identity_{r_2}
        \nonumber
        \\
        \big( t_{r_1}^a \,\otimes\, \Identity_{r_2} \big)
        \big( \Identity_{r_1} \,\otimes\, t_{r_2}^b \big) 
        & = & t_{r_1}^a \,\otimes\, t_{r_2}^b
\label{Eq_matrix_multiplication_in_tensor_product_space}
\eea
and the vanishing commutator 
\vspace{0.2cm}
\be
        \left[t_{r_1}^a \otimes \Identity_{r_2}, 
        \Identity_{r_1} \otimes t_{r_2}^b\right] 
        = 0
\label{Eq_[t_x_1,1_x_t]}
\vspace{0.2cm}
\ee
the two exponentials in (\ref{Eq_trace_trick_<W[C1]W[C2]>}) commute
and can be written as one exponential
\vspace{0.2cm}
\be
        \Big\langle W[C_{1}] W[C_{2}] \Big\rangle_G =
        \Big\langle 
        \nronexrtwoTr\,\Ps \exp\!
            \left[-i\,\frac{g}{2} 
                \int_{S} \! d\sigma^{\mu\nu}(x) 
                \GG_{\mu\nu}(o,x;C_{xo}) 
            \right]
        \Big\rangle_G    
\label{Eq_<W[C1]W[C2]>_analogous_to_<W[C]>}
\vspace{0.2cm}
\ee
with the following gluon field strength tensor acting in the tensor
product space
\vspace{0.2cm}
\be
        \GG_{\mu\nu}(o,x;C_{xo})
        := \left\{ \begin{array}{cc}
            \G_{\mu\nu}^a(o,x;C_{xo})
                \big( t_{r_1}^a \,\otimes\, \Identity_{r_2} \big)
            & \mbox{for $\,\,\, x\,\,\, \in \,\,\, S_1$} \\
            \G_{\mu\nu}^a(o,x;C_{xo})
                \big( \Identity_{r_1} \,\otimes\, t_{r_2}^a \big)
            & \mbox{for $\,\,\, x\,\,\, \in \,\,\, S_2$}
        \end{array}\right. \ .
\label{Eq_GG} 
\vspace{0.2cm}
\ee
In Eq.~(\ref{Eq_<W[C1]W[C2]>_analogous_to_<W[C]>}), the surface
integrals over $S_1$ and $S_2$ are written as one integral over the
combined surface $S = S_1 + S_2$ so that the right-hand side (rhs)
of~(\ref{Eq_<W[C1]W[C2]>_analogous_to_<W[C]>}) becomes very similar to
the rhs of~(\ref{Eq_Non-Abelian_Stokes_<W[C]>}). This allows us to
proceed analogously to the computation of $\langle W_r[C] \rangle_G$
in the previous section: With the linearity of the functional
integral, the matrix cumulant expansion, the color-neutrality of the
vacuum, and the Gaussian approximation now in the color components of
the gluon field strength tensor $\GG_{\mu\nu}(o,x;C_{xo})$, only the
$n=2$ term of the matrix cumulant expansion survives, which leads to
\bea
&& \!\!\!\!\!\!\!\!\!\!\!
        \Big\langle W_{r_1}[C_{1}] W_{r_2}[C_{2}] \Big\rangle_G 
        \\
&& \!\!\!\!\!\!\!\!\!\!\!
        = \nronexrtwoTr
        \exp\!\left[\!-\frac{g^2}{8} \!\!
          \int_{S} \!\! d\sigma^{\mu\nu}(x_1) \!
          \int_{S} \!\! d\sigma^{\rho\sigma}(x_2) 
          \Big\langle\!\Ps 
          [\GG_{\mu\nu}(o,x_1;C_{x_1 o}) 
          \GG_{\rho\sigma}(o,x_2;C_{x_2 o})] 
          \!\Big\rangle_{\!\!G} 
        \right] \ .
        \nonumber  
\label{Eq_matrix_cumulant_expansion_<W[C1]W[C2]>}
\eea

Note that the Gaussian approximation on the level of the color
components of the gluon field strength tensor (component
factorization) differs from the one on the level of the gluon field
strength tensor (matrix factorization) used to compute $\langle
W_{r}[C] \rangle$ in the original version of the
SVM~\cite{Dosch:1987sk+X}. Nevertheless, with the additional ordering
rule~\cite{Rueter:1994cn} explained in detail in Sec.~2.4
of~\cite{Dosch:2000va}, a modified component factorization is obtained
that leads to the same area law as the matrix factorization.

\vskip 0.3cm

Using definition~(\ref{Eq_GG}) and relations~(\ref{Eq_matrix_multiplication_in_tensor_product_space}), we now redivide the exponent in~(\ref{Eq_matrix_cumulant_expansion_<W[C1]W[C2]>}) into integrals of the ordinary parallel transported gluon field strengths over the separate surfaces $S_{1}$ and $S_{2}$

\bea
        && \Big\langle W_{r_1}[C_{1}] W_{r_2}[C_{2}] \Big\rangle_G = 
        \nronexrtwoTr
        \exp \Bigg[             \\ &&\hspace{-0.8cm}
          -\frac{g^2}{8} \!\!
          \int_{S_1} \!\!\! d\sigma^{\mu\nu}(x_1) \!\! 
          \int_{S_2} \!\!\! d\sigma^{\rho\sigma}(x_2)\, 
          \Ps \!\left [ \Big\langle\! 
          \G^a_{\mu\nu}(o,x_1;C_{x_1 o}) 
          \G^b_{\rho\sigma}(o,x_2;C_{x_2 o})\!\Big\rangle_{\!\!G}
                \big(t_{r_1}^a  \,\otimes\, t_{r_2}^b\big) \right ] 
          \nonumber \\
        &&\hspace{-0.8cm}
          -\frac{g^2}{8} \!\!
          \int_{S_2} \!\!\! d\sigma^{\mu\nu}(x_1) \!\! 
          \int_{S_1} \!\!\! d\sigma^{\rho\sigma}(x_2)\, 
          \Ps \!\left [ \Big\langle\! 
            \G^a_{\mu\nu}(o,x_1;C_{x_1 o}) 
            \G^b_{\rho\sigma}(o,x_2;C_{x_2 o})\!\Big\rangle_{\!\!G}
                \big(t_{r_1}^a  \,\otimes\, t_{r_2}^b\big) \right ] 
            \nonumber \\
        &&\hspace{-0.8cm}
          -\frac{g^2}{8} \!\!
          \int_{S_1} \!\!\! d\sigma^{\mu\nu}(x_1) \!\! 
          \int_{S_1} \!\!\! d\sigma^{\rho\sigma}(x_2)\, 
          \Ps \!\left [ \Big\langle \!
            \G^a_{\mu\nu}(o,x_1;C_{x_1 o}) 
            \G^b_{\rho\sigma}(o,x_2;C_{x_2 o})\!\Big\rangle_{\!\!G}
                \big(t_{r_1}^a t_{r_1}^b \,\otimes\, \Identity_{r_2}\big)\right ] 
        \nonumber \\
        &&\hspace{-0.8cm}
        \left.
          -\frac{g^2}{8}\!\! 
          \int_{S_2} \!\!\! d\sigma^{\mu\nu}(x_1) \!\! 
          \int_{S_2} \!\!\! d\sigma^{\rho\sigma}(x_2)\, 
           \Ps \!\left [ \Big\langle\! 
             \G^a_{\mu\nu}(o,x_1;C_{x_1 o}) 
             \G^b_{\rho\sigma}(o,x_2;C_{x_2 o})\!\Big\rangle_{\!\!G}
                \big(\Identity_{r_1} \,\otimes\, t_{r_2}^a t_{r_2}^b\big) \right ]
              \!\right] \ .
        \nonumber
\label{Eq_exponent_decomposition_<W[C1]W[C2]>}
\eea
Here the surface ordering $\Ps$ is again irrelevant due to the
color-neutrality of the vacuum~(\ref{Eq_Ansatz}), and (\ref{Eq_exponent_decomposition_<W[C1]W[C2]>}) becomes
\bea
        && \Big\langle W_{r_1}[C_{1}] W_{r_2}[C_{2}] \Big\rangle_G 
        = \nronexrtwoTr
        \exp\!\Bigg[
                - i\,\inv{2}\left\{\left(\chi_{S_1 S_2}+\chi_{S_2 S_1}\right)\,
                \big(t_{r_1}^a \,\otimes\, t_{r_2}^a\big) \right. 
        \nonumber \\
        && \vspace*{3cm}\quad\quad\quad\quad\quad\quad  \left.  
        + \,\chi_{S_1 S_1}
                \big(t_{r_1}^a t_{r_1}^a\,\otimes\,\rtwoIdentity\big) 
            + \chi_{S_2 S_2}
                \big(\roneIdentity\,\otimes\,t_{r_2}^a t_{r_2}^a\big) 
             \right\} \Bigg]
\label{Eq_eikonal_functions_<W[C1]W[C2]>}
\eea
with 
\be
        \chi_{S_i S_j}
        := -\,i\,\frac{\pi^2}{4} 
        \int_{S_i} \! d\sigma^{\mu\nu}(x_1) 
        \int_{S_j} \! d\sigma^{\rho\sigma}(x_2)
        F_{\mu\nu\rho\sigma}(x_1,x_2,o;C_{x_1 o},C_{x_2 o}) 
        \ .
\label{Eq_chi_Si_Sj}        
\ee
The symmetries in the tensor structure of $F_{\mu\nu\rho\sigma}$ in
Minkowski space-time -- see~(\ref{Eq_F_decomposition}),
(\ref{Eq_MSV_Ansatz_F}), and~(\ref{Eq_PGE_Ansatz_F}) -- lead to
$\chi_{S_1 S_1} = \chi_{S_2 S_2} = 0$ for light-like loops as
explained in Sec.~\ref{Sec_Chi_Computation}, and also to $\chi_{S_1
  S_2} = \chi_{S_2 S_1} =: \chi$.\footnote{In Euclidean space-time a different
result is obtained since $\chi_{S_i S_i} \neq 0$ as shown in~\cite{Shoshi:2002rd}.} Our
final Minkowskian result for general $SU(N_c)$ representations $r_1$ and
$r_2$ becomes
\be
        \Big\langle W_{r_1}[C_{1}] W_{r_2}[C_{2}] \Big\rangle_{\!G} 
         = \nronexrtwoTr
        \exp\!\Bigg[
                - i\,\chi\,
                \big(t_{r_1}^a \,\otimes\, t_{r_2}^a\big)\Bigg] \ .
\label{Eq_final_general_Euclidean_result_<W[C1]W[C2]>}
\ee
After specifying the representations $r_1$ and $r_2$, the tensor product
$\tensor:=t_{r_1}^a \,\otimes\, t_{r_2}^a$ can be expressed as a sum
of projection operators $\Projector_i$ with the property $\Projector_i
\,\tensor = \lambda_i \,\Projector_i$ where  
\be
        \tensor = \sum \lambda_i\,\Projector_i
        \quad\quad \mbox{with} \quad\quad 
        \lambda_i = \frac{\nronexrtwoTr\big(\Projector_i \,\tensor\big)}{\nronexrtwoTr \big(\Projector_i\big)}
        \ ,
\label{Eq_tensor_decomposition}
\ee
which corresponds to the decomposition of the tensor product space into irreducible representations.

For two \WW-loops in the {\em fundamental representation} of $SU(N_c)$, $r_1 = r_2 = \Fundamental$, that describe the trajectories of two quark-antiquark dipoles, the decomposition~(\ref{Eq_tensor_decomposition}) becomes trivial
\be
        t_{\fundamental}^a \,\otimes\, t_{\fundamental}^a
        = \frac{N_c-1}{2N_c} \Projector_s - \frac{N_c+1}{2N_c} \Projector_a
        \ ,
\label{Eq_projector_ta_x_ta_relation}
\ee
with the projection operators
\bea
        &&
        (\Projector_s)_{(\alpha_1 \alpha_2)( \beta_1 \beta_2)} =
        \frac{1}{2}
        (\delta_{\alpha_1 \beta_1} \delta_{\alpha_2 \beta_2} 
        +\delta_{\alpha_1 \beta_2} \delta_{\alpha_2 \beta_1})
        \\
        &&
        (\Projector_a)_{(\alpha_1 \alpha_2)( \beta_1 \beta_2)} =
        \frac{1}{2}
        (\delta_{\alpha_1 \beta_1} \delta_{\alpha_2 \beta_2} 
        -\delta_{\alpha_1 \beta_2} \delta_{\alpha_2 \beta_1})
\label{Eq_projectors}
\eea
that decompose the direct product space of two fundamental $SU(N_c)$
representations into the irreducible representations
\be
        \mbox{N}_c \,\otimes\, \mbox{N}_c 
        = (\mbox{N}_c + 1)\mbox{N}_c/2 \,\oplus\, \overline{\mbox{N}_c(\mbox{N}_c - 1)/2}
        \ .
\label{Eq_tensor_product_fundamental_decomposition}
\ee
Using $\Tr_{\fundamental\otimes\fundamental}\,\Identity_{\fundamental\otimes\fundamental} = N_c^2$ and the projector properties
\be
        \Projector^2_{s,a} = \Projector_{s,a}
        \ , \quad 
        \Tr_{\fundamental\otimes\fundamental} \,\Projector_s = (N_c + 1)N_c/2
        \ , \quad \mbox{and} \quad
        \Tr_{\fundamental\otimes\fundamental} \,\Projector_a = (N_c - 1)N_c/2
        \ ,
\label{Eq_projector_properties_fundamental}
\ee 
we find a simple expression for the loop-loop correlation function with both loops in the fundamental $SU(N_c)$ representation
\be
        \Big\langle W_{\fundamental}[C_{1}] W_{\fundamental}[C_{2}] \Big\rangle_G 
        = \frac{N_c+1}{2N_c}\exp\!\left[-i\,\frac{N_c-1}{2N_c}\chi\right]
        + \frac{N_c-1}{2N_c}\exp\!\left[ i\,\frac{N_c+1}{2N_c}\chi\right]
        \nonumber 
\label{Eq_final_Euclidean_result_<W[C1]W[C2]>_fundamental}
\ee
and recover, of course, for $N_c = 3$ the result
from~\cite{Berger:1999gu}. 

For one \WW-loop in the {\em fundamental} and one in the {\em adjoint
  representation} of $SU(N_c)$, $r_1 = \Fundamental$ and $r_2 =
\Adjoint$, that can be used to describe the scattering of a
quark-antiquark dipole with a glueball, the
decomposition~(\ref{Eq_tensor_decomposition}) reads
\be
        t_{\fundamental}^a \,\otimes\, t_{\adjoint}^a
        \,\,=\,\, 
        -\,\frac{N_c}{2}\,\Projector_1
        \,+\, \inv{2}\,\Projector_2 
        \,-\, \inv{2}\,\Projector_3 
\label{Eq_projector_tf_x_ta_relation}
\ee
with the projection operators\footnote{The explicit form of the
  projection operators $\Projector_1$, $\Projector_2$, and
  $\Projector_3$ can be found in~\cite{Cvitanovic_1984} but note that
  we use the Gell-Mann (conventional) normalization of the gluons. The
  eigenvalues, $\lambda_i$, of the projection operators
  in~(\ref{Eq_projector_tf_x_ta_relation}) can be evaluated
  conveniently with the computer program
  ``Colour''~\cite{Hakkinen:1996bb}.} $\Projector_1$, $\Projector_2$,
and $\Projector_3$ that decompose the direct product space of one
fundamental and one adjoint $SU(N_c)$ representation into the the
irreducible representations
\be
        \Fundamental\,\otimes\,\Adjoint
        \,\,=\,\,
        \mbox{N}_c
        \,\,\oplus\,\,\inv{2}\,\mbox{N}_c(\mbox{N}_c-1)(\mbox{N}_c+2)
        \,\,\oplus\,\,\inv{2}\,\mbox{N}_c(\mbox{N}_c+1)(\mbox{N}_c-2)
\label{Eq_tensor_product_f_x_a_decomposition}
\ee
which reduces for $N_c = 3$ to the well-known $SU(3)$ decomposition
\be
        3\,\otimes\,8 
        = 3\,\oplus\,15\,\oplus\,6
        \ .
\label{Eq_tensor_product_f_x_a_SU(3)_decomposition}
\ee
With $\Tr_{\fundamental\otimes\adjoint}\,\Identity_{\fundamental\otimes\adjoint} = N_c(N_c^2 - 1)$ and projector properties analogous to~(\ref{Eq_projector_properties_fundamental}), we obtain the loop-loop correlation function for one loop in the fundamental and one in the adjoint representation of $SU(N_c)$
\bea
        && \!\!\!\!\!\!\!\!
        \Big\langle W_{\fundamental}[C_{1}] W_{\adjoint}[C_{2}] \Big\rangle_G 
\label{Eq_final_Euclidean_result_<Wf[C1]Wa[C2]>}\\
        && = \,
        \,\inv{N_c^2\!-\!1}\,\exp\!\Big[i\,\frac{N_c}{2}\,\chi\Big]
        +\frac{N_c\!+\!2}{2(N_c\!+\!1)}\exp\!\Big[\!-i\,\inv{2}\chi\Big]
        +\frac{N_c\!-\!2}{2(N_c\!-\!1)}\exp\!\Big[i\,\inv{2}\chi\Big]
        \ .
\nonumber
\eea

Note that our results given in
Eqs.~(\ref{Eq_final_Euclidean_result_<W[C1]W[C2]>_fundamental})
and~(\ref{Eq_final_Euclidean_result_<Wf[C1]Wa[C2]>}) are quite general
since they are obtained directly from the color-neutrality of the QCD
vacuum and the Gaussian approximation of the functional integrals. For
the explicit computation of $\chi$-function in
Eqs.~(\ref{Eq_final_Euclidean_result_<W[C1]W[C2]>_fundamental})
and~(\ref{Eq_final_Euclidean_result_<Wf[C1]Wa[C2]>}) one has to
specify the gluon field strength correlator $F_{\mu\nu\rho\sigma}$ and
the surfaces $S_{1,2}$ with the restriction $\partial S_{1,2} =
C_{1,2}$ that appear in the $\chi$-function~(\ref{Eq_chi_Si_Sj}). This
we do in the following sections.  Furthermore, we concentrate in the
following on the dipole-dipole
scattering~(\ref{Eq_final_Euclidean_result_<W[C1]W[C2]>_fundamental})
as the relevant case and postpone the dipole-glueball
scattering~(\ref{Eq_final_Euclidean_result_<Wf[C1]Wa[C2]>}) for future
investigations.
\section[Perturbative and Non-Perturbative QCD Components]{\myfonta \hspace{-0.4cm}\letterspace to .92\naturalwidth{Perturbative and Non-Perturbative QCD Components}}
\label{Sec_Non-pert_Pert_Cont}

We decompose the gauge-invariant bilocal gluon field strength
correlator~(\ref{Eq_Ansatz}) into a perturbative ($\pert$) and
non-perturbative ($\nprt$) component
\be
        F_{\mu\nu\rho\sigma} 
        = F_{\mu\nu\rho\sigma}^{\pert} + F_{\mu\nu\rho\sigma}^{\nprt} 
        \ ,
\label{Eq_F_decomposition}
\ee
where $F_{\mu\nu\rho\sigma}^{\pert}$ gives the perturbative physics
(short-range correlations) described by {\em perturbative gluon
  exchange} and $F_{\mu\nu\rho\sigma}^{\nprt}$ the non-perturbative
physics (long-range correlations) modelled by the {\em stochastic
  vacuum model} (SVM)~\cite{Dosch:1987sk+X}. This combination allows
us to describe long and short distance correlations in agreement with
lattice calculations of the Euclidean gluon field strength
correlator~\cite{DiGiacomo:1992df+X,Meggiolaro:1999yn}. Moreover, this
two component ansatz leads to the static quark-antiquark potential
with color-Coulomb behavior for small and confining linear rise for
large source separations in good agreement with lattice data as shown
in our recent work~\cite{Shoshi:2002rd}. Note that
besides our two component ansatz an ongoing effort to reconcile the
non-perturbative SVM with perturbative gluon exchange that has led to
complementary
methods~\cite{Simonov:kt,Shevchenko:1998ej,Shevchenko:2002xi}.

We compute the perturbative gluon field strength correlator
$F_{\mu\nu\rho\sigma}^{\pert}$ from the gluon propagator
in Feynman-'t~Hooft gauge
\be
        \Big\langle  \G^a_{\mu}(x_1)\G^b_{\nu}(x_2) \Big\rangle
        = \int
        \frac{d^4k}{(2\pi)^4}
        \,\frac{-i\delta^{ab}g_{\mu\nu}}{k^2-m_G^2}
        \, e^{-ik(x_1-x_2)}
        \ ,
\label{Eq_massive_gluon_propagator}
\ee
where we introduce an {\em effective gluon mass} $m_G$ to limit the
range of the perturbative interaction in the infrared (IR) region. In
leading order in the strong coupling $g$, the perturbative correlator
is gauge-invariant already without the parallel transport to a common
reference point and depends only on the difference $z:= x_1 - x_2$, 
\bea
        F_{\mu\nu\rho\sigma}^{\pert}(z)
        \!\!&=&\!\!\frac{g^2}{\pi^2}\, \inv{2}\Bigl[
                       \frac{\partial}{\partial z_\nu}
                         \left(z_\sigma g_{\mu\rho}
                         -z_\rho g_{\mu\sigma}\right)
                       +\frac{\partial}{\partial z_\mu}
                         \left(z_\rho g_{\nu\sigma}
                         -z_\sigma g_{\nu\rho}\right)\Bigr]\,
              D_{\pert}(z^2)
        \nonumber \\
        \!\!&=&\!\! -\,\frac{g^2}{\pi^2}\!
                \int \!\!\frac{d^4k}{(2\pi)^4} \,e^{-ikz}\,\Bigl[
                k_\nu k_\sigma g_{\mu\rho}  - k_\nu k_\rho   g_{\mu\sigma}
              + k_\mu k_\rho  g_{\nu\sigma} - k_\mu k_\sigma g_{\nu\rho} \Bigr]\,
           \tilde{D}_{\pert}^{\prime}(k^2)
        \nonumber\\
\label{Eq_PGE_Ansatz_F}
\eea
with the {\em perturbative correlation function}
\be
        \tilde{D}_{\pert}^{\prime}(k^2) 
        := \frac{d}{dk^2} \int d^4z \,D_{\pert}(z^2)\, e^{ikz} 
         = \frac{i}{k^2 - m_G^2}
         \ .
\label{Eq_massive_D_pge_prime}
\ee

We take into account radiative corrections in perturbative
correlations by replacing the constant coupling $g^2$ with the {\em
  running coupling}
%
\be
        g^2(\vec{z}_{\!\perp})
        = 4 \pi \alphaS(\vec{z}_{\!\perp})
        = \frac{48 \pi^2}
        {(33-2 N_f) 
        \ln\left[
                (|\vec{z}_{\!\perp}|^{-2} + M^2)/\Lambda_{QCD}^2
        \right]}
\label{Eq_g2(z_perp)}
\ee
in the final step of the computation of the eikonal function $\chi$. Here
the renormalization scale is provided by $|\vec{z}_{\!\perp}|$
that represents the spatial separation of the interacting dipoles in
transverse space\footnote{Only transverse separations appear in the
  final expression of $\chi$ as explained in Sec.~\ref{Sec_Chi_Computation}.}.
In~(\ref{Eq_g2(z_perp)}) $N_f$ denotes the number of dynamical quark
flavors, which is set to $N_f = 0$ in agreement with the quenched
approximation, $\Lambda_{QCD} = 0.25\;\GeV$, and $M^2$ allows us to
freeze $g^2$ for $|\vec{z}_{\!\perp}| \rightarrow \infty$.

If the path connecting the points $x_1$ and $x_2$ is a straight line,
the non-perturbative correlator $F_{\mu\nu\rho\sigma}^{\nprt}$ depends
also only on the difference $z:= x_1 - x_2$. Then, the most general
form of the correlator in four-dimensional Minkowski space-time that
respects translational, Lorentz, and parity invariance
reads~\cite{Kramer:1990tr,Dosch:1994ym}
\bea
        \!\!\!\!
      F_{\mu\nu\rho\sigma}^{\nprt}(z) 
        &\!\!\!:=\!\!\!& F_{\mu\nu\rho\sigma}^{\nprt_{(nc)}}(z) +  
               F_{\mu\nu\rho\sigma}^{\nprt_{(c)}}(z)
\label{Eq_MSV_Ansatz_F}\\
\vspace*{-1cm}\nonumber\\
        \!\!\!\!
        F_{\mu\nu\rho\sigma}^{\nprt_{(nc)}}(z)
        &\!\!\!=\!\!\!&  \frac{G_2\,(1\!-\!\kappa)}{6(N_c^2-1)} 
        \Bigl(\frac{\partial}{\partial z_\nu}
        \left(z_\sigma g_{\mu\rho}
          \!-\!z_\rho g_{\mu\sigma}\right)
        \!+\!\frac{\partial}{\partial z_\mu}
        \left(z_\rho g_{\nu\sigma}
          \!-\!z_\sigma g_{\nu\rho}\right)\Bigr)\,
        D_1(z^2) 
\label{Eq_MSV_Ansatz_F_nc}\\
        && \hspace{-2.cm}=-\,\frac{G_2\,(1-\kappa)}{6(N_c^2-1)} 
        \int \frac{d^4k}{(2\pi)^4} \,e^{-ikz} \Bigl(
        k_\nu k_\sigma g_{\mu\rho}   - k_\nu k_\rho   g_{\mu\sigma}
        + k_\mu k_\rho  g_{\nu\sigma} - k_\mu k_\sigma g_{\nu\rho} \Bigr)\,
               \tilde{D}_{1}^{\prime}(k^2) 
\nonumber\\
\vspace*{1cm}\nonumber\\
        \!\!\!\!
        F_{\mu\nu\rho\sigma}^{\nprt_{(c)}}(z)
        &\!\!\!=\!\!\!& \frac{G_2\,\kappa}{3(N_c^2-1)}\, 
        \left(g_{\mu\rho}g_{\nu\sigma}
          -g_{\mu\sigma}g_{\nu\rho}\right) \,
        D(z^2)                                
        \nonumber\\
        &\!\!\!=\!\!\!& \frac{G_2\,\kappa}{3(N_c^2-1)}\,
        \left(g_{\mu\rho}g_{\nu\sigma}
          -g_{\mu\sigma}g_{\nu\rho}\right)\, 
        \int \frac{d^4k}{(2\pi)^4} \,e^{-ikz}\,
        \tilde{D}(k^2) \ .
\label{Eq_MSV_Ansatz_F_c}
\eea
and was originally constructed in Euclidean
space-time~\cite{Dosch:1987sk+X}. In all previous applications of the
SVM, this form depending only on $x_1$ and $x_2$ has been used. New
lattice results on the path dependence of the
correlator~\cite{DiGiacomo:2002mq} show a dominance of the shortest
path. This result is effectively incorporated in the model since the
straight paths dominate in the average over all paths.

In~(\ref{Eq_MSV_Ansatz_F_nc}) and~(\ref{Eq_MSV_Ansatz_F_c}), $a$ is
the {\em correlation length}, $G_2 := \langle \frac{g^2}{4\pi^2}
\G^a_{\mu\nu}(0) \G^a_{\mu\nu}(0) \rangle$ is the {\em gluon
  condensate}~\cite{Shifman:1979bx+X}, $\kappa$ determines the
relative weight of the two different tensor structures, $D$ and $D_1$
are the {\em non-perturbative correlation functions} in four
dimensional Minkowski space-time, and
\be
        \tilde{D}_{1}^{\prime}(k^2) 
        := \frac{d}{dk^2} \int d^4z D_{1}(z^2/a^2) \ e^{ikz} 
        \ .
\label{Eq_D1_prime}
\ee
Euclidean correlation functions are accessible together with the
Euclidean correlator in lattice
QCD~\cite{DiGiacomo:1992df+X,Meggiolaro:1999yn}.  We adopt for our
calculations the simple {\em exponential correlation functions}
specified in four dimensional Euclidean space-time
\be
        D^{E}(Z^2/a^2) = D^{E}_1(Z^2/a^2) = \exp(-|Z|/a)
        \ ,
\label{Eq_MSV_correlation_functions}
\ee
that are motivated by lattice QCD measurements of the gluon field
strength correlator
$F_{\mu\nu\rho\sigma}^{\nprt}(Z)$~\cite{DiGiacomo:1992df+X,Meggiolaro:1999yn}.
This correlation function stays positive for all Euclidean distances
$Z$ and, thus, is compatible with a spectral representation of the
correlation function~\cite{Dosch:1998th}.  This means a conceptual
improvement as compared with the correlation function used in earlier
applications of the \SVM which becomes negative at large
distances~\cite{Dosch:1994ym,Rueter:1996yb+X,Dosch:1998nw,Rueter:1998up,D'Alesio:1999sf,Berger:1999gu,Dosch:2001jg}.
By analytic continuation of (\ref{Eq_MSV_correlation_functions}) we
obtain the Minkowski correlation functions
in~(\ref{Eq_MSV_Ansatz_F_nc}) and~(\ref{Eq_MSV_Ansatz_F_c}) as shown
in Appendix~\ref{Sec_Correlation_Functions}.


The perturbative and non-perturbative gluon field strength correlators
can be analytically continued from Euclidean to Minkowski space-time
by the substitution $\delta_{\mu\rho} \rightarrow - g_{\mu\rho}$ and
the analytic continuation of the Euclidean correlation functions to
real time, $D^{E}_x(Z^2) \rightarrow
D^M_x(z^2)$,~\cite{Kramer:1990tr,Dosch:1994ym}. An alternative
analytic continuation has been recently proposed by
Meggiolaro~\cite{Meggiolaro:1996hf+X}. In our resent
work~\cite{Shoshi:2002rd} we have generalized
Meggiolaro's analytic continuation from parton-parton to
gauge-invariant dipole-dipole scattering. In this approach one
considers the correlation of two Wegner-Wilson loops tilted by an
angle $\Theta$ with respect to each other in Euclidean space-time and
obtaines -- after Maggiolaro's prescription for the analytic
continuation of the angle $\Theta$ -- the $S$-matrix element for
dipole-dipole scattering in Minkowski
space-time~\cite{Shoshi:2002rd}. Both analytic
continuations give, of course, the same results and allow us to study
the effect of confinement examined in Euclidean space-time on
high-energy reactions computed in Minkowski
space-time~\cite{Shoshi:2002in,Shoshi:2002fq,Shoshi:2002rd}.
In
chapter~\ref{Chapt_string_decomposition} we show the important
features of the confining QCD string and its manifestation in
high-energy scattering.

\subsubsection*{The Static Color-Dipole Potential} 
To illustrate the meaning of the different contributions
$F_{\mu\nu\rho\sigma}^{\nprt_{(nc)}}(z)$ and
$F_{\mu\nu\rho\sigma}^{\nprt_{(c)}}(z)$ in the non-perturbative
correlator~(\ref{Eq_MSV_Ansatz_F}), let us consider the potential of a
static color-dipole which is obtained from the expectation
value of one Wegner-Wilson loop\footnote{Here the subtraction of the
  self-energy of the color-sources is understood.}  computed in
Euclidean space-time (E)~\cite{Wilson:1974sk,Brown:1979ya}
\be
        V_r(R) 
        = - \lim_{T \to \infty} \inv{T} 
        \ln \langle W^E_r[C] \rangle
        \ .
\label{Eq_static_potential}
\ee
The computation of $\langle W^E_r[C] \rangle$ follows the same line as
the one for the Minkowskian case shown in
Sec.~\ref{Sec_VEV_Of_One_Wegner-Wilson_Loop}: The only difference is
the non-vanishing Euclidean $\chi^E_{SS}$ in the final step of the
computation which has to be computed now with the Euclidean correlator
$F^E_{\mu\nu\rho\sigma}$. For a loop in the fundamental
representation, $V_r = V_{N_c}$ describes the static
quark-antiquark potential and for a loop in the adjoint
respresentation, $V_r = V_{N_c^2 - 1}$ gives the static
potential of a gluino pair (adjoint dipole).

Considered in Euclidean space-time, the perturbative
correlator~(\ref{Eq_PGE_Ansatz_F}) leads to the non-confining {\em
  color-Yukawa potential} of a static
color-dipole~\cite{Shoshi:2002rd},
\be
        V_r^{\pert}(R) 
        = - \frac{C_2(r) g^2(R)}{4 \pi R} \exp[-m_g R] \ ,
\label{Eq_Vr(R)_color-Yukawa}
\ee
where $C_2(r)$ is the quadratic Casimir operator defined
in~(\ref{Eq_quadratic_Casimir_operator}) and $r$ denotes the
representation of $SU(N_c)$. The perturbative contribution dominates
the static potential at small dipole sizes $R$. In the limit $m_g \to
0$, Eq.~(\ref{Eq_Vr(R)_color-Yukawa}) reduces to the well-known {\em
  color-Coulomb potential} \cite{Kogut:1979wt}.

The tensor
structure of $F_{\mu\nu\rho\sigma}^{\nprt_{(nc)}}(z)$ given in~(\ref{Eq_MSV_Ansatz_F_nc}) is characteristic for Abelian
gauge theories, coincides with the tensor structure of the perturbative
correlator~(\ref{Eq_PGE_Ansatz_F}), and does not lead to confinement
when considered in Euclidean space-time since it gives the static
color-dipole potential~\cite{Shoshi:2002rd},
\be
        V_r^{\nprt\,\,nc}(R) =  
        -\,C_2(r)\,\,
        \frac{\pi^2 G_2\,(1-\kappa)\,a}{3 (N_c^2-1)}\,
        R^2\,K_2[R/a] \ ,
\label{Eq_Vr(R)_NP_nc}
\ee
which vanishes exponentially for large dipole sizes $R$. 

In contrast, the tensor structure of
$F_{\mu\nu\rho\sigma}^{\nprt_{(c)}}(z)$ given
in~(\ref{Eq_MSV_Ansatz_F_c}) can only occur in non-Abelian gauge
theories and Abelian gauge-theories with monopoles. Its Euclidean
version leads to the following static color-dipole
potential~\cite{Shoshi:2002rd}
\be
        V_r^{\nprt\,\,c}(R) 
         =  
        C_2(r)\,\,
        \frac{2\pi^2 G_2 \kappa}{3(N_c^2-1)}\,\,
        R^2
        \int_0^R \!\! d\rho\,
        (R-\rho)\,
        \rho\,K_1[\rho/a]
\label{Eq_Vr(R)_NP_c}
\ee
when computed with the {\em minimal surface}, i.e., the planar surface
bounded by the loop as indicated by the shaded area in
Fig.~\ref{Fig_loop_loop_scattering_surfaces}. For large dipole sizes,
$R \gtsim 0.5\ \fm$, Eq.~(\ref{Eq_Vr(R)_NP_c}) reduces to a linearly
increasing static potential
%
\be
        V_r^{\nprt\,\,c}(R)\Big|_{R\,\gtsim\,0.5\,\mbox{\scriptsize fm}} 
        = \sigma_r R  + \mbox{const.} \ ,
\label{Eq_Vr(R)_NP_c_linear}
\ee
which leads to {\em
  confinement}~\cite{Dosch:1987sk+X,Shoshi:2002rd}.
Therefore, we call the tensor structure in~(\ref{Eq_MSV_Ansatz_F_nc})
containing $(1-\kappa)$ non-confining ($nc$) and the tensor
structurein~(\ref{Eq_MSV_Ansatz_F_c}) containing $\kappa$ confining
($c$).

Color-confinement~\cite{Dosch:1987sk+X} is realized in our model by
the formation of a flux-tube of color-electric fields between the
color-sources in the
dipole~\cite{DelDebbio:1994zn,Rueter:1995cn,Shoshi:2002rd}.
We study the intrinsic structure of this flux-tube or confining
QCD string and the interaction between QCD strings in
chapter~\ref{Chapt_string_decomposition}.

The QCD {\em string tension} $\sigma_r$ is given by the
non-perturbative confining component~(\ref{Eq_MSV_Ansatz_F_c}) and
reads for a color-dipole in the representation $r$ of $SU(N_c)$
\cite{Dosch:1987sk+X,Shoshi:2002rd}
\be
        \sigma_r 
        = C_2(r)\,\,\frac{\pi^3 G_2 \kappa}{48} 
          \int_0^\infty dZ^2 D(Z^2,a^2) 
        = C_2(r)\,\,\frac{\pi^3 \kappa G_2 a^2}{24} \ ,
\label{Eq_string_tension}
\ee
with the exponential correlation
function~(\ref{Eq_MSV_correlation_functions}) used in the final step.
Since the string tension can be computed from first principles within
lattice QCD~\cite{Bali:2000gf}, relation~(\ref{Eq_string_tension})
puts an important constraint on the three fundamental parameters of
the non-perturbative QCD vacuum $a$, $G_2$, and $\kappa$. With the
values for $a$, $G_2$, and $\kappa$ given in
Sec.~\ref{Sec_Model_Parameters}, that are used throughout this work,
one obtains for the string tension of the $SU(3)$ quark-antiquark
potential ($r=3$) a value of $\sigma_3 = 0.22\,\GeV^2 \equiv 1.12
\,\GeV/\fm$ which coincides with those obtained from hadron
spectroscopy~\cite{Kwong:1987mj}, Regge
theory~\cite{Goddard:1973qh+X}, and lattice QCD
investigations~\cite{Bali:2000gf}.


The {\em Casimir scaling} of the static potential can be directly seen
from~(\ref{Eq_Vr(R)_color-Yukawa})-(\ref{Eq_Vr(R)_NP_c}). It emerges
trivially in our approach as a consequence of the Gaussian
approximation explained in
Sec.~\ref{Sec_VEV_Of_One_Wegner-Wilson_Loop}. Since the Casimir
scaling hypothesis of the static potential has been verified to high
accuracy for $SU(N_c)$ on the lattice~\cite{Bali:2000un}, this result
has been interpreted as a strong hint towards Gaussian dominance in
the QCD vacuum and, thus, as evidence for a strong suppression of
higher cumulant
contributions~\cite{Shevchenko:2000du,Shevchenko:2001ij}. In
comparison to our model, the instanton model can neither describe
Casimir scaling~\cite{Shevchenko:2001ij} nor the linear rise of the
confining potential~\cite{Chen:1999ct}.

The static potential of an adjoint dipole differs from that of a
fundamental dipole only in the eigenvalue of the corresponding
quadratic Casimir operator: $C_2(r) = C_2(N^2_c-1) = N_c$ for adjoint
and $C_2(r) = C_2(N_c) = (N_c^2-1)/(2 N_c)$ for fundamental dipoles.
Our model working in the quenched approximation has a shortcoming at
large dipole sizes: {\em string breaking} cannot occur in fundamental
dipoles as dynamical quark-antiquark production is excluded but should
be present for adjoint dipoles because of dynamical gluon production.
From Eqs.~(\ref{Eq_Vr(R)_NP_c_linear}) and~(\ref{Eq_string_tension}) it is
clear that string breaking is neither described for fundamental nor
for adjoint dipoles in our model. Interestingly, even on the lattice
there has been no striking evidence for adjoint quark screening in
quenched QCD~\cite{Kallio:2000jc}. It is even conjectured that the
{\WW} loop operator is not suited to studies of string
breaking~\cite{Gusken:1997sa+X}.

\section[Evaluation of the $\chi$-Function with Minimal
Surfaces]{\myfonta \hspace{-0.4cm}\letterspace to
  .93\naturalwidth{Evaluation of the} \boldmath$\chi$-\myfonta \letterspace to .93\naturalwidth{Function with Minimal Surfaces}}
\label{Sec_Chi_Computation}

For the computation of the $\chi$-function~(\ref{Eq_chi_SS})
\bea    
        \chi 
        &:=& \chi^{\pert} + \chi^{\nprt}_{nc} + \chi^{\nprt}_{c} 
        \nonumber \\
        &=& - \,i\,\frac{\pi^2}{4}\!
        \int_{S_1} \! d\sigma^{\mu\nu}(x_1) 
        \int_{S_2} \! d\sigma^{\rho\sigma}(x_2)
        \left( 
          F_{\mu\nu\rho\sigma}^{\pert} 
        + F_{\mu\nu\rho\sigma}^{\nprt_{(nc)}} 
        + F_{\mu\nu\rho\sigma}^{\nprt_{(c)}} 
        \right) \ ,
\label{chi_amplitude}
\eea
one has to specify surfaces $S_{1,2}$ with the restriction $\partial
S_{1,2} = C_{1,2}$ according to the non-Abelian Stokes' theorem. As
illustrated in Fig.~\ref{Fig_loop_loop_scattering_surfaces}, we put
the reference point $o$ at the origin of the coordinate system and
choose for $S_{1,2}$ the {\em minimal surfaces} that are built from
the areas spanned by the corresponding loops $C_{1,2}$ and the
infinitesimally thin tube which connects the two surfaces $S_1$ and
$S_2$.  The thin tube allows us to compare the field strengths in
surface $S_1$ with the field strengths in surface $S_2$.

Due to the truncation of the cumulant expansion or, equivalently, the
Gaussian approximation, the non-perturbative confining component
$\chi_{c}^{\nprt}$ depends on the specific surface choice. This is not
the case for the perturbative and the non-perturbative non-confining
component. A different and more complicated result for
$\chi_{c}^{\nprt}$ was obtained with the pyramid mantle choice for the
surfaces $S_{1,2}$ in earlier applications of the \SVM\ to high-energy
scattering~\cite{Dosch:1994ym,Rueter:1996yb+X,Dosch:1998nw,Rueter:1998up,D'Alesio:1999sf,Berger:1999gu,Dosch:2001jg}.
In this work we use the minimal surfaces because of the following
reasons. Minimal surface are usually used to obtain Wilson's area
law~\cite{Dosch:1987sk+X,Shoshi:2002rd}.  The minimal
surfaces are also favored by other complementary approaches such as
the strong coupling expansion in lattice QCD, where plaquettes cover
the minimal surface, or large-$N_c$ investigations, where the planar
gluon diagrams dominate in the large-$N_c$ limit. Within bosonic
string theory, our minimal surfaces represent the worldsheets of the
{\em rigid} strings: Our model does not describe fluctuations or
excitations of the string and thus cannot reproduce the L\"uscher term
which has recently been confirmed with unprecedented precision by
L\"uscher and Weisz~\cite{Luscher:2002qv}.

Internal consistencies of our model favorize also the minimal surface:
Since our results for the vacuum expectation value (VEV) of a
rectangular {\WW} loop lead to a static quark-antiquark potential that
is in good agreement with lattice
data~\cite{Shoshi:2002rd}, we are led to conclude that
the choice of the minimal surface is required by the Gaussian
approximation in the gluon field strengths. Furthermore, we have shown
in our recent work~\cite{Shoshi:2002rd} that the
minimal surfaces are actually needed to ensure the consistency of
our results for the VEV of one loop $\langle W_{r}[C] \rangle$ and the
loop-loop correlation function $\langle W_{r_1}[C_1] W_{r_2}[C_2]
\rangle$. Phenomenologically, in comparison with pyramid mantles, the
description of the slope parameter $B(s)$, the differential elastic
cross section $d\sigma^{el}/dt(s,t)$, and the elastic cross section
$\sigma^{el}(s)$ can be improved with minimal surfaces as shown in
chapter~\ref{Sec_Comparison_Data}.

The simplicity of the minimal surfaces is appealing: It allows us to
show for the first time the structure of non-perturbative
dipole-dipole interactions in momentum space, to represent the QCD
string as a collection of stringless dipoles and to compute
unintegrated gluon distributions of hadrons and photons. This is shown
in chapter~\ref{Chapt_string_decomposition} and~\cite{Shoshi:2002fq}.

In applications of the model to high-energy
scattering~\cite{Shoshi:2002in,Shoshi:2002ri,Shoshi:2002fq} the
minimal surfaces are the most natural choice to examine the scattering
of two rigid strings without any fluctuations or excitations. Our
model does unfortunately not choose the surface dynamically and, thus,
cannot describe string flips between two non-perturbative
color-dipoles. This could generate also the energy dependence in
non-perturbative interactions. Recently, new developments towards a
dynamical surface choice and a theory for the dynamics of the
confining strings have been reported~\cite{Shevchenko:2002xi}.

%
\section*{\large Parametrization of the Minimal Surfaces}
%

The minimal surfaces $S_1$ and $S_2$ shown as shaded areas in
Fig.~\ref{Fig_loop_loop_scattering_surfaces} can be parametrized with
the upper (lower) subscripts and signs referring to $S_1$ ($S_2$) as
follows
\be
        S_{\!\!{1 \atop (2)}} =  
        \left\{ 
            \Big(x^{\mu}_{\!\!\!\!{1\atop(2)}}(u,v)\Big) 
            = \Big(
            r^{\mu}_{\!\!\!\!{1q\atop(2q)}}
            + u\,n^{\mu}_{\!\!\!{\oplus\atop(\ominus)}} 
            + v\,r^{\mu}_{\!\!\!{1\atop(2)}}
            \Big),
            \;u \in [-T,T], \;v \in [0,1] \right\}
        \ ,
\label{Eq_S1(S2)_parameterization}
\ee
where 
\be
        \Big(n^{\mu}_{\!\!\!{\oplus\atop(\ominus)}}\Big) 
        := \left( \barray{c} 1 \\ \vec{0} \\ {\scriptscriptstyle{+\atop(-)}}\!1 \earray \right)
        , \quad
        \Big( r^{\mu}_{\!\!\!\!{1q\atop(2q)}} \Big)
        := \left( \barray{c} 0 \\ \vec{r}_{\!\!\!{1q\atop(2q)}} \\ 0 \earray \right)
        , \quad \mbox{and} \quad
        \Big( r^{\mu}_{\!\!\!{1\atop(2)}} \Big)
        := \left( \barray{c} 0 \\ \vec{r}_{\!\!\!{1\atop(2)}}   \\ 0 \earray \right)
        \ .
\label{Eq_C1(C2)_four_vectors}
\ee
The infinitesimally thin tube is neglected since it does not
contribute to the $\chi$-function. The computation of the
$\chi$-function requires only the parametrized parts of the minimal
surfaces~(\ref{Eq_S1(S2)_parameterization}), the corresponding
infinitesimal surface elements
\be
        d\sigma^{\mu\nu} 
        = \left( \frac{\partial x^{\mu}}{\partial u} 
                 \frac{\partial x^{\nu}}{\partial v}
               - \frac{\partial x^{\mu}}{\partial v} 
                 \frac{\partial x^{\nu}}{\partial u} \right)\,du\,dv
        = \left( n^{\mu}_{\!\!\!{\oplus\atop(\ominus)}} 
                 r^{\nu}_{\!\!\!{1\atop(2)}}
               - r^{\mu}_{\!\!\!{1\atop(2)}} 
                 n^{\nu}_{\!\!\!{\oplus\atop(\ominus)}} \right)\,du\,dv
        \ ,
\label{Eq_S1(S2)_surface_element}
\ee
and the limit $T \to \infty$ which is appropriate since the
correlation length $a$ is much smaller (see
Sec.~\ref{Sec_Model_Parameters}) than the longitudinal extension of
the loops.

\section*{\large \boldmath$\chi_{c}^{\nprt}$-Computation}

Starting with the confining component 
\bea
        \!\!\!\!\!\!\!\!\!\!&&\!\!\!\!\! 
        \chi_{c}^{\nprt} 
        := - \, i \,\frac{\pi^2}{4} 
                \int_{S_1} \!\! d\sigma^{\mu\nu}(x_1) 
                \int_{S_2} \!\! d\sigma^{\rho\sigma}(x_2)\,
                F_{\mu\nu\rho\sigma}^{\nprt_{(c)}}(z = x_1 - x_2)
        \nonumber \\ 
        \!\!\!\!\!\!\!\!\!\!&& 
        = -\,\frac{\pi^2 G_2 \kappa}{12(N_c^2-1)}
        \int_{S_1} \!\! d\sigma^{\mu\nu}(x_1) 
        \int_{S_2} \!\! d\sigma^{\rho\sigma}(x_2)
        \left(g_{\mu\rho}g_{\nu\sigma}
                -g_{\mu\sigma}g_{\nu\rho}\right)
        iD(z^2/a^2) 
        \ ,
\label{Eq_chi_MSV_confining_definition}
\eea
one exploits the anti-symmetry of the surface elements,
$d\sigma^{\mu\nu} = - d\sigma^{\nu\mu}$, and applies the surface
parametrization~(\ref{Eq_S1(S2)_parameterization}) with the
corresponding surface elements~(\ref{Eq_S1(S2)_surface_element}) to
obtain
\be
        \chi_{c}^{\nprt} = 
        \frac{\pi^2 G_2 \kappa}{3(N_c^2-1)}\,2
        \left(\vec{r}_1\cdot\vec{r}_2\right)
        \int_0^1 \!\! dv_1 \int_0^1 \!\! dv_2
        \lim_{T\to\infty} \int_{-T}^T\!\!du_1 \int_{-T}^T\!\!du_2\,
        iD(z^2/a^2)
        \ ,
\label{Eq_chi_MSV_confining_intermediate}
\ee
where
\be
        z^{\mu} = x_1^{\mu} - x_2^{\mu} 
        = u_1 n^{\mu}_{\oplus} - u_2 n^{\mu}_{\ominus}
        + r^{\mu}_{1q} - r^{\mu}_{2q}
        + v_1 r^{\mu}_{1} - v_2 r^{\mu}_{2}
        \ ,
\label{Eq_z_S1_S2}
\ee
and the identities $n_{\oplus}\cdot r_{2} = r_{1}\cdot n_{\ominus} =
0$ and $n_{\oplus}\cdot n_{\ominus} = 2$, evident
from~(\ref{Eq_C1(C2)_four_vectors}), have been used. Next, one Fourier
transforms the correlation function and performs the $u_1$ and $u_2$
integrations in the limit $T \to \infty$
\bea
        && \lim_{T\to\infty} 
                \int_{-T}^T \!\! du_1 
                \int_{-T}^T \!\! du_2\,iD(z^2/a^2)
        \nonumber \\
        && = \int \frac{d^4k}{(2\pi)^4}\,i\tilde{D}(k^2) 
                \lim_{T\to\infty} \int_{-T}^T \!\! du_1 
                \int_{-T}^T \!\! du_2 \,e^{-ikz}
        \nonumber \\
        && = \int \frac{d^4k}{(2\pi)^2}\,i\tilde{D}(k^2)\, 
        \exp[-ik_{\mu}
        (r^{\mu}_{1q}-r^{\mu}_{2q}+v_1 r^{\mu}_{1}-v_2 r^{\mu}_{2})]
        \,\delta(k^0-k^3)\,\delta(k^0+k^3)
        \nonumber \\
        && = \inv{2} \, iD^{(2)}
        \left(\vec{r}_{1q}+v_1\vec{r}_1-\vec{r}_{2q}-v_2\vec{r}_2\right)
        \ ,
\label{Eq_light-cone_coordinates_integrated_out}
\eea
where $iD^{(2)}$ is the confining correlation function in the
two-dimensional transverse space
(cf.~Appendix~\ref{Sec_Correlation_Functions})
\be
        D^{(2)}(\vec{z}_{\!\perp}) 
        =  \int \frac{d^2k_{\!\perp}}{(2\pi)^2} 
        e^{i\vec{k}_{\!\perp}\vec{z}_{\!\perp}}
        \tilde{D}^{(2)}(\vec{k}_{\!\perp}) \ .
\label{Eq_transverse_Fourier_transform}
\ee
The contributions along the light-cone coordinates have been
integrated out so that $\chi_{c}^{\nprt}$ is completely determined by
the transverse projection of the minimal surfaces. Inserting
(\ref{Eq_light-cone_coordinates_integrated_out}) into
(\ref{Eq_chi_MSV_confining_intermediate}), one finally obtains
\be
      \chi_{c}^{\nprt} = 
        \frac{\pi^2 G_2}{3(N_c^2-1)}\,\kappa
        \left(\vec{r}_1\cdot\vec{r}_2\right) \,
        \int_0^1 \! dv_1 \int_0^1 \! dv_2 \, 
        iD^{(2)}\left(\vec{r}_{1q} + v_1\vec{r}_1 
        - \vec{r}_{2q} - v_2\vec{r}_2\right)
        \ .
\label{Eq_chi_MSV_confining} 
\ee
With $\tilde{D}^{(2)}(\vec{k}_{\!\perp})$ obtained from the
exponential correlation function~(\ref{Eq_MSV_correlation_functions}),
cf.~Appendix~\ref{Sec_Correlation_Functions}, we find
\be
        iD^{(2)}(\vec{z}_{\!\perp})
        =  2 \pi \, a^2 
        \left[1+(|\vec{z}_{\!\perp}|/a)\right] 
        \exp\!\left( -|\vec{z}_{\!\perp}|/a\right)
\label{Eq_F2[i_D_confining]}
\ee
which is positive for all transverse distances. 


\section*{\large \boldmath$\chi_{nc}^{\nprt}$-Computation}

Continuing with the computation of the non-confining component
\bea
        \chi_{nc}^{\nprt}\!\!\!\!\!\!  
        &&:= -\, i \,\frac{\pi^2}{4} 
                \int_{S_1} \!\! d\sigma^{\mu\nu}(x_1) 
                \int_{S_2} \!\! d\sigma^{\rho\sigma}(x_2)\,
                F_{\mu\nu\rho\sigma}^{\nprt_{(nc)}}(z = x_1 - x_2)
        \nonumber \\ 
        && = \frac{\pi^2 G_2 (1-\kappa)}{12(N_c^2-1)} \,
                \int_{S_1} \!\! d\sigma^{\mu\nu}(x_1) \,
                \int_{S_2} \!\! d\sigma^{\rho\sigma}(x_2) 
        \\
\label{Eq_chi_MSV_non-confining_definition}
        && \hphantom{=} 
        \times \int \!\! \frac{d^4k}{(2\pi)^4} \,e^{-ikz}\,
        \Bigl[k_\nu k_\sigma g_{\mu\rho} 
        - k_\nu k_\rho g_{\mu\sigma}
        + k_\mu k_\rho  g_{\nu\sigma} 
        - k_\mu k_\sigma g_{\nu\rho} \Bigr]\,
        i\tilde{D}_{1}^{\prime}(k^2) \nonumber \ ,
\eea
we exploit again the anti-symmetry of both surface elements to obtain
\bea
        \chi_{nc}^{\nprt} 
        & = & \frac{\pi^2 G_2 (1-\kappa)}{3(N_c^2-1)} \,
                \int_0^1 \!\! dv_1 \int_0^1 \!\! dv_2 \,
                \int \!\! \frac{d^4k}{(2\pi)^4} \,
                \lim_{T\to\infty}\,
                \int_{-T}^T\!\!du_1\int_{-T}^T\!\!du_2\,e^{-ikz}
        \nonumber \\ 
        && \times
        \Bigl[2\,(r_1\cdot k)\,(r_2\cdot k)\,
        -\,(\vec{r}_1\cdot\vec{r}_2)\,(k^0 - k^3)(k^0 + k^3)\Bigr]\,
        i\tilde{D}_{1}^{\prime}(k^2) 
\label{Eq_chi_MSV_non-confining_intermediate_1}
\eea
with $z$ as given in~(\ref{Eq_z_S1_S2}). Again the identities
$n_{\oplus}\cdot r_{2} = r_{1}\cdot n_{\ominus} = 0$ and
$n_{\oplus}\cdot n_{\ominus} = 2$ have been used. Performing the $u_1$
and $u_2$ integrations in the limit $T \rightarrow \infty$, one
obtains --- as in~(\ref{Eq_light-cone_coordinates_integrated_out}) ---
two $\delta$-functions which allow us to carry out the integrations
over $k^0$ and $k^3$ immediately. This leads to
\bea
        \chi_{nc}^{\nprt} 
        & \!\!\!\! = \!\!\! & \frac{\pi^2 G_2 (1-\kappa)}{3(N_c^2-1)}
                \!\int_0^1 \!\! dv_1 \!\int_0^1 \!\! dv_2
                \! \int \!\! \frac{d^2k_{\!\perp}}{(2\pi)^2}
                i\tilde{D}_{1}^{\prime\,(2)}(\vec{k}_{\!\perp}^2)
        (\vec{r}_1\cdot\vec{k}_{\!\perp})\,(\vec{r}_2\cdot\vec{k}_{\!\perp})
        e^{i\vec{k}_{\!\perp}(\vec{r}_{1q}+v_1\vec{r}_1-\vec{r}_{2q}-v_2\vec{r}_2)}
        \nonumber \\
        & \!\!\!\! = \!\!\! & \frac{\pi^2 G_2 (1-\kappa)}{3(N_c^2-1)} 
                \int_0^1 \!\! dv_1 \frac{\partial}{\partial v_1} 
                \int_0^1 \!\! dv_2 \frac{\partial}{\partial v_2}\, 
                iD_{1}^{\prime\,(2)}
                (\vec{r}_{1q} + v_1\vec{r}_1 - \vec{r}_{2q} - v_2\vec{r}_2) 
        \ , 
\label{Eq_chi_MSV_non-confining_intermediate_2}
\eea
where $iD_{1}^{\prime\,(2)}$ is the non-confining correlation function
in transverse space defined analogously
to~(\ref{Eq_transverse_Fourier_transform}). The $v_1$ and $v_2$
integrations are trivial and lead (cf.\ 
Fig.~\ref{Fig_loop_loop_scattering_surfaces}b) to
\bea
     &&\!\!\!\!\!\!\!\!\! 
        \chi_{nc}^{\nprt} = 
        \frac{\pi^2 G_2}{3(N_c^2-1)}\,(1-\kappa) 
        \left[ 
        iD^{\prime\,(2)}_1
        \left(\vec{r}_{1q}-\vec{r}_{2q}\right) 
        +iD^{\prime\,(2)}_1
        \left(\vec{r}_{1\qbar}-\vec{r}_{2\qbar}\right)
        \right.
        \nonumber \\
     &&\!\!\!\!\!\!\!\!\! 
     \hphantom{\chi_{nc}^{\nprt}=\frac{\pi^2 G_2}{3(N_c^2-1)}\,(1-\kappa)}
        \left.
        -\,iD^{\prime\,(2)}_1
        \left(\vec{r}_{1q}-\vec{r}_{2\qbar}\right)
        -iD^{\prime\,(2)}_1
        \left(\vec{r}_{1\qbar}-\vec{r}_{2q}\right)
        \right] \ .
\label{Eq_chi_MSV_non-confining}
\eea
Using $\tilde{D}_{1}^{\prime\,(2)}(\vec{k}_{\!\perp}^2)$, derived from
the exponential correlation
function~(\ref{Eq_MSV_correlation_functions}) in
Appendix~\ref{Sec_Correlation_Functions}, we obtain
\be
         iD^{\prime\,(2)}_1(\vec{z}_{\!\perp})
         =  
        \pi \, a^4  \left[3 + 3(|\vec{z}_{\!\perp}|/a) + (|\vec{z}_{\!\perp}|/a)^2 \right]
        \exp\!\left( -|\vec{z}_{\!\perp}|/a\right)
        \ .
\label{Eq_F2[i_D_non-confining_prime]}
\ee

The non-perturbative components, $\chi^{\nprt}_c$ and
$\chi^{\nprt}_{nc}$, lead to {\em color transparency} for small
dipoles, i.e.\ a dipole-dipole cross section with
$\sigma_{DD}(\vec{r}_1,\vec{r}_2) \propto |\vec{r}_1|^2|\vec{r}_2|^2$
for $|\vec{r}_{1,2}| \to 0$, as known for the perturbative
case~\cite{Nikolaev:1991ja}. This will be shown in
Sec.~\ref{Sec_The_Decomposition_of_the_String_into_Dipoles_and_the_Unintegrated_Gluon_Distribution}.

\section*{\large \boldmath$\chi^{\pert}$-Computation}
%
The perturbative component $\chi^{\pert}$ is defined as
\bea
        \chi^{\pert} 
        &:=& -\,i\,\frac{\pi^2}{4} 
                \int_{S_1} \!\! d\sigma^{\mu\nu}(x_1) 
                \int_{S_2} \!\! d\sigma^{\rho\sigma}(x_2)\,
                F_{\mu\nu\rho\sigma}^{\pert}(z = x_1 - x_2) 
        \hphantom{i\tilde{D}_{\pert}^{\prime\,(4)}k}
        \nonumber \\ 
        & =& \,\frac{g^2}{4}
                \int_{S_1} \!\! d\sigma^{\mu\nu}(x_1)\,
                \int_{S_2} \!\! d\sigma^{\rho\sigma}(x_2) 
        \\ && 
        \times \int \!\! \frac{d^4k}{(2\pi)^4} \,e^{-ikz}\,
                \Bigl[k_\nu k_\sigma g_{\mu\rho}   
                - k_\nu k_\rho   g_{\mu\sigma}
                + k_\mu k_\rho  g_{\nu\sigma} 
                - k_\mu k_\sigma g_{\nu\rho} \Bigr]\,
                i\tilde{D}_{\pert}^{\prime}(k^2) 
        \ , \nonumber
\label{Eq_chi_PGE_definition}
\eea
and shows a structure identical to the one of $\chi_{nc}^{\nprt}$ given
in~(\ref{Eq_chi_MSV_non-confining_definition}).  Accounting for the
different prefactors and the different correlation function, the
result for $\chi_{nc}^{\nprt}$~(\ref{Eq_chi_MSV_non-confining}) can be
used to obtain
\bea
     &&\!\!\!\!\!\!\!\!\!
        \chi^{\pert} = 
        \left[ 
        g^2\!\left(\vec{r}_{1q}-\vec{r}_{2q}\right)
        iD^{\prime\,(2)}_{\pert}
        \left(\vec{r}_{1q}-\vec{r}_{2q}\right)
        +g^2\!\left(\vec{r}_{1\qbar}-\vec{r}_{2\qbar}\right)
        iD^{\prime\,(2)}_{\pert}
        \left(\vec{r}_{1\qbar}-\vec{r}_{2\qbar}\right)
        \right.
        \nonumber \\
     &&\!\!\!\!\!\!\!\!\!\!\!\!\!\!\!\!\!\!\!\!\!
        \hphantom{\chi^{\pert} = }
        \left.
        -\,g^2\!\left(\vec{r}_{1q}-\vec{r}_{2\qbar}\right)
        iD^{\prime\,(2)}_{\pert}
        \left(\vec{r}_{1q}-\vec{r}_{2\qbar}\right)
        -g^2\!\left(\vec{r}_{1\qbar}-\vec{r}_{2q}\right)
        iD^{\prime\,(2)}_{\pert}
        \left(\vec{r}_{1\qbar}-\vec{r}_{2q}\right)
        \right]
        \ , 
\label{Eq_chi_PGE}
\eea
where the running coupling $g^2(\vec{z}_{\!\perp})$ is understood as
given in~(\ref{Eq_g2(z_perp)}). With~(\ref{Eq_massive_D_pge_prime})
one obtains the perturbative correlation function in transverse space
\bea
        iD^{\prime\,(2)}_{\pert}
        (\vec{z}_{\!\perp})
        = \inv{2\pi} K_0\left(m_G |\vec{z}_{\!\perp}|\right)
        \ ,
\label{Eq_F2[i_massive_D_pge_prime]}
\eea
where $K_0$ denotes the $0^{th}$ modified Bessel function (McDonald
function).

In contrast to the confining component $\chi_{c}^{\nprt}$, the
non-confining components, $\chi_{nc}^{\nprt}$ and $\chi^{\pert}$, depend
only on the transverse position between the quark and antiquark of the
two dipoles and are therefore independent of the surface choice.

Finally, we explain that the vanishing of $\chi_{S_1S_1}$ and
$\chi_{S_2S_2}$ anticipated in Sec.~\ref{Sec_loop_loop_func} results
from the light-like loops and the tensor structures in
$F_{\mu\nu\rho\sigma}$. Concentrating --- without loss of generality
--- on $\chi_{S_1 S_1}$, the appropriate infinitesimal surface
elements~(\ref{Eq_S1(S2)_surface_element}) and the
$F_{\mu\nu\rho\sigma}$--ansatz given in~(\ref{Eq_F_decomposition}),
(\ref{Eq_MSV_Ansatz_F}), and~(\ref{Eq_PGE_Ansatz_F}) are inserted
into~(\ref{Eq_chi_SS}). Having simplified the resulting expression by
exploiting the anti-symmetry of the surface elements, one finds only
terms proportional to $n_{\oplus}^2$, $n_{\oplus}\cdot r_{1}$, and
$n_{\oplus}\cdot z$ with $z^{\mu} = x_1^{\mu} - x_2^{\mu} = (u_1 -
u_2) n^{\mu}_{\oplus} + (v_1 - v_2) r^{\mu}_{1}$. Since $n_{\oplus}^2
= 0$ and $n_{\oplus}\cdot r_{1} = 0$, which is evident
from~(\ref{Eq_C1(C2)_four_vectors}), all terms vanish and $\chi_{S_1
  S_1} = 0$ is derived.

Note that $\chi = \chi^{\pert} + \chi_{nc}^{\nprt} + \chi_{c}^{\nprt}$ is
a real-valued function. Since, in addition, the wave functions
$|\psi_i(z_i,\vec{r}_i)|^2$ used in this work (cf.\ 
Appendix~\ref{Sec_Wave_Functions}) are invariant under the replacement
$(\vec{r}_i \rightarrow -\vec{r}_i, z_i \rightarrow 1-z_i)$, the
$T$-matrix element~(\ref{Eq_model_T_amplitude}) upon inserting the
results~(\ref{VEV_WWL_1}) and~(\ref{Eq_final_Euclidean_result_<W[C1]W[C2]>_fundamental})  becomes purely imaginary and reads for $N_c=3$
\bea
        \!\!\!\!\!\!\!\!\!\!\!\!\!\!\!\!\!
        T(s,t) 
        & = & 2is \int \!\!d^2b_{\!\perp} 
                e^{i {\vec q}_{\!\perp} {\vec b}_{\!\perp}}
                \int \!\!dz_1 d^2r_1 \!
                \int \!\!dz_2 d^2r_2 \,\,
                |\psi_1(z_1,\vec{r}_1)|^2   \,\,
                |\psi_2(z_2,\vec{r}_2)|^2       
        \nonumber \\    
        && \!\!\!\!\!\!\!\!\!\!\!\!\!\!
        \times 
        \left[1-\frac{2}{3} 
        \cos\!\left(\frac{1}{3}
        \chi({\vec b}_{\!\perp},z_1,\vec{r}_1,z_2,\vec{r}_2)\!\right)
        - \frac{1}{3}
        \cos\!\left(\frac{2}{3}
        \chi({\vec b}_{\!\perp},z_1,\vec{r}_1,z_2,\vec{r}_2)\!\right)
        \right] \ .
\label{Eq_model_purely_imaginary_T_amplitude}
\eea
The real part averages out in the integration over ${\vec r}_i$ and
$z_i$ since the $\chi$-function changes sign
\be
        \chi(\vec{b}_{\!\perp},1-z_1,-\vec{r}_1,z_2,\vec{r}_2)
        = - \chi(\vec{b}_{\!\perp},z_1,\vec{r}_1,z_2,\vec{r}_2)
        \ ,
\label{Eq_odd_eikonal_function}
\ee
which can be seen directly
from~(\ref{Eq_chi_MSV_confining}),(\ref{Eq_chi_MSV_non-confining}) and
(\ref{Eq_chi_PGE}) as $(\vec{r}_1 \rightarrow -\vec{r}_1, z_1
\rightarrow 1-z_1)$ implies $\vec{r}_{1q} \rightarrow
\vec{r}_{1\qbar}$. In physical terms, $(\vec{r}_i \rightarrow
-\vec{r}_i, z_i \rightarrow 1-z_i)$ corresponds to {\em charge
  conjugation}\, i.e.\ the replacement of each parton with its
antiparton and the associated reversal of the loop direction.

Consequently, the
$T$-matrix~(\ref{Eq_model_purely_imaginary_T_amplitude}) describes
only charge conjugation $C = +1$ exchange. Since in our quenched
approximation purely gluonic interactions are modelled,
(\ref{Eq_model_purely_imaginary_T_amplitude}) describes only
pomeron\footnote{Odderon $C = -1$ exchange is excluded in our model.
  It would survive in the following cases: (a) Wave functions are used
  that are not invariant under the transformation $(\vec{r}_i
  \rightarrow -\vec{r}_i, z_i \rightarrow 1-z_i)$. (b) The proton is
  described as a system of three quarks with finite separations
  modelled by three loops with one common light-like line. (c) The
  Gaussian approximation that enforces the truncation of the cumulant
  expansion is relaxed and additional higher cumulants are taken into
  account.}  but not reggeon exchange.

\section{Energy Dependence}
\label{Sec_Energy_Dependence}

Up to now the $T$-matrix
element~(\ref{Eq_model_purely_imaginary_T_amplitude}) leads to
energy-independent total cross sections in contradiction to the
experimental observation. This is disappointing from the
phenomenological point of view but not surprising since our formalism
does not describe gluon radiation which would generate the energy
dependence.  Nevertheless, we introduce the energy dependence in a
phenomenological way inspired by other successful models so that a
unified description of hadron-hadron, photon-proton, and photon-photon
reactions is achieved. The powerlike ansatz used for the energy
dependence is crucial to guarantee the Froissart
bound~\cite{Froissart:1961ux+X} as shown in chapter~\ref{$S$-Matrix
  Unitarity and Gluon Saturation}.

Most models for high-energy scattering are constructed to describe
either hadron-hadron or photon-hadron reactions.  For example,
Kopeliovich et al.~\cite{Kopeliovich:2001pc} as well as Berger and
Nachtmann~\cite{Berger:1999gu} focus on hadron-hadron scattering. In
contrast, Golec-Biernat and W\"usthoff~\cite{Golec-Biernat:1999js,Golec-Biernat:1999qd}
and Forshaw, Kerley, and Shaw~\cite{Forshaw:1999uf} concentrate on
photon-proton reactions. A model that describes the energy dependence
in both hadron-hadron and photon-hadron reactions up to large photon
virtualities is the two-pomeron model of Donnachie and
Landshoff~\cite{Donnachie:1998gm+X}. Based on Regge theory, they find
hard pomeron trajectory with intercept $1 + \epsilon_{hard} \approx
1.4$ that governs the strong energy dependence of $\gamma^*p$
reactions with high $Q^2$ and a soft pomeron trajectory with intercept
$ 1 + \epsilon_{soft} \approx 1.08$ that governs the weak energy
dependence of hadron-hadron or $\gamma^{*} p$ reactions with low
$Q^2$.  Similarly, we aim at a simultaneous description of
hadron-hadron, photon-proton, and photon-photon reactions involving
real and virtual photons as well.

In line with other two-component (hard $+$ soft)
models~\cite{Donnachie:1998gm+X,D'Alesio:1999sf,Forshaw:1999uf,Rueter:1998up,Donnachie:2001wt}
and the different hadronization mechanisms in hard and soft
collisions, our physical ansatz demands that the perturbative and
non-perturbative contributions do not interfere. Therefore, we modify the
cosine-summation in~(\ref{Eq_model_purely_imaginary_T_amplitude})
allowing only even numbers of soft and hard correlations, $\left
  (\chi^{\pert} \right)^{2n} \left ( \chi^{\nprt} \right)^{2m}$ with
$n,m \in I\!\!N$.  Interference terms with odd numbers of hard and
soft correlations are subtracted by the replacement
\be
        \cos\left[ c\chi \right] =  
        \cos\left[c\left( \chi^{\pert} + \chi^{\nprt} \right)\right] 
        \rightarrow 
        \cos\left[c\chi^{\pert}\right]\cos\left[c\chi^{\nprt}\right] 
        \ ,
\label{Eq_interference_term_subtraction}
\ee
where $c = 1/3$ or $2/3$. This prescription leads to the following
factorization of soft and hard physics in the $T$-matrix
element~(\ref{Eq_model_purely_imaginary_T_amplitude}),
\bea
        && \!\!\!\!\!\!\!\!\!\!
        T(s,t) 
        = 2is \int \!\!d^2b_{\!\perp} 
        e^{i {\vec q}_{\!\perp} {\vec b}_{\!\perp}}
        \int \!\!dz_1 d^2r_1 \!\int \!\!dz_2 d^2r_2\,\,
        |\psi_1(z_1,\vec{r}_1)|^2 \,\, 
        |\psi_2(z_2,\vec{r}_2)|^2       
        \nonumber \\    
        &&\!\!\!\!\!\!\!\!\!\!\!\! 
        \times \left[ 1 - \frac{2}{3} 
        \cos\!\left(\!\frac{1}{3}\chi^{\pert}\!\right)
        \cos\!\left(\!\frac{1}{3}\chi^{\nprt}\!\right)         
        - \frac{1}{3}
        \cos\!\left(\!\frac{2}{3}\chi^{\pert}\!\right)
        \cos\!\left(\!\frac{2}{3}\chi^{\nprt}\!\right)
        \right] \ . 
\label{Eq_model_purely_imaginary_T_amplitude_almost_final_result}
\eea
In the limit of small $\chi$-functions, $|\chi^{\pert}|
\ll 1$ and $|\chi^{\nprt}| \ll 1$, one gets
\bea
        T(s,t) 
        & = & 2is \!\int \!\!d^2b_{\!\perp} 
        e^{i {\vec q}_{\!\perp} {\vec b}_{\!\perp}}
        \!\int \!\!dz_1 d^2r_1 \!\int \!\!dz_2 d^2r_2\,\,
        |\psi_1(z_1,\vec{r}_1)|^2 \,\,
        |\psi_2(z_2,\vec{r}_2)|^2       
        \nonumber \\  
        && \times \frac{1}{9}\left[
        \left(\chi^{\pert}\right)^2
        +\left(\chi^{\nprt}\right)^2
        \right].
\label{Eq_model_purely_imaginary_T_amplitude_small_chi_limit}
\eea
In this limit, the $T$-matrix element evidently becomes a sum of a
perturbative and a non-perturbative component. We show in the
following chapter that the perturbative correlations,
$(\chi^{\pert})^2$, describe the well-known {\em two-gluon
  exchange}~\cite{Low:1975sv+X,Gunion:iy} to dipole-dipole scattering
and the non-perturbative correlations, $(\chi^{\nprt})^2$, the
corresponding non-perturbative two-point interactions.

As the two-component structure of
(\ref{Eq_model_purely_imaginary_T_amplitude_small_chi_limit}) reminds
of the two-pomeron model of Donnachie and
Landshoff~\cite{Donnachie:1998gm+X}, we adopt the powerlike energy
increase and ascribe a strong energy dependence to the perturbative
component $\chi^{\pert}$ and a weak one to the non-perturbative
component $\chi^{\nprt}$
\bea
        \left(\chi^{\pert}\right)^2 \quad & \rightarrow & \quad 
        \left(\chi^{\pert}(s)\right)^2 := \left(\chi^{\pert}\right)^2
        \left(\frac{s}{s_0} 
        \frac{\vec{r}_1^{\,2}\,\vec{r}_2^{\,2}}{R_0^4}\right)^{\epsilon^{\pert}}
        \nonumber \\
        \left(\chi^{\nprt}\right)^2 \quad & \rightarrow & \quad 
        \left(\chi^{\nprt}(s)\right)^2 := \left(\chi^{\nprt}\right)^2 
        \left(\frac{s}{s_0}
        \frac{\vec{r}_1^{\,2}\,\vec{r}_2^{\,2}}{R_0^4}\right)^{\epsilon^{\nprt}}
\label{Eq_energy_dependence}
\eea
with the scaling factor $s_0 R_0^4$. The powerlike energy
dependence with the exponents $0\approx \epsilon^{\nprt} <
\epsilon^{\pert} < 1$ guarantees Regge type behavior at moderately high
energies, where the small-$\chi$
limit~(\ref{Eq_model_purely_imaginary_T_amplitude_small_chi_limit}) is
appropriate. In~(\ref{Eq_energy_dependence}), the energy variable
$s$ is scaled by the factor $\vec{r}_1^{\,2}\,\vec{r}_2^{\,2}$ that
allows to rewrite the energy dependence in photon-hadron scattering in
terms of the appropriate Bjorken scaling variable $x$
\be
        s\,\vec{r}_1^{\,2} \propto \frac{s}{Q^2} = \inv{x}
        \ ,
\label{Eq_x_Bj_<->_s}
\ee
where $|\vec{r}_1|$ is the transverse extension of the $q\qbar $
dipole in the photon. A similar factor has been used before in the
dipole model of Forshaw, Kerley, and Shaw~\cite{Forshaw:1999uf} and
also in the model of Donnachie and Dosch~\cite{Donnachie:2001wt} in
order to respect the scaling properties observed in the structure
function of the proton.\footnote{In the model of Donnachie and
  Dosch~\cite{Donnachie:2001wt}, $s\,|\vec{r}_1|\,|\vec{r}_2|$ is used
  as the energy variable if both dipoles are small, which is in
  accordance with the choice of the typical BFKL energy scale but
  leads to discontinuities in the dipole-dipole cross section. In
  order to avoid such discontinuities, we use the energy
  variable~(\ref{Eq_energy_dependence}) also for the scattering of two
  small dipoles.} In the dipole-proton cross section of Golec-Biernat
and W\"usthoff~\cite{Golec-Biernat:1999js,Golec-Biernat:1999qd},
Bjorken $x$ is used directly as energy variable which is important for
the success of the model. In fact, also in our model, the
$\vec{r}_1^{\,2}\,\vec{r}_2^{\,2}$ factor improves the description of
$\gamma^{*} p$ reactions at large $Q^2$.

The powerlike Regge type energy
dependence~(\ref{Eq_energy_dependence}) can be derived in more
theoretical frameworks: A powerlike energy dependence is found for
hadronic reactions by Kopeliovich et al.~\cite{Kopeliovich:2001pc} and
for hard photon-proton reactions from the BFKL equation~\cite{BFKL}.
However, these approaches need unitarization since their powerlike
energy dependence will ultimately violate $S$-matrix unitarity at
asymptotic energies. In our model, we use the following $T$-matrix
element as the basis for chapter~\ref{$S$-Matrix Unitarity and Gluon
  Saturation} and~\ref{Sec_Comparison_Data}
\bea
        &&\!\!\!\!\!\!\!\!\!\!
        T(s,t)  
        =  2is \int \!\!d^2b_{\!\perp} 
        e^{i {\vec q}_{\!\perp} {\vec b}_{\!\perp}}
        \int \!\!dz_1 d^2r_1 \!\int \!\!dz_2 d^2r_2\,\, 
        |\psi_1(z_1,\vec{r}_1)|^2 \,\, 
        |\psi_2(z_2,\vec{r}_2)|^2       
        \nonumber \\    
        &&\!\!\!\!\!\!\!\!\!\!\!\! 
        \times \left[1 - \frac{2}{3} 
        \cos\!\left(\!\frac{1}{3}\chi^{\pert}(s)\!\right)
        \cos\!\left(\!\frac{1}{3}\chi^{\nprt}(s)\!\right)         
        - \frac{1}{3}
        \cos\!\left(\!\frac{2}{3}\chi^{\pert}(s)\!\right)
        \cos\!\left(\!\frac{2}{3}\chi^{\nprt}(s)\!\right)
        \right] 
        \ . \nonumber \\
\label{Eq_model_purely_imaginary_T_amplitude_final_result}
\eea
Here the cosine functions in combination with the powerlike energy
ansatz ensure the $S$-matrix unitarity in impact parameter space and
the Froissart bound~\cite{Froissart:1961ux+X} as shown in
chapter~\ref{$S$-Matrix Unitarity and Gluon Saturation}. Furthermore,
the multiple gluonic interactions associated with the higher order
terms in the expansion of the cosine functions provide the mechanism
which leads to saturation effects in cross sections at ultra-high
energies.

Having ascribed the energy dependence to the $\chi$-function, the
energy behavior of hadron-hadron, photon-hadron, and photon-photon
scattering results exclusively from the {\em universal } loop-loop
correlation function $S_{DD}$. In this way a unified description of
hadronic and photonic interactions is achieved.

\section{Model Parameters}
\label{Sec_Model_Parameters}

Lattice QCD simulations provide important information and constraints
on the model parameters. The fine tuning of the parameters was,
however, directly performed on the high-energy scattering data for
hadron-hadron, photon-hadron, and photon-photon reactions where an
error ($\chi^2$) minimization was not feasible because of the
non-trivial multi-dimensional integrals in the $T$-matrix
element~(\ref{Eq_model_purely_imaginary_T_amplitude_final_result}).

The parameters $a$, $\kappa$, $G_2$, $m_G$, $M^2$, $s_0R^4_0$,
$\epsilon^{\nprt}$ and $\epsilon^{\pert}$ determine the dipole-dipole
scattering and are universal for all reactions described. In addition,
there are reaction-dependent parameters associated with the wave
functions which are provided in Appendix~\ref{Sec_Wave_Functions}.

The perturbative component involves the gluon mass $m_G$ as IR
regulator (or inverse ``perturbative correlation length'') and the
parameter $M^2$ that freezes the running
coupling~(\ref{Eq_g2(z_perp)}) for large distance scales.
We adopt the parameters
\be
        m_G =  m_{\rho} = 0.77\,\GeV 
        \quad \mbox{and }\quad 
        M^2 = 1.04\,\GeV^2
        \ .
\label{Eq_PGE_scattering_fit_parameter_results}
\ee
The value of the gluon mass is important for the interplay between the
perturbative and non-perturbative contribution. Using our perturbative
component, it gives a reasonable ``perturbative glueball mass'' (GB)
of $M_{GB}^{\pert} = 2m_G = 1.54\,\GeV$.  The value of the
parameter $M^2$ in~(\ref{Eq_PGE_scattering_fit_parameter_results})
that fixes the running coupling
at large distances at $\alpha_s = 0.4$
is taken from~\cite{D'Alesio:1998sf}.

The non-perturbative component involves the correlation length $a$,
the gluon condensate $G_2$, and the parameter $\kappa$ indicating the
non-Abelian character of the correlator. With the simple exponential
correlation functions specified in Euclidean space-time
(\ref{Eq_MSV_correlation_functions}), we obtain the following values
for the parameters of the non-perturbative
correlator~(\ref{Eq_MSV_Ansatz_F})
\be
        a =  0.302\,\fm, \quad 
        \kappa = 0.74, \quad 
        G_2 = 0.074\,\GeV^4
        \ ,
\label{Eq_MSV_scattering_fit_parameter_results}
\ee
and, correspondingly, the string tension for the fundamental
representation of $SU(3)$ becomes
\be
        \sigma_3 = 
        \frac{\pi^3 \kappa\,G_2\,a^2}{18} 
        = 0.22\,\GeV^2 \equiv 1.12 \,\GeV/\fm
        \ ,
\label{Eq_sting_tension_from_exp_correlation}
\ee 
which is consistent with hadron spectroscopy~\cite{Kwong:1987mj},
Regge theory~\cite{Goddard:1973qh+X}, and lattice QCD
investigations~\cite{Bali:2000gf}. Using the above value for the
correlation length, the non-perturbative component generates a
``non-perturbative glueball mass'' of $M_{GB}^{\nprt} = 2/a =
1.31\,\GeV$ that is smaller than the ``perturbative glueball mass''
$M_{GB}^{\pert}$ and thus governs the long-range interactions as
expected.


Lattice QCD computations of the gluon field strength correlator down
to distances of $0.4\,\fm$ have obtained the following values with the
exponential correlation
function~(\ref{Eq_MSV_correlation_functions})~\cite{Meggiolaro:1999yn}:
$a = 0.219\,\fm$, $\kappa = 0.746$, $G_2 = 0.173\,\GeV^4$. This value
for $\kappa$ is in agreement with the one
in~(\ref{Eq_MSV_scattering_fit_parameter_results}), while the fit to
high-energy scattering data clearly requires a larger value for $a$
and a smaller value for $G_2$.

The energy dependence of the model is associated with the energy
exponents $\epsilon^{\nprt}$ and $\epsilon^{\pert}$, and the scaling
parameter $s_0R^4_0$ 
\be
        \epsilon^{\nprt} = 0.125, \quad 
        \epsilon^{\pert} = 0.73, \quad \mbox{and} \quad 
        s_0 R_0^4 = (\,47\,\GeV\,\fm^2\,)^2
        \ .
\label{Eq_energy_dependence_scattering_fit_parameter_results}
\ee
In comparison with the energy exponents of Donnachie and
Landshoff~\cite{Donnachie:1992ny,Donnachie:1998gm+X}, $\epsilon_{soft}
\approx 0.08$ and $\epsilon_{hard} \approx 0.4$, our exponents are
larger. However, the cosine functions in our $T$-matrix
element~(\ref{Eq_model_purely_imaginary_T_amplitude_final_result})
reduce the large exponents so that the energy dependence of the cross
sections agrees with the experimental data as illustrated in
chapter~\ref{Sec_Comparison_Data}.

We have fixed the above parameters by trying to reproduce as many
experimental observables as possible. The numerous observables can, of
course, only narrow the range of the parameters. Additional contraints
have been lattice QCD simulations and other models. Nonetheless, since
an error ($\chi^2$) minimization has not been possible, the given
parameter values cannot be mandatory.




%% file: string_decomposition.tex
%
\chapter{Anatomy of QCD Strings and Unintegrated Gluon Distributions \
}
\label{Chapt_string_decomposition}
\vspace{0.5cm}

In this chapter we investigate the QCD structure of the perturbative and
non-perturbative dipole-dipole interaction in momentum
space within our loop-loop correlation model. We reproduce the known
results for perturbative interactions between the dipoles and give new
insights into non-perturbative interactions between QCD strings which
lead to quark-confinement in dipoles.
Non-perturbative string-string interactions show a new structure
different from perturbative dipole-dipole interactions. The confining
QCD string exhibits a very nice feature: It can be exactly
represented as a collection of stringless quark-antiquark dipoles.
This outstanding result allows us to extract the microscopic structure
of unintegrated gluon distributions of hadrons and photons. Our
unintegrated gluon distribution of the proton is compared with those
obtained in other approaches. Confining QCD string manifestations are
shown explictly in dipole-hadron cross sections and unintegrated gluon
distributions. The $|\vec{k}_{\!\perp}|$-factorization, known in
perturbative physics, is found to be valid also in non-perturbative
interactions within our model.

\vspace{0.2cm}

The structural aspects exposed in this chapter have been possible
because of the simple minimal surfaces (see
Fig.~\ref{Fig_loop_loop_scattering_surfaces}) used in our model.
The previous complicated pyramid mantle choice of the 
surfaces~\cite{Dosch:1994ym,Berger:1999gu,Rueter:1996yb+X,Dosch:1997ss,Dosch:1998nw,Rueter:1998up,D'Alesio:1999sf,Dosch:2001jg}
did not allow such studies.

\vspace{0.2cm}

We consider throughout this chapter only the lowest order
contribution to dipole-dipole interactions, i.e., the small-$\chi$
limit of the $T$-matrix
element~(\ref{Eq_model_purely_imaginary_T_amplitude_small_chi_limit}).
With a phenomenological energy dependence assigned to the
$\chi$-functions as discussed in Sec.~\ref{Sec_Energy_Dependence},
this limit can be used to describe experimental observables for $x =
Q^2/s \geq 10^{-4}$. For lower Bjorken-$x$ (higher c.m. energy)
values, the exact $T$-matrix
element~(\ref{Eq_model_purely_imaginary_T_amplitude_final_result}) is
needed to ensure $S$-matrix unitarity conditions  and to observe
saturation effects in high-energy reactions as discussed in chapter~\ref{$S$-Matrix Unitarity and Gluon Saturation}
and~\ref{Sec_Comparison_Data}.

%
\section{QCD Structure of Dipole-Dipole Scattering in Momentum Space}
\label{Momentum-Space Structure of Dipole-Dipole Scattering}

The total dipole-dipole cross section in the small-${\chi}$
limit is obtained from
Eq. (\ref{Eq_model_purely_imaginary_T_amplitude_small_chi_limit}) via the
optical theorem
%
\bea
\sigma^{tot}_{\!\mbox{\tiny\it D}\mbox{\tiny\it D}}(s_0) 
            &=& \inv{s_0}\,\im\,T(s_0, t=0)
            \nonumber \\
            &=&2 \int \!\!d^2b_{\!\perp} \int \!\!dz_1d^2r_1 
            \int \!\!dz_2 d^2r_2 
            |\psi_{\!\mbox{\tiny\it D}_1}(z_1,\vec{r}_1)|^2 
            |\psi_{\!\mbox{\tiny\it D}_2}(z_2,\vec{r}_2)|^2 \nonumber \\
            && \times \frac{1}{9} 
            \left[\left(\chi^{\pert}\right)^2 +
            \left(\chi^{\nprt}_{nc} + \chi^{\nprt}_{c} \right)^2  
            \right ] \ , 
\label{sigtot_DD}
\eea  
where $\sqrt{s_0} \approx 20\,\GeV$ denotes the c.m. energy at which
our
model~(\ref{Eq_model_purely_imaginary_T_amplitude_almost_final_result})
reproduces the experimental observables. The dipoles
in Eq.~(\ref{sigtot_DD}) have fixed $z_i$ and $|\vec{r}_i|$ values
%
\be
|\psi_{\!\mbox{\tiny\it D}_i}(z_i,\vec{r}_i)|^2 =
     \inv{2\pi|\vec{r}_{\!\mbox{\tiny\it D}_i}|}\,\delta
     (|\vec{r}_i|-|\vec{r}_{\!\mbox{\tiny\it D}_i}|)\,\delta
     (z_i-z_{\!\mbox{\tiny\it D}_i}) \ 
\label{dip_wf}
\ee
but are averaged over all orientations. Being interested in the
momentum-space structure of the scattering, we
use~(\ref{Eq_transverse_Fourier_transform}) to writte the
${\chi}$-functions computed with minimal surfaces in the previous
chapter, see~(\ref{Eq_chi_MSV_confining}),
(\ref{Eq_chi_MSV_non-confining}), (\ref{Eq_chi_PGE}), in the form
\bea
\!\!\!\!\!\!
\chi^{\pert}&=&
         4\pi \int \! \frac{d^2k_{\!\perp}}{(2\pi)^2} \ 
         \alpha_s(k_{\!\perp}^2)\,i\tilde{D}_{\pert}^{\prime \,(2)}(k_{\!\perp}^2)\,
          \left [ e^{i\vec{k}_{\!\perp}( 
          \vec{r}_{1q}-\vec{r}_{2q})} + e^{i\vec{k}_{\!\perp}( 
          \vec{r}_{1\bar{q}}-\vec{r}_{2\bar{q}})} \right . 
          \nonumber \\
          &&
          \hphantom{\chi^{\pert} = 4\pi \int \! \frac{d^2k_{\!\perp}}{(2\pi)^2} \ 
          \alpha_s(k_{\!\perp}^2)\,iD} 
          \left .
          - e^{i\vec{k}_{\!\perp}( 
          \vec{r}_{1q}-\vec{r}_{2\bar{q}})} - e^{i\vec{k}_{\!\perp}( 
          \vec{r}_{1\bar{q}}-\vec{r}_{2q})} \right ]  
\label{chi_p}  \\ \vspace{2cm} \nonumber \\
\!\!\!\!\!\!
\chi_{nc}^{\nprt}&=&
         \frac{\pi^2\,G_2\,(1-\kappa)}{24}  
         \int \! \frac{d^2k_{\!\perp}}{(2\pi)^2} \ 
         i\tilde{D}^{\prime \,(2)}_{1}(k_{\!\perp}^2)\,
         \left [ e^{i\vec{k}_{\!\perp} (\vec{r}_{1q}-\vec{r}_{2q})} + 
         e^{i\vec{k}_{\!\perp} (\vec{r}_{1\bar{q}}-
         \vec{r}_{2\bar{q}})} \right . 
         \nonumber \\
         &&\!\!\!\!\!\!\!\!\!
         \hphantom{\chi_{nc}^{\nprt} = \frac{\pi^2\,G_2\,(1-\kappa)}{24} 
          \int \! \frac{d^2k_{\!\perp}}{(2\pi)^2} 
          i\tilde{D}D}
         \left .
         - e^{i\vec{k}_{\!\perp} ( 
         \vec{r}_{1q}-\vec{r}_{2\bar{q}})} - e^{i \vec{k}_{\!\perp} ( 
         \vec{r}_{1\bar{q}}-\vec{r}_{2q})} \right ]          
\label{chi_nc} \\ \vspace{2cm} \nonumber \\
\!\!\!\!\!\!
\chi_{c}^{\nprt}&=& 
       \frac{\pi^2\,G_2\,\kappa}{24}
       \left(\vec{r}_1\cdot\vec{r}_2\right) \, 
       \int_0^1 \! dv_1 \int_0^1 \! dv_2 \, \int \!
       \frac{d^2k_{\!\perp}}
       {(2\pi)^2} \ 
       i\tilde{D}^{(2)}(k_{\!\perp}^2)\, e^{i\vec{k}_{\!\perp} \left
       (\vec{r}_{1q}+ v_1\vec{r}_1-\vec{r}_{2q}-v_2\vec{r}_2\right ) }
       \  \nonumber \\
\label{chi_c}
\eea
with the Minkowskian correlation functions in transverse space
\bea
        i\tilde{D}_{\pert}^{\prime \,(2)}(k_{\!\perp}^2) 
         &=& \frac{1}{k^2_{\!\perp} + m_G^2} \ , 
\label{Eq_massive_D_pge_prime_trans} \\ \vspace{1cm} \nonumber \\
        i\tilde{D}^{\prime \,(2)}_1(k^2_{\!\perp}) 
        &=&  \frac{30\,\pi^2}{a(k^2_{\!\perp} +
        a^{-2})^\frac{7}{2}} \ , 
\label{Eq_D_(prime)(k^2)_for_exp_correlation_perp}\\
        i\tilde{D}^{(2)}(k^2_{\!\perp}) 
        &=&  \frac{12\,\pi^2}{a\,(k^2_{\!\perp} +
        a^{-2})^\frac{5}{2}} \ .
\label{Eq_D(k^2)_for_exp_correlation_perp} 
\eea
The perturbative correlation function, $\tilde{D}_{\pert}^{\prime
  \,(2)}(k_{\!\perp}^2)$, results from Eq.~(\ref{Eq_massive_D_pge_prime})
for $k^0 = k^3 = 0$, i.e., $k^2=-{\vec k}^2_{\!\perp}$, which is
enforced in the computation of $\chi$ - functions. The derivation of
the non-perturbative correlation functions, $\tilde{D}^{\prime
  \,(2)}_1(k^2_{\!\perp})$ and $\tilde{D}^{(2)}(k^2_{\!\perp}) $,
from Eq.~(\ref{Eq_MSV_correlation_functions}) is shown in Appendix
\ref{Sec_Correlation_Functions}.  


The component $\chi^{\pert}$ describes the perturbative interaction of
the quark and antiquark of one dipole with the quark and antiquark of
the other dipole as evident from the $\vec{r}_{iq}$ and
$\vec{r}_{i{\bar q}}$ dependence of~(\ref{chi_p}) and
Fig.~\ref{Fig_loop_loop_scattering_surfaces}b. The component
$\chi_{nc}^{\nprt}$ has the same structure as $\chi^{\pert}$ and gives
the non-perturbative interaction between the quarks and antiquarks of
the dipoles. With the term quark used generically for quarks and
antiquarks in the following, we refer to $\chi^{\pert}$ and
$\chi_{nc}^{\nprt}$ as {\em quark-quark interactions}.


The component $\chi_{c}^{\nprt}$ given in Eq.~(\ref{chi_c}) shows a
different structure. Here the integrations over $v_1$ and $v_2$ sum
non-perturbative interactions between the gluon field strengths
connecting the quark and antiquark in each of the two dipoles. These
connections are manifestations of the strings that confine the
corresponding quark and antiquark in the dipole and are visualized in
Fig.~\ref{Fig_loop_loop_scattering_surfaces}b.  Indeed, the
non-perturbative component $\chi_{c}^{\nprt}$ shows a flux tube (string)
between a static quark-antiquark pair in Euclidean
space-time~\cite{DelDebbio:1994zn,Rueter:1995cn,Shoshi:2002rd}
which confines the quark and antiquark in the dipole as explained in
chapter~\ref{Sec_The_Model}.  Therefore, we understand the confining
component $\chi_{c}^{\nprt}$ as a {\em string-string interaction}.


The mixed contribution $\chi_{nc}^{\nprt}\,\chi_{c}^{\nprt}$ that
occurs in the dipole-dipole cross section~(\ref{sigtot_DD}) gives the
non-perturbative interaction of the quark and antiquark of one dipole
with the string of the other dipole, i.e., it represents the {\em
  quark-string interaction}.


To be able to examine the $|\vec{k}_{\!\perp}|$ - structure of
dipole-dipole interactions, we carry out all integrations
in Eq.~(\ref{sigtot_DD}) except, of course, the one over transverse
momentum $|\vec{k}_{\!\perp}|$ that enters through the
${\chi}$-functions~(\ref{chi_p})--(\ref{chi_c}). The resulting
perturbative ($\pert$) and non-perturbative ($\nprt$) integrand of the
total dipole-dipole cross section 
\bea
\sigma^{tot}_{\!\mbox{\tiny\it D}\mbox{\tiny\it D}}(s_0) 
     &=& 
     \int d|\vec{k}_{\!\perp}|\, \left[ 
     I^{\pert}(s_0, |\vec{k}_{\!\perp}|) +
     I^{\nprt}(s_0, |\vec{k}_{\!\perp}|)\right]
   \\
     &=& 
     \int d|\vec{k}_{\!\perp}|\, \left[ 
     I^{\pert}(s_0, |\vec{k}_{\!\perp}|) +
     I^{\nprt}_{qq}(s_0, |\vec{k}_{\!\perp}|) +
     I^{\nprt}_{ss}(s_0, |\vec{k}_{\!\perp}|)
    \right] \ ,
\label{sigt}
\eea
show the following momentum-space structure:
\bea
I^{\pert}(s_0, |\vec{k}_{\!\perp}|)&=& 
      \frac{2}{9}\frac{1}{2\pi} |\vec{k}_{\!\perp}| 
      \left (4\pi \alphaS(k^2_{\!\perp})\right)^2 
      \left [i\tilde{D}_{\pert}^{\prime \,(2)}(k^2_{\!\perp})\right ]^2
\label{Ip} \\
      &&\times
      \left [2\ \langle \psi_{\!\mbox{\tiny\it D}_1}|1-e^{i\vec{k}_{\!\perp}\vec{r}_1}
      |\psi_{\!\mbox{\tiny\it D}_1} \rangle \ 
      2\ \langle \psi_{\!\mbox{\tiny\it D}_2}|1-e^{i\vec{k}_{\!\perp}\vec{r}_2}
      |\psi_{\!\mbox{\tiny\it D}_2} \rangle \right ] 
\nonumber\\ \vspace{0.3cm} \nonumber \\
I^{\nprt}_{qq}(s_0, |\vec{k}_{\!\perp}|)
      &=& 
      \frac{2}{9}\frac{1}{2\pi} |\vec{k}_{\!\perp}| 
      \left(\frac{\pi^2 G_2}{24}\right)^2
      \left [ 
        (1-\kappa)\,i \tilde{D}^{\prime \,(2)}_1(k^2_{\!\perp})\!
        + \frac{\kappa}{k_{\!\perp}^2} \,i\tilde{D}^{(2)}(k^2_{\!\perp})
      \right]^2
\label{Eq_I_NP_qq}     
      \\
      &&\times
      \left [2\ \langle \psi_{\!\mbox{\tiny\it D}_1}
      |1-e^{i\vec{k}_{\!\perp}\vec{r}_1} 
      |\psi_{\!\mbox{\tiny\it D}_1}\rangle \  
      2\ \langle \psi_{\!\mbox{\tiny\it D}_2}
      |1-e^{i\vec{k}_{\!\perp}\vec{r}_2} 
      |\psi_{\!\mbox{\tiny\it D}_2}\rangle\right]
      \nonumber\\ \vspace{0.3cm} \nonumber \\
I^{\nprt}_{ss}(s_0, |\vec{k}_{\!\perp}|)
      &=& 
      \frac{2}{9}\frac{1}{2\pi} |\vec{k}_{\!\perp}| 
      \left(\frac{\pi^2 G_2}{24}\right)^2
      \left[\frac{\kappa}{k_{\!\perp}^2} \,i\tilde{D}^{(2)}(k^2_{\!\perp})\right]^2 
\label{Eq_I_NP_ss}
      \\
      &&\times
      \left [2\ \langle \psi_{\!\mbox{\tiny\it D}_1}|\tan^2\!\!\phi_1 
      (1-e^{i\vec{k}_{\!\perp}\vec{r}_1})
      |\psi_{\!\mbox{\tiny\it D}_1}\rangle \  
      2\ \langle \psi_{\!\mbox{\tiny\it D}_2}|\tan^2\!\!\phi_2 
      (1-e^{i\vec{k}_{\!\perp}\vec{r}_2}) 
      |\psi_{\!\mbox{\tiny\it D}_2}\rangle \right]
      \nonumber
\eea
where the brackets $\langle \psi_{\!\mbox{\tiny\it D}_i}| ...
|\psi_{\!\mbox{\tiny\it D}_i}\rangle$ denote the averages
\be
\langle \psi_{\!\mbox{\tiny\it D}_i}|A_i|\psi_{\!\mbox{\tiny\it D}_i}\rangle = 
      \int \!dz_id^2r_i |\psi_{\!\mbox{\tiny\it D}_i}(z_i, \vec{r}_i)|^2 A_i
\label{average}
\ee
and the dipole orientation $\phi_i$ is defined as the angle between
transverse momentum $\vec{k}_{\!\perp}$ and dipole vector $\vec{r}_i$.
With Eq.~(\ref{dip_wf}) the integration over the dipole orientations
$\phi_i$ leads, respectively, to the Bessel function
$J_0(|\vec{k}_{\!\perp}||\vec{r}_{\!\mbox{\tiny\it D}_i}|)$ and the
generalized hypergeometric function\footnote{A review of generalized
  hypergeometric functions can be found in~\cite{MOS}. In the computer
  program Mathematica~\cite{Math} ``HypergeometricPFQ$[\{ {-1/2}\},\{
  {1/2,1}\}, -k^2_{\!\perp} r^2_{\!\mbox{\tiny\it D}}/4]$'' denotes
  this function.} $_1F_2 (-1/2;1/2,1;-k_{\!\perp}^2
r_{\!\mbox{\tiny\it D}_i}^2/4)$
\bea
\!\!\!\!\!\!\!\!\!\!\!
    \langle \psi_{\!\mbox{\tiny\it
    D}_i}|1-e^{i\vec{k}_{\!\perp}\vec{r}_i}|\psi_{\!\mbox{\tiny\it
    D}_i}\rangle 
&=& 
       \frac{1}{2\pi} \int_0^{2\pi} d\phi_{i}
       (1-e^{i\vec{k}_{\!\perp}
       \vec{r}_{\!\mbox{\tiny\it D}_i}})
\nonumber \\
&=& 
        1- J_0(|\vec{k}_{\!\perp}||\vec{r}_{\!\mbox{\tiny\it D}_i}|)
          \ ,
\label{average_dip}\\
\vspace{1cm}
\nonumber\\
\!\!\!\!\!\!\!\!\!\!\!
     \langle \psi_{\!\mbox{\tiny\it D}_i}|\tan^2\!\!\phi_i 
     (1-e^{i\vec{k}_{\!\perp}\vec{r}_i}) 
     |\psi_{\!\mbox{\tiny\it D}_i}\rangle &=& 
            \frac{1}{2\pi} \int_0^{2\pi} d\phi_{i}
            \tan^2\!\!\phi_{i} (1-e^{i\vec{k}_{\!\perp}
            \vec{r}_{\!\mbox{\tiny\it D}_i}}) \nonumber \\
            &=& -1+\ \!\! _1F_2 (-\frac{1}{2};\frac{1}{2},1; 
            \frac{-k_{\!\perp}^2 r_{\!\mbox{\tiny\it D}_i}^2}{4}) \  
        \ .
\label{average_string}
\eea
The important implications will be discussed in the following section.

\begin{figure}[]
\setlength{\unitlength}{1.cm}
\begin{center}
\epsfig{file=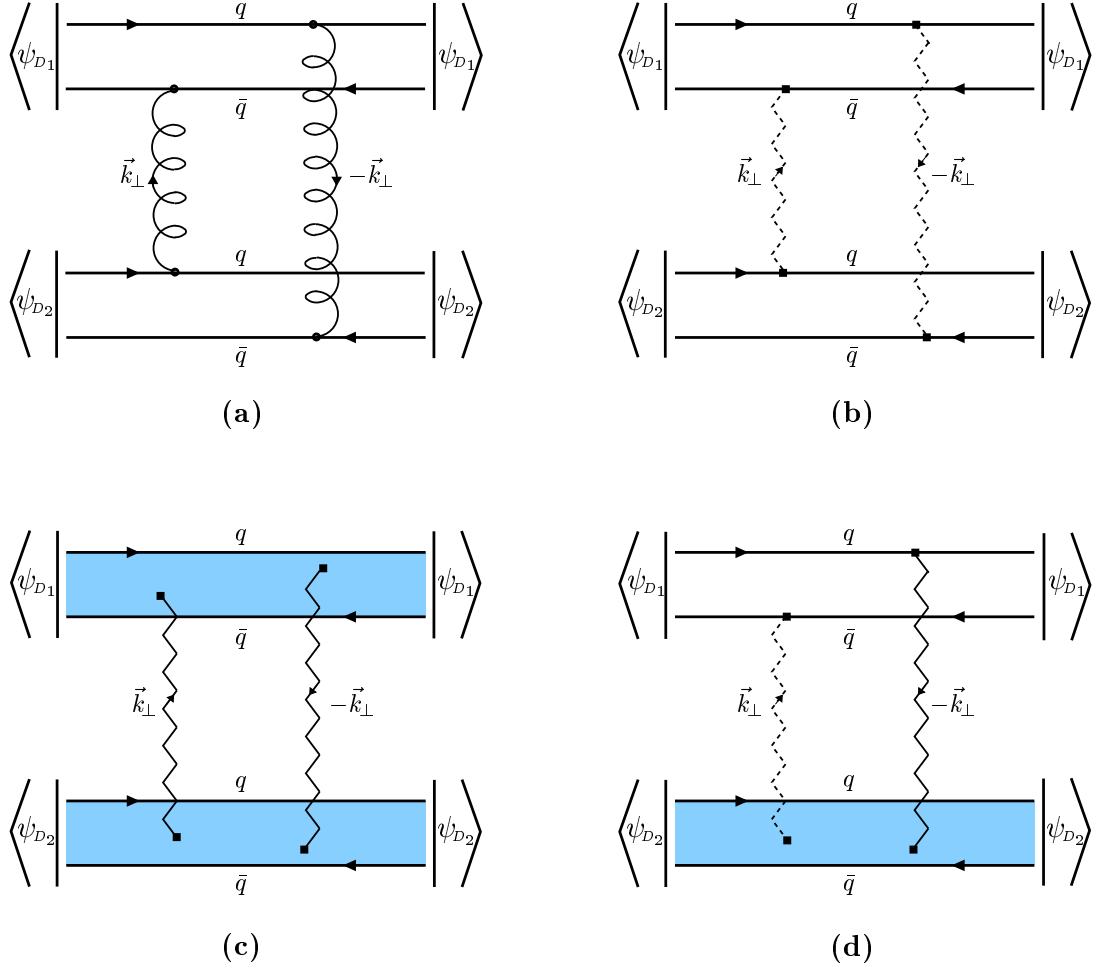, width=14.5cm}
\end{center}
\caption{\small Perturbative and non-perturbative contributions to dipole-dipole scattering:
  ${\bf (a)}$ perturbative quark-quark interaction and
  non-perturbative ${\bf (b)}$ quark-quark, ${\bf (c)}$ string-string,
  and ${\bf (d)}$ quark-string interactions. The term quark is used
  genuinely for quarks and antiquarks. Only one diagram is shown for
  each type of interaction. Paired horizontal lines represent
  quark-antiquark dipoles, surrounding brackets indicate the averages
  over the dipole orientations~(\ref{average}), shaded areas visualize
  strings, curly lines denote exchanged perturbative gluons, and
  dashed and solid zig-zag lines symbolize, respectively, the
  non-perturbative non-confining and confining correlation functions.}
\label{Fig_dipole_dipole_interactions}
\end{figure}

The integrand $I^{\pert}$ given in Eq.~(\ref{Ip}) describes the known
perturbative two-gluon\footnote{The exact $T$-matrix
  element~(\ref{Eq_model_purely_imaginary_T_amplitude_final_result}), 
  used to describe scattering processes at ultra-high energies in
  chapter~\ref{Sec_Comparison_Data}, goes beyond two-gluon exchange due
  to the higher orders in the cosine expansion.}
exchange~\cite{Low:1975sv+X,Gunion:iy} between the quarks and
antiquarks of the two dipoles. The ingredients of $I^{\pert}$ are
visualized for one combination of gluon exchanges in
Fig.~\ref{Fig_dipole_dipole_interactions}a: the paired horizontal
lines represent the {\em dipole factors}
$(1-e^{i\vec{k}_{\!\perp}\vec{r_i}})$ that describe the phase
difference between the quark and antiquark at separate transverse
positions, the surrounding brackets indicate the average over the
dipole orientations~(\ref{average}), the two curly lines illustrate
the gluon propagator squared $[i\tilde{D}_{\pert}^{\prime \,(2)}]^2$,
and the four vertices (dots) correspond to the strong coupling to the
fourth power $g^4=(4\pi\alphaS)^2$.


The integrand $I^{\nprt} = I^{\nprt}_{qq} + I^{\nprt}_{ss}$ given in
Eqs.~(\ref{Eq_I_NP_qq}) and~(\ref{Eq_I_NP_ss}) describes the
non-perturbative interactions: the quark-quark, string-string, and
quark-string interactions identified by the appropriate correlation
functions $[i\tilde{D}^{\prime \,(2)}_1]^2$, $[i\tilde{D}^{(2)}]^2$,
and $[i\tilde{D}^{\prime \,(2)}_1\,i\tilde{D}^{(2)}]$. These
interactions are illustrated in
Figs.~\ref{Fig_dipole_dipole_interactions}b,
\ref{Fig_dipole_dipole_interactions}c and
\ref{Fig_dipole_dipole_interactions}d, respectively. Analogous to the
perturbative interaction in
Fig.~\ref{Fig_dipole_dipole_interactions}a, the dashed and solid
zig-zag lines represent, respectively, the non-confining
$(1-\kappa)\,i\tilde{D}^{\prime \,(2)}_1$ and the confining
$(\kappa/k^2_{\!\perp})\,i\tilde{D}^{(2)}$ non-perturbative
correlations, the shaded areas symbolize the strings, and the four
vertices (squares) in each figure indicate the ''non-perturbative
coupling'' to the fourth power $g^4_{\nprt} :=
\left(\pi^2G_2/24\right)^2$.
%


The integrand $I^{\nprt}_{qq}$ describes the non-perturbative {\em
  interactions between the quarks and antiquarks} of the two dipoles
and exhibits the same dipole factors
$(1-e^{i\vec{k}_{\!\perp}\vec{r_i}})$ that appear in the perturbative
integrand~(\ref{Ip}). $I^{\nprt}_{qq}$ contains three components: the
non-confining component, the confining component, and their
interference term. While the non-confining component visualized in
Fig.~\ref{Fig_dipole_dipole_interactions}b has the same structure as the
perturbative contribution, see also Eqs.~(\ref{chi_p}) and~(\ref{chi_nc}),
the confining component shown in its more general form in
Fig.~\ref{Fig_dipole_dipole_interactions}c comes from the interaction between
the quarks and antiquarks at the endpoints of the strings, which will
be further discussed below Eq.~(\ref{ang_dep}). The interference term
describes the quark-string interaction as illustrated in
Fig.~\ref{Fig_dipole_dipole_interactions}d. Note that it reduces entirely to an
interaction between the quarks and antiquarks of the dipoles with the
additional denominator $1/k^2_{\!\perp}$ generated by the integrations
over the variables $v_1$ and $v_2$ in the confining component
$\chi^{\nprt}_c$ given in Eq.~(\ref{chi_c}).


The integrand $I^{\nprt}_{ss}$ describes the non-perturbative {\em
  string-string interaction}\,  shown in
Fig.~\ref{Fig_dipole_dipole_interactions}c.  The new angular dependencies in
the string-string interaction, the modified dipole factors
$\tan^2\!\!\phi_{i} (1-e^{i\vec{k}_{\!\perp} \vec{r}_{\!\mbox{\tiny\it
      D}_i}})$ in Eq.~(\ref{Eq_I_NP_ss}), are obtained as follows. The
integrations over $v_1$, $v_2$, $v_1^{\prime}$, and $v_2^{\prime}$ in
$(\chi^{\nprt}_c)^2$ produce the dipole factors
$(1-e^{i\vec{k}_{\!\perp}\vec{r_i}})$ and the denominator
$1/[(\vec{k}_{\!\perp}\vec{r}_1)^2 (\vec{k}_{\!\perp}\vec{r}_2)^2]$.
This denominator multiplied with the additional factor
$(\vec{r_1}\vec{r_2})^2$ from $(\chi_c^{\nprt})^2$, see Eq.~(\ref{chi_c}),
gives the total angular dependence
\bea
\frac{(\vec{r}_1\vec{r}_2)^2}
{(\vec{k}_{\!\perp}\vec{r}_1)^2
(\vec{k}_{\!\perp}\vec{r}_2)^2} 
&=& 
   \frac{r^2_1r^2_2\cos^2(\phi_1-\phi_2)}
   {\left(k^2_{\!\perp}r^2_1\cos^2\!\phi_1\right)
   \left(k^2_{\!\perp}r^2_2\cos^2\!\phi_2\right)} \nonumber\\
&=& 
   \frac{\left(\cos\phi_1\cos\phi_2+\sin\phi_1\sin\phi_2\right)^2}
   {k^4_{\!\perp}\cos^2\!\phi_1\cos^2\!\phi_2} \nonumber\\
&=&
   \frac{1}{k^4_{\!\perp}} \left(1+2\tan\!\phi_1\tan\!\phi_2+
   \tan^2\!\phi_1\tan^2\!\phi_2 \right) \ .
\label{ang_dep}
\eea
The first term in Eq.~(\ref{ang_dep}) explains the interaction between the
endpoints of the strings with the additional factor $1/k^4_{\!\perp}$
in the first contribution $I^{\nprt}_{qq}$ already mentioned above.
The product of the second term in Eq.~(\ref{ang_dep}),
$2\tan\!\phi_1\tan\!\phi_2$, with the dipole factors vanishes after
the integration over the dipole orientations $\phi_1$ and $\phi_2$.
The third term in Eq.~(\ref{ang_dep}), $\tan^2\!\phi_1\tan^2\!\phi_2$,
weights the different orientations of the strings and is
characteristic for the string-string interaction~(\ref{Eq_I_NP_ss}) in
our model. Due to this factor, the string-string interaction differs
significantly from the interaction between the quarks and antiquarks
of the dipoles known from perturbative two-gluon exchange~(\ref{Ip}).

\begin{figure}[htp]
\setlength{\unitlength}{1.cm}
\begin{center}
\epsfig{file=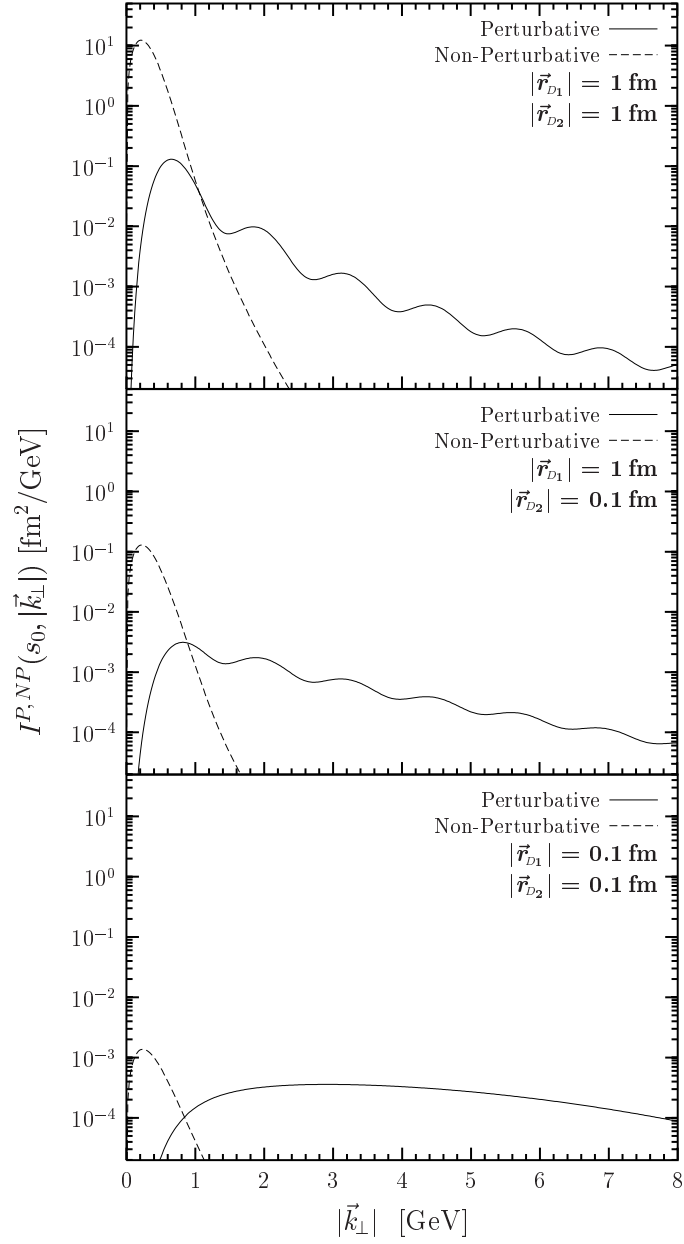, width=9.cm}
\end{center}
\caption{\small The perturbative 
  integrand $I^{\pert}$ (solid line) and the non-perturbative
  integrand $I^{\nprt}$ (dashed line) of the total dipole-dipole cross
  section~(\ref{sigt}) as a function of transverse momentum
  $|\vec{k}_{\!\perp}|$ for various dipole sizes
  $|\vec{r}_{\!\mbox{\tiny\it D}_1}|$ and $|\vec{r}_{\!\mbox{\tiny\it
      D}_2}|$.  The integrands are shown for the parameters given in
  Sec.~\ref{Sec_Model_Parameters} that allow a good description of the
  experimental data of hadron and photon reactions at $\sqrt{s_0}
  \approx 20\,\GeV$ with the wave functions given in
  Appendix~\ref{Sec_Wave_Functions} and the exact $T$-matrix
  element~(\ref{Eq_model_purely_imaginary_T_amplitude_almost_final_result}).
  The oscillations of the perturbative integrand $I^{\pert}$ originate
  from the Bessel function
  $J_0(|\vec{k}_{\!\perp}||\vec{r}_{\!\mbox{\tiny\it D}_i}|)$.  }
\label{integrand}
\end{figure}
In Fig.~\ref{integrand} we show the perturbative integrand
$I^{\pert}$ (solid line) and the non-perturbative integrand
$I^{\nprt}$ (dashed line) of the total dipole-dipole cross
section (\ref{sigt}) as a function of transverse momentum
$|\vec{k}_{\!\perp}|$ for various dipole sizes. The integrands have
been calculated with the model parameters given in
Sec.~\ref{Sec_Model_Parameters} that allow a good
description of the experimental data of hadron and photon reactions at
the c.m.\ energy of $\sqrt{s_0} \approx 20\,\GeV$ with the wave
functions given in Appendix~\ref{Sec_Wave_Functions} and the exact
$T$-matrix
element~(\ref{Eq_model_purely_imaginary_T_amplitude_almost_final_result}).
Evidently, the non-perturbative integrand $I^{\nprt}$ governs the low
momenta and the perturbative integrand $I^{\pert}$ the high momenta
exchanged in the dipole-dipole scattering. This behavior is a direct
consequence of the correlation functions: The non-perturbative
correlation functions~(\ref{Eq_D_(prime)(k^2)_for_exp_correlation_perp})
and~(\ref{Eq_D(k^2)_for_exp_correlation_perp}) favor low momenta and
suppress high momenta in comparison to the perturbative correlation
function~(\ref{Eq_massive_D_pge_prime}). For fixed dipole sizes
$|\vec{r}_{\!\mbox{\tiny\it D}_1}|$ and $|\vec{r}_{\!\mbox{\tiny\it
    D}_2}|$ the absolute values of the perturbative and
non-perturbative integrand are controlled, respectively, by the
parameters ($m_g$, $M$) and ($G_2$, $a$, $\kappa$) given in
Sec.~\ref{Sec_Model_Parameters}. With decreasing dipole
sizes the ratio $I^{\pert}/I^{\nprt}$ increases: for large dipole
sizes $|\vec{r}_{\!\mbox{\tiny\it D}_1}| = |\vec{r}_{\!\mbox{\tiny\it
    D}_2}| = 1\, \fm$, the non-perturbative contribution gives the
main contribution to the total dipole-dipole cross section; for small
dipole sizes $|\vec{r}_{\!\mbox{\tiny\it D}_1}| =
|\vec{r}_{\!\mbox{\tiny\it D}_2}| = 0.1\, \fm$, the perturbative
contribution dominates.

As can be seen from the analytic results in
Eqs.~(\ref{Ip})--(\ref{average_string}), the total dipole-dipole cross
section~(\ref{sigt}) or the forward scattering amplitude $T(s_0,t=0)$
does not depend on the longitudinal quark momentum fractions
$z_{\!\mbox{\tiny\it D}_i}$ of the dipoles. For $t=0$ the parameter
$z_{\!\mbox{\tiny\it D}_i}$ disappears upon the integration over $z_i$
since only $|\psi_{\!\mbox{\tiny\it D}_i}(z_i,\vec{r}_i)|^2$ given in
Eq.~(\ref{dip_wf}) depends on $z_i$.

The structure presented for dipole-dipole scattering remains in
reactions involving hadrons and photons: the hadronic and photonic
total cross sections are obtained from the total dipole-dipole cross
section~(\ref{sigtot_DD}) by replacing $|\psi_{\!\mbox{\tiny\it
    D}_i}(z_i,\vec{r}_i)|^2$ given in Eq.~(\ref{dip_wf}) with the hadron
and photon wave functions given in Appendix~\ref{Sec_Wave_Functions}.
As the total dipole-dipole cross section, the total hadronic and
photonic cross sections are independent of the parameters which
control the $z_i$\,-\,distribution in the wave functions due to the
normalization of the $z_i$\,-\,distributions.  The independence of the
total hadronic cross section on the widths $\Delta z_h$ can be seen
immediately with the Gaussian hadron wave
functions~(\ref{Eq_hadron_wave_function})
\bea
\!\!\!\!\!\!\!\!\!\!\!
\langle \psi_{h}|1-e^{i\vec{k}_{\!\perp}\vec{r}_i}|\psi_{h}\rangle &=& 
      1-e^{-\frac{1}{2} k_{\!\perp}^2 S_h^2} \ ,
\label{averages_explicit_qq}\\
\!\!\!\!\!\!\!\!\!\!\!
      \langle \psi_{h}|\tan^2\!\!\phi_i (1-e^{i\vec{k}_{\!\perp}\vec{r}_i}) 
      |\psi_{h}\rangle &=& 
      -1+e^{-\frac{1}{2} k_{\!\perp}^2 S_h^2}
      + \sqrt{\frac{\pi}{2}} \,|\vec{k}_{\!\perp}| S_h\, 
      \mbox{Erf}\left (\frac{|\vec{k}_{\!\perp}| S_h}{\sqrt{2}}\right)
\label{averages_explicit_ss}
\eea
with the error function $\mbox{Erf}(z)=\sqrt{2/\pi}\int_0^z dt
\exp(-t^2)$. These analytical results confirm the $\Delta z_h$ and
$z_{\!\mbox{\tiny\it D}}$\,-\,independence of the total dipole-proton
cross section assumed in phenomenological
models~\cite{Forshaw:1999uf,Golec-Biernat:1999js,Golec-Biernat:1999qd}.  However, the
non-forward hadronic scattering amplitude $T(s_0, t\neq 0)$ depends on
the parameter $\Delta z_h$ as shown explicitly in
Appendix~\ref{App_T_tneq0}.  Thus, the differential elastic cross
section $d\sigma^{el}/dt(s,t)$~(\ref{Eq_dsigma_el_dt}) and its
logarithmic slope $B(s,t)$~(\ref{Eq_elastic_local_slope}) are $\Delta
z_h$\,-\,dependent. In fact, this $\Delta z_h$\,-\,dependence is
essential for the agreement with experimental
data~\cite{Shoshi:2002in} as discussed in chapter~\ref{Sec_Comparison_Data}.

The $|\vec{k}_{\!\perp}|$\,-\,dependence of the perturbative and
non-perturbative integrand for hadron-hadron, hadron-photon, and
photon-photon cross sections at high photon virtualities is similar to
the one of the perturbative and non-perturbative integrand of the
dipole-dipole cross section shown in Fig.~\ref{integrand} for
($|\vec{r}_{\!\mbox{\tiny\it D}_1}| = |\vec{r}_{\!\mbox{\tiny\it
    D}_2}| = 1\, \fm$), ($|\vec{r}_{\!\mbox{\tiny\it D}_1}| = 1\,
\fm$, $|\vec{r}_{\!\mbox{\tiny\it D}_2}| = 0.1\, \fm$), and
($|\vec{r}_{\!\mbox{\tiny\it D}_1}| = |\vec{r}_{\!\mbox{\tiny\it
    D}_2}| = 0.1\, \fm$), respectively. Of course, the absolute values
differ and the oscillations of the perturbative integrand caused by
the Bessel functions disappear.
\vspace{5cm}
\section[Decomposition of the QCD String into Dipoles \\ and 
Unintegrated Gluon Distributions]{\letterspace to
  1.06\naturalwidth{Decomposition of the QCD String into} \\ 
  \letterspace to 0.96\naturalwidth{Dipoles and Unintegrated Gluon
    Distributions}}
\label{Sec_The_Decomposition_of_the_String_into_Dipoles_and_the_Unintegrated_Gluon_Distribution}


The {\em unintegrated gluon distribution} of hadrons\footnote{The word
  hadron and the subscript $h$ is used genuinely for hadrons and
  photons in this section.} ${\cal F}_{h}(x,k_{\!\perp}^2)$ is a
basic, universal quantity convenient for the computation of many
scattering observables at small $x$. It is the central object in the
BFKL~\cite{Kuraev:fs+X} and CCFM~\cite{Ciafaloni:1987ur+X} evolution
equations. Upon integration over the transverse gluon momentum
$|\vec{k}_{\!\perp}|$ it leads to the conventional gluon distribution
$xG_{h}(x,Q^2)$ used in the DGLAP evolution
equation~\cite{Gribov:ri+X}. The unintegrated gluon distribution is
crucial to describe processes in which transverse momenta are
explicitly exposed such as dijet~\cite{Nikolaev:1994cd+X} or vector
meson~\cite{Nemchik:1997xb} production at HERA. Its explicit
$|\vec{k}_{\!\perp}|$ dependence is particularly suited to study the
interplay between soft and hard physics. In this section an exact
representation of the string as a collection of stringless
quark-antiquark dipoles is presented that allows us to extract the
perturbative and non-perturbative contributions to ${\cal
  F}_{h}(x,k_{\!\perp}^2)$ from our total dipole-hadron cross section
via $|\vec{k}_{\!\perp}|$\,-\,factorization.


We calculate the unintegrated gluon distribution ${\cal F}_{h}(x,k_{\!\perp}^2)$ with the $T$-matrix element in the limit of
small
${\chi}$-functions~(\ref{Eq_model_purely_imaginary_T_amplitude_small_chi_limit}).
Following Sec.~\ref{Sec_Energy_Dependence}, we give a strong energy
dependence to the perturbative contribution ${\chi^{\pert}}$ and a
weak one to the non-perturbative contribution ${\chi^{\nprt}}$, see
also Eqs.~(\ref{Eq_energy_dependence}) and~(\ref{Eq_x_Bj_<->_s}),
\bea 
        ({\chi^{\pert}})^2 & \to & 
        ({\chi^{\pert}})^2 \left( \frac{x_0}{x}\right)^{\epsilon^{\pert}}
\nonumber \\
        ({\chi^{\nprt}})^2 & \to & ({\chi^{\nprt}})^2
        \left(\frac{x_0}{x}\right)^{\epsilon^{\nprt}} 
\label{x_dep}
\eea
where the values of the exponents ${\epsilon^{\pert}}$ and
${\epsilon^{\nprt}}$ are given in Sec.~\ref{Sec_Model_Parameters} and
$x_0 = 2.4\times10^{-3}$ is adjusted to reproduce at $Q^2=1\,\GeV^2$
the integrated gluon distribution of the proton $xG_{\!p}(x,Q^2)$
extracted from the HERA data~\cite{Abramowicz:1998ii+X}.  The
small-${\chi}$
limit~(\ref{Eq_model_purely_imaginary_T_amplitude_small_chi_limit})
considered here is applicable only for $x = Q^2/s \ge 10^{-4}$. For $x
\le 10^{-4}$, the full $T$-matrix
element~(\ref{Eq_model_purely_imaginary_T_amplitude_almost_final_result})
has to be used which satisfies the $S$-matrix unitarity and leads to a
successful description of many reactions as shown in
chapter~\ref{$S$-Matrix Unitarity and Gluon Saturation}
and~\ref{Sec_Comparison_Data}. The structure of the interactions
worked out in this section, however, is independent of the
phenomenological energy dependence considered.


The total dipole-hadron cross section
$\sigma_{\!Dh}(x,|\vec{r}_{\!\mbox{\tiny\it D}}|)$ is obtained from
the total dipole-dipole cross section~(\ref{sigtot_DD}) by replacing
$|\psi_{\!\mbox{\tiny\it D}_2}(z_2,\vec{r}_2)|^2$ with a squared
hadron wave function $|\psi_{\!\mbox{\tiny\it h}}(z_2,\vec{r}_2)|^2$.
Accordingly, the $x$-dependent total dipole-hadron cross section reads
\bea
&&\!\!\!\!\!\!\!\!\!\!
\sigma_{\!Dh}(x, |\vec{r}_{\!\mbox{\tiny\it D}}|) = 
    \frac{8}{9}\frac{1}{4\pi} \int dk_{\!\perp}^2 
\label{sigdip}\\
    &&\!\!\!\!\!\!\!\!\!\!\times
    \left [
    \left (4\pi \alphaS(k^2_{\!\perp})\right)^2
    \left \{ \left [i\tilde{D}_{\pert}^{\prime \,(2)}(k^2_{\!\perp})\right ]^2
    \left(1-J_0(|\vec{k}_{\!\perp}||\vec{r}_{\!\mbox{\tiny\it D}}|)
    \right) 
    \langle \psi_{h}|1-e^{i\vec{k}_{\!\perp}\vec{r}_2}|\psi_{h}\rangle
    \right \} \left ( \frac{x_0}{x} \right )^{\!\!\epsilon^{\pert}} \right .
    \nonumber \\
    &&\!\!\!\!\!\!\!\!\!\!\!
    + \!\left(\!\frac{\pi^2G_2}{24}\!\right)^{\!\!2}\!
    \left \{ \!\!
    \left[\frac{\kappa}{k_{\!\perp}^2}\, 
    i\tilde{D}^{(2)}(k_{\!\perp}^2)
    + (1 - \kappa)\,
    i\tilde{D}^{\prime \,(2)}(k_{\!\perp}^2)\! 
    \right]^2\!\!
    \left(1-J_0(|\vec{k}_{\!\perp}||\vec{r}_{\!\mbox{\tiny\it D}}|)
    \right) 
    \langle \psi_{h}|1-e^{i\vec{k}_{\!\perp}\vec{r}_2}|\psi_{h}\rangle
    \right .
    \nonumber \\
    &&\!\!\!\!\!\!\!\!\!\!\!\!
    \left . \left . 
    +\! 
    \left[\frac{\kappa}{k_{\!\perp}^2}\,i\tilde{D}^{(2)}(k_{\!\perp}^2)\right]^2 
    \!\!\left(-1\!+\ \!\! _1F_2 (-\frac{1}{2};\frac{1}{2},1; 
    \frac{-k_{\!\perp}^2 r_{\!\mbox{\tiny\it D}}^2}{4})\right)
    \!\langle \psi_{h}|\tan^2\!\!\phi_2(1-e^{i\vec{k}_{\!\perp}\vec{r}_2}) 
    |\psi_{h}\rangle\! 
    \right \}\!\left ( \frac{x_0}{x} \right )^{\!\!\epsilon^{\nprt}} \right
    ] \nonumber
\eea
with the Bessel function $J_0(|\vec{k}_{\!\perp}||\vec{r}_{\!\mbox{\tiny\it D}}|)$ and the
generalized hypergeometric function $_1F_2 (-1/2;1/2,1;-k_{\!\perp}^2
r_{\!\mbox{\tiny\it D}}^2/4)$ derived in the previous section.

For dipole sizes $|\vec{r}_{\!\mbox{\tiny\it D}}| \to 0$, the
perturbative contribution to $\sigma_{\!Dh}(x,
|\vec{r}_{\!\mbox{\tiny\it D}}|)$ is known to vanish quadratically
with decreasing dipole size,
$\sigma_{\!Dh}(x,|\vec{r}_{\!\mbox{\tiny\it D}}|) \propto
r^2_{\!\mbox{\tiny\it D}}$. This behavior reflects the weak absorption
of a small color-singlet dipole in the hadron and is known as {\em
  color transparency}. It can be seen immediately from Eq.~(\ref{sigdip}) as
\be
\left (1-J_0(|\vec{k}_{\!\perp}||\vec{r}_{\!\mbox{\tiny\it D}}|)\right) \approx 
\frac{k_{\!\perp}^2 r_{\!\mbox{\tiny\it D}}^2}{4} \quad \mbox{for
$|\vec{r}_{\!\mbox{\tiny\it D}}| \to 0$ and finite $|\vec{k}_{\!\perp}|$}. 
\label{bess_small_r}
\ee
The non-perturbative contribution to $\sigma_{\!Dh}(x,
|\vec{r}_{\!\mbox{\tiny\it D}}|)$ gives color transparency as well
since the generalized hypergeometric function behaves as
\be
\left(-1+\ \!\!_1F_2
(-\frac{1}{2};\frac{1}{2},1;\frac{-k_{\!\perp}^2 r_{\!\mbox{\tiny\it D}}^2}{4})\right) \approx 
\frac{k_{\!\perp}^2 r_{\!\mbox{\tiny\it D}}^2}{4} \quad \mbox{for
$|\vec{r}_{\!\mbox{\tiny\it D}}| \to 0$ and finite $|\vec{k}_{\!\perp}|$}.
\ee

For large dipole sizes, $|\vec{r}_{\!\mbox{\tiny\it D}}|\,\gtsim\,1\,
\fm$, the perturbative contribution to $\sigma_{\!Dh}(x,
|\vec{r}_{\!\mbox{\tiny\it D}}|)$ describing interactions of the quark
and antiquark of the dipole with the hadron saturates since
\be
\left (1-J_0(|\vec{k}_{\!\perp}||\vec{r}_{\!\mbox{\tiny\it D}}|)\right) \approx 
1 \quad \mbox{for
large $|\vec{k}_{\!\perp}||\vec{r}_{\!\mbox{\tiny\it D}}|$} \ .
\label{bess_large_r}
\ee
In contrast, the non-perturbative contribution to $\sigma_{\!Dh}(x,
|\vec{r}_{\!\mbox{\tiny\it D}}|)$ increases linearly with increasing
dipole size, $\sigma_{\!Dh}(x, |\vec{r}_{\!\mbox{\tiny\it D}}|)
\propto |\vec{r}_{\!\mbox{\tiny\it D}}|$.  This linear increase is
generated by the interaction of the string of the dipole with the
hadron: The string elongates linearly with the dipole size
$|\vec{r}_{\!\mbox{\tiny\it D}}|$ and, thus, has a linearly increasing
geometric cross section with the hadron. Indeed, this feature of the
string can be seen analytically since
\be
\left(-1+\ \!\!_1F_2
(-\frac{1}{2};\frac{1}{2},1;\frac{-k_{\!\perp}^2 r_{\!\mbox{\tiny\it D}}^2}{4})\right) \propto |\vec{k}_{\!\perp}||\vec{r}_{\!\mbox{\tiny\it D}}|
\quad \mbox{for
large $|\vec{k}_{\!\perp}||\vec{r}_{\!\mbox{\tiny\it D}}|$} \ .
\ee
When considered in Euclidean space-time, the same string gives also
the linear confining potential between a static quark and antiquark at
large $q{\bar q}$
separations~\cite{Rueter:1995cn,Shoshi:2002rd} as shown
briefly at the end of Sec.~\ref{Sec_Non-pert_Pert_Cont}.  Thus, the
behavior of the total dipole-hadron cross section is related to the
confining potential. Furthermore, as we are working in the quenched
approximation, there is no string breaking through dynamical
quark-antiquark production at large dipole sizes. String breaking is
expected to stop the linear increase of the total dipole-hadron cross
section at dipole sizes of $|\vec{r}_{\!\mbox{\tiny\it D}}| \,\gtsim\,
1\, \fm$ analogous to the saturation of the static $q{\bar q}$
potential seen for large $q{\bar q}$ separations on the lattice in
full QCD~\cite{Laermann:1998gm,Bali:2000gf}.

\begin{figure}[h!]
\setlength{\unitlength}{1.cm}
\begin{center}
\epsfig{file=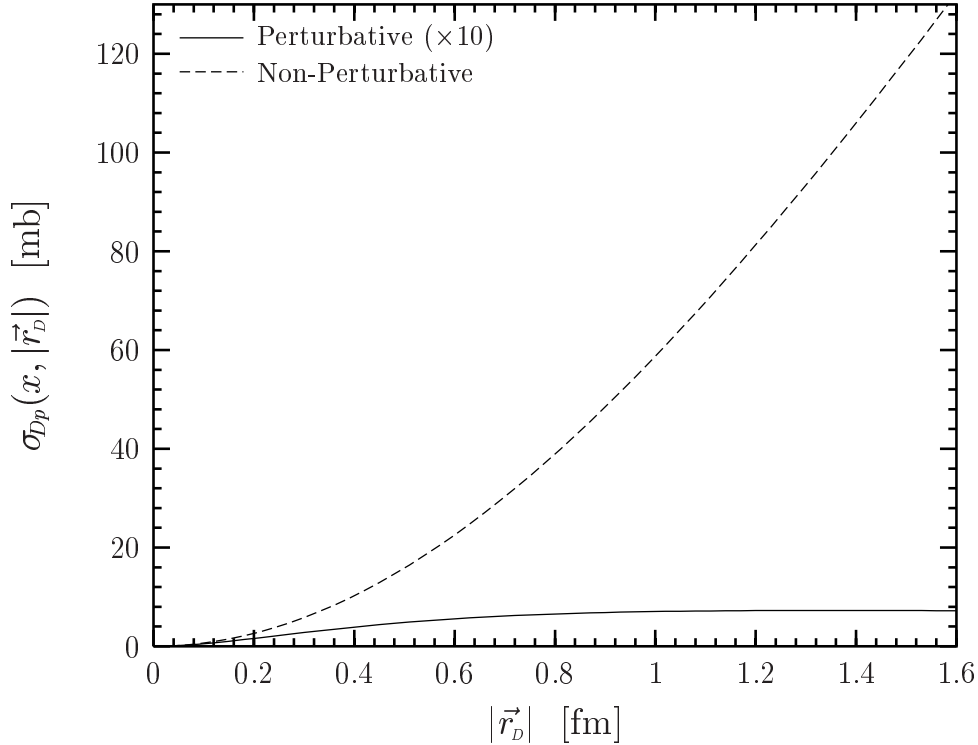,width=13.cm}
\end{center}
\vspace{-0.5cm}
\caption{\small Perturbative (solid line) and
  non-perturbative (dashed line) contributions to the total
  dipole-proton cross section $\sigma_{\!Dp}(x,
  |\vec{r}_{\!\mbox{\tiny\it D}}|)$ as a function of the dipole size
  $|\vec{r}_{\!\mbox{\tiny\it D}}|$ at $x = x_0 = 2.4\cdot 10^{-3}$.
  with the perturbative contribution multiplied by a factor of $10$.
  The perturbative contribution shows color transparency at small
  $|\vec{r}_{\!\mbox{\tiny\it D}}|$ and saturates at large
  $|\vec{r}_{\!\mbox{\tiny\it D}}|$. The non-perturbative contribution
  shows also color transparency at small
  $|\vec{r}_{\!\mbox{\tiny\it D}}|$ but increases linearly with
  increasing $|\vec{r}_{\!\mbox{\tiny\it D}}|$.}
\label{sig_dipp}
\end{figure}

In Fig.~\ref{sig_dipp} we show the perturbative (solid line) and
non-perturbative (dashed line) contributions to the total
dipole-proton cross section $\sigma_{\!Dp}(x,
|\vec{r}_{\!\mbox{\tiny\it D}}|)$ as a function of the dipole size
$|\vec{r}_{\!\mbox{\tiny\it D}}|$ at $x = x_0 = 2.4\cdot 10^{-3}$,
where the perturbative contribution is multiplied by a factor of $10$.
The dipole-proton cross section is computed with the simple Gaussian
proton wave function~(\ref{Eq_hadron_wave_function}) and illustrates
the general features discussed above: The perturbative contribution
shows color transparency at small $|\vec{r}_{\!\mbox{\tiny\it D}}|$
and saturates at large $|\vec{r}_{\!\mbox{\tiny\it D}}|$. The
non-perturbative contribution shows also color
transparency at small $|\vec{r}_{\!\mbox{\tiny\it D}}|$ but increases
linearly with increasing $|\vec{r}_{\!\mbox{\tiny\it D}}|$.  

Our result~(\ref{sigdip}) shows that the
$|\vec{k}_{\!\perp}|$\,-\,dependence of the hadron constituents
factorizes from the rest of the process in both the perturbative and
the non-perturbative contribution to the total dipole-hadron cross
section. This factorization -- known in perturbative QCD as
$|\vec{k}_{\!\perp}|$\,-\,factorization~\cite{Catani:1990xk+X} --
allows to define the unintegrated gluon distribution ${\cal
  F}_{h}(x,k^2_{\!\perp})$ as follows~\cite{Nikolaev:1991ja,Nikolaev:ce,Golec-Biernat:1999qd}
\be 
\sigma_{\!Dh}(x,|\vec{r}_{\!\mbox{\tiny\it D}}|) =
     \frac{4\pi^2r^2_{\!\mbox{\tiny\it D}}}{3}\int \!\!dk_{\!\perp}^2\,
     \frac{\left(1-J_0(|\vec{k}_{\!\perp}||\vec{r}_{\!\mbox{\tiny\it D}}|)
     \right)} {(|\vec{k}_{\!\perp}||\vec{r}_{\!\mbox{\tiny\it D}}|)^2}
     \,\alphaS(k^2_{\!\perp}) {\cal F}_{h}(x,k_{\!\perp}^2) \ .
\label{unint_gl_distr_def}
\ee 
For small dipole sizes
$|\vec{r}_{\!\mbox{\tiny\it D}}|$, this equation together
with~(\ref{bess_small_r}) and the integrated gluon
distribution
\be 
        xG_{h}(x,Q^2)=
        \int_0^{Q^2} \!dk_{\!\perp}^2 {\cal F}_{h}(x,k_{\!\perp}^2) 
\label{Eq_xG(x,Q^2)}   
\ee 
leads to the widely used perturbative QCD relation~\cite{Blaettel:rd}
\be
\sigma_{\!Dh}(x,|\vec{r}_{\!\mbox{\tiny\it D}}|) =
     \frac{\pi^2r^2_{\!\mbox{\tiny\it D}}}{3} 
     \left [ \alphaS(Q^2) xG_{h}(x,Q^2) \right ]_
     {\,Q^2 \,=\, c/r^2_{\!\mbox{\tiny\it D}}} \ ,
\label{rel_sigdip_xG}
\ee
where $c \approx 10$ is estimated from the properties of the Bessel
function $J_0(|\vec{k}_{\!\perp}||\vec{r}_{\!\mbox{\tiny\it
    D}}|)$~\cite{Nikolaev:ce+X}.

\begin{figure}[htb]
\setlength{\unitlength}{1.cm}
\begin{center}
\epsfig{file=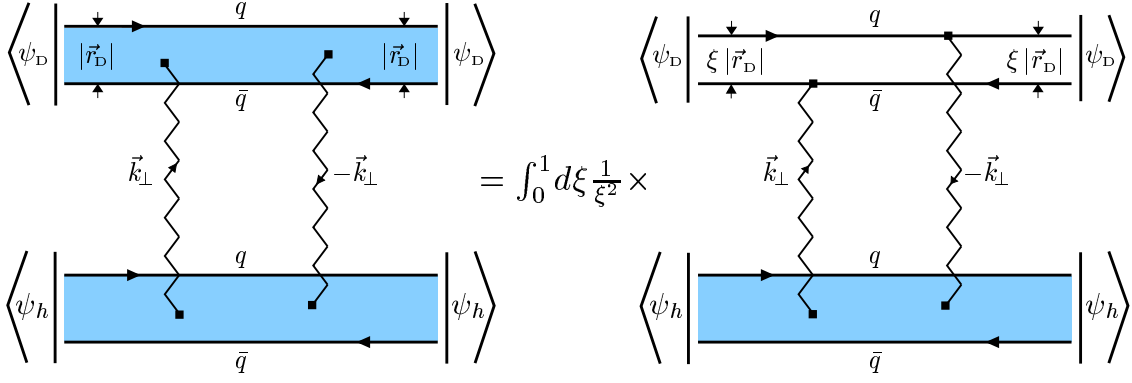,width=15.cm}
\end{center}
\caption{\small The string of length
  $|\vec{r}_{\!\mbox{\tiny\it D}}|$ is made up of stringless dipoles
  of size $\xi|\vec{r}_{\!\mbox{\tiny\it D}}|$ with $0\leq\xi\leq 1$
  and dipole number density $n(\xi) = 1/\xi^2$. The string-hadron
  scattering process reduces to an incoherent superposition of
  stringless dipole-hadron scattering processes.}
\label{sstoqs}
\end{figure}

To extract the unintegrated gluon distribution ${\cal
  F}_{h}(x,k^2_{\!\perp})$, we compare~(\ref{unint_gl_distr_def})
with (\ref{sigdip}) using the following mathematical identity
\be
\left(-1+\ \!\! _1F_2 (-\frac{1}{2};\frac{1}{2},1; 
\frac{-k_{\!\perp}^2 r_{\!\mbox{\tiny\it D}}^2}{4}) \right) = 
      \int_0^1\,d\xi \,\frac{1}{\xi^2}\left(1-J_0(|\vec{k}_{\!\perp}|
      |\vec{r}_{\!\mbox{\tiny\it D}}|\xi)\right) \ . 
\label{mathident}
\ee
As discussed in the previous section, the lhs of~(\ref{mathident})
results from the string averaged over all
orientations~(\ref{average_string}). Thus, the string confining the
quark-antiquark dipole of length $|\vec{r}_{\!\mbox{\tiny\it D}}|$ can
be represented as an integral over stringless dipoles of sizes
$\xi|\vec{r}_{\!\mbox{\tiny\it D}}|$ with $0\leq\xi\leq1$ and a dipole
number density of $n(\xi) = 1/\xi^2$.  As visualized in
Fig.~\ref{sstoqs}, the string-hadron scattering process reduces to an
incoherent superposition of stringless dipole-hadron scattering
processes with dipole sizes $0\leq \xi|\vec{r}_{\!\mbox{\tiny\it
    D}}|\leq|\vec{r}_{\!\mbox{\tiny\it D}}|$ and dipole number
density $n(\xi) = 1/\xi^2$. The decomposition of the string into many
smaller stringless dipoles via~(\ref{mathident}) behaves similar to
the wave function of a $q{\bar q}$ onium state in the large $N_c$
limit~\cite{Mueller:1994rr,Mueller:1994jq}: The numerous gluons
emitted inside the onium state can be considered as many $q{\bar q}$
dipoles in the large $N_c$ limit.

Inserting~(\ref{mathident}) into~(\ref{sigdip}) and rescaling the
momentum variable $|\vec{k}_{\!\perp}^{\prime}|=
\xi|\vec{k}_{\!\perp}|$, the string-hadron ($sh$) contribution to the
total dipole-hadron cross section~(\ref{sigdip}) becomes
\bea
\sigma_{\!Dh}^{\,sh}
(x, |\vec{r}_{\!\mbox{\tiny\it D}}|)&=& 
      \frac{8}{9}\frac{1}{4\pi} \int dk_{\!\perp}^{\prime \,2}
      \left(1-J_0(|\vec{k}_{\!\perp}^{\prime}|
      |\vec{r}_{\!\mbox{\tiny\it D}}|)\right)
\label{sigdip_sp} \\
      &&
      \hphantom{\hspace{-3cm}}
      \times 
      \left( \frac{\pi^2G_2}{24} \right)^2 \left \{ \frac{\kappa^2}
      {k_{\!\perp}^{\prime \,4}}
      \int_0^1 d\xi
      \left[i\tilde{D}^{(2)}(\frac{k_{\!\perp}^{\prime \,2}}{\xi^2})\right]^2 
      \langle \psi_{h}|\tan^2\!\!\phi_2
      (1-e^{i(\vec{k}_{\!\perp}^{\prime}/
      {\xi})\vec{r}_2})|\psi_{h} 
      \rangle \right \}\left ( \frac{x_0}{x} 
      \right )^{\!\!\epsilon^{\nprt}}. \nonumber
\eea
The dipole factor $(1-J_0(|\vec{k}_{\!\perp}^{\prime}|
|\vec{r}_{\!\mbox{\tiny\it D}}|)$ indicates that the string-hadron
interaction has been rewritten into a stringless dipole-hadron
interaction. The string confining the dipole has been shifted into the
hadron. Comparing (\ref{unint_gl_distr_def}) with our new expression for the total dipole-hadron cross section,
\bea
\!\!\!\!\!\!\!\!
\sigma_{\!Dh}(x, |\vec{r}_{\!\mbox{\tiny\it D}}|) &=& 
    \frac{8}{9}\frac{1}{4\pi} \int dk_{\!\perp}^2 
    \left(1-J_0(|\vec{k}_{\!\perp}||\vec{r}_{\!\mbox{\tiny\it D}}|)\right)
    \nonumber \\
    &&
    \hphantom{\hspace{-2cm}}
    \times
    \left [
    \left (4\pi\alphaS(k^2_{\!\perp})\right)^2
    \left \{ \left [i\tilde{D}_{\pert}^{\prime \,(2)}(k^2_{\!\perp})\right ]^2
    \langle \psi_{h}|1-e^{i\vec{k}_{\!\perp}\vec{r}_2}|\psi_{h} \rangle
    \right \} \left( \frac{x_0}{x} \right )^{\!\!\epsilon^{\pert}} \right .
    \nonumber \\
    &&
    \hphantom{\hspace{-2cm}}
    + \left(\frac{\pi^2G_2}{24}\right)^2
    \left \{ \!\!
    \left[(1-\kappa)
    \left[i\tilde{D}_1^{\prime \,(2)}(k_{\!\perp}^2)\right]+
    \frac{\kappa}{k_{\!\perp}^2} 
    \left[i\tilde{D}^{(2)}(k_{\!\perp}^2)\right]
    \right]^2
    \langle \psi_{h}|1-e^{i\vec{k}_{\!\perp}\vec{r}_2}|\psi_{h}\rangle
    \right .
    \nonumber \\
    &&
    \hphantom{\hspace{-2cm}}
    \left . \left .
    +\, \frac{\kappa^2}{k_{\!\perp}^4} \int_0^1 d\xi
    \left[i\tilde{D}^{(2)}(\frac{k_{\!\perp}^2}{\xi^2})\right]^2 
    \langle
    \psi_{h}|\tan^2\!\!\phi_2(1-e^{i(\vec{k}_{\!\perp}/{\xi})\vec{r}_2
    }) |\psi_{h}\rangle 
    \right \} \left( \frac{x_0}{x} \right )^{\!\!\epsilon^{\nprt}}
    \right ] \ ,
\label{sigdipxi}
\eea
one obtains the unintegrated gluon distribution 
\bea
\!\!\!\!\!
{\cal F}_{h}(x,k_{\!\perp}^2) &=& 
    \frac{k_{\!\perp}^2}{6\pi^3\alphaS(k^2_{\!\perp})}
    \nonumber \\
    &&
    \hphantom{\hspace{-2cm}}
    \times
    \left [
    \left (4\pi\alphaS(k^2_{\!\perp})\right)^2
    \left \{ \left [i\tilde{D}_{\pert}^{\prime \,(2)}(k^2_{\!\perp})\right ]^2
    \langle \psi_{h}|1-e^{i\vec{k}_{\!\perp}\vec{r}_2}|\psi_{h}\rangle
    \right \} \left( \frac{x_0}{x} \right )^{\!\!\epsilon^{\pert}} \right .
    \nonumber \\
    &&
    \hphantom{\hspace{-2cm}}
    +\left(\frac{\pi^2G_2}{24}\right)^2
    \left \{ \!\!
    \left[(1-\kappa)
    \left[i\tilde{D}_1^{\prime \,(2)}(k_{\!\perp}^2)\right] +
    \frac{\kappa}{k_{\!\perp}^2} 
    \left[i\tilde{D}^{(2)}(k_{\!\perp}^2)\right]
    \right]^2
    \langle \psi_{h}|1-e^{i\vec{k}_{\!\perp}\vec{r}_2}|\psi_{h}\rangle
    \right .
    \nonumber \\
    &&
    \hphantom{\hspace{-2cm}}
    \left . \left .
    + \frac{\kappa^2}{k_{\!\perp}^4}\int_0^1 d\xi
    \left[i\tilde{D}^{(2)}(\frac{k_{\!\perp}^2}{\xi^2})\right]^2 
    \langle
    \psi_{h}|\tan^2\!\!\phi_2(1-e^{i(\vec{k}_{\!\perp}/{\xi})
    \vec{r}_2})|\psi_{h}\rangle 
    \right \} \left( \frac{x_0}{x} \right )^{\!\!\epsilon^{\nprt}}
    \right ] \ .
\label{lf}
\eea
This result shows explicitly the microscopic structure of the
perturbative and non-perturbative contribution to ${\cal
  F}_{h}(x,k^2_{\!\perp})$. It is valid for any hadron wave
function. To get numerical values for the unintegrated gluon
distribution ${\cal F}_{h}(x,k^2_{\!\perp})$, the hadron wave
functions must be specified.

\section{Numerical Results for Unintegrated Gluon \\ Distributions in Hadrons and Photons}
\label{Sec_Numerical_Results_of_Unintegrated_Gluon_Distributions_of_Hadrons_and_Photons} 

In this section we present the unintegrated gluon
distribution~(\ref{lf}) for protons, pions, kaons, and photons
computed with the Gaussian hadron wave
function~(\ref{Eq_hadron_wave_function}) and the perturbatively
derived photon wave
functions~(\ref{Eq_photon_wave_function_T_squared})
and~(\ref{Eq_photon_wave_function_L_squared}). To account for
non-perturbative effects at low photon virtuality $Q^2$ in the photon
wave functions, quark masses $m_f(Q^2)$ are used that interpolate
between the current quarks at large $Q^2$ and the constituent quarks
at small $Q^2$~\cite{Dosch:1998nw} as discussed in
Appendix~\ref{Sec_Wave_Functions}.

\begin{figure}[htb]
\setlength{\unitlength}{1.cm}
\begin{center}
\epsfig{file=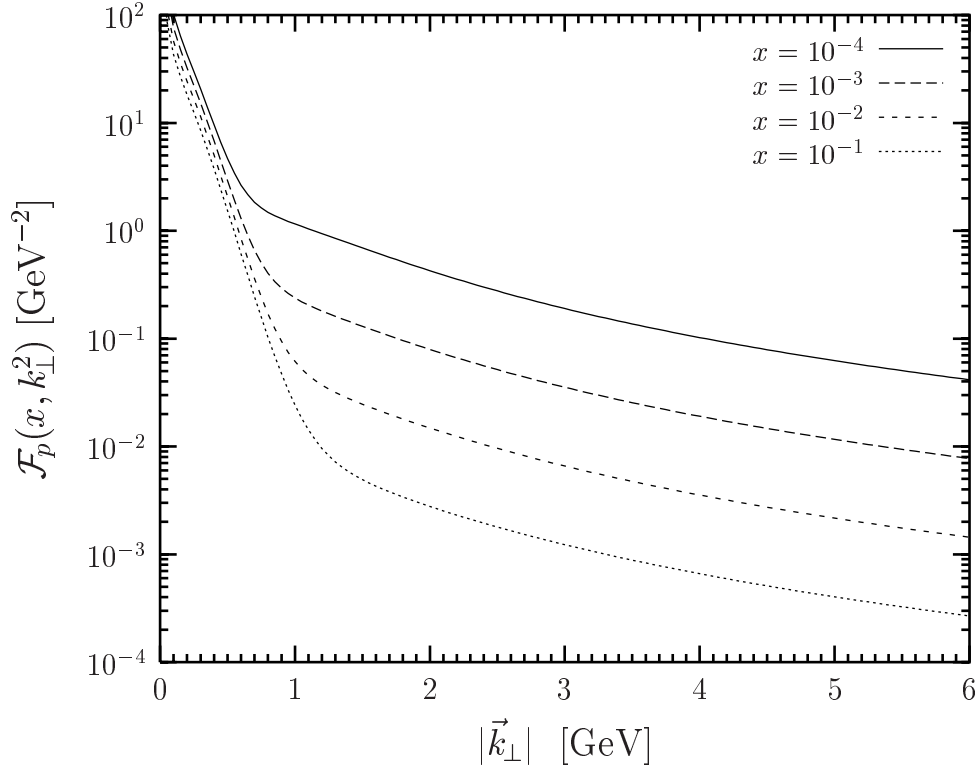,width=13.cm}
\end{center}
\caption{\small The unintegrated gluon distribution of the proton
  ${\cal F}_{\!p}(x,k_{\!\perp}^2)$ as a function of transverse
  momentum $|\vec{k}_{\!\perp}|$ at Bjorken\,-\,$x$ values of
  $10^{-1}$, $10^{-2}$, $10^{-3}$ and $10^{-4}$.}
\label{fg_k.eps}
\end{figure}

\begin{figure}[htb]
\setlength{\unitlength}{1.cm}
\begin{center}
\epsfig{file=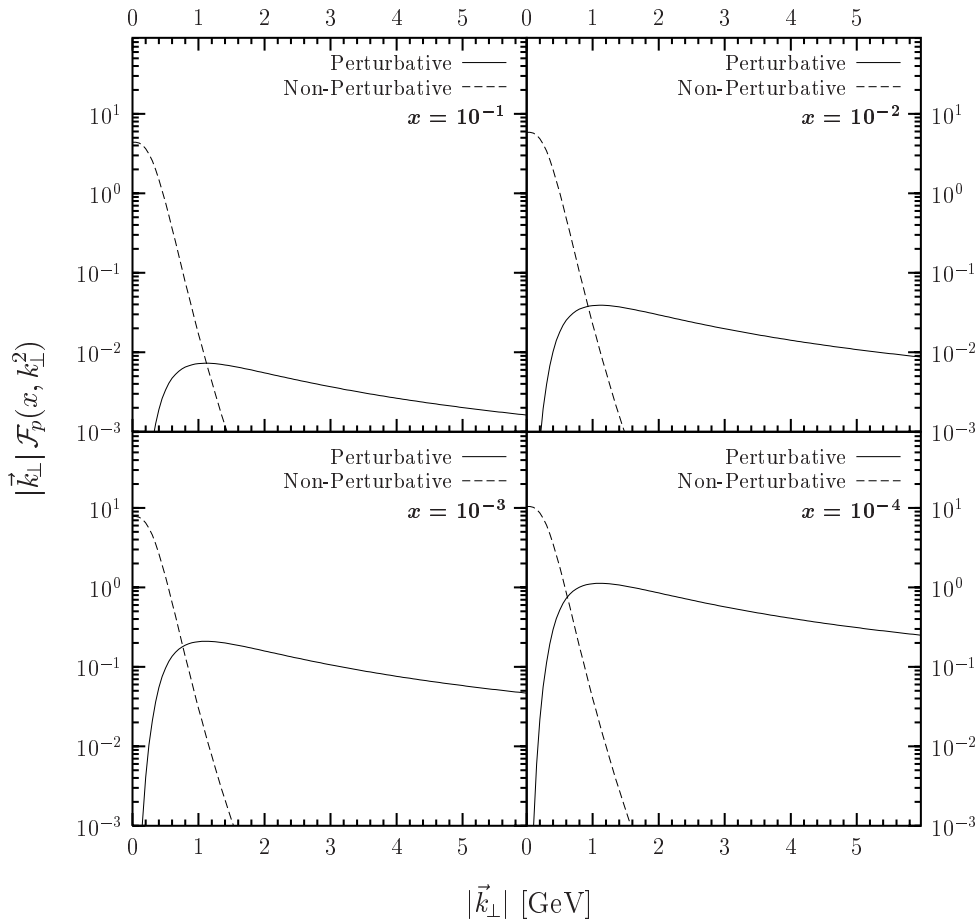,width=13cm}
\end{center}
\caption{\small The unintegrated gluon distribution of the proton ${\cal
    F}_{\!p}(x,k_{\!\perp}^2)$ times the tranverse momentum
  $|\vec{k}_{\!\perp}|$ as a function of $|\vec{k}_{\!\perp}|$ at
  Bjorken\,-\,$x$ values of $10^{-1}$, $10^{-2}$, $10^{-3}$ and
  $10^{-4}$.}
\label{p_fg_k_p_vs_np.eps}
\end{figure}

The unintegrated gluon distribution of the proton ${\cal
  F}_{\!p}(x,k_{\!\perp}^2)$ is shown as a function of transverse
momentum $|\vec{k}_{\!\perp}|$ at $x = 10^{-1}$, $10^{-2}$, $10^{-3}$,
and $10^{-4}$ in Fig.~\ref{fg_k.eps}. Figure~\ref{p_fg_k_p_vs_np.eps}
illustrates the interplay of the perturbative (solid line) and
non-perturbative (dashed line) contributions to
$|\vec{k}_{\!\perp}|{\cal F}_{\!p}(x,k_{\!\perp}^2)$ as a function of
transverse momentum $|\vec{k}_{\!\perp}|$ for the same values of $x$.
At small momenta $|\vec{k}_{\!\perp}|$, the unintegrated gluon
distribution is dominated by the non-perturbative contribution and
behaves as $1/|\vec{k}_{\!\perp}|$. This behavior is a string
manifestation as it reflects the linear increase of the total
dipole-proton cross section at large dipole sizes. In contrast, the
saturation model of Golec-Biernat and
W{\"u}sthoff~\cite{Golec-Biernat:1999qd} shows the behavior ${\cal
  F}^{\mbox{\tiny $GBW$}}_{\!p}(x,k_{\!\perp}^2) \propto
k_{\!\perp}^2$ for small momenta. With increasing
$|\vec{k}_{\!\perp}|$, the non-perturbative contribution to ${\cal
  F}_{\!p}(x,k_{\!\perp}^2)$ decreases rapidly which results from the
strong suppression of large momenta by the non-perturbative
correlation functions $i\tilde{D}_1^{\prime \,(2)}(k_{\!\perp}^2)$ and
$i\tilde{D}^{(2)}(k_{\!\perp}^2)$ given
in~(\ref{Eq_D_(prime)(k^2)_for_exp_correlation_perp})
and~(\ref{Eq_D(k^2)_for_exp_correlation_perp}). For
$|\vec{k}_{\!\perp}| \,\gtsim\, 1\,\GeV$, the perturbative
contribution dominates the unintegrated gluon distribution. It drops
as $1/k_{\!\perp}^2$ in accordance with the perturbative correlation
function $i\tilde{D}_{\pert}^{\prime \,(2)}(k_{\!\perp}^2)$ given
in~(\ref{Eq_massive_D_pge_prime_trans}).  This perturbative QCD result
is not reproduced by the phenomenological model of Golec-Biernat and
W{\"u}sthoff~\cite{Golec-Biernat:1999qd} which predicts a Gaussian
decrease of ${\cal F}_{\!p}(x,k_{\!\perp}^2)$ with increasing
$|\vec{k}_{\!\perp}|$.

The $x$\,-\,dependence of ${\cal F}_{\!p}(x,k_{\!\perp}^2)$ can be
seen in Figs.~\ref{fg_k.eps} and~\ref{p_fg_k_p_vs_np.eps}: With
decreasing $x$, the perturbative contribution increases much stronger
than the non-perturbative contribution which results from the energy
exponents $\epsilon^{\pert} \gg \epsilon^{\nprt}$ in~(\ref{x_dep})
necessary to describe the experimental data within our model as
discussed in Sec.~\ref{Sec_Energy_Dependence}.
Moreover, the perturbative contribution extends into the
small\,-\,$|\vec{k}_{\!\perp}|$ region as $x$ decreases. Indeed, the
soft-hard transition point moves towards smaller momenta with
decreasing $x$ as shown in Fig.~\ref{p_fg_k_p_vs_np.eps}. Such a
hard-to-soft diffusion is observed also in~\cite{Ivanov:2000cm} where
the unintegrated gluon distribution has been parametrized to reproduce
the experimental data for the proton structure function $F_2(x,Q^2)$
at small $x$. The opposite behavior is obtained in the approach of the
color glass condensate~\cite{Iancu:2002xk}: With decreasing $x$, gluons
are produced predominantely in the high\,-\,$|\vec{k}_{\!\perp}|$
region of lower density and weaker repulsive interactions.

\begin{figure}[htb]
\setlength{\unitlength}{1.cm}
\begin{center}
\epsfig{file=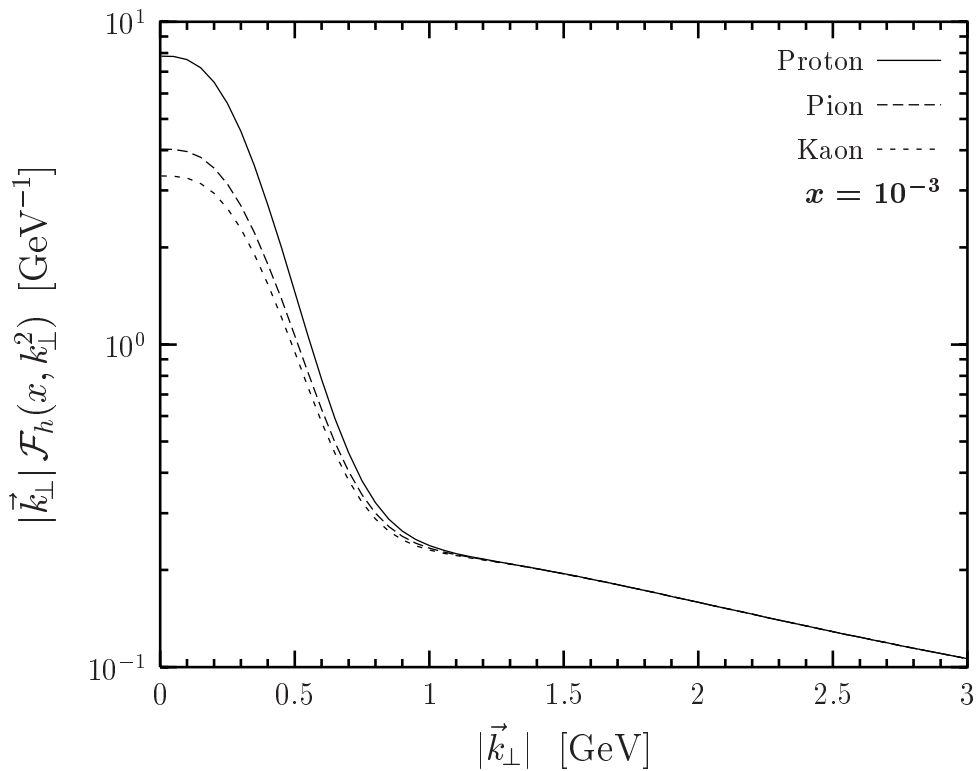,width=13.cm}
\end{center}
\caption{\small The unintegrated gluon distribution of the proton,
  pion, and kaon ${\cal F}_h(x,k_{\!\perp}^2)$ times the transverse
  momentum $|\vec{k}_{\!\perp}|$ as a function of
  $|\vec{k}_{\!\perp}|$ at Bjorken-variable $x=10^{-3}$.}
\label{p_pi_k_k_fg_k.eps}
\end{figure}
In Fig.~\ref{p_pi_k_k_fg_k.eps}, the unintegrated gluon distribution
of the proton, pion, and kaon ${\cal F}_h(x,k_{\!\perp}^2)$ times the
transverse momentum $|\vec{k}_{\!\perp}|$ is shown as a function of
$|\vec{k}_{\!\perp}|$ at $x=10^{-3}$. The hadrons are characterized by
different values for $\Delta z_h$ and $S_h$ in the hadron wave
function~(\ref{Eq_hadron_wave_function}). However, ${\cal
  F}_h(x,k_{\!\perp}^2)$ depends only on $S_h$. Due to the
normalization of the hadron wave functions, $\Delta z_h$ disappears
upon the integration over $z_i$ as can be seen directly
from~(\ref{averages_explicit_qq}) and~(\ref{averages_explicit_ss}). At
small momenta, ${\cal F}_h(x,k_{\!\perp}^2) \propto S_h^2$ is found
from~(\ref{averages_explicit_qq}), (\ref{averages_explicit_ss}),
and~(\ref{lf}). It becomes visible in Fig.~\ref{p_pi_k_k_fg_k.eps} for
the chosen hadron extensions: $S_p = 0.86\,\fm$, $S_{\pi} =
0.607\,\fm$, and $S_{K} = 0.55\,\fm$.  At large momenta where the
perturbative contribution dominates, the dependence on $S_h$ vanishes
as can be seen from~(\ref{averages_explicit_qq}) and the unintegrated
gluon distributions of protons, pions, and kaons become identical. Of
course, this behavior results from the wave function normalization
being identical for protons, pions, and kaons with two valence
constituents which are the quark and antiquark in the pion and kaon
and the quark and diquark in the proton. At large
$|\vec{k}_{\!\perp}|$, i.e., high resolution, the realistic
description of protons as three-quark systems becomes necessary. In
fact, the three-quark description of protons leads to a different
perturbative contribution in~(\ref{lf}): the quark counting factors of
$2$ (appropriate for mesons) in the square brackets in~(\ref{Ip}) are
substituted by factors of $3$ (appropriate for
baryons)~\cite{Low:1975sv+X,Gunion:iy}. At other values of $x$, the
unintegrated gluon distributions of protons, pions and kaons show the
same features.

\begin{figure}[htb]
\setlength{\unitlength}{1.cm}
\begin{center}
\epsfig{file=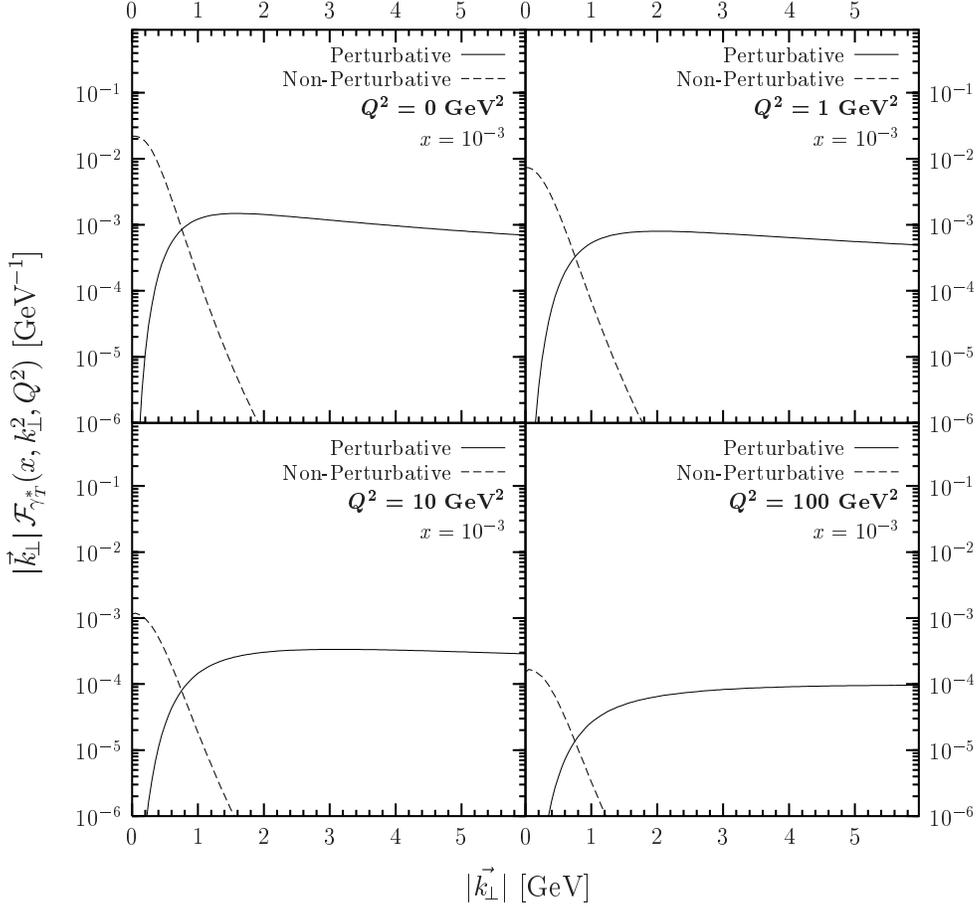,width=13.cm}
\end{center}
\caption{\small The unintegrated gluon distribution of the transverse
  polarized photon ${\cal F}_{\!\gamma^*_T}(x,k_{\!\perp}^2,Q^2)$
  times the tranverse momentum $|\vec{k}_{\!\perp}|$ as a function of
  $|\vec{k}_{\!\perp}|$ at photon virtualities of $Q^2 = 0$, $1$, $10$,
  and $100\,\GeV^2$ and Bjorken-variable $x=10^{-3}$.}
\label{gT_k_fg_k.eps}
\end{figure}
\begin{figure}[htb]
\setlength{\unitlength}{1.cm}
\begin{center}
\epsfig{file=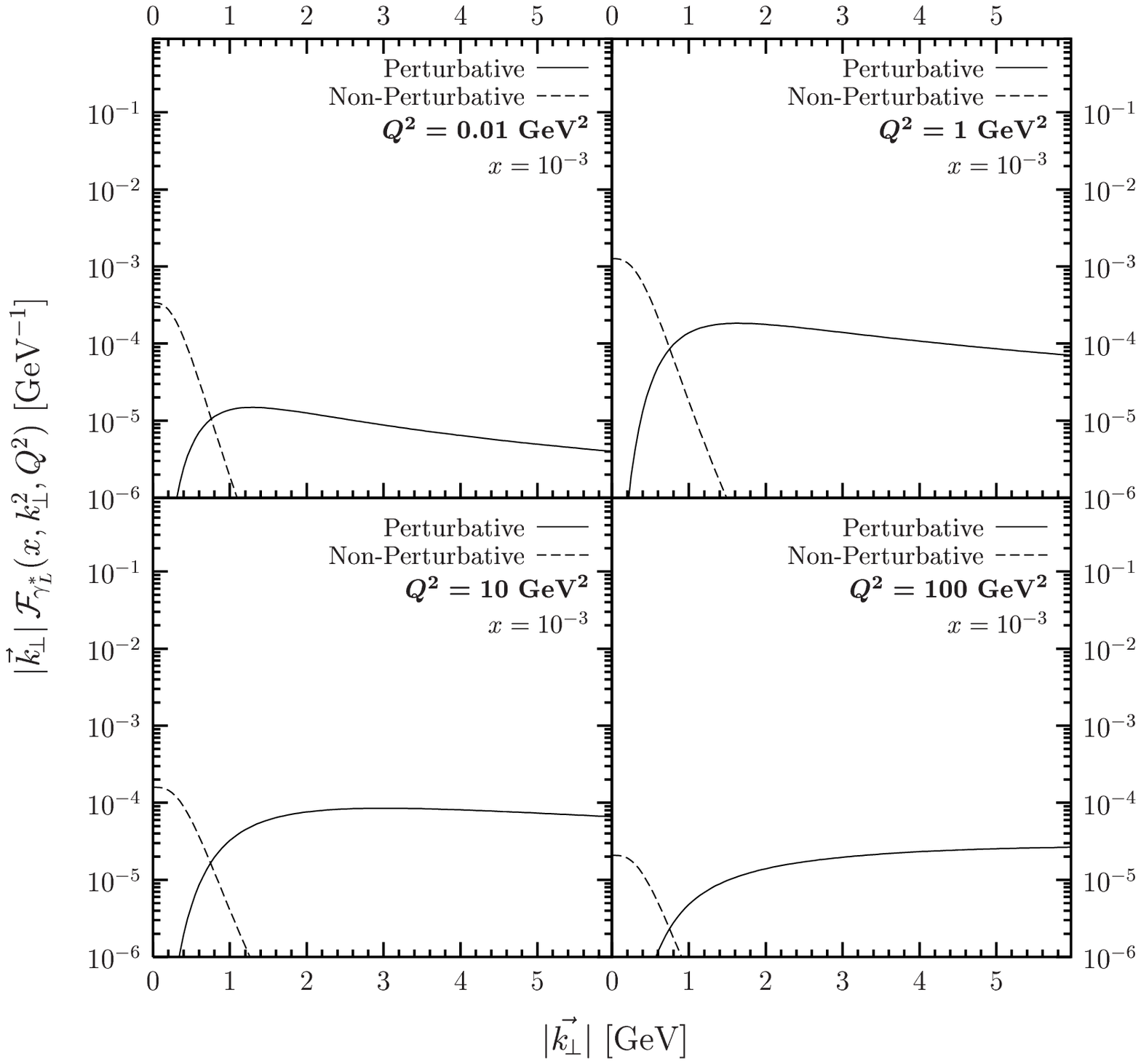,width=13.cm}
\end{center}
\caption{\small The unintegrated gluon distribution of the
  longitudinally polarized photon ${\cal
    F}_{\!\gamma^*_L}(x,k_{\!\perp}^2,Q^2)$ times the tranverse
  momentum $|\vec{k}_{\!\perp}|$ as a function of
  $|\vec{k}_{\!\perp}|$ at photon virtualities of $Q^2 = 0.01$, $1$,
  $10$, and $100\,\GeV^2$ and Bjorken-variable $x=10^{-3}$.}
\label{gL_k_fg_k.eps}
\end{figure}

Photons are particularly interesting because the
transverse size of the quark-antiquark dipole into which a photon
fluctuates is controlled by the photon virtuality~$Q^2$ (cf.
Appendix~\ref{Sec_Wave_Functions})
\be 
        |\vec{r}_{\gamma^*_{T,L}}| \approx
                         \frac{2}{Q^2+4m^2_u(Q^2)} \ , 
\label{rQ}
\ee 
where $m_u(Q^2)$ is the running $u$-quark mass given in
Appendix~\ref{Sec_Wave_Functions}. In Figs.~\ref{gT_k_fg_k.eps}
and~\ref{gL_k_fg_k.eps}, the unintegrated gluon distribution of
transverse~($T$) and longitudinal~($L$) photons ${\cal
  F}_{\!\gamma^*_{T,L}}(x,k_{\!\perp}^2,Q^2)$ times the transverse
momentum $|\vec{k}_{\!\perp}|$ is shown as a function of
$|\vec{k}_{\!\perp}|$ for various photon virtualities $Q^2$ at
$x=10^{-3}$.
With increasing $Q^2$, i.e., decreasing ``photon size''
$|\vec{r}_{\gamma^*_{T,L}}|$, the ratio of the perturbative to the
non-perturbative contribution to ${\cal
  F}_{\!\gamma^*_{T,L}}(x,k_{\!\perp}^2,Q^2)$ increases. This behavior
has already been discussed in Sec.~\ref{Momentum-Space Structure of
  Dipole-Dipole Scattering} on the level of dipole-dipole scattering,
where decreasing dipole sizes increase the ratio of the perturbative
to non-perturbative contribution, see
Fig.~\ref{Fig_dipole_dipole_interactions}. Due to the different
$Q^2$\,-\,dependence in the transverse and longitudinally polarized
photon wave functions, ${\cal
  F}_{\!\gamma^*_{T}}(x,k_{\!\perp}^2,Q^2)$ decreases continously with
increasing $Q^2$ while ${\cal
  F}_{\!\gamma^*_{L}}(x,k_{\!\perp}^2,Q^2)$ increases for $Q^2 \ltsim
1\,\GeV^2$ and decreases for $Q^2 \,\gtsim\, 1\,\GeV^2$.

\begin{figure}[htb]
\setlength{\unitlength}{1.cm}
\begin{center}
\epsfig{file=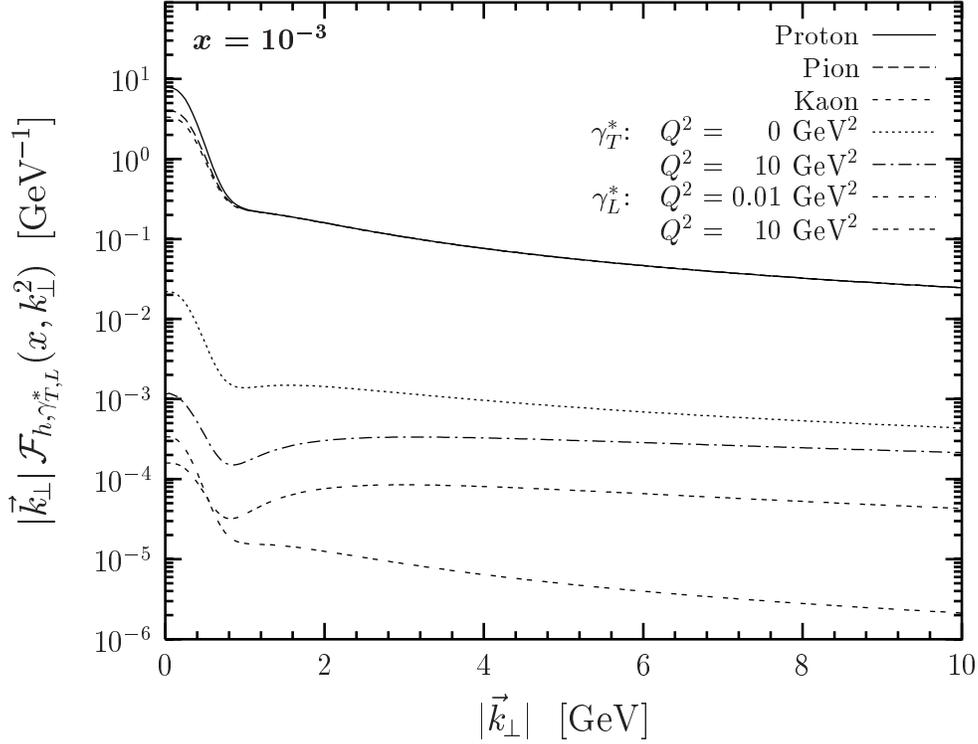,width=13.cm}
\end{center}
\caption{\small The unintegrated gluon distribution of the proton,
  pion, kaon, and transverse and longitudinally polarized photon
  ${\cal F}_{h,\gamma^*_{T,L}}(x,k_{\!\perp}^2)$ times the transverse
  momentum $|\vec{k}_{\!\perp}|$ as a function of
  $|\vec{k}_{\!\perp}|$ at Bjorken-variable $x=10^{-3}$. Results for
  transverse polarized photons are shown for photon virtualities of
  $Q^2=0$ and $10\,\GeV^2$ and results for longitudinally polarized
  photons for photon virtualities of $Q^2=0.01$ and $10\,\GeV^2$.}
\label{h_g_k_fg_k.eps}
\end{figure}
In Fig.~\ref{h_g_k_fg_k.eps}, the unintegrated gluon distributions of
hadrons and photons discussed above are compared. The unintegrated
gluon distribution of real ($Q^2=0$) photons ($|\vec{r}_{\gamma^{}_T}|
\approx S_h$) is suppressed by a factor of order $\alpha$ in
comparison to the one of the hadrons otherwise its shape is very
similar. The suppression factor comes from the photon\,-\,dipole
transition described by the photon wave functions given in
Appendix~\ref{Sec_Wave_Functions}. The ratio of the perturbative to
the non-perturbative contribution of unintegrated gluon distributions
increases as one goes from hadrons to photons with high virtuality
$Q^2$. Since the wave functions of protons, pions, and kaons are
normalized to the same value, the unintegrated gluon distributions of
these hadrons are, as mentioned above, identical at large
$|\vec{k}_{\!\perp}|$ and do not depend on the hadron size. In
contrast, the $Q^2$\,-\,dependence of the photon wave functions leads
to $Q^2$\,-\,dependent, i.e., ``photon size''\,-\,dependent,
unintegrated gluon distributions at large $|\vec{k}_{\!\perp}|$. With
increasing $|\vec{k}_{\!\perp}|$, the unintegrated gluon distributions
of hadrons and photons become parallel in line with the vanishing
dependence on the specific form of the wave function.

\begin{figure}[htb]
\setlength{\unitlength}{1.cm}
\begin{center}
\epsfig{file=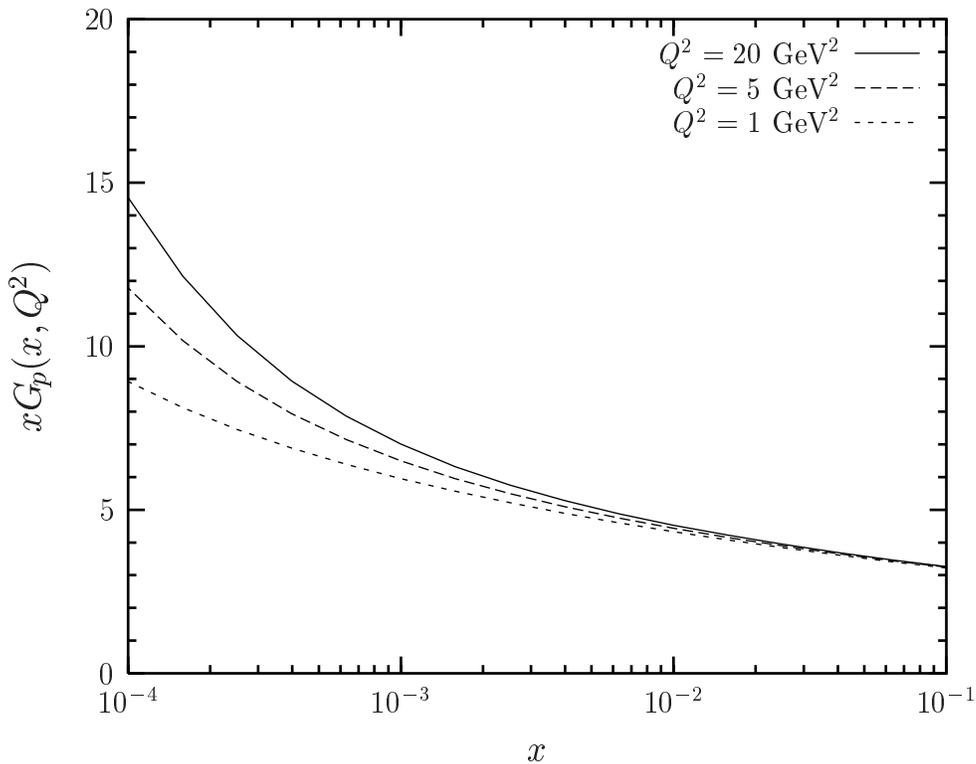,width=13.cm}
\end{center}
\caption{\small The gluon distribution of the proton
  $xG_p(x,Q^2)$ as a function of Bjorken\,-\,$x$ at photon
  virtualities of $Q^2=1$, $5$, and $20\, \GeV^2$.}
\label{xG_x.eps}
\end{figure}
In Fig.~\ref{xG_x.eps}, the integrated gluon distribution of the
proton $xG_p(x,Q^2)$ is shown as a function of Bjorken\,-\,$x$ at
photon virtualities of $Q^2 = 1, 5$, and $20\,\GeV^2$.  Recall that
the parameter $x_0=2.4 \cdot 10^{-3}$ has been adjusted in the
previous section such that the experimental data of $xG_p(x,Q^2)$ at
$Q^2=1\,\GeV^2$~\cite{Abramowicz:1998ii+X} are reproduced. For $x
\,\gtsim\, 10^{-3}$, $xG_p(x,Q^2)$ is mainly determined by
non-perturbative physics as can be seen from Figs.~\ref{fg_k.eps}
and~\ref{p_fg_k_p_vs_np.eps}. Perturbative physics becomes relevant
for $x\,\ltsim\,10^{-3}$ and generates the steep increase of
$xG_p(x,Q^2)$ with decreasing $x$ at fixed $Q^2$. Also the rise of
$xG_p(x,Q^2)$ with increasing $Q^2$ at fixed $x$ results from the
perturbative contribution. For $x \ll 10^{-4}$, we show
explicitly in Sec.~\ref{Sec_Gluon_Saturation} that
multiple gluonic exchanges contained in the full $T$-matrix
element~(\ref{Eq_model_purely_imaginary_T_amplitude_final_result})
slow down the powerlike increase of $xG_p(x,Q^2)$ with decreasing $x$
in accordance with $S$-matrix unitarity constraints.

\newpage
\section{Comparison with Other Work}
\label{Sec_Comparison_with_Other_Work}

In this section, we compare the unintegrated gluon distribution of the
proton extracted from our loop-loop correlation model (LLCM) with
those obtained from the saturation model of Golec-Biernat and
W{\"u}sthoff (GBW)~\cite{Golec-Biernat:1999qd}, the derivative of the
Gl{\"u}ck, Reya, and Vogt (GRV) parametrization of
$xG_p(x,Q^2)$~\cite{Gluck:1998xa}, and the approach of Ivanov and
Nikolaev (IN)~\cite{Ivanov:2000cm}.

\medskip

In the approach of Golec-Biernat and
W{\"u}sthoff~\cite{Golec-Biernat:1999qd}, the unintegrated gluon
distribution of the proton is extracted from the total dipole-proton
cross section by inverting~(\ref{unint_gl_distr_def}),
\be
{\cal F}_{p}(x,k_{\!\perp}^2) = 
        \frac{3\,\sigma_0}{4\,\pi^2\,\alphaS}\,
        R_0^2(x)\,k^2_{\!\perp}\,\exp\left(-R_0^2(x)\,k^2_{\!\perp}\right) 
\quad \mbox{with} \quad
        R_0^2(x) = \frac{1}{Q_0^2}\left(\frac{x}{x_0}\right)^{\lambda}
\label{GBW}
\ee
where the parameter $\sigma_0 = 29.12\,\mb$, $\alphaS = 0.2$, $Q_0 =
1\,\GeV$, $\lambda = 0.277$, and $x_0 = 0.41 \cdot 10^{-4}$ are
obtained from a fit to the proton structure function $F_2(x,Q^2)$
including charm quarks in the photon wave function. Note, however that
the GBW approach uses the dipole-proton cross section of the GBW model
on the lhs of~(\ref{unint_gl_distr_def}) which implies multiple gluon
exchanges while the rhs of~(\ref{unint_gl_distr_def}) describes only
two-gluon exchange as discussed in Sec.~\ref{Momentum-Space Structure
  of Dipole-Dipole Scattering}. Moreover, as demonstrated below, the
large $k^2_{\!\perp}$\,-\,behavior of the unintegrated gluon
distribution~(\ref{GBW}) deviates significantly from the DGLAP
results. This mismatch motivated the recent modifications of the
saturation model~\cite{Bartels:2002cj}.

\medskip

The unintegrated gluon distribution of the proton is also computed
from the integrated gluon distribution $xG_p(x,Q^2)$ by
inverting~(\ref{Eq_xG(x,Q^2)})
\be
{\cal F}_{p}(x,k_{\!\perp}^2) =
        \left.\frac{dxG_p(x,Q^2)}{dQ^2}\right|_{Q^2=k^2_{\!\perp}} \ .
\label{GRV}
\ee
Here, we use for the integrated gluon distribution $xG_p(x,Q^2)$ the
leading order (LO) parametrization of Gl{\"u}ck, Reya, and Vogt
(GRV)~\cite{Gluck:1998xa}, which covers the kinematic region $10^{-9}
< x < 1$ and $0.8\,\GeV^2 < Q^2 < 10^6\,\GeV^2$.

\medskip

Ivanov and Nikolaev~\cite{Ivanov:2000cm} have constructed a
two-component (soft $+$ hard) ansatz for the unintegrated gluon
distribution of the proton
\be
{\cal F}_p(x,k_{\!\perp}^2) 
        = {\cal F}_{\soft}(x,k_{\!\perp}^2)\,\frac{k^2_s}{k^2_{\!\perp}+k^2_s} 
        + {\cal
         F}_{\hard}(x,k_{\!\perp}^2)\,\frac{k^2_{\!\perp}}{k^2_{\!\perp} +
         k^2_h}
\label{IN}
\ee
with the soft and hard component
\bea
\!\!\!\!\!\!\!\!
{\cal F}_{\soft}(k_{\!\perp}^2) 
        & = &
       a_{\soft}\,C_F\,N_c\,\frac{\alphaS(k^2_{\!\perp})}{\pi}\,
       \frac{k^2_{\!\perp}}{\left(k^2_{\!\perp}+\mu_{\soft}^2\right)^2}\,
       \left [1-
         \left(1+\frac{3\,k^2_{\!\perp}}{\Lambda^2}\right)^{-2}\right ]
\label{F_IN_soft}\\
\!\!\!\!\!\!\!\!
{\cal F}_{\hard}(x,k_{\!\perp}^2) 
        & = &
      {\cal F}^{(B)}_{\pt}(k_{\!\perp}^2)\,\frac{{\cal
      F}_{\pt}(x,Q^2_c)}{{\cal
      F}_{\pt}^{(B)}(Q^2_c)}\,\Theta(Q^2_c-k^2_{\!\perp}) +
      {\cal F}_{\pt}(x,k^2_{\!\perp})\,\Theta(k^2_{\!\perp}-Q^2_c) 
\label{F_IN_hard}
\eea
where ${\cal F}^{(B)}_{\pt}(k_{\!\perp}^2)$ has the same form
as~(\ref{F_IN_soft}) with the parameters $a_{\pt}$ and $\mu_{\pt}$
instead of $a_{\soft}$ and $\mu_{\soft}$ and ${\cal
  F}_{\pt}(x,k^2_{\!\perp})$ is the derivative of the integrated gluon
distribution~(\ref{GRV}). In the IN approach, the running coupling is
given by $\alphaS(k^2_{\!\perp}) = \mbox{min}\{0.82,
(4\,\pi)/(\beta_0\,\log[k^2_{\!\perp}/\Lambda^2_{QCD}])\}$. With the
GRV-parametrization~\cite{Gluck:1998xa} for the integrated gluon
distribution, the structure function of the proton $F_2(x,Q^2)$ has
been described successfully using the following parameters: $k^2_s =
3\,\GeV^2$, $k^2_h = (1+0.0018\log(1/x)^4)^{0.5}$, $a_{\soft}=2$,
$a_{\pt}=1$, $\mu_{\soft}=0.1\,\GeV$, $\mu_{\pt}=0.75\,\GeV$, $Q^2_c =
0.895\,\GeV^2$, $\beta_0 = 9$, and $\Lambda_{QCD} = 0.2\,\GeV$.

\medskip

\begin{figure}[p!]
  \setlength{\unitlength}{1.cm}
\begin{center}
\epsfig{file=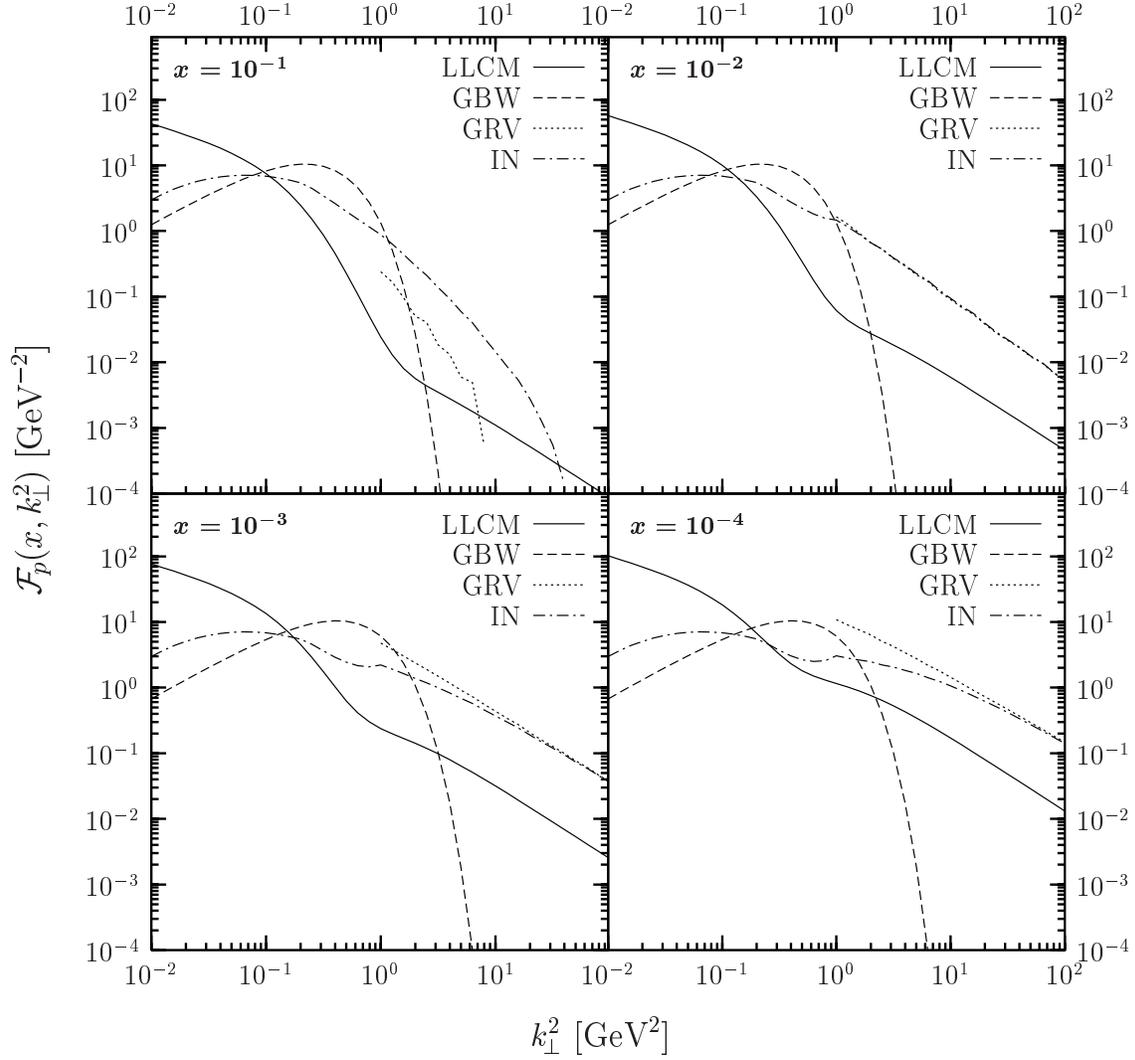,width=15.cm}
\end{center}
\caption{\small The unintegrated gluon distribution of the
  proton ${\cal F}_{p}(x,k_{\!\perp}^2)$ as a function of transverse
  momentum squared $k^2_{\!\perp}$ at Bjorken\,-\,$x$ values of
  $10^{-1}$, $10^{-2}$, $10^{-3}$, and $10^{-4}$. The different curves
  are obtained from our loop-loop correlation model (LLCM), the
  Golec-Biernat and W{\"u}sthoff (GBW)
  model~\cite{Golec-Biernat:1999qd}, the derivative of the Gl{\"u}ck,
  Reya, and Vogt (GRV) parametrization of
  $xG_p(x,Q^2)$~\cite{Gluck:1998xa}, and the Ivanov and Nikolaev (IN)
  approach~\cite{Ivanov:2000cm}.}
\label{fg_SSDP_GBW_IN_GRV_k2_4plots.eps}
\end{figure}
In Fig.~\ref{fg_SSDP_GBW_IN_GRV_k2_4plots.eps}, we show the LLCM, GBW,
GRV, and IN results for the unintegrated gluon distribution of the
proton ${\cal F}_{p}(x,k_{\!\perp}^2)$ as a function of transverse
momentum squared $k^2_{\!\perp}$ for $x=10^{-1}$, $10^{-2}$,
$10^{-3}$, and $10^{-4}$. At small transverse momenta,
$k_{\!\perp}^2\,\ltsim\,0.1\,\GeV^2$, our model gives the largest
values for ${\cal F}_{p}(x,k_{\!\perp}^2)$. As mentioned in the
previous section, our LLCM unintegrated gluon distribution increases
as $1/\sqrt{k_{\!\perp}^2}$ with decreasing $k_{\!\perp}^2$ as a
consequence of the linear increase of the total dipole-proton cross
section at large dipole sizes. In contrast, for $k_{\!\perp}^2 \to 0$,
the unintegrated gluon distribution of GBW decreases as
$k_{\!\perp}^2$ and the one of IN as $k_{\!\perp}^4$. In the
perturbative region, $k_{\!\perp}^2\,\gtsim\,1\,\GeV^2$, the
unintegrated gluon distribution of the LLCM becomes smaller than the
one of GRV and IN but is still larger than the one of GBW. Moreover,
the LLCM, GRV, and IN unintegrated gluon distributions become parallel
for $x\,\ltsim\,10^{-2}$ and drop as $1/k_{\!\perp}^2$ for large
$k_{\!\perp}^2$. This perturbative QCD behavior is not reproduced by
the GBW unintegrated gluon distribution which decreases exponentially
with increasing $k_{\!\perp}^2$.

\medskip

\begin{figure}[p!]
\setlength{\unitlength}{1.cm}
\begin{center}
\epsfig{file=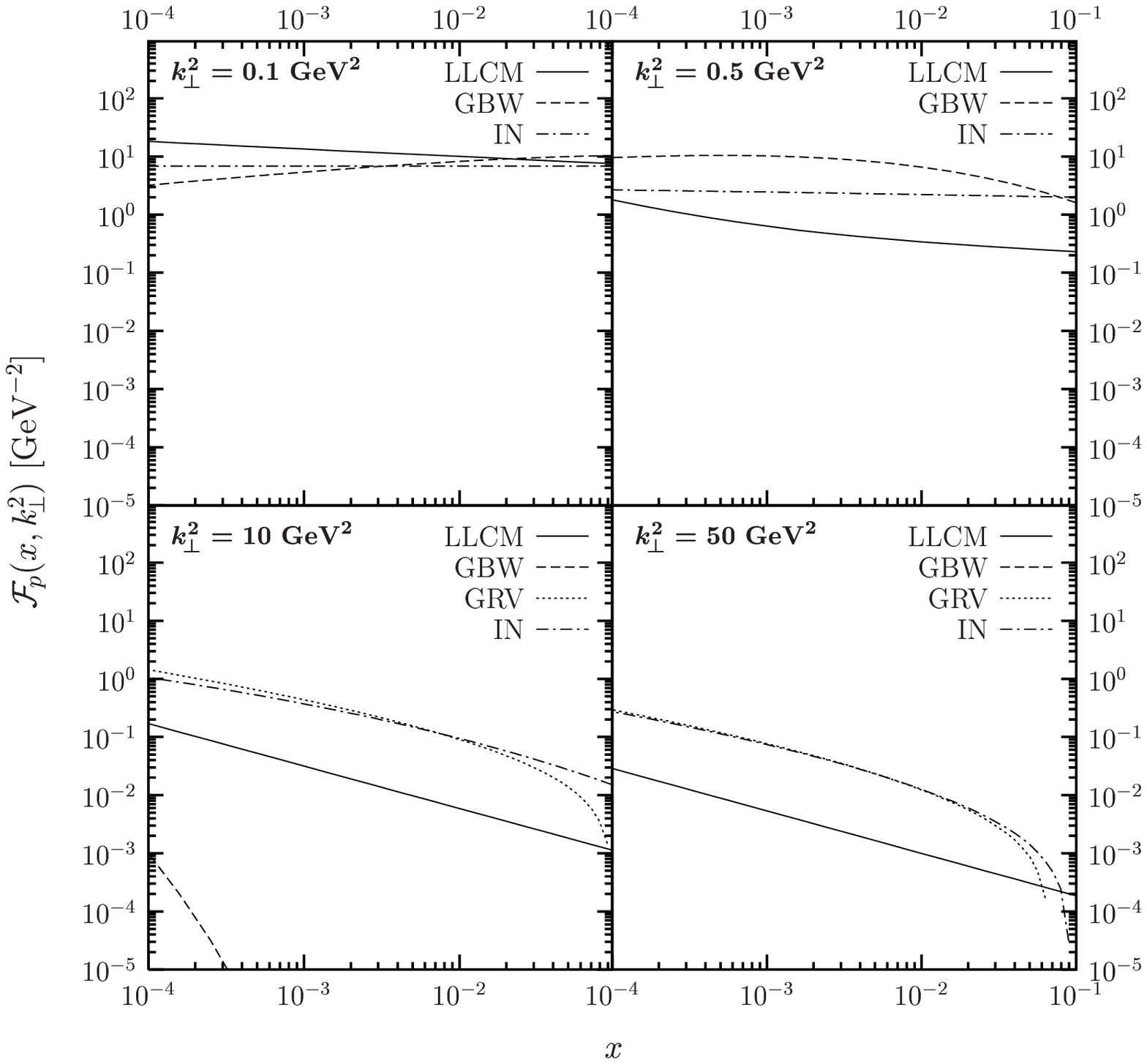,width=15.cm}
\end{center}
\caption{\small The unintegrated gluon distribution of the
  proton ${\cal F}_{p}(x,k_{\!\perp}^2)$ as a function of
  Bjorken\,-\,$x$ at transverse momenta squared $k^2_{\!\perp}=0.1$,
  $0.5$, $10$, and $50\,\GeV^2$. The different curves are obtained
  from our loop-loop correlation model (LLCM), the Golec-Biernat and
  W{\"u}sthoff (GBW) model~\cite{Golec-Biernat:1999qd}, the derivative
  of the Gl{\"u}ck, Reya, and Vogt (GRV) parametrization of
  $xG_p(x,Q^2)$~\cite{Gluck:1998xa}, and the Ivanov and Nikolaev (IN)
  approach~\cite{Ivanov:2000cm}. Note that the GRV parametrization is
  only available for $k^2_{\!\perp} \ge 0.8\,\GeV^2$. Moreover, the
  GBW result for ${\cal F}_{p}(x,k_{\!\perp}^2)$ is below $10^{-5}$
  for $k^2_{\!\perp} = 50\,\GeV^2$.}
\label{fg_SSDP_GBW_IN_GRV_x_4plots.eps}
\end{figure}
The $x$-dependence of the LLCM, GBW, GRV, and IN unintegrated gluon
distributions ${\cal F}_{p}(x,k_{\!\perp}^2)$ is shown for transverse
momenta squared $k^2_{\!\perp}=0.1$, $0.5$, $10$, and $50\,\GeV^2$ in
Fig.~\ref{fg_SSDP_GBW_IN_GRV_x_4plots.eps}.  In the non-perturbative
region, $k_{\!\perp}^2\,=\,0.1\,\GeV^2$ and
$k_{\!\perp}^2\,=\,0.5\,\GeV^2$, the LLCM, GBW, and IN unintegrated
gluon distributions show a weak $x$-dependence.  In the perturbative
region, $k_{\!\perp}^2\,=\,10\,\GeV^2$ and
$k_{\!\perp}^2\,=\,50\,\GeV^2$, the $x$-dependence of the unintegrated
gluon distributions becomes stronger. The LLCM, GRV, and IN
unintegrated gluon distributions show nearly the same rise with
decreasing $x$. In contrast, the GBW unintegrated gluon distribution
increases much faster as $x$ decreases.

In addition to the one-scale unintegrated gluon distributions, ${\cal
  F}(x,k_{\!\perp}^2)$, discussed in this work, there exist also
two-scale unintegrated gluon distributions, ${\cal
  F}(x,k_{\!\perp}^2,\mu^2)$. In the CCFM evolution
equation~\cite{Ciafaloni:1987ur+X}, this additional scale $\mu^2$ is
related to the maximum angle allowed in the gluon emission. Two-scale
unintegrated gluon distributions are obtained in the approach of
Bl\"umlein~\cite{Blumlein:1995eu}, Jung and
Salam~\cite{Jung:1998mi+X}, Kimber, Martin, and
Ryskin~\cite{Kimber:2001sc}, and in the linked dipole chain (LDC)
model~\cite{Andersson:1995ju+X,Gustafson:2002jy}. A comparison of
their results can be found in~\cite{Anderson:2002cf,Gustafson:2002jy}
where also the one-scale unintegrated gluon distributions of
Kwiecinski, Martin and Stasto~\cite{Kwiecinski:1997ee}, and Ryskin and
Shabelski~\cite{Ryskin:wz} are discussed.

%% file: Phenomenology.tex
%
\chapter[Impact Parameter Profiles and Gluon Saturation]{Impact
  Parameter Profiles \\ and Gluon Saturation}
\label{$S$-Matrix Unitarity and Gluon Saturation}

In this chapter we study saturation effects that manifest the
unitarity of the $S$-matrix at high c.m. energies. We investigate the
scattering amplitude in impact parameter space where the $S$-matrix
unitarity limit becomes most explicit. In fact, the $S$-matrix
unitarity imposes the {\em black disc limit} on the height of such
impact parameter profiles.  We show explicitly that our model respects
the black disc limit. Furthermore, the width of the impact parameter
profiles is shown to increase logarithmically at asymptotic energies
as needed to guarantee the Froissart bound~\cite{Froissart:1961ux+X}.

We compute the impact parameter profiles for hadron-hadron and
longitudinal photon-proton scattering. Concrete energy values are
determined at which the profiles saturate at the black disc limit for
small impact parameters. The impact parameter profiles provide an
intuitive geometrical picture for the energy dependence of the
scattering process as they illustrate the evolution of the opacity and
size of the interacting particles with increasing c.m. energy. We use
the explicit black disc limit of the profiles in the following chapter
to localize saturation effects in experimental observables.

Using a leading twist, next-to-leading order DGLAP relation, we
estimate the {\em impact parameter dependent gluon distribution} of
the proton $xG(x,Q^2,|\vec{b}_{\perp}|)$ from the profile function for
longitudinal photon-proton scattering. We find a low-$x$ saturation of
$xG(x,Q^2,|\vec{b}_{\perp}|)$ as a manifestation of the $S$-matrix
unitarity. The implications on the integrated gluon distribution
$xG(x,Q^2)$ are studied and compared with complementary investigations
of gluon saturation.

The results shown in this chapter are obtained with the $T$-matrix
element given in~(\ref{Eq_model_purely_imaginary_T_amplitude_final_result}), the model parameters explained in
Sec.~\ref{Sec_Model_Parameters}, and the hadron and photon wave
functions discussed extensively in Appendix~\ref{Sec_Wave_Functions}.

\section[$S$-Matrix Unitarity and Impact Parameter Profiles]%
{\hspace{-0.3cm}\boldmath$S$-\letterspace to 0.893\naturalwidth{Matrix Unitarity and Impact Parameter Profiles}}
\label{Sec_Impact_Parameter}

In this section we show that our model respects the $S$-matrix
unitarity limit in impact parameter space of the
scattering amplitude.

\hspace{-0.3cm}The impact parameter dependence of the scattering amplitude is given
by $\impactT(s,|\vec{b}_{\!\perp}|)$, \quad 
\be
        T(s,t=-{\vec q}_{\!\perp}^{\,\,2}) \;=\;
        4s\!\int \!\!d^2b_{\!\perp}\,
        e^{i {\vec q}_{\!\perp} {\vec b}_{\!\perp}}\,
        \impactT(s,|\vec{b}_{\!\perp}|)
\label{Eq_Fourier_transformed_T-matrix_element}
\ee
and in particular by the {\em profile function}
\be
        J(s,|\vec{b}_{\!\perp}|) 
        = 2\,\im\impactT(s,|\vec{b}_{\!\perp}|)
        \ ,
\label{Eq_profile_function_def}
\ee 
which describes the {\em blackness} or {\em opacity} of the
interacting particles as a function of the impact parameter $|{\vec
  b}_{\!\perp}|$ and the c.m.\ energy $\sqrt{s}$. In fact, the profile
function~(\ref{Eq_profile_function_def}) determines every observable
if the $T$-matrix is --- as in our model --- purely imaginary.

The $S$-matrix unitarity, $SS^{\dagger} = S^{\dagger}S = \Identity$,
leads directly to the {\em unitarity condition} in impact parameter
space~\cite{Amaldi:1976gr,Castaldi:1985ft}
\be
        \im\impactT(s,|\vec{b}_{\!\perp}|)
        = |\impactT(s,|\vec{b}_{\!\perp}|)|^2 + G_{inel}(s,|\vec{b}_{\!\perp}|)
        \ ,
\label{Eq_unitarity_condition}
\ee 
where $G_{inel}(s,|\vec{b}_{\!\perp}|) \ge 0$ is the inelastic overlap
function~\cite{VanHove:1964rp}.\footnote{Integrating
  (\ref{Eq_unitarity_condition}) over the impact parameter space and
  multiplying by a factor of $4$ one obtains the relation
  $\sigma^{tot}(s) = \sigma^{el}(s) + \sigma^{inel}(s)$.} This
unitarity condition imposes an absolute limit on the profile function
\be
        0 \;\;\leq\;\;
        2\,|\impactT(s,|\vec{b}_{\!\perp}|)|^2
        \;\;\leq\;\; 
        J(s,|\vec{b}_{\!\perp}|) 
        \;\;\leq\;\; 2
\label{Eq_absolute_unitarity_limit}
\ee 
and the inelastic overlap function, $G_{inel}(s,|\vec{b}_{\!\perp}|)
\le 1/4$.  At high energies, however, the elastic amplitude is
expected to be purely imaginary.  Consequently, the solution
of~(\ref{Eq_unitarity_condition}) reads
\be
        J(s,|\vec{b}_{\!\perp}|) = 1 \pm \sqrt{1-4\,G_{inel}(s,|\vec{b}_{\!\perp}|)}
\label{Eq_solution_unitarity_condition}
\ee
and leads with the minus sign corresponding to the physical situation
to the {\em reduced unitarity bound}
\be
        0 \;\;\leq\;\;
        J(s,|\vec{b}_{\!\perp}|) 
        \;\;\leq\;\; 1
        \ .
\label{Eq_reduced_unitarity_bound}
\ee 
Reaching the {\em black disc limit} or {\em maximum opacity} at a
certain impact parameter $|\vec{b}_{\!\perp}|$,
$J(s,|\vec{b}_{\!\perp}|) = 1$, corresponds to maximal inelastic
absorption $G_{inel}(s,|\vec{b}_{\!\perp}|) = 1/4$ and equal elastic
and inelastic contributions to the total cross section at that impact
parameter.

In our model, every reaction is reduced to dipole-dipole scattering
with well defined dipole sizes $|{\vec r}_i|$ and longitudinal quark
momentum fractions $z_i$. The unitarity condition in our model
becomes, therefore, most explicit in the profile function
\be
        J_{DD}(s,|\vec{b}_{\!\perp}|,z_1,|\vec{r}_1|,z_2,|\vec{r}_2|)  
        = \int \frac{d\phi_1}{2\pi}  \int \frac{d\phi_2}{2\pi} 
        \left[1 - S_{DD}(s,\vec{b}_{\!\perp},z_1,{\vec r}_1,z_2,{\vec r}_2)\right]
        \ ,
\label{Eq_DD_profile_function}
\ee
where $\phi_i$ describes the dipole orientation, i.e.\ the angle
between ${\vec r}_i$ and $\vec{b}_{\!\perp}$, and $S_{DD}$ describes
{\em elastic dipole-dipole scattering}
\be
        S_{DD}
        = \frac{2}{3} 
        \cos\!\left(\frac{1}{3}\chi^{\nprt}(s)\right)
        \cos\!\left(\frac{1}{3}\chi^{\pert}(s)\right)         
        + \frac{1}{3}
        \cos\!\left( \frac{2}{3}\chi^{\nprt}(s)\right)
        \cos\!\left( \frac{2}{3}\chi^{\pert}(s)\right)
\label{Eq_S_DD_final_result}
\ee
with the purely real-valued eikonal functions $\chi^{\nprt}(s)$ and
$\chi^{\pert}(s)$ defined in~(\ref{Eq_energy_dependence}). Because of
$|S_{DD}| \leq 1$, a consequence of the cosine functions
in~(\ref{Eq_S_DD_final_result}) describing multiple gluonic
interactions, $J_{DD}$ respects the absolute
limit~(\ref{Eq_absolute_unitarity_limit}). Thus, the elastic
dipole-dipole scattering respects the unitarity
condition~(\ref{Eq_unitarity_condition}).  At high energies, the
arguments of the cosine functions in $S_{DD}$ become so large that
these cosines average to zero in the integration over the dipole
orientations. This leads to the black disc limit $J_{DD}^{max} = 1$
reached at high energies first for small impact parameters.

If one considers the scattering of two dipoles with fixed orientation,
the inelastic overlap function obtained from the unitarity
constraint~(\ref{Eq_unitarity_condition}),
\bea
        &&\!\!\!\!\!\!\!\!
        G_{inel}^{DD}(s,|\vec{b}_{\!\perp}|) 
\\
        &&\!\!\!\!\!\!\!\!
        = \inv{4} \left( 1 - \left[
            \frac{2}{3} \cos\!\left(\frac{1}{3}\chi^{\nprt}(s)\right)
            \cos\!\left(\frac{1}{3}\chi^{\pert}(s)\right) + \frac{1}{3}
            \cos\!\left( \frac{2}{3}\chi^{\nprt}(s)\right) \cos\!\left(
              \frac{2}{3}\chi^{\pert}(s)\right) \right]^2\right) \ , 
\nonumber
\eea
shows nonphysical behavior with increasing energy. This behavior is a
consequence of aritifically fixing the orientations of the dipoles. If
one averages over the dipole orientations as in all high-energy
reactions considered in this work, no unphysical behavior is observed.
\vspace{-0.5cm}
\section[Profile Function for Hadron-Hadron Scattering]
{\letterspace to .94\naturalwidth{Profile Function for Hadron-Hadron Scattering}}
\label{Sec_PP_Profile_Function}
In this section we show the c.m. energy and impact parameter
dependence of the hadron-hadron scattering amplitude, determine the
energy values at which saturation effects set in as a manifestation of
the $S$-matrix unitarity, and compute the logarithmic rise of the
profile width at asymptotic c.m. energies which is equivalent to the
Froissart bound~\cite{Froissart:1961ux+X}.

The profile function for hadron-hadron (hh) scattering
\be
        J_{hh}(s,|\vec{b}_{\!\perp}|) \!=\! 
        \int \!\!dz_1 d^2r_1 \!\int \!\!dz_2 d^2r_2      
        |\psi_h(z_1,\vec{r}_1)|^2 |\psi_h(z_2,\vec{r}_2)|^2\!
        \left[1-S_{DD}(s,\vec{b}_{\!\perp},z_1,{\vec r}_1,z_2,{\vec r}_2)\right]
\label{Eq_model_pp_profile_function}
\ee
is obtained from~(\ref{Eq_DD_profile_function}) by weighting the
dipole sizes $|{\vec r}_i|$ and longitudinal quark momentum fractions
$z_i$ with the hadron wave functions $|\psi_h(z_i,\vec{r}_i)|^2$ from
Appendix~\ref{Sec_Wave_Functions}.


The profile functions for proton-proton, pion-proton, and kaon-proton
scattering are shown in Fig.~\ref{Fig_J_pp(b,s)} for c.m.\ energies
$\sqrt{s}$ from $10\,\GeV$ to $10^8\,\GeV$. The profiles are obtained with
model parameters from Sec.~\ref{Sec_Model_Parameters} and the hadron
extensions $S_p = 0.86\,\fm$, $S_{\pi} = 0.607\,\fm$, and $S_{K} =
0.55\,\fm$ used in the Gaussian hadron wave functions. Up to $\sqrt{s}
\approx 100\,\GeV$, the profiles have approximately a Gaussian shape.
Above $\sqrt{s}=1\,\TeV$, they significantly develop into broader and
higher profiles until the black disc limit is reached for $\sqrt{s}
\approx 10^6\,\GeV$ and $|\vec{b}_{\!\perp}|=0$.  At this point, the
cosine functions in $S_{DD}$ average to zero
\be
        \int \!\!dz_1 d^2r_1 \!\int \!\!dz_2 d^2r_2  
        |\psi_h(z_1,{\vec r}_1)|^2|\psi_h(z_2,{\vec r}_2)|^2
        S_{DD}(\sqrt{s}\gtsim10^6\,\GeV,|\vec{b}_{\!\perp}|=0,\dots) 
        \approx 0
\label{Eq_pp_black_disc_limit_explained}
\ee
so that the hadron wave function normalizations determine the maximum
opacity
\be
        J_{hh}^{max}
        =\int \!\!dz_1 d^2r_1 \!\int \!\!dz_2 d^2r_2\,        
        |\psi_h(z_1,{\vec r}_1)|^2\,|\psi_h(z_2,{\vec r}_2)|^2
        = 1
        \ .
\label{Eq_pp_black_disc_limit}
\ee
Once the maximum opacity is reached at a certain impact parameter, the
profile function saturates at that $|\vec{b}_{\!\perp}|$ and extends
towards larger impact parameters with increasing energy. Thus, the
multiple gluonic interactions important to respect the $S$-matrix
unitarity constraint~(\ref{Eq_unitarity_condition}) lead to saturation
for $\sqrt{s} \gtsim 10^6\,\GeV$.
\befig[h!]  
\vspace{-0.2cm}
\begin{center}
  \epsfig{figure=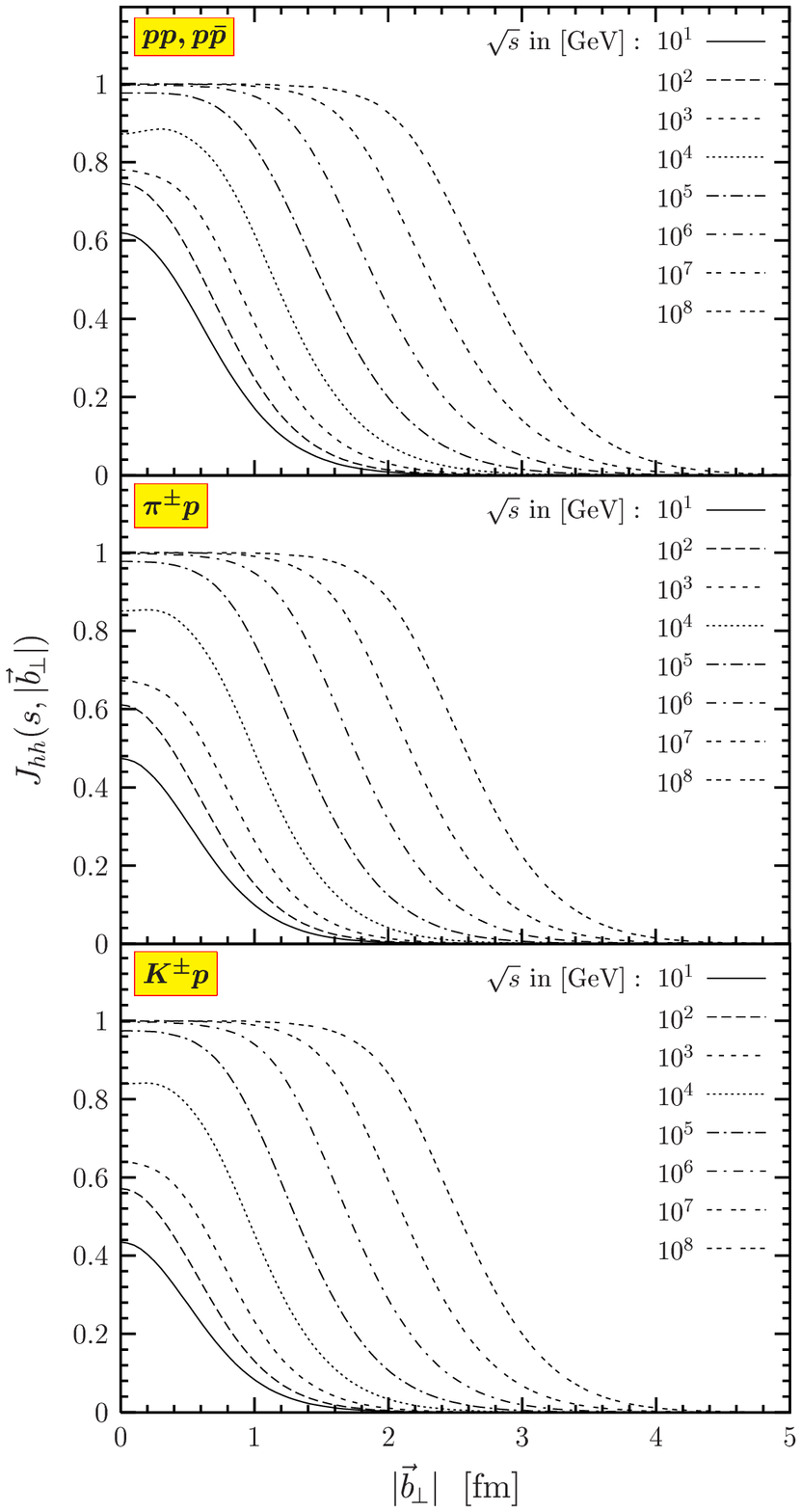,
    width=10.4cm}
\end{center} 
\vspace{-0.8cm} 
\protect\caption{\small The profile function for proton-proton,
  pion-proton, and kaon-proton scattering
  $J_{hh}(s,|\vec{b}_{\!\perp}|)$ is shown versus the impact
  parameter $|\vec{b}_{\!\perp}|$ for c.m.\ energies $\sqrt{s}$ from
  $10\,\GeV$ to $10^8\,\GeV$. The unitarity
  limit~(\ref{Eq_absolute_unitarity_limit}) corresponds to
  $J_{hh}(s,|\vec{b}_{\!\perp}|) = 2$ and the black disc
  limit~(\ref{Eq_reduced_unitarity_bound}) to
  $J_{hh}(s,|\vec{b}_{\!\perp}|) = 1$.}
\label{Fig_J_pp(b,s)}
\vspace{-1.2cm}
\efig
%
The above behavior of the profile functions illustrates the evolution
of the hadron with increasing c.m.\ energy. The hadron is gray and of
small transverse size at small $\sqrt{s}$ but becomes blacker and more
transversally extended with increasing $\sqrt{s}$ until it reaches the
black disc limit in its center at $\sqrt{s} \approx 10^6\,\GeV$.
Beyond this energy, the hadron cannot become blacker in its central
region but in its periphery with continuing transverse growth.
Furthermore, the hadron boundary seems to stay diffusive as claimed also
in~\cite{Frankfurt:2001nt+X}.


The differences between the profile functions for $pp$ ($p\pbar$),
$\pi^{\pm}p$, and $K^{\pm}p$ scattering especially at low energies
result from the different transverse hadron extensions, $S_p =
0.86\,\fm > S_{\pi} = 0.607\,\fm > S_{K} = 0.55\,\fm$, cf.\ 
Appendix~\ref{Sec_Wave_Functions}.  For the smaller pion and kaon
extensions the short distance physics described by the perturbative
component becomes more important and leads to a stronger energy
increase ($\epsilon^{\pert} = 0.73 > \epsilon^{\nprt} = 0.125$) of the
$\pi^{\pm}p$ and $K^{\pm}p$ profile functions as compared to the one
of the $pp$ ($p\pbar$) scattering in the c.m. energy range $10\,\GeV
\le \sqrt{s} \le 10^6\,\GeV$.  The black disc
limit~(\ref{Eq_pp_black_disc_limit}) is identical for $\pi p$, $K p$,
and $pp$ ($p\pbar$) scattering due to the same pion, kaon, and proton
wave function
normalization~(\ref{Eq_hadron_wave_function_normalization}). The
smaller size of pions and kaons in comparison to the one of protons
demands, however, slightly higher c.m.\ energies to reach the black
disc limit and yields narrower $\pi^{\pm}p$ and $K^{\pm}p$ profile
widths for energies somewhat larger than $\sqrt{s} = 10^6\,\GeV$. At
asymptotic energies, however, the profile widths become universal
(hadron size independent) as shown below.

\vspace{0.25cm}

According to our model the black disc limit will not be reached at
LHC. Our prediction of $\sqrt{s} \approx 10^6\,\GeV = 10^3\,\TeV$ for
the onset of the black disc limit in proton-proton collisions is about
two orders of magnitude beyond the LHC energy $\sqrt{s} = 14\,\TeV$.
This is in contrast, for example, with~\cite{Desgrolard:1999pr}, where
the value predicted for the onset of the black disc limit is $\sqrt{s}
= 2\,\TeV$, i.e.\ small enough to be reached at LHC. However, we feel
confidence in our LHC prediction since our profile function
$J_{pp}(s,|\vec{b}_{\!\perp}|)$ yields good agreement with
experimental data for cross sections up to the highest energies as
shown in chapter~\ref{Sec_Comparison_Data}.

\vspace{0.25cm}

The discussion in the previous section, the black disc limit in
Eq.~(\ref{Eq_pp_black_disc_limit}), and Fig.~\ref{Fig_J_pp(b,s)} show
that our model respects the $S$-matrix unitarity in impact parameter
space of the scattering amplitude. This, however, is not enough to
guarantee the Froissart bound~\cite{Froissart:1961ux+X} for total
hadronic cross sections at asymptotic c.m.  energies. Since the latter
are obtained by integrating the impact parameter profiles over all
impact parameters, difficulties with the Froissart bound can arise if
the black disc radius $R(s)$, i.e., the impact parameter range for
which $J_{hh}(s,|\vec{b}_{\!\perp}|) \approx 1$, increases with
growing c.m. energy faster than $\ln(s)$.  A calculation of the energy
dependence of $R(s)$, which agrees with the numerical results shown in
chapter~\ref{Sec_Comparison_Data}, shows below that the Froissart
bound is respected in our model.

\vspace{0.25cm}

For impact parameters much larger than the dipole sizes involved in
the interaction, $|\vec{b}_{\!\perp}| \gg |\vec{r}_1|$, $|\vec{r}_2|$,
the leading term of the $\chi^{\pert}$-function~(\ref{Eq_chi_PGE})
which is needed in the computation of $R(s)$ reads
\bea
        \chi^{\pert}(s, \vec{r}_1, \vec{r}_2, \vec{b}_{\!\perp}) 
        &\!\!\!\!\!=\!\!\!\!\!&
        \left(\frac{s}{s_0} 
        \frac{\vec{r}_1^{\,2}\,\vec{r}_2^{\,2}}{R_0^4}
        \right)^{\!\epsilon^{\pert}/2}\!
        \frac{g^2}{2\pi} \sqrt{\frac{\pi}{2}} 
        \frac{e^{-m_G\,|\vec{b}_{\perp}|}}{\sqrt{m_G\,|\vec{b}_{\!\perp}|}}
        \left[\!
        \left( \inv{4|\vec{b}_{\!\perp}|^2} + 
        \frac{m_G}{2|\vec{b}_{\!\perp}|} \right)\!
        |\vec{r}_1||\vec{r}_2| \cos(\phi) \right.
        \nonumber \\
        && \left. \hspace{2.3cm}
        -
        \left( \frac{5}{4|\vec{b}_{\!\perp}|^2} + 
        \frac{2m_G}{|\vec{b}_{\!\perp}|} + m_G^2 \right) 
        |\vec{r_1}||\vec{r_2}| \cos(\phi_1) \cos(\phi_2) 
        \right] \ , \nonumber \\
\label{chi_p_analytic}    
\eea 
where $\phi$ denotes the angle betweeen the dipoles $\vec{r}_1$ and
$\vec{r}_2$ and $\phi_i$ the angle between the impact parameter
$\vec{b}_{\!\perp}$ and dipole $\vec{r}_i$. To obtain the above result
we have used the longitudinal momentum fractions $z_1=z_2=0.5$. These
values, however, do not affect the asymptotic behaviour of $R(s)$.

With the above energy and impact parameter dependence of the
$\chi^{\pert}$-function, the black disc radius $R(s)$ can now be
obtained as follows: For c.m.  energies $\sqrt{s} \gg 10^6\,\GeV$,
perturbative correlations $\chi^{\pert}$ dominate the dipole-dipole
scattering amplitude $S_{DD}$ in~(\ref{Eq_model_pp_profile_function})
because of the stronger energy dependence ($\epsilon^{\pert} = 0.73 >
\epsilon^{\nprt} = 0.125$) as compared with non-perturbative
correlations $\chi^{\nprt}$.  Consequently, $\cos(c \chi^{\pert})$
oscillates much more than $\cos(c \chi^{\nprt})$ in the integration
over the dipole sizes and orientations ($c =1/3$ or $2/3$), i.e., only
$\cos(c\chi^{\pert})$ needs to be considered here.  The condition for
the black disc radius $R(s)$ results from the fact that for
$|\vec{b}|\le R(s)$ the functions $\cos(c\chi^{\pert})$ have to
average to zero when integrated over the dipole orientations in the
profile function~(\ref{Eq_model_pp_profile_function}). This is the
case when the main contribution to the
$\chi^{\pert}$-function~(\ref{chi_p_analytic}) at large impact
parameters fulfilles the relation
\be
        \left(\frac{s}{s_0} 
        \frac{\vec{r}_1^{\,2}\,\vec{r}_2^{\,2}}{R_0^4}
        \right)^{\epsilon^{\pert}/2}
        \frac{g^2}{2\pi}\,\sqrt{\frac{\pi}{2}}\, 
        \frac{e^{-m_G\,R(s)}}{\sqrt{m_G\,R(s)}}
        \,m_G^2\,|\vec{r_1}|\,|\vec{r_2}|\  = C_0 \ \gg \ \pi \ . \nonumber \\
\label{chi_p_R(s)}    
\ee 
Taking the logarithm on both sides of Eq~(\ref{chi_p_R(s)}), one
obtaines for the leading contribution to the black disc radius $R(s)$
at ultra-high energies
\be
   R(s) = 
     \frac{\epsilon^{\pert}}{2\,m_G}\,\ln\left(\frac{s}{{\bar
   s}_0}\right) 
     = \frac{\epsilon^{\pert}}{M^P_{GB}}\,\ln\left(\frac{s}{{\bar s}_0}\right)\ ,
\label{Eq_R(s)}
\ee
where ${\bar s}_0$ denotes the reference energy, $m_G$ the gluon
mass, and $M^P_{GB} = 2 m_G$ the ``glueball mass'' obtained from our
perturbative contribution~\cite{Shoshi:2002rd}.
The logarithmic increase of $R(s)$ is equivalent to the Froissart
bound~\cite{Froissart:1961ux+X} since $\sigma^{tot}_{hh}(s) = 2 \pi
R(s)^2$ at asymptotic energies (see the following chapter).

For the derivation of the Froissart bound the powerlike energy
dependence $s^{\epsilon^{\pert}}$ as well as the exponential fall off
$\exp(-m_G\,b)$ are crucial. This provides an a posteriori
justification for the powerlike energy ansatz used in this work. The
exponential $|\vec{b}_{\perp}|$ - decrease exists because of the
non-zero gluon mass $m_G$ which has been introduced in the gluon
propagator as a cutoff to suppress the effect of perturbative physics in the
non-perturbative region of small momenta.

The coefficient $\epsilon^{\pert}/(2 m_G)$ of the $\ln(s)$ term
contains the energy exponent $\epsilon^{\pert}$ of the perturbative
contribution and the semiperturbative parameter $m_G$ which determines
the interplay between perturbative and non-perturbative physics. This
coefficient is {\em universal} for all hadron interactions computed
within our model. Different coefficients have been used in the
literature: Lukaszuk and Martin~\cite{LukaszukMartin:1967} have used
the factor $\sqrt{\pi}/m_{\pi}$ ($m_{\pi}$ is the pion mass),
Heisenberg~\cite{Heisenberg:1952} has obtained by considering the
dynamics of the scattering $\sqrt{\pi}/(2 m_{\pi})$, Ferreiro et
al.~\cite{Ferreiro:2002kv} have deduced from the Color Glass
Condensate model the result $\sqrt{\pi/2}\,(4 \ln(2) \alphaS N_c)/(\pi
m_{\pi})$ with the BFKL-exponent $4 \ln(2) \alphaS N_c/\pi$, and Dosch
et al.~\cite{Dosch:2002pg} have updated Heisenberg's approach to get
$\sqrt{\pi}/(2 M_{GB})$. So far the coefficient of the $\ln(s)$ term
is not conclusive.

A universal black disc radius $R(s)$ means that all hadrons are of the
same size at asymptotic energies. This behavior starts becoming
visible in the profile functions shown in Fig.~\ref{Fig_J_pp(b,s)} at
the highest energies. The universal increase exhibits itself most
explicitly in total hadronic cross sections which we discuss
extensively in chapter~\ref{Sec_Comparison_Data}.

\section[Profile Function for Photon-Proton Scattering]
{\letterspace to .94\naturalwidth{Profile Function for Photon-Proton Scattering}}
\label{Sec_GP_Profile_Function}
In this section we compute the photon-proton scattering amplitude
which is very appropriate to study the dependence of profile functions
on the size of the scattered particles by varying the photon
virtuality.

The profile function for a longitudinal photon $\gamma_L^*$ scattering
off a proton $p$
\bea
        J_{\gamma_L^* p}(s,|\vec{b}_{\!\perp}|,Q^2)  
        & = & 
        \int \!\!dz_1 d^2r_1 \!\int \!\!dz_2 d^2r_2\,
        |\psi_{\gamma_L^*}(z_1,\vec{r}_1,Q^2)|^2 \,
        |\psi_p(z_2,\vec{r}_2)|^2
        \nonumber \\
        && 
        \times
        \left[1-S_{DD}(\vec{b}_{\!\perp},s,z_1,{\vec r}_1,z_2,{\vec r}_2)\right]
\label{Eq_model_gp_profile_function}
\eea
is calculated with the longitudinal photon wave function
$|\psi_{\gamma_L^*}(z_i,\vec{r}_i,Q^2)|^2$ given
in~(\ref{Eq_photon_wave_function_L_squared}). In this way, the profile
function~(\ref{Eq_model_gp_profile_function}) is ideally suited for
the investigation of dipole size effects since the photon virtuality
$Q^2$ determines the transverse size of the dipole~(\ref{rQ}) into which the
photon fluctuates before it interacts with the proton.

Figure~\ref{Fig_J_gp_(b,s,Q^2)} shows the $|\vec{b}_{\!\perp}|$
dependence of the profile function $J_{\gamma_L^*
  p}(s,|\vec{b}_{\!\perp}|,Q^2)$ divided by $\alphaEM/{\pi}$ for c.m.\ 
energies $\sqrt{s}$ from $10\,\GeV$ to $10^9\,\GeV$ and photon
virtualities $Q^2 = 1, 10$, and $100\,\GeV^2$, where $\alphaEM$ is the
fine-structure constant.
\befig[h!] 
\vspace{-0.2cm}
\begin{center}
\epsfig{figure=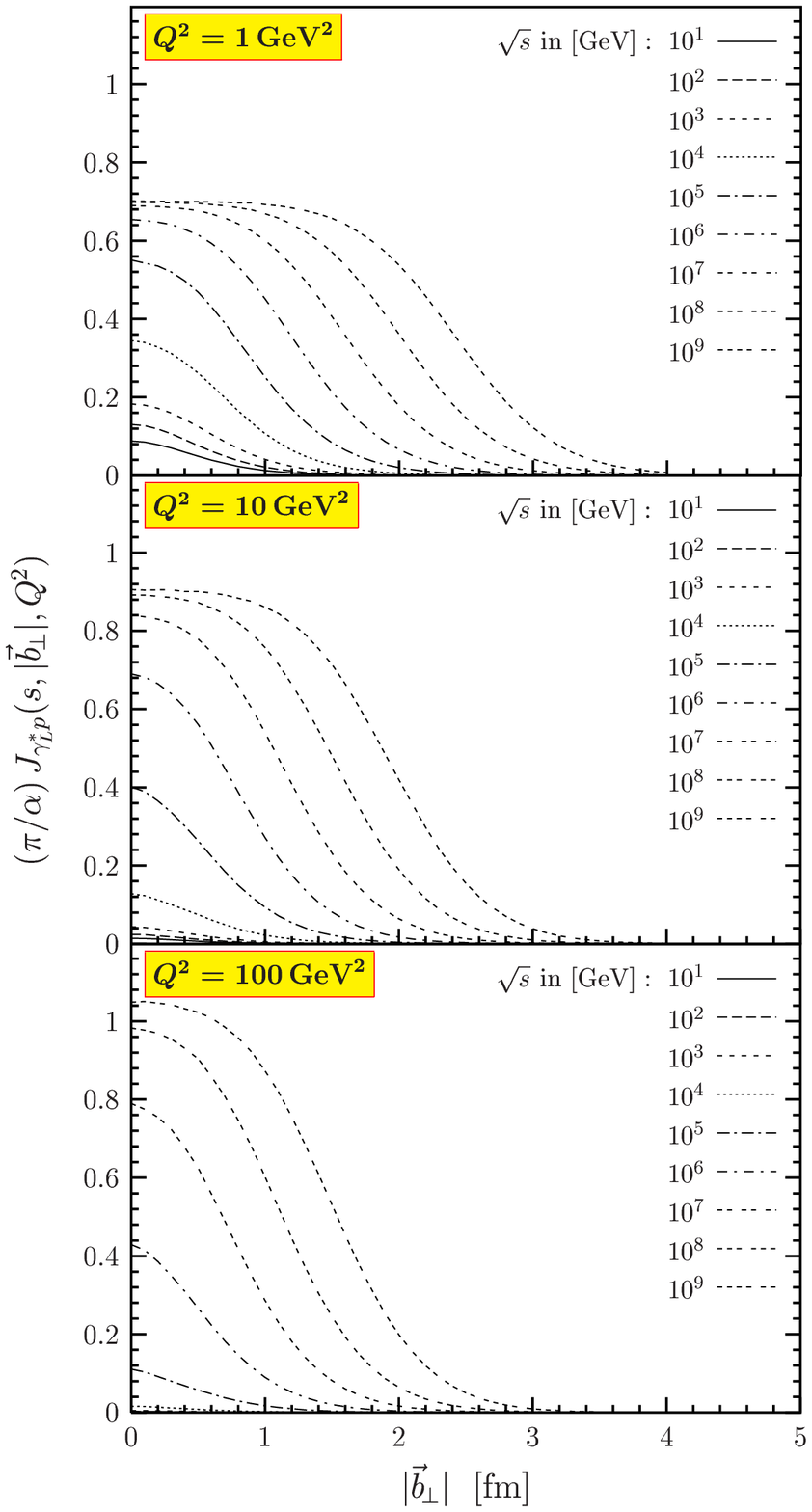, width = 10.4cm}
\end{center}
\vspace{-0.8cm} 
\protect\caption{\small The profile function for a longitudinal photon
  scattering off a proton $J_{\gamma_L^*
    p}(s,|\vec{b}_{\!\perp}|,Q^2)$ divided by $\alphaEM/{\pi}$ is
  shown versus the impact parameter $|\vec{b}_{\!\perp}|$ for photon
  virtualities $Q^2 = 1, 10$, and $100\,\GeV^2$ and c.m.\ energies
  $\sqrt{s}$ from $10\,\GeV$ to $10^9\,\GeV$. The value of the black disc
  limit $J_{\gamma_L^*p}^{max}(Q^2)$ and the width of the profiles depend on $Q^2$.}
\label{Fig_J_gp_(b,s,Q^2)}
\vspace{-1.2cm}
\end{figure}
One clearly sees that the qualitative behavior of this rescaled
profile function is similar to the one for hadron-hadron scattering.
However, the black disc limit induced by the underlying dipole-dipole
scattering depends on the photon virtuality $Q^2$ and is given by the
normalization of the longitudinal photon wave function
\bea
        J_{\gamma_L^* p}^{max}(Q^2)  
        = \int \!\!dz d^2r |\psi_{\gamma_L^*}(z,\vec{r},Q^2)|^2
\label{Eq_gp_black_disc_limit}
\eea
since the proton wave function is normalized to one.

The photon virtuality $Q^2$ does not only determine the absolute value
of the black disc limit and the c.m.\ energy at which it is reached
but also the width of the profiles at fixed $\sqrt{s}$ as illustrated
in Fig.~\ref{Fig_J_gp_(b,s,Q^2)}. With increasing resolution $Q^2$,
i.e.\ decreasing dipole sizes, $|\vec{r}_{\gamma_L^*}|^2 \propto
1/Q^2$, the absolute value of the black disc limit grows and higher
energies are needed to reach this limit.\footnote{Note that the
  Bjorken $x$ at which the black disc limit is reached decreases with
  increasing photon virtuality $Q^2$.  (See also
  Fig.~\ref{Fig_xg(x,Q^2,b=0)_vs_x})} This can be seen most explicitly
in Fig.~\ref{Fig_J_gp_(b=0,s,Q^2)} where the $\sqrt{s}$ dependence of
$J_{\gamma_L^* p}(s,|\vec{b}_{\!\perp}|=0,Q^2)$ divided by
$\alphaEM/{\pi}$ is presented for $Q^2 =
1,\,10,\,\mbox{and}\,100\,\GeV^2$. 
\befig[htb]
\centerline{\epsfig{figure=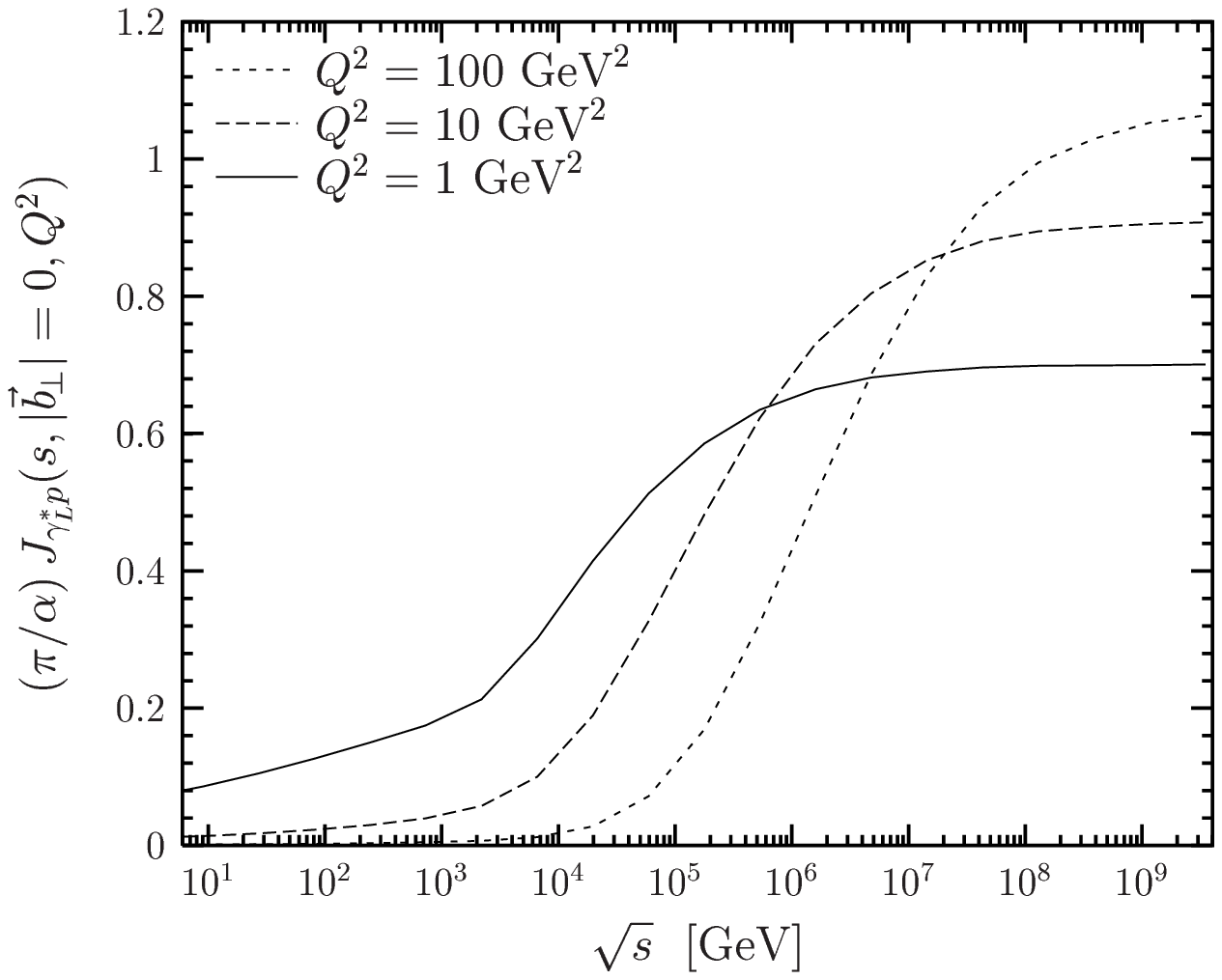}}
\protect\caption{\small The profile function for a longitudinal photon
  scattering off a proton $J_{\gamma_L^*p}(s,|\vec{b}_{\!\perp}|,Q^2)$
  divided by $\alphaEM/{\pi}$ is shown versus the c.m.\ energy
  $\sqrt{s}$ at zero impact parameter ($|\vec{b}_{\!\perp}|=0$) for
  photon virtualities $Q^2 = 1,\,10,\,\mbox{and}\,100\,\GeV^2$.}
\label{Fig_J_gp_(b=0,s,Q^2)}
\end{figure}
The growth of the absolute value of
the black disc limit is simply due to the normalization of the
longitudinal photon wave function while the requirement of higher
energies to reach this limit is due to the decreasing interaction
strength with decreasing dipole size. The latter explains also why the
energies needed to reach the black disc limit in $\pi^{\pm} p$ and
$K^{\pm} p$ scattering are slightly higher than in $pp\ (\bar{p}p)$
scattering. In comparison to the hadron-hadron profile function,
Fig.~\ref{Fig_J_gp_(b,s,Q^2)} shows explicitly that the width of
the $\gamma_L^*p$ profiles at energies somewhat beyond the black disc limit
decreases with decreasing photon size $|\vec{r}_{\gamma_L^*}|^2
\propto 1/Q^2$.  Furthermore, comparing $\gamma_L^*p$ scattering at
$Q^2=1\,\GeV^2$ with $pp$ scattering quantitatively, the black disc
limit $J_{\gamma_L^*p}^{max}(Q^2=1\,\GeV^2) = 0.00164$ is about three
orders of magnitude smaller because of the photon wave function
normalization ($\propto \alphaEM/{\pi}$). At $|\vec{b}_{\!\perp}|=0$
it is reached at an energy of $\sqrt{s} \approx 10^8\,\GeV$, which is
about two orders of magnitude higher because of the smaller dipoles
involved.

The way in which the profile function $J_{\gamma_L^*
  p}(s,|\vec{b}_{\!\perp}|,Q^2)$ approaches the black disc limit at
high energies depends on the shape of the proton and longitudinal
photon wave function at small dipole sizes $|\vec{r}_{1,2}|$. At high
energies, dipoles of typical sizes $0 \leq |\vec{r}_{1,2}| \leq
R_0\,(s_0/s)^{1/4}$ give the main contribution to
$S_{\gamma_L^*p} = 1 - J_{\gamma_L^*p}$ because
of~(\ref{Eq_energy_dependence}) and the fact that the contribution of
the large dipole sizes averages to zero upon integration over the
dipole orientations,
cf.~also~(\ref{Eq_pp_black_disc_limit_explained}). Since
$S_{\gamma_L^*p}$ is a measure of the proton transmittance, this means
that only small dipoles can penetrate the proton at high energies.
Increasing the energy further, even these small dipoles are absorbed
and the black disc limit is reached. However, the dependence of the
profile function on the short distance behavior of normalizable wave functions is
weak which can be understood as follows. Because of color
transparency, the eikonal functions $\chi^{\nprt}(s)$ and
$\chi^{\pert}(s)$ are small for small dipole sizes $0 \leq
|\vec{r}_{1,2}| \leq R_0\,(s_0/s)^{1/4}$ at large
$\sqrt{s}$. Consequently, $S_{DD} \approx 1$ and
\bea
        \!\!\!\!\!\!\!\!
        && 
        \!\!\!\!\!\!\!\!\!\!\!\!\!\!\!\!
        J_{\gamma_L^* p}(s,|\vec{b}_{\!\perp}|,Q^2)  
\nonumber \\
        \!\!\!\!\!\!\!\!
        &&
        \!\!\!\!\!\!\!\!\!\!\!\!\!\!\!\!
         \approx 
        J_{\gamma_L^* p}^{max}(Q^2) - 4\pi^2\!\!
        \int\limits_0^1 \!\!dz_1 \!\!
        \int\limits_0^{r_c(s)}\!\!\!dr_1 r_1 
        |\psi_{\gamma_L^*}(z_1,r_1,Q^2)|^2 
        \int\limits_0^1 \!\!dz_2 \!\!
        \int\limits_0^{r_c(s)}\!\!\!dr_2 r_2 
        |\psi_p(z_2,r_2)|^2
\label{Eq_model_gp_profile_function_wavefunction_independence}
\eea
where $r_c(s) \approx R_0\,(s_0/s)^{1/4}$. Clearly, the
linear behavior from the phase space factors $r_{1,2}$ dominates over
the $r_{1,2}$-dependence of normalizable wave functions.\footnote{For our
  choice of the wave functions
  in~(\ref{Eq_model_gp_profile_function_wavefunction_independence}),
  one sees very explicitly that the specific Gaussian behavior of
  $|\psi_p(z_2,r_2)|^2$ and the logarithmic short distance behavior of
  $|\psi_{\gamma_L^*}(z_1,r_1,Q^2)|^2$ is dominated by the phase space
  factors $r_{1,2}$.} More generally, for any profile function
involving normalizable wave functions, the way in which the black disc
limit is approached depends only weakly on the short distance behavior
of the wave functions.
\vspace*{2cm}
%
\section{A Scenario for Gluon Saturation}
\label{Sec_Gluon_Saturation}

In this section we estimate the {\em impact parameter dependent gluon
  distribution} of the proton $xG(x,Q^2,|\vec{b}_{\!\perp}|)$.  Using
a leading twist, next-to-leading order QCD relation between
$xG(x,Q^2)$ and the longitudinal structure function $F_L(x,Q^2)$, we
relate $xG(x,Q^2,|\vec{b}_{\!\perp}|)$ to the profile function
$J_{\gamma_L^* p}(s=Q^2/x,|\vec{b}_{\!\perp}|,Q^2)$ and find low-$x$
saturation of $xG(x,Q^2,|\vec{b}_{\!\perp}|)$ as a manifestation of
$S$-matrix unitarity. The resulting $xG(x,Q^2,|\vec{b}_{\!\perp}|)$
is, of course, only an estimate since our profile function contains
also higher twist contributions. Furthermore, in the considered low-$x$
region, the leading twist, next-to-leading order QCD formula may be
inadequate as higher twist contributions~\cite{Martin:1998kk+X} and
higher order QCD corrections~\cite{Gribov:1983tu,Mueller:1986wy} are
expected to become important. Nevertheless, still assuming a close
relation between $F_L(x,Q^2)$ and $xG(x,Q^2)$ at low $x$, we think
that our approach provides some insight into the gluon distribution as
a function of the impact parameter and into its saturation.

The {\em gluon distribution}\ of the proton $~xG(x,Q^2)~$ has the
following meaning: $xG(x,Q^2)dx$ gives the momentum fraction of the
proton which is carried by the gluons in the interval $[x, x+dx]$ as
seen by probes of virtuality $Q^2$. The {\em impact parameter
  dependent gluon distribution} $xG(x,Q^2,|\vec{b}_{\!\perp}|)$ is the
gluon distribution $xG(x,Q^2)$ at a given impact parameter
$|\vec{b}_{\!\perp}|$ so that
\be
        xG(x,Q^2) = \int
        \!\!d^2b_{\!\perp}\,xG(x,Q^2,|\vec{b}_{\!\perp}|) \ .
\label{Eq_def_xg(x,Q^2)}
\ee

In leading twist, next-to-leading order QCD, the gluon distribution
$xG(x,Q^2)$ is related to the structure functions $F_L(x,Q^2)$ and
$F_2(x,Q^2)$ of the proton~\cite{Martin:1988vw}
\be
        F_L(x, Q^2) 
        = \frac{\alphaS}{\pi}\!
        \left[
        \frac{4}{3}\int_x^1 \!
        \frac{dy}{y}\!\left(\frac{x}{y}\right)^{\!\!2} \!F_2(y,Q^2)
        + 2 \sum_f e_f^2\!\int_x^1 \!
        \frac{dy}{y}\!\left(\frac{x}{y}\right)^{\!\!2} \!\!
        \left(\!1-\frac{x}{y}\right) yG(y,Q^2)
        \right]
\label{Eq_FL_QCD_prediction}
\ee
where $\sum_f e_f^2$ is a flavor sum over the quark charges squared.
For four active flavors and $x \ltsim 10^{-3}$,
(\ref{Eq_FL_QCD_prediction}) can be approximated as
follows~\cite{Cooper-Sarkar:1988ds+X}
\be
        xG(x,Q^2) 
        \approx \frac{3}{5}\,5.8\, 
        \left[ 
        \frac{3\pi}{4\alphaS}\, F_L(0.417 x, Q^2)
        - \inv{1.97}\, F_2(0.75 x, Q^2) 
        \right] 
        \ .
\label{Eq_xg(x,Q^2)_approximation}
\ee
For typical $\Lambda_{QCD} = 100-300\,\MeV$ and $Q^2 = 50 -
100\,\GeV^2$, the coefficient of $F_L$ in
(\ref{Eq_xg(x,Q^2)_approximation}), $3\pi/(4\alphaS) = {\cal{O}}(10)$,
is large compared to the one of $F_2$. Taking into account also the
values of $F_2$ and $F_L$, in this $Q^2$ region and for $x \ltsim
10^{-3}$, the gluon distribution is mainly determined by the
longitudinal structure function. The latter can be expressed in terms
of the profile function for longitudinal photon-proton scattering
using the optical theorem (cf.~(\ref{Eq_optical_theorem}))
\be
        F_L(x,Q^2) 
        = \frac{Q^2}{4\,\pi^2\,\alphaEM}\,
        \sigma^{tot}_{\gamma^*_L p}(x,Q^2) 
        = \frac{Q^2}{4\,\pi^2\,\alphaEM}\, 
        2\!\int \!\!d^2b_{\!\perp}\,
        J_{\gamma_L^*p}(x,|\vec{b}_{\!\perp}|,Q^2) 
        \ ,
\label{fl}
\ee
where the $s$-dependence of the profile function is rewritten in terms
of the Bjorken scaling variable, $x = Q^2/s$. Neglecting the $F_2$
term in~(\ref{Eq_xg(x,Q^2)_approximation}), consequently, the gluon
distribution reduces to
\be
        xG(x,Q^2) 
        \approx
        1.305\,\frac{Q^2}{\pi^2 \alphaS}\,\frac{\pi}{\alphaEM}
        \int \!\!d^2b_{\!\perp}\,
        J_{\gamma_L^*p}(0.417 x,|\vec{b}_{\!\perp}|,Q^2)
        \ . 
\label{Eq_xg(x,Q^2)-J_gLp(x,b,Q^2)_connection}
\ee
Comparing (\ref{Eq_def_xg(x,Q^2)}) with
(\ref{Eq_xg(x,Q^2)-J_gLp(x,b,Q^2)_connection}), it seems natural to
relate the integrand of (\ref{Eq_xg(x,Q^2)-J_gLp(x,b,Q^2)_connection})
to the impact parameter dependent gluon distribution
\be
        xG(x,Q^2,|\vec{b}_{\!\perp}|) 
        \approx
        1.305\,\frac{Q^2}{\pi^2 \alphaS}\,\frac{\pi}{\alphaEM}\,
        J_{\gamma_L^*p}(0.417 x,|\vec{b}_{\!\perp}|,Q^2)
        \ .
\label{Eq_xg(x,Q^2,b)-J_gLp(x,b,Q^2)_relation}
\ee

The black disc limit of the profile function for longitudinal
photon-proton scattering~(\ref{Eq_gp_black_disc_limit}) imposes
accordingly an upper bound on $xG(x,Q^2,|\vec{b}_{\!\perp}|)$
\be
        xG(x,Q^2,|\vec{b}_{\!\perp}|)\ \leq \ 
        xG^{max}(Q^2)
        \approx 
        1.305\,\frac{Q^2}{\pi^2 \alphaS}\,\frac{\pi}{\alphaEM}\,
        J_{\gamma^*_L p}^{max}(Q^2)
        \ ,
\label{Eq_low_x_saturation} 
\ee
which is the low-$x$ saturation value of the gluon distribution
$xG(x,Q^2,|\vec{b}_{\!\perp}|)$ in our approach. With $\pi
J_{\gamma^*_L p}^{max}(Q^2)/\alphaEM \approx 1$ as shown in
Fig.~\ref{Fig_J_gp_(b=0,s,Q^2)}, a compact approximation
of~(\ref{Eq_low_x_saturation}) is obtained
\be
        xG(x,Q^2,|\vec{b}_{\!\perp}|)\ \leq \ 
        xG^{max}(Q^2)
        \approx 
        \frac{Q^2}{\pi^2 \alphaS}
        \ ,
\label{Eq_low_x_saturation_approximation} 
\ee
which is consistent with the results
in~\cite{Mueller:1986wy,Mueller:1999wm,Iancu:2001md} and indicates
strong color-field strengths $G^a_{\mu \nu} \sim 1/ \sqrt{\alphaS}$ as
well.

According to our
relations~(\ref{Eq_xg(x,Q^2,b)-J_gLp(x,b,Q^2)_relation})
and~(\ref{Eq_low_x_saturation}), the {\em blackness} described by the
profile function is a measure for the gluon density and the {\em black
  disc limit} corresponds to the maximum gluon density reached at the
impact parameter under consideration. In accordance with the behavior
of the profile function $J_{\gamma_L^*p}$, see
Fig.~\ref{Fig_J_gp_(b,s,Q^2)}, the gluon distribution
$xG(x,Q^2,|\vec{b}_{\!\perp}|)$ decreases with increasing impact
parameter for given values of $x$ and $Q^2$. The gluon density,
consequently, has its maximum in the geometrical center of the proton,
i.e.\ at zero impact parameter, and decreases towards the periphery.
With decreasing $x$ at given $Q^2$, the gluon distribution
$xG(x,Q^2,|\vec{b}_{\!\perp}|)$ increases and extends towards larger
impact parameters just as the profile function $J_{\gamma_L^*p}$ for
increasing $s$.  The saturation of the gluon distribution
$xG(x,Q^2,|\vec{b}_{\!\perp}|)$ sets in first in the center of the
proton ($|\vec{b}_{\!\perp}|=0$) at very small Bjorken $x$.

In Fig.~\ref{Fig_xg(x,Q^2,b=0)_vs_x}, the gluon distribution
$xG(x,Q^2,|\vec{b}_{\!\perp}|=0)$ is shown as a function of $x$ for
$Q^2 = 1,\,10,\,\mbox{and}\,100\,\GeV^2$, where the
relation~(\ref{Eq_xg(x,Q^2,b)-J_gLp(x,b,Q^2)_relation}) has been used
also for low photon virtualities.
\begin{figure}[htb]
  \centerline{\epsfig{figure=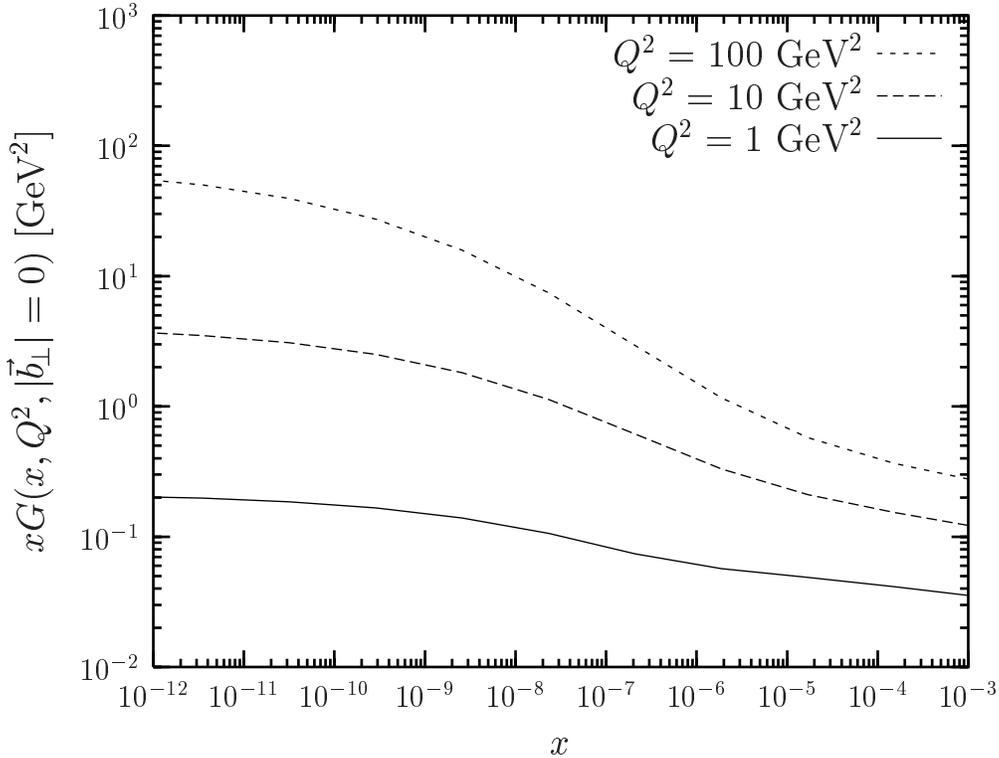}}
  \protect\caption{\small The gluon distribution of the proton at zero
    impact parameter $xG(x,Q^2,|\vec{b}_{\!\perp}|=0)$ is shown as a
    function of $x$ for $Q^2 = 1,\,10,\,\mbox{and}\,100\,\GeV^2$. The
    results are obtained within the
    approximation~(\ref{Eq_xg(x,Q^2,b)-J_gLp(x,b,Q^2)_relation}).}
\label{Fig_xg(x,Q^2,b=0)_vs_x}
\end{figure}
Evidently, the gluon distribution $xG(x,Q^2,|\vec{b}_{\!\perp}|=0)$
saturates at very low values of $x \ltsim 10^{-10}$ for $Q^2 \gtsim
1\,\GeV^2$. The photon virtuality $Q^2$ determines the saturation
value~(\ref{Eq_low_x_saturation}) and the Bjorken-$x$ at which it is
reached (cf. also Fig.~\ref{Fig_J_gp_(b,s,Q^2)}).  For larger $Q^2$,
the low-$x$ saturation value is larger and is reached at smaller
values of $x$, as claimed also in \cite{Gotsman:2001ku}.  Moreover, the
growth of $xG(x,Q^2,|\vec{b}_{\!\perp}|=0)$ with decreasing $x$ becomes
stronger with increasing $Q^2$. This results from the stronger energy
increase of the perturbative component, $\epsilon^{\pert} = 0.73$, that
becomes more important with decreasing dipole size.

According to our approach, the onset of the
$xG(x,Q^2,|\vec{b}_{\!\perp}|)$-saturation appears for $Q^2 \gtsim
1\,\GeV^2$ at $x \ltsim 10^{-10}$, which is far below the $x$-region
accessible at HERA ($x \gtsim 10^{-6}$). Even for THERA ($x\gtsim
10^{-7}$), gluon saturation is not predicted for $Q^2 \gtsim 1
\,\GeV^2$. Munier et al. have also shown in a recent
publication~\cite{Munier:2001gj} that the black disc limit is not
reached in the energy range of HERA. However, since the HERA data can
be described by models with and without saturation
embedded~\cite{Gotsman:2001ku}, the present situation is not
conclusive.\footnote{So far, the most striking hint for saturation in
  the present HERA data at $x\approx 10^{-4}$ and $Q^2 < 2\,\GeV^2$
  has been the turnover of $dF_2(x,Q^2)/d\ln(Q^2)$ towards small $x$
  in the Caldwell plot~\cite{Abramowicz:1999ii}, which is still a
  controversial issue due to the correlation of $Q^2$ and $x$ values.}

Note that the $S$-matrix unitarity
condition~(\ref{Eq_unitarity_condition}) together
with~(\ref{Eq_xg(x,Q^2,b)-J_gLp(x,b,Q^2)_relation}) requires the
saturation of the impact parameter dependent gluon distribution
$xG(x,Q^2,|\vec{b}_{\!\perp}|)$ but not the saturation of the
integrated gluon distribution $xG(x,Q^2)$. Due to multiple gluonic
interactions in our model, this requirement is fulfilled, as can be
seen from Fig.~\ref{Fig_J_gp_(b,s,Q^2)} and
relation~(\ref{Eq_xg(x,Q^2,b)-J_gLp(x,b,Q^2)_relation}). Indeed,
approximating the gluon distribution $xG(x,Q^2,|\vec{b}_{\!\perp}|)$
in the saturation regime of very low $x$ by a step-function
\be
        xG(x,Q^2,|\vec{b}_{\!\perp}|) 
        \approx xG^{max}(Q^2)\,
        \Theta(\,R(x,Q^2)-|\vec{b}_{\!\perp}|\,)
        \ ,
\label{Eq_J_gp_(x,b,Q^2)_Theta-approximation}
\ee
where $R(x,Q^2)$ denotes the black disc radius (or the full width at
half maximum of the profile function), one obtains
with~(\ref{Eq_def_xg(x,Q^2)}), (\ref{Eq_low_x_saturation}) and
(\ref{Eq_low_x_saturation_approximation}) the integrated gluon
distribution
\be
        xG(x,Q^2) 
        \;\approx\;
        1.305\,\frac{Q^2\,R^2(x,Q^2)}{\pi \alphaS}\,
        \frac{\pi}{\alphaEM}\,
        J_{\gamma^*_L p}^{max}(Q^2)
        \;\approx\;
        \frac{Q^2\,R^2(x,Q^2)}{\pi\alphaS}
        \ ,
\label{Eq_xg(x,Q^2)_saturation_regime}
\ee
which does not saturate because of the increase of the effective
proton radius $R(x,Q^2)$ with decreasing $x$. Nevertheless, although
$xG(x,Q^2)$ does not saturate, the saturation of
$xG(x,Q^2,|\vec{b}_{\!\perp}|)$ leads to a slow-down from powerlike to
squard logarithmic (see~(\ref{Eq_R(s)})) growth of $xG(x,Q^2)$ towards small
$x$.  \footnote{This is analogous to the slow down of the total $pp$
  cross section at high c.m.\ energy as soon as the corresponding
  profile function $J_{pp}(s,|\vec{b}_{\!\perp}|)$ reaches its black
  disc limit as shown in Sec.~\ref{Sec_Total_Cross_Sections}.}
Interestingly, our result~(\ref{Eq_xg(x,Q^2)_saturation_regime})
coincides with the result of Mueller and Qiu~\cite{Mueller:1986wy}.

Finally, it must be emphasized that the low-$x$ saturation of
$xG(x,Q^2,|\vec{b}_{\!\perp}|)$, required in our approach by the
$S$-matrix unitarity, is realized by {\em multiple gluonic
  interactions}. In other approaches that describe the evolution of
the gluon distribution with varying $x$ and $Q^2$, {\em gluon
  recombination} leads to gluon
saturation~\cite{Gribov:1983tu,Mueller:1986wy,McLerran:1994ni+X,Jalilian-Marian:1999dw+X,Iancu:2001hn+X},
which is reached when the probability of a gluon splitting up into two
is equal to the probability of two gluons fusing into one. A more
phenomenological understanding of saturation is attempted
in~\cite{Golec-Biernat:1999js,Capella:2001hq+X}.

%
\chapter[Comparison with Data and Saturation
    Effects in Observables ]
{Comparison with Data and Saturation
    Effects in Observables}
\label{Sec_Comparison_Data}

In this chapter we show the phenomenological performance of our
model. We compute total, differential, and elastic cross sections,
structure functions, and diffractive slopes for hadron-hadron,
photon-proton, and photon-photon scattering, compare the results with
experimental data including cosmic ray data, and provide predictions
for future experiments. Making use of the explicit saturation of the
impact parameter profiles at the black disc limit, the corresponding
energy values, and the logarithmic rise of the black disc radius
studied in the preceding chapter, we show explicitly manifestations of
$S$-matrix unitarity limits in experimental observables.

Using the $T$-matrix element given
in~(\ref{Eq_model_purely_imaginary_T_amplitude_final_result}), 
we compute the {\em pomeron} contribution to $pp$, $p\pbar$,
$\pi^{\pm}p$, $K^{\pm}p$, $\gamma^{*} p$, and $\gamma \gamma$
reactions in terms of the universal dipole-dipole scattering amplitude
$S_{DD}$. This allows one to compare reactions induced by hadrons and
photons in a systematic way. In fact, it is our aim to provide a
unified description of all these reactions and to show in this way
that the pomeron contribution to the above reactions is universal and
can be traced back to the dipole-dipole scattering amplitude $S_{DD}$.

Our model describes pomeron ($C=+1$ gluon exchange) but neither
odderon ($C=-1$ gluon exchange) nor reggeon exchange (quark-antiquark
exchange) as discussed in Sec.~\ref{Sec_Chi_Computation}. Only in the
computation of the hadronic total cross sections the reggeon
contribution is added~\cite{Donnachie:1992ny,Donnachie:2000kp}. This
improves the agreement with the data for $\sqrt{s} \ltsim 100\,\GeV$
and describes exactly the differences between hadron-hadron and
antihadron-hadron reactions.

The model parameters have been adjusted in fits to the high-energy
scattering data shown in this chapter. The resulting parameter set
given in Sec.~\ref{Sec_Model_Parameters} and
Appendix~\ref{Sec_Wave_Functions} is used throughout this work.

\section{Total Cross Sections}
\label{Sec_Total_Cross_Sections}

The total cross section for the high-energy reaction $ab \to X$ is
related via the {\em optical theorem} to the imaginary part of the
forward elastic scattering amplitude and can also be expressed in
terms of the profile function~(\ref{Eq_profile_function_def})
\be
        \sigma^{tot}_{ab}(s) 
        \;=\; \inv{s}\,\im\,T(s, t=0) 
        \;=\; 2 \int \!d^2b_{\!\perp}\,J_{ab}(s,|\vec{b}_{\!\perp}|)
        \ ,  
\label{Eq_optical_theorem}
\ee
where $a$ and $b$ label the initial particles whose masses were
neglected as they are small in comparison to the c.m.\ energy
$\sqrt{s}$. 

We compute the pomeron contribution to the total cross section,
$\sigma^{tot, \Pomeron}_{ab}(s)$, from the
$T$-matrix~(\ref{Eq_model_purely_imaginary_T_amplitude_final_result}),
as explained above, and add only here a reggeon contribution of the
form~\cite{Donnachie:1992ny,Donnachie:2000kp}
\be
        \sigma^{tot, \Reggeon}_{ab}(s)
        = X_{ab}\, \left(  \frac{s}{1\,\GeV^2} \right)^{-0.4525} 
        \ ,
\label{Eq_DL_reggeon_contribution}
\ee
where $X_{ab}$ depends on the reaction considered: $X_{pp} =
56.08\,\mb$, $X_{p\pbar} = 98.39\,\mb$, $X_{\pi^+p} = 27.56\,\mb$,
$X_{\pi^-p} = 36.02\,\mb$, $X_{K^+p} = 8.15\,\mb$, $X_{K^-p} =
26.36\,\mb$, $X_{\gamma p} = 0.129\,\mb$, and $X_{\gamma \gamma} =
605\,\nb$. Accordingly, we obtain the total cross section
\be
        \sigma^{tot}_{ab}(s)
        = \sigma^{tot, \Pomeron}_{ab}(s) 
        + \sigma^{tot, \Reggeon}_{ab}(s)
\label{Eq_total_cross_section_final_result}
\ee
for $pp$, $p\pbar$, $\pi^{\pm}p$, $K^{\pm}p$, $\gamma p$ and $\gamma
\gamma$ scattering.

\begin{figure}[p]
\setlength{\unitlength}{1.cm}
\begin{center}
\epsfig{file=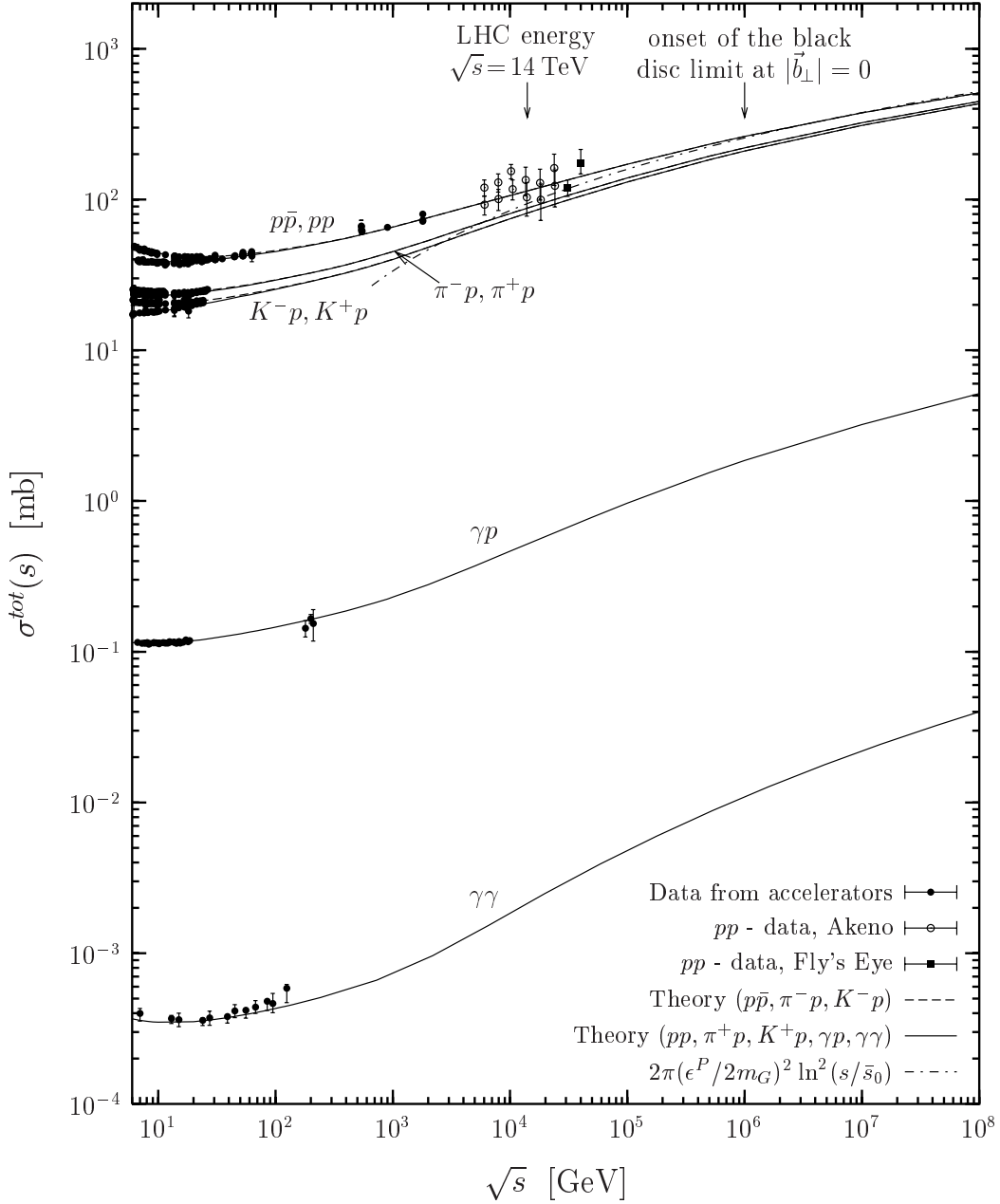,width=14.3cm}
\end{center}
\vspace{-0.5cm}
\caption{ \small The total cross section $\sigma^{tot}$ is shown as a
  function of the c.m.\ energy $\sqrt{s}$ for $pp$, $p\pbar$,
  $\pi^{\pm}p$, $K^{\pm}p$, $\gamma p$ and $\gamma \gamma$ scattering.
  The solid lines represent the model results for $pp$, $\pi^+p$,
  $K^+p$, $\gamma p$ and $\gamma \gamma$ scattering and the dashed
  lines the ones for $p\pbar$, $\pi^-p$, and $K^-p$ scattering. The
  dot-dashed line represents
  $2\pi(\epsilon^{\pert}/2m_G)^2\ln^2(s/{\bar s}_0)$ with $\sqrt{{\bar
      s}} = 20\,\GeV$. The $pp$, $p\pbar$, $\pi^{\pm}p$, $K^{\pm}p$,
  $\gamma p$~\cite{Groom:2000in} and $\gamma \gamma$
  data~\cite{Abbiendi:2000sz+X} taken at accelerators are indicated by
  the closed circles while the closed squares (Fly's eye
  data)~\cite{Baltrusaitis:1984ka+X} and the open circles (Akeno
  data)~\cite{Honda:1993kv+X} indicate cosmic ray data.  The arrows at
  the top point to the LHC energy, $\sqrt{s} = 14\,\TeV$, and to the
  onset of the black disc limit in $pp$ ($p\pbar$) reactions,
  $\sqrt{s} \approx 10^6\,\GeV$.}
\label{Fig_sigma_tot}
\end{figure}

The good agreement of the computed total cross sections with the
experimental data is shown in Fig.~\ref{Fig_sigma_tot}.
Here the solid lines represent the theoretical results for $pp$,
$\pi^+p$, $K^+p$, $\gamma p$, and $\gamma \gamma$ scattering and the
dashed lines the ones for $p\pbar$, $\pi^-p$, and $K^-p$ scattering.
The dot-dashed line represents
$2\pi(\epsilon^{\pert}/2m_G)^2\ln^2(s/{\bar s}_0)$ with $\sqrt{{\bar
    s}} = 20\,\GeV$.  The $pp$, $p\pbar$, $\pi^{\pm}p$, $K^{\pm}p$,
$\gamma p$~\cite{Groom:2000in} and $\gamma \gamma$
data~\cite{Abbiendi:2000sz+X} taken at accelerators are indicated by
the closed circles while the closed squares (Fly's eye
data)~\cite{Baltrusaitis:1984ka+X} and the open circles (Akeno
data)~\cite{Honda:1993kv+X} indicate cosmic ray data. Concerning the
photon-induced reactions, only real photons are considered which are,
of course, only transverse polarized.

The prediction for the total $pp$ cross section at LHC ($\sqrt{s} =
14\,\TeV$) is $\sigma^{tot}_{pp} = 114.2\,\mb$ in good agreement with
the cosmic ray data. Compared with other works, our LHC prediction is
close to the one of Block et al.~\cite{Block:1999hu},
$\sigma^{tot}_{pp} = 108 \pm 3.4\,\mb$, but considerably larger than
the one of Donnachie and Landshoff~\cite{Donnachie:1992ny},
$\sigma^{tot}_{pp} = 101.5\,\mb$.

The differences between $ab$ and $\bar{a}b$ reactions for $\sqrt{s}
\ltsim 100\,\GeV$ result solely from the different reggeon
contributions which die out rapidly as the energy increases. The
pomeron contribution to $ab$ and $\bar{a}b$ reactions is, in
contrast, identical and increases as the energy increases. It thus
governs the total cross sections for $\sqrt{s} \gtsim 100\,\GeV$ where
the results for $ab$ and $\bar{a}b$ reactions coincide.

The differences between $pp$ ($p\pbar$), $\pi^{\pm}p$, and $K^{\pm}p$
scattering result from the different transverse extension parameters,
$S_p = 0.86\,\fm > S_{\pi} = 0.607\,\fm > S_{K} = 0.55\,\fm$, cf.\ 
Appendix~\ref{Sec_Wave_Functions}.  Since a smaller transverse
extension parameter favors smaller dipoles, the total cross section
becomes smaller, and the short distance physics described by the
perturbative component becomes more important and leads to a stronger
energy growth due to $\epsilon^{\pert} = 0.73 > \epsilon^{\nprt} =
0.125$. In fact, the ratios $\sigma^{tot}_{pp}/\sigma^{tot}_{\pi p}$
and $\sigma^{tot}_{pp}/\sigma^{tot}_{Kp}$ converge slowly towards
unity with increasing energy as can already be seen in
Fig.~\ref{Fig_sigma_tot}.

For real photons, the transverse size is governed by the constituent
quark masses $m_f(Q^2=0)$, cf.\ Appendix~\ref{Sec_Wave_Functions},
where the light quarks have the strongest effect, i.e.\ 
$\sigma^{tot}_{\gamma p} \propto 1/m_{u,d}^2$ and
$\sigma^{tot}_{\gamma \gamma} \propto 1/m_{u,d}^4$. Furthermore, in
comparison with hadron-hadron scattering, there is the additional
suppression factor of $\alphaEM$ for $\gamma p$ and $\alphaEM^2$ for
$\gamma \gamma$ scattering coming from the photon-dipole transition.
In the $\gamma \gamma$ reaction, also the box diagram
contributes~\cite{Budnev:1975zs,Donnachie:2000kp} but is neglected
since its contribution to the total cross section is less than
1\%~\cite{Donnachie:2001wt}.

It is worthwhile mentioning that total cross sections for $pp$
($p\pbar$), $\pi^{\pm}p$, and $K^{\pm}p$ scattering do not depend on
the width $\Delta z_h$ of the longitudinal quark momentum distribution
in the hadrons~(\ref{Eq_hadron_wave_function}) since the underlying
dipole-dipole scattering process is independent of the longitudinal
quark momentum fraction $z_i$ for $t = 0$. On the two-gluon-exchange
level, this is shown analytically in
chapter~\ref{Chapt_string_decomposition}.

Saturation effects as a manifestation of the $S$-matrix unitarity can
be seen in Fig.~\ref{Fig_sigma_tot}. Having investigated the profile
function for hadron-hadron scattering, we know that this profile
function becomes higher and broader with increasing energy until it
saturates the black disc limit first for zero impact parameter
($|\vec{b}_{\!\perp}|=0$) at $\sqrt{s} \approx 10^6\,\GeV$.  Beyond
this energy, the profile function cannot become higher but expands
towards larger values of $|\vec{b}_{\!\perp}|$. Consequently, the
total cross section~(\ref{Eq_optical_theorem}) increases no longer due
to the growing blackness at the center but only due to the transverse
expansion of the hadrons. This tames the growth of the total hadronic
cross sections as can be seen for c.m. energies beyond $\sqrt{s}
\approx 10^6\,\GeV$ in Fig.~\ref{Fig_sigma_tot}.

At energies far beyond the onset of the black disc limit at zero impact
parameter, the profile function can be approximated by
\be 
      J_{ab}^{approx}(s,|\vec{b}_{\!\perp}|) =
      N_a\,N_b\,\Theta\left(R(s)-|\vec{b}_{\!\perp}|\right) \,
\label{Eq_J_ab_asymptotic_energies}
\ee
where $N_{a,b}$ denotes the normalization of the wave functions of the
scattered particles and $R(s)$ the black disc radius defined in
Sec.~\ref{Sec_PP_Profile_Function} that reflects the effective radii
of the interacting particles. Thus, the energy dependence of the total
cross section~(\ref{Eq_optical_theorem}) is driven exclusively by the
increase of the transverse extension of the particles $R(s)$
\be
        \sigma^{tot}_{ab}(s) = 2 \pi N_a N_b R(s)^2
        \ ,
\label{Eq_sigma_tot_ab_asymptotic_energies}
\ee
which is known as {\em geometrical
  scaling}~\cite{Amaldi:1980kd,Castaldi:1983ft}. Introducing the
  analytical result for the black disc radius $R(s)$ at
  asymptotic energies~(\ref{Eq_R(s)}) in the
  above equation, one obtaines the $\ln^2$-growth
\be
        \sigma^{tot}_{ab}(s) = 
        2 \pi N_a N_b \left(\frac{\epsilon^{\pert}}{2\,m_G}\right)^2\,
        \ln^2\left(\frac{s}{{\bar s}_0}\right)
        \ ,
\label{Eq_sigma_tot_logarithmic}
\ee
which coincides with the Froissart bound~\cite{Froissart:1961ux+X}.
For hadron wave function normalizations $N_{a,b}=1$, the total cross
section~(\ref{Eq_sigma_tot_logarithmic}) is independent of the hadron
species involved in the hadron-hadron scattering at asymptotic
energies. The hadronic cross sections start joining already at the
highest energies shown in Fig.~\ref{Fig_sigma_tot}. Also for photons
of different virtuality $Q_1^2$ and $Q_2^2$ one can check that the
ratio of the total cross sections $\sigma^{tot}_{\gamma^*
  p}(Q_1^2)/\sigma^{tot}_{\gamma^* p}(Q_2^2)$ converges to unity at
asymptotic energies in agreement with the conclusion
in~\cite{Schildknecht:2001qe}.

The total cross section~(\ref{Eq_sigma_tot_logarithmic}) obtained with
the values for $\epsilon^{\pert}$ and $m_G$ from
Sec.~\ref{Sec_Model_Parameters}, $N_{a,b}=1$, and the c.m. energy
$\sqrt{{\bar s}_0} = 20\,\GeV$ is shown as a dot-dashed line for
$\sqrt{s} \ge 600\,\GeV$ in Fig.~\ref{Fig_sigma_tot}. Evidently, the
so obtained logarithmic growth coincides with the numerical result for
the $pp$ cross section at high energies. In fact, a transition from a
powerlike to an $\ln^2$-increase of total cross sections seems to set
in at about $\sqrt{s} = 10^6\,\GeV$ as visible in
Fig.~\ref{Fig_sigma_tot}. Following the recent
publications~\cite{Gauron:2000ri+X,Dosch:2002pg}, already fits to available
forward scattering data support the universal
$\ln^2(s)$-dependence of the cross sections.

\section{Proton Structure Function}
\label{Sec_Structure_Functions}

The total cross section for the scattering of a transverse ($T$) and
longitudinally ($L$) polarized photon off the proton,
$\sigma_{\gamma^*_{T\!,L}p}^{tot}(x,Q^2)$, at photon virtuality $Q^2$ and
c.m.\ energy\footnote{Here, $\sqrt{s}$ refers to the c.m.\ energy in
  the $\gamma^* p$ system.} squared, $s=Q^2/x$, is equivalent to the
{\em structure functions} of the proton 
\be
        F_{T,L}(x,Q^2) 
        = \frac{Q^2}{4\pi^2\alphaEM} 
        \sigma_{\gamma^*_{T\!,L}p}^{tot}(x,Q^2)
\label{Eq_FTL}
\ee
and
\be
        F_2(x,Q^2) = F_{T}(x,Q^2) + F_{L}(x,Q^2)
        \ .
\label{Eq_F2}
\ee

Reactions induced by virtual photons are particularly interesting
because the transverse separation of the quark-antiquark pair that
emerges from the virtual photon decreases as the photon virtuality
increases, cf.\ Eq.~(\ref{rQ}). With increasing virtuality, one probes
therefore smaller transverse distance scales of the proton.

\begin{figure}[p]
  \centerline{\psfig{figure=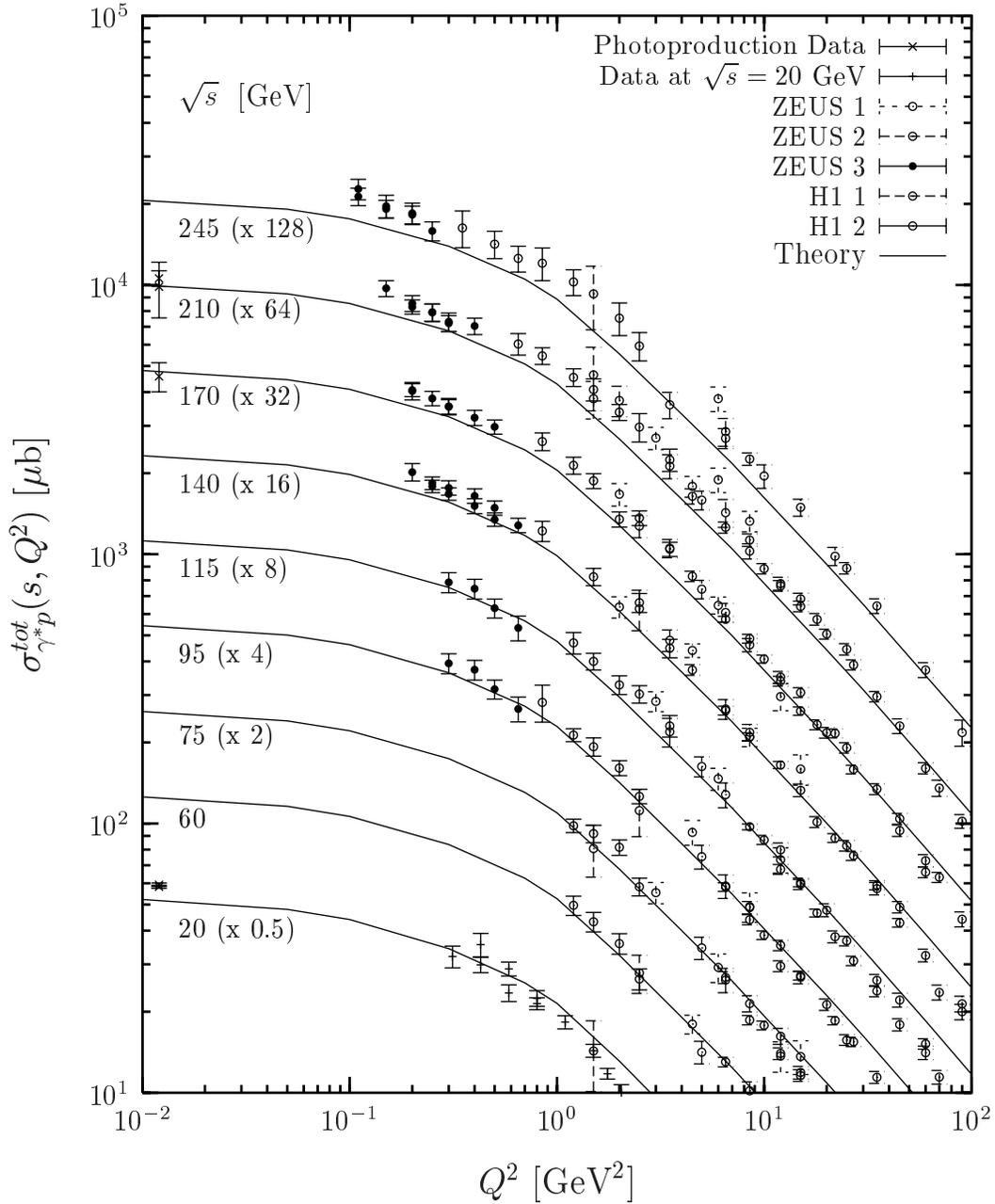,
      width=14cm}} \protect\caption{\small The total $\gamma^* p$
    cross section, $\sigma^{tot}_{\gamma^* p}(s,Q^2)$, is shown as a
    function of the photon virtuality $Q^2$ for c.m.\ energies from
    $\sqrt{s} = 20\,\GeV$ to $\sqrt{s} = 245\,\GeV$, where the model
    results (solid lines) and the experimental data at different
    $\sqrt{s}$ values are scaled with the indicated factors. The low
    energy data at $\sqrt{s} = 20\,\GeV$ are
    from~\cite{Benvenuti:1989rh+X}, the data at higher energies from
    the H1~\cite{Aid:1996au+X} and ZEUS
    collaboration~\cite{Derrick:1996ef+X}. The photoproduction
    ($Q^2=0$) data from~\cite{Caldwell:1978yb+X} are displayed at $Q^2
    = 0.012\,\GeV^2$.}
\label{Fig_sigma_tot_gp_vs_Q^2}
\end{figure}

In Fig.~\ref{Fig_sigma_tot_gp_vs_Q^2}, the $Q^2$-dependence of the
total $\gamma^* p$ cross section
\be
        \sigma^{tot}_{\gamma^*p}(s,Q^2)
        = \sigma^{tot}_{\gamma_T^*p}(s,Q^2)
        + \sigma^{tot}_{\gamma_L^*p}(s,Q^2)
\label{Eq_sigma_tot_gp_=_sigma_T_+_sigma_L}
\ee
is presented, where the model results (solid lines) are compared with
the experimental data for c.m.\ energies from $\sqrt{s} = 20\,\GeV$ up
to $\sqrt{s} = 245\,\GeV$. Note the indicated scaling factors at
different $\sqrt{s}$ values. The low energy data at $\sqrt{s} =
20\,\GeV$ are from~\cite{Benvenuti:1989rh+X} while the data at higher
energies have been measured at HERA by the H1~\cite{Aid:1996au+X} and
ZEUS collaboration~\cite{Derrick:1996ef+X}. At $Q^2 = 0.012\,\GeV^2$,
also the photoproduction ($Q^2=0$) data from~\cite{Caldwell:1978yb+X}
are displayed.

In the window shown in Fig.~\ref{Fig_sigma_tot_gp_vs_Q^2}, the model
results are in reasonable agreement with the experimental data. The
total $\gamma^* p$ cross section levels off towards small values of
$Q^2$ as soon as the photon size $|\vec{r}_\gamma|$, i.e\ the
resolution scale, becomes comparable to the proton size. Our model
reproduces this behavior by using the perturbative photon wave
functions with $Q^2$-dependent quark masses, $m_f(Q^2)$, that
interpolate between the current (large $Q^2$) and the constituent
(small $Q^2$) quark masses as explained in detail in
Appendix~\ref{Sec_Wave_Functions}. The decrease of
$\sigma^{tot}_{\gamma^* p}$ with increasing $Q^2$ results from the
decreasing dipole sizes since small dipoles do not interact as
strongly as large dipoles.

\begin{figure}[htb]
\centerline{\psfig{figure=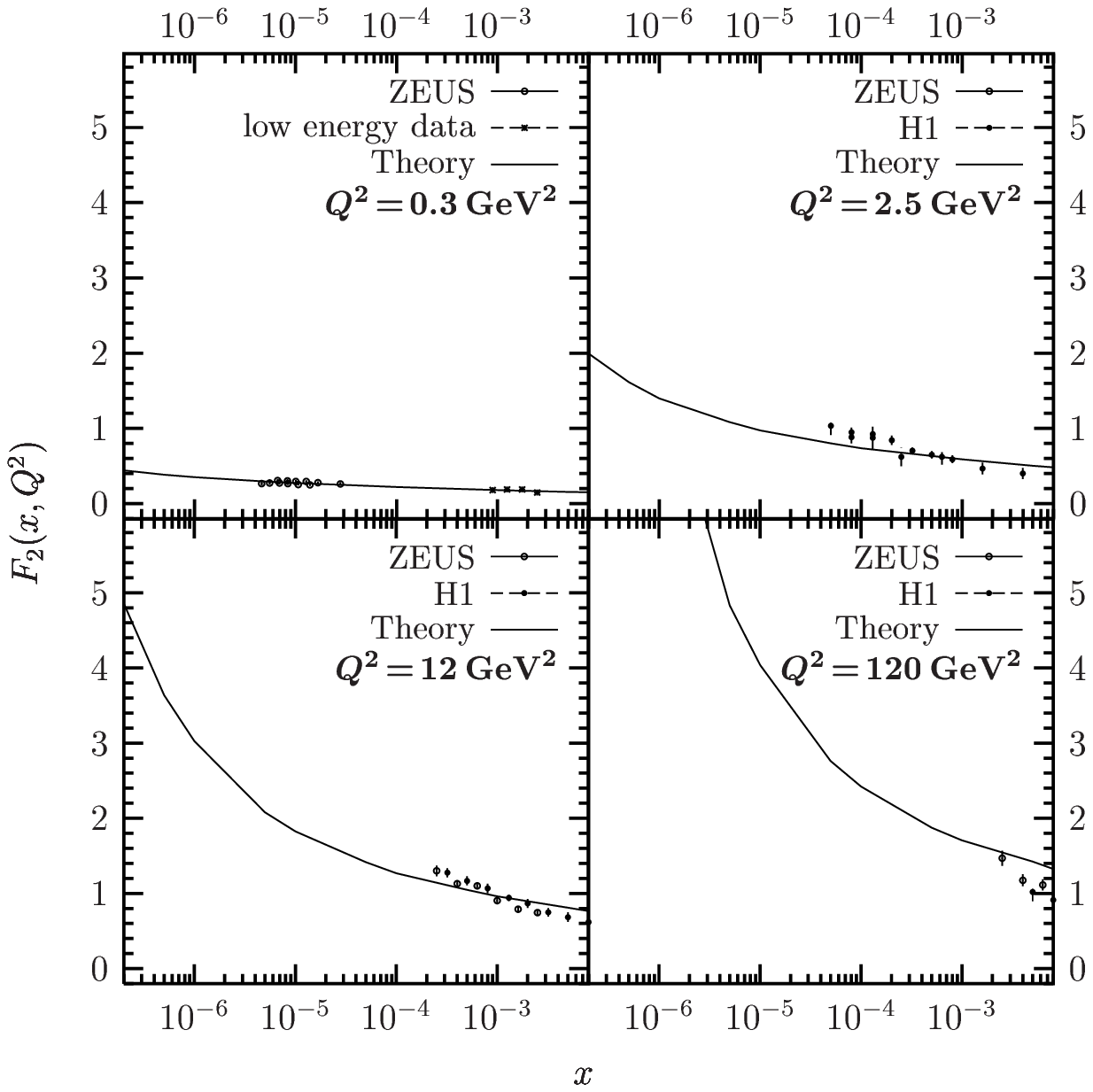, width=12.cm}}
\caption{\small The $x$-dependence of the computed proton structure
  function $F_2(x,Q^2)$ (solid line) is shown for $Q^2 =
  0.3,\,2.5,\,12$ and $120\,\GeV^2$ in comparison to the data measured
  by the H1~\cite{Abt:1993cb+X} and ZEUS~\cite{Derrick:1993ft+X}
  detector, and the low energy data at $\sqrt{s} = 20\,\GeV$
  from~\cite{Benvenuti:1989rh+X}.}
\label{F2_p}
\end{figure}

The $x$-dependence of the computed proton structure function
$F_2(x,Q^2)$ is shown in Fig.~\ref{F2_p} for $Q^2 = 0.3,\,2.5,\,12$
and $120\,\GeV^2$ in comparison to the data measured by the
H1~\cite{Abt:1993cb+X} and ZEUS~\cite{Derrick:1993ft+X} detector.
Within our model, the increase of $F_2(x,Q^2)$ towards small Bjorken
$x$ becomes stronger with increasing $Q^2$ in agreement with the
trend in the HERA data. This behavior results from the fast energy
growth of the perturbative component that becomes more important with
increasing $Q^2$ due to the smaller dipole sizes involved.

As can be seen in Fig.~\ref{F2_p}, the data show a stronger increase
with decreasing $x$ than the model outside the low-$Q^2$ region. This
results from the weak energy boost of the non-perturbative component
that dominates $F_2(x,Q^2)$ in our model. In fact, even for large
$Q^2$ the non-perturbative contribution overwhelms the perturbative
one, which explains also the overestimation of the data for $x \gtsim
10^{-3}$.
 
This problem is typical for the \SVM\ model applied to the scattering
of a small size dipole off a proton. In an earlier application by
R\"uter~\cite{Rueter:1998up}, an additional cut-off was introduced to
switch from the non-perturbative to the perturbative contribution as
soon as one of the dipoles is smaller than $r_{cut} = 0.16\,\fm$. This
yields a better agreement with the data also for large $Q^2$ but leads
to a discontinuous dipole-proton cross section. In the model of
Donnachie and Dosch~\cite{Donnachie:2001wt}, a similar \SVM-based
component is used also for dipoles smaller than $R_c = 0.22\,\fm$ with
a strong energy boost instead of a perturbative component.
Furthermore, their \SVM-based component is tamed for large $Q^2$ by an
additional $\alphaS(Q^2)$ factor.

We did not follow these lines in order to keep a continuous,
$Q^2$-independent dipole-proton cross section and, therefore, cannot
improve the agreement with the $F_2(x,Q^2)$ data without losing
quality in the description of hadronic observables. Since our
non-perturbative component relies on lattice QCD, we are more
confident in describing non-perturbative physics and, thus, put more
emphasis on the hadronic observables. Admittedly, our model misses
details of the proton structure that become visible with increasing
$Q^2$. In comparison, most other existing models provide neither the
profile functions nor a simultaneous description of hadronic and
$\gamma^*$-induced processes. For example, the phenomenological model
of Golec-Biernat and
W\"usthoff~\cite{Golec-Biernat:1999js,Golec-Biernat:1999qd} provides a
successful and economical description of the $\gamma^* p$ reactions
but cannot be applied to hadron-hadron reactions.

\section[Slope Parameter $B$ of Elastic Forward
Scattering]{\letterspace to .918\naturalwidth{Slope Parameter}
    \boldmath$B$ \letterspace to .918\naturalwidth{of Elastic Forward Scattering}}
\label{Sec_Slope_B}

The {\em local slope} of elastic scattering $B(s,t)$ is defined as
\be
        B(s,t) := 
        \frac{d}{dt} \left( \ln \left[ \frac{d\sigma^{el}}{dt}(s,t) \right] \right)
\label{Eq_elastic_local_slope}
\ee
and, thus, characterizes the diffractive peak of the differential
elastic cross section $d\sigma^{el}/dt(s,t)$ discussed below. Here, we
concentrate on the slope for elastic forward ($t=0$) scattering also
called {\em slope parameter}
\be
        B(s) 
        := B(s,t=0) 
        = \inv{2} 
        \frac{\int\!d^2b_{\!\perp}\,|\vec{b}_{\!\perp}|^2\,J(s,|\vec{b}_{\!\perp}|)}
        {\int\!d^2b_{\!\perp}\,J(s,|\vec{b}_{\!\perp}|)}
        = \inv{2} \langle b^2 \rangle \ ,
\label{Eq_elastic_forward_slope}
\ee
which measures the rms interaction radius $\langle b^2 \rangle$ of the
scattered particles, and does not depend on the opacity. 

We compute the slope parameter with the profile function from the
$T$-matrix (\ref{Eq_model_purely_imaginary_T_amplitude_final_result})
and neglect the reggeon contributions, which are relevant only at
small c.m.\ energies, so that the same result is obtained for $ab$ and
$\bar{a}b$ scattering.

\begin{figure}[htb] 
\centerline{\epsfig{figure=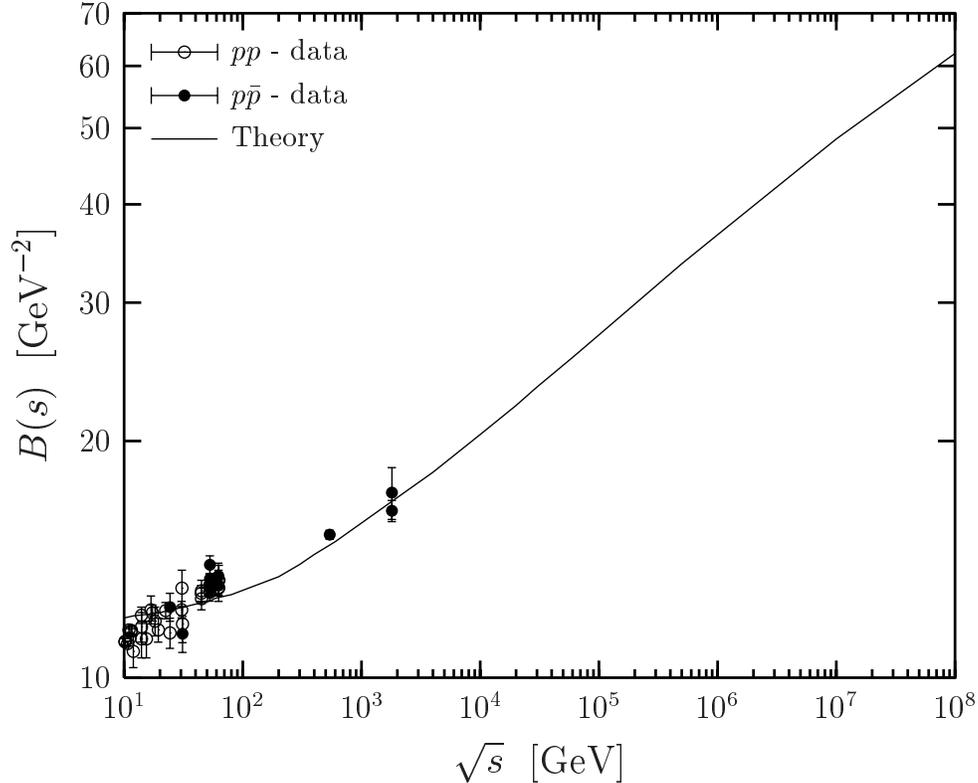, width=13.cm}}
\caption{The elastic slope parameter $B(s)$ is shown as a function of 
  the c.m.\ energy $\sqrt{s}$ for $pp$ and $p\pbar$ forward ($t=0$)
  scattering. The solid line represents the model result that is
  compared with the data for $pp$ (open circles) and $p\pbar$ (closed
  circles) reactions
  from~\cite{Amaldi:1971kt+X,Bozzo:1984ri,Amos:1989at}.}
\label{Fig_B_pp}
\end{figure}

In Fig.~\ref{Fig_B_pp}, the resulting slope parameter $B(s)$ is shown
as a function of $\sqrt{s}$ for $pp$ and $p\pbar$ scattering (solid
line) and compared with the $pp$ (open circles) and $p\pbar$ (closed
circles) data from~\cite{Amaldi:1971kt+X,Bozzo:1984ri,Amos:1989at}.
As expected from the opacity independence of the slope parameter
(\ref{Eq_elastic_forward_slope}), saturation effects as seen in the
total cross sections do not occur. Indeed, one observes an approximate
$B(s) \propto R^2(s) \propto \ln^2(s)$ growth for
$\sqrt{s} \gtsim 10^4\,\GeV$.  This behavior agrees, of course, with
the transverse expansion of $J_{pp}(s,|\vec{b}_{\!\perp}|)$ for
increasing $\sqrt{s}$ shown in Fig.~\ref{Fig_J_pp(b,s)}. Analogous
results are obtained also for $\pi p$ and $Kp$ scattering.

For the good agreement of our model with the data, the finite width of
the longitudinal quark momentum distribution in the hadrons, i.e.\ 
$\Delta z_p,\,\Delta z_{\pi},\,\mbox{and}\,\Delta z_{K}\neq 0$
in~(\ref{Eq_hadron_wave_function}), is important as the numerator in
(\ref{Eq_elastic_forward_slope}) depends on this width. In fact,
$B(s)$ comes out more than 10\% smaller with $\Delta z_p,\,\Delta
z_{\pi},\,\mbox{and}\,\Delta z_{K}= 0$. The $\Delta z_h$-dependence of
the non-forward scattering amplitude computed analytically on the
two-gluon exchange level is shown explicitly in
Appendix~\ref{App_T_tneq0}. Furthermore, a strong growth of the
perturbative component, $\epsilon^{\pert} = 0.73$, is important to
achieve the increase of $B(s)$ for $\sqrt{s} \gtsim 500\,\GeV$
indicated by the data.

It must be emphasized that only the simultaneous fit of the total
cross section and the slope parameter provides the correct shape of
the profile function $J(s,|\vec{b}_{\!\perp}|)$. This shape leads then
automatically to a good description of the differential elastic cross
section $d\sigma^{el}/dt(s,t)$ as demonstrated below. Astonishingly,
only few phenomenological models provide a satisfactory description of
both quantities~\cite{Block:1999hu,Kopeliovich:2001pc}. In the
approach of~\cite{Berger:1999gu}, for example, the total cross section
is described correctly while the slope parameter exceeds the data by
more than 20\% already at $\sqrt{s} = 23.5\,\GeV$ and, thus, indicates
deficiencies in the form of $J(s,|\vec{b}_{\!\perp}|)$.

\section{Differential Elastic Cross Sections}
\label{Sec_Diff_El_Cross_Section}

The {\em differential elastic cross section} obtained from the squared
absolute value of the $T$-matrix element
\be
        \frac{d\sigma^{el}}{dt}(s,t) 
        = \inv{16 \pi s^2}|T(s,t)|^2
\label{Eq_dsigma_el_dt}
\ee
can be expressed for our purely imaginary
$T$-matrix~(\ref{Eq_model_purely_imaginary_T_amplitude_final_result})
in terms of the profile function
\be
        \frac{d\sigma^{el}}{dt}(s,t) 
        = \inv{4\pi} \left[ 
        \int \!\!d^2b_{\!\perp}\,
        e^{i {\vec q}_{\!\perp} {\vec b}_{\!\perp}}\,
        J(s,|\vec{b}_{\!\perp}|)
        \right ]^2
        \ .
\label{Eq_dsigma_el_dt_model}
\ee
and is, thus, very sensitive to the transverse extension {\em and}
opacity of the scattered particles.
Equation~(\ref{Eq_dsigma_el_dt_model}) reminds of optical diffraction,
where $J(s,|\vec{b}_{\!\perp}|)$ describes the distribution of an
absorber that causes the diffraction pattern observed for incident
plane waves.

\begin{figure}[p]
  \centerline{\psfig{figure=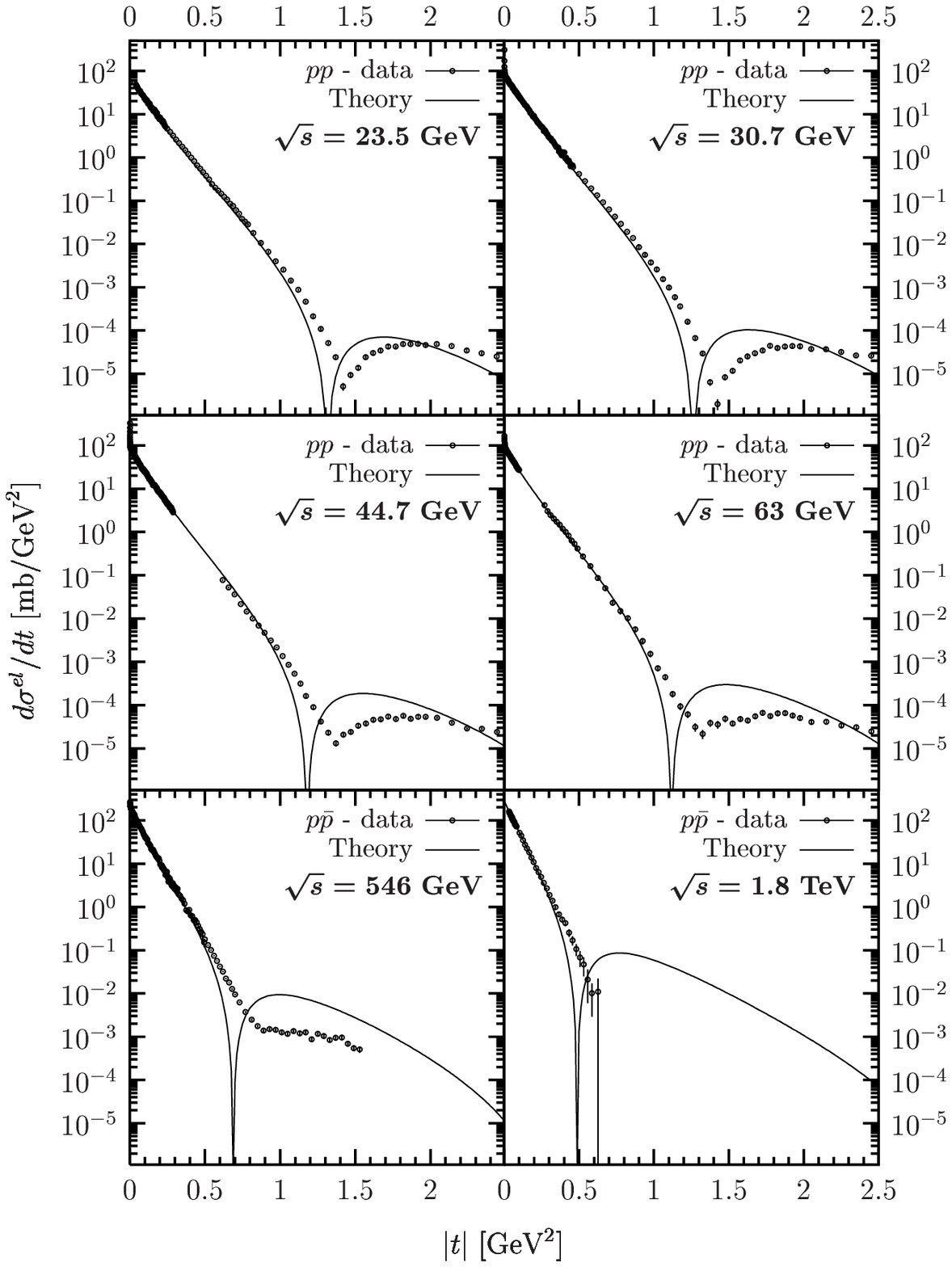,
      width=13.cm}} \protect\caption{ \small The differential elastic
    cross section for $pp$ and $p\pbar$ scattering is shown as a
    function of $|t|$ up to $2.5\,\GeV^2$. The result of our model,
    indicated by the solid line, is compared for $\sqrt{s} =
    23.5,\,30.7,\,44.7,\,\mbox{and}\,63\,\GeV$ with the CERN ISR $pp$
    data~\cite{Amaldi:1980kd}, for $\sqrt{s} = 546\,\GeV$ with the
    CERN $Sp{\pbar}S$ $p\pbar$ data~\cite{Bozzo:1984ri}, and for
    $\sqrt{s} = 1.8\,\TeV$ with the Fermilab Tevatron $p\pbar$
    data~\cite{Amos:1989at,Amos:1990jh}, all indicated by the open
    circles with error bars.}
\label{Fig_dsigma_el_dt_pp}
\end{figure}

\begin{figure}[tb]
  \centerline{\psfig{figure=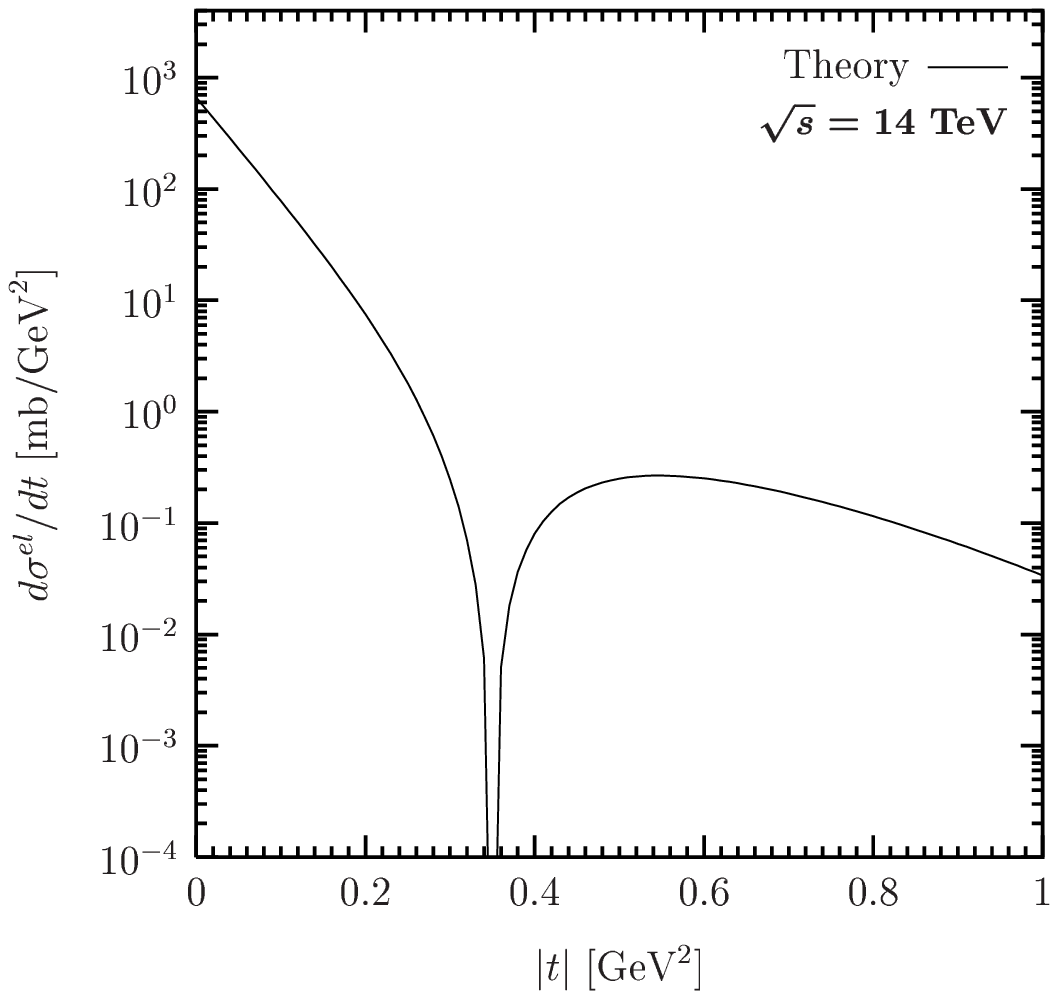,
      width=11.cm}} \protect\caption{ \small The prediction of our
    model for the $pp$ differential elastic cross section at LHC
    ($\sqrt{s} = 14\,\TeV$) is shown as a function of momentum
    transfer $|t|$ up to $1\,\GeV^2$.}
\label{Fig_dsigma_el_dt_pp_LHC}
\end{figure}

In Fig.~\ref{Fig_dsigma_el_dt_pp}, the differential elastic cross
section computed for $pp$ and $p\pbar$ scattering (solid line) is
shown as a function of $|t|=\vec{q}_{\!\perp}^2$ at $\sqrt{s} =
23.5,\,30.7,\,44.7,\,63,\,546$, and $1800\,\GeV$ and compared with
experimental data (open circles), where the $pp$ data at $\sqrt{s} =
23.5,\,30.7,\,44.7,\,\mbox{and}\,63\,\GeV$ were measured at the CERN
ISR~\cite{Amaldi:1980kd}, the $p\pbar$ data at $\sqrt{s} = 546\,\GeV$
at the CERN $Sp{\pbar}S$~\cite{Bozzo:1984ri}, and the $p\pbar$ data at
$\sqrt{s} = 1.8\,\TeV$ at the Fermilab
Tevatron~\cite{Amos:1989at,Amos:1990jh}. The prediction of our model
for the $pp$ differential elastic cross section at the CERN LHC,
$\sqrt{s} = 14\,\TeV$, is given in Fig.~\ref{Fig_dsigma_el_dt_pp_LHC}.

For all energies, the model reproduces the experimentally observed
diffraction pattern, i.e\ the characteristic {\em diffraction peak} at
small $|t|$ and the {\em dip} structure at medium $|t|$. As the energy
increases, also the {\em shrinking of the diffraction peak} is
described which reflects the rise of the slope parameter $B(s,t=0)$
already discussed above. The shrinking of the diffraction peak comes
along with a dip structure that moves towards smaller values of
$|t|$ as the energy increases. This motion of the dip is also
described approximately.

The dip in the theoretical curves reflects a change of sign in the
$T$-matrix
element~(\ref{Eq_model_purely_imaginary_T_amplitude_final_result}). As
the latter is purely imaginary, it is not surprising that there are
deviations from the data in the dip region. Here, the real part is
expected to be important~\cite{Amos:1990jh} which is in the small
$|t|$ region negligible in comparison to the imaginary part.

The difference between the $pp$ and $p\pbar$ data, a deep dip for $pp$
but only a bump or shoulder for $p\pbar$ reactions, requires a $C = -
1$ contribution. Besides the reggeon contribution at small
energies,\footnote{Zooming in on the result for $\sqrt{s} =
  23.5\,\GeV$, one finds further an underestimation of the data for
  all $|t|$ before the dip, which is correct as it leaves room for the
  reggeon contribution being non-negligible at small energies.} one
expects here an additional perturbative $C=-1$ contribution such as
three-gluon exchange~\cite{Fukugita:1979fe,Donnachie:1984hf+X} or an
odderon~\cite{Lukaszuk:1973nt+X,Rueter:1999gj,Dosch:2002ai}. In fact, allowing a
finite size diquark in the (anti-)proton an odderon appears that
supports the dip in $pp$ but leads to the shoulder in $p\pbar$
reactions~\cite{Dosch:2002ai}.

For the differential elastic cross section at the LHC energy,
$\sqrt{s} = 14\,\TeV$, the above findings suggest an accurate
prediction in the small-$|t|$ region but a dip at a position smaller
than the predicted value at $|t| \approx 0.35\,\GeV^2$. Our confidence
in the validity of the model at small $|t|$ is supported additionally
by the total cross section that fixes $d\sigma^{el}/dt(s,t=0)$ and is
in agreement with the cosmic ray data shown in
Fig.~\ref{Fig_sigma_tot}. Concerning our prediction for the dip
position, it is close to the value $|t| \approx 0.41\,\GeV^2$
of~\cite{Block:1999hu} but significantly below the value $|t| \approx
0.55\,\GeV^2$ of~\cite{Berger:1999gu}. Beyond the dip position, the
height of the computed shoulder is always above the data and, thus,
very likely to exceed also the LHC data. In comparison with other
works, the height of our shoulder is similar to the one
of~\cite{Block:1999hu} but almost one order of magnitude above the one
of~\cite{Berger:1999gu}.

Considering Figs.~\ref{Fig_dsigma_el_dt_pp}
and~\ref{Fig_dsigma_el_dt_pp_LHC} more quantitatively in the
small-$|t|$ region, one can use the well known parametrization of the
differential elastic cross section in terms of the slope parameter
$B(s)$ and the {\em curvature} $C(s)$
\be
        d\sigma^{el}/dt(s,t) 
        = d\sigma^{el}/dt(s,t=0)\,\exp\left[B(s)t+C(s)t^2\right]  
        \ .
\label{Eq_dsigma_el_dt_exp_parameterization}
\ee
Using $B(s)$ from the preceding section and assuming for the moment
$C(s) = 0$, one achieves a good description at small momentum
transfers and energies, which is consistent with the approximate
Gaussian shape of $J_{pp}(s,|\vec{b}_{\!\perp}|)$ at small energies
shown in Fig.~\ref{Fig_J_pp(b,s)}. The dip, of course, is generated by
the deviation from the Gaussian shape at small impact parameters.
According to (\ref{Eq_dsigma_el_dt_exp_parameterization}), the
shrinking of the diffraction peak with increasing energy reflects
simply the increasing interaction radius described by $B(s)$.

For small energies $\sqrt{s}$, our model reproduces the experimentally
observed change in the slope at $|t| \approx
0.25\,\GeV^2$~\cite{Barbiellini:1972ua+X} that is characterized by a
positive curvature. For LHC, we find clearly a negative value for the
curvature in agreement with~\cite{Block:1999hu} but in contrast
to~\cite{Berger:1999gu}. The change of sign in the curvature reflects
the transition of $J(s,|\vec{b}_{\!\perp}|)$ from the approximate
Gaussian shape at low energies to the approximate step-function
shape~(\ref{Eq_J_ab_asymptotic_energies}) at high energies.

Important for the good agreement with the data is the longitudinal
quark momentum distribution in the proton. Besides the slope
parameter, which characterizes the diffraction peak, also the dip
position is very sensitive to the distribution width $\Delta z_p$,
i.e.\ with $\Delta z_p= 0$ the dip position appears at more than 10\%
lower values of $|t|$. In the earlier \SVM\ 
approach~\cite{Berger:1999gu}, the reproduction of the correct dip
position was possible without the $z$-dependence of the hadronic wave
functions but a deviation from the data in the low-$|t|$ region had to
be accepted. In this low-$|t|$ region, we achieved a definite
improvement with the new correlation
functions~(\ref{Eq_MSV_correlation_functions}) and the minimal
surfaces used in our model.
\begin{figure}[htb]
  \centerline{\psfig{figure=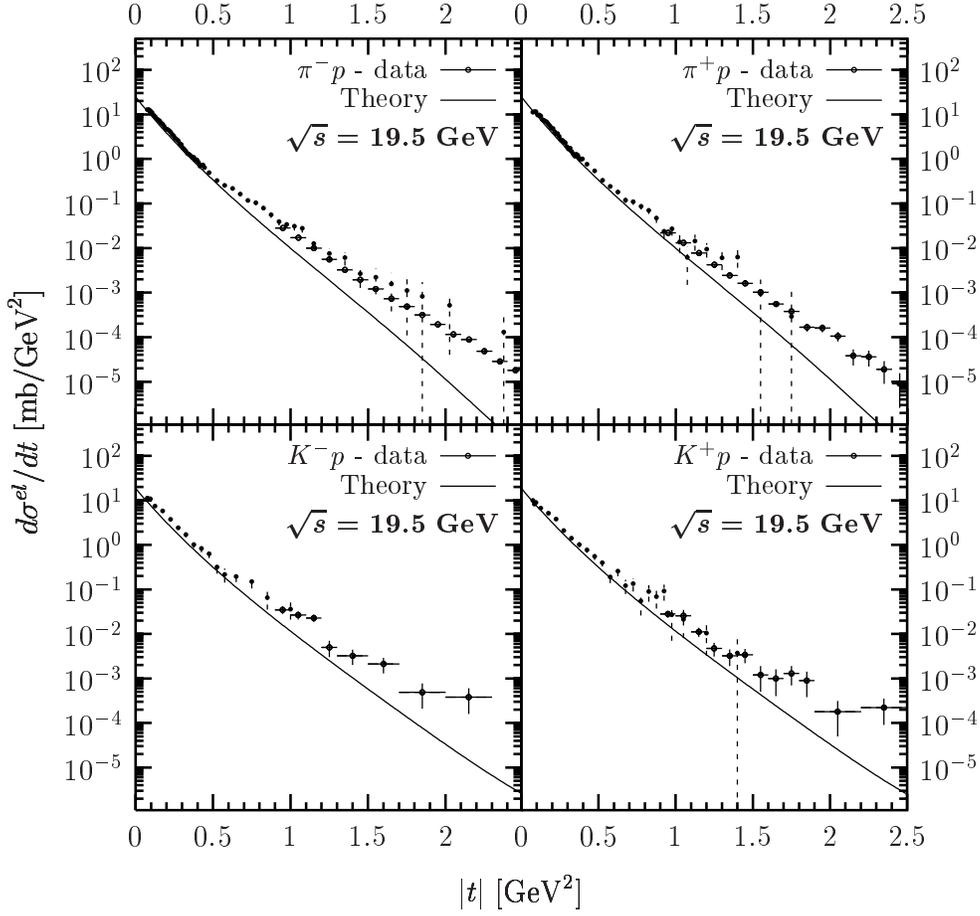,
      width=13.cm}} \protect\caption{ \small The differential elastic
    cross section $d\sigma^{el}/dt(s,t)$ is shown versus the momentum
    transfer $|t|$ for $\pi^{\pm}p$ and $K^{\pm}p$ reactions at the
    c.m.\ energy $\sqrt{s} = 19.5$ GeV. The model results (solid line)
    are compared with the data (closed circles with error bars)
    from~\cite{Akerlof:1976gk+X}.}
\label{Fig_dsigdt_kp_pip}
\end{figure} 

The differential elastic cross section computed for $\pi^{\pm}p$ and
$K^{\pm}p$ reactions has the same behavior as the one for $pp$
($p\pbar$) reactions. The only difference comes from the different
$z$-distribution widths, $\Delta z_{\pi}$ and $\Delta z_{K}$, and the
smaller extension parameters, $S_{\pi}$ and $S_{K}$, which shift the
dip position to higher values of $|t|$. This is illustrated in
Fig.~\ref{Fig_dsigdt_kp_pip}, where the model results (solid line) for
the $\pi^{\pm}p$ and $K^{\pm}p$ differential elastic cross section as
a function of $|t|$ are shown at $\sqrt{s} = 19.5\,\GeV$ in comparison
with experimental data (closed circles) from~\cite{Akerlof:1976gk+X}.
The deviations from the data towards large $|t|$ leave room for
odderon and reggeon contributions. Indeed, with a finite diquark size
in the proton, an odderon occurs that improves the description of the
data at large values of $|t|$~\cite{Berger:PhDthesis:1999}.

\section[Elastic Cross Sections $\sigma^{el}$, $\sigma^{el}/ \sigma^{tot}$, and $\sigma^{tot}/B$]{\hspace{-0.3cm}\letterspace to .9\naturalwidth{Elastic Cross Sections} \boldmath$\sigma^{el}$, \boldmath$\sigma^{el}\!/\!\sigma^{tot}$, \letterspace to .9\naturalwidth{and} \boldmath$\sigma^{tot}\!/\!B$}

The {\em elastic cross section} obtained by integrating the
differential elastic cross section
\be
        \sigma^{el}(s) 
        = \int_0^{-\infty}\!dt\,\frac{d\sigma^{el}}{dt}(s,t) 
        = \int_0^{-\infty}\!dt\,\inv{16 \pi s^2}|T(s,t)|^2
\label{Eq_total_elastic_cross_section}
\ee
reduces for our purely imaginary
$T$-matrix~(\ref{Eq_model_purely_imaginary_T_amplitude_final_result})
to
\be
        \sigma^{el}(s) 
        = \int \!\!d^2b_{\!\perp}\,|J(s,|\vec{b}_{\!\perp}|)|^2 
        \ .
\label{Eq_total_elastic_cross_section_J}
\ee
Due to the squaring, it exhibits the saturation of
$J(s,|\vec{b}_{\!\perp}|)$ at the black disc limit more clearly than
$\sigma^{tot}(s)$. Even more transparent is the saturation in the
following ratios given here for a purely imaginary $T$-matrix
\bea
        \frac{\sigma^{el}}{\sigma^{tot}}(s) 
        & = & 
        \frac
        {\int\!d^2b_{\!\perp}\,|J(s,|\vec{b}_{\!\perp}|)|^2}
        {2\int\!d^2b_{\!\perp}\,J(s,|\vec{b}_{\!\perp}|)}
        \ ,
\label{Eq_sigma_el/sigma_tot} \\
        \frac{\sigma^{tot}}{B}(s) 
        & = & 
        \frac
        {\left(2\int\!d^2b_{\!\perp}\,J(s,|\vec{b}_{\!\perp}|)\right)^2}
        {\int\!d^2b_{\!\perp}\,|\vec{b}_{\!\perp}|^2\,J(s,|\vec{b}_{\!\perp}|)} 
\label{Eq_sigma_tot/B}
        \ ,
\eea
which are directly sensitive to the opacity of the particles. This
sensitivity can be illustrated within the approximation 
\be
        T(s,t) = i\, s\, \sigma^{tot}(s)\, \exp[B(s) t/2]
\label{Eq_T_matrix_exp_parameterization}
\ee
that leads to the differential cross
section~(\ref{Eq_dsigma_el_dt_exp_parameterization}) with $C(s) = 0$,
i.e.\ an exponential decrease over $|t|$ with a slope $B(s)$. As the
purely imaginary $T$-matrix
element~(\ref{Eq_T_matrix_exp_parameterization}) is equivalent to
\be
        J(s,|\vec{b}_{\!\perp}|)
        =(\sigma^{tot}/4\pi B)\,\exp[-|\vec{b}_{\!\perp}|^2/2B] 
        =(4\sigma^{el}/\sigma^{tot})\,\exp[-|\vec{b}_{\!\perp}|^2/2B]
        \ ,
\label{Eq_J(b,s)_exp_parameterization}
\ee
one finds that the above ratios are a direct measure for the opacity
at zero impact parameter
\be
        J(s,|\vec{b}_{\!\perp}|=0) 
        = (\sigma^{tot}/4\pi B)
        = (4\sigma^{el}/\sigma^{tot}) 
        \ .
\label{Eq_J(b=0,s)_exp_parameterization}
\ee
For a general purely imaginary $T$-matrix, $T(s,t) =
i\,s\,\sigma^{tot}\,g(|t|)$ with an arbitrary real-valued function
$g(|t|)$, $J(s,|\vec{b}_{\!\perp}|=0)$ is given by
$(\sigma^{el}/\sigma^{tot})$ times a pure number which depends on the
shape of $g(|t|)$.

\begin{figure}[p]
\centerline{\psfig{figure=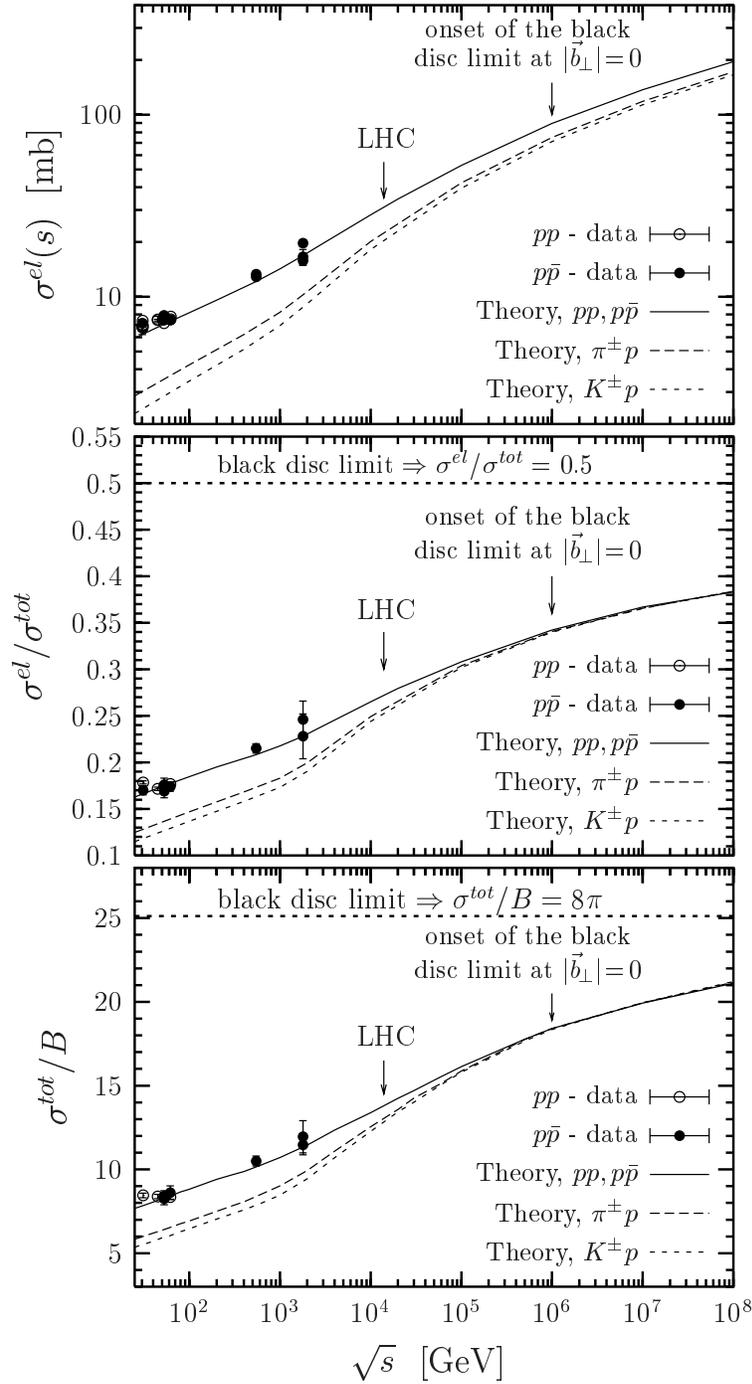, width=10.cm}}
\protect\caption{ \small The elastic cross section $\sigma^{el}$ and
  the ratios $\sigma^{el}/ \sigma^{tot}$ and $\sigma^{tot}/B$ are
  shown as a function of the c.m.\ energy $\sqrt{s}$. The model
  results for $pp$ ($p\pbar$), $\pi p$, and $K p$ scattering are
  represented by the solid, dashed and dotted lines, respectively. The
  experimental data for the $pp$ and $p\pbar$ reactions are indicated
  by the open and closed circles, respectively. The data for the
  elastic cross section are taken from~\cite{Groom:2000in} and the
  data for $\sigma^{tot}$ and $B$ from the references given in
  previous sections.}
\label{Fig_sigtot_el_and_ratios}
\end{figure}

We compute the elastic cross section $\sigma^{el}$ and the ratios
$\sigma^{el}/ \sigma^{tot}$ and $\sigma^{tot}/B$ in our model without
taking into account reggeons. In Fig.~\ref{Fig_sigtot_el_and_ratios},
the results for $pp$ and $p\pbar$ reactions (solid lines) are compared
with the experimental data (open and closed circles). The data for the
elastic cross section are taken from~\cite{Groom:2000in} and the data
for $\sigma^{tot}$ and $B$ from the references given in previous
sections.
For $pp$ ($p\pbar$) scattering, we indicate explicitly the prediction
for LHC at $\sqrt{s}=14\,\TeV$ and the onset of the black disc limit
at $\sqrt{s} = 10^6\,\GeV$. The model results for $\pi p$ and $K p$
reactions are presented as dashed and dotted line, respectively. For
the ratios, the asymptotic limits are indicated: Since the maximum
opacity or black disc limit governs the $\sqrt{s} \to \infty$
behavior, $\sigma^{el}/\sigma^{tot}$ ($\sigma^{tot}/B$) cannot exceed
$0.5$ ($8\pi$) in hadron-hadron scattering.

In the investigation of $pp$ ($p\pbar$) scattering, our theoretical
curves confront successfully the experimental data for the elastic
cross section and both ratios. At low energies, the data are
underestimated as reggeon contributions are not taken into account.
Again, the agreement is comparable to the one achieved
in~\cite{Block:1999hu} and better than in the approach
of~\cite{Berger:1999gu}, where $\sigma^{el}$ comes out too small due
to an underestimation of $d\sigma^{el}/dt$ in the low-$|t|$ region.

Concerning the energy dependence, $\sigma^{el}$ shows a similar
behavior as $\sigma^{tot}$ but with a more pronounced flattening
around $\sqrt{s} \gtsim 10^6\,\GeV$. This flattening is even stronger
for the ratios, drawn on a linear scale, and reflects very clearly the
onset of the black disc limit. As expected from the simple
approximation~(\ref{Eq_J(b=0,s)_exp_parameterization}),
$\sigma^{el}/\sigma^{tot}$ and $\sigma^{tot}/B$ show a similar
functional dependence on $\sqrt{s}$. At the highest energy shown,
$\sqrt{s} = 10^8\,\GeV$, both ratios are still below the indicated
asymptotic limits, which reflects that $J(s,|\vec{b}_{\!\perp}|)$
still deviates from the step-function
shape~(\ref{Eq_J_ab_asymptotic_energies}). The ratios
$\sigma^{el}/\sigma^{tot}$ and $\sigma^{tot}/B$ reach their upper
limits $0.5$ and $8\pi$, respectively, at asymptotic energies,
$\sqrt{s} \to \infty$, where the hadrons become infinitely large,
completely black discs.

In comparision to the $pp$ ($p\pbar$) result for $\sigma^{el}$, the
$\pi p$ and $Kp$ elastic cross sections are smaller due to the smaller
pion and kaon sizes or the corresponding narrower profile functions
shown in Fig.~\ref{Fig_J_pp(b,s)}. For $pp$ ($p\pbar$), $\pi p$, and
$Kp$ scattering, the ratios $\sigma^{el}/\sigma^{tot}$ and
$\sigma^{tot}/B$ representing the opacity of the interacting hadrons
in their center (cf.\ Fig.~\ref{Fig_J_pp(b,s)} at $|\vec{b}_{\!\perp}|
= 0$) converge for $\sqrt{s} \geq 10^6\,\GeV$ as shown in
Fig.~\ref{Fig_sigtot_el_and_ratios}. This follows from the identical
normalizations of the hadron wave functions that lead to an identical
black disc limit for hadron-hadron reactions approached in the center
of the hadrons at $\sqrt{s} \approx 10^6\,\GeV$.

%% file: SectionConclusion.tex
%
\chapter{Conclusions and Outlook}
\label{Sec_Conclusion}

%
%
We have developed a model that combines perturbative and
non-perturbative QCD to compute high-energy reactions of hadrons and
photons. We have investigated the QCD structure of non-perturbative
interactions in momentum space. A decomposition of the QCD string into
stringless dipoles has been found, confinement effects have been shown
explicitly in high-energy scattering and a microscopic structure of
unintegrated gluon distributions of hadrons and photons has been
obtained. Saturation effects that manifest the unitarity of the
$S$-matrix have been studied in impact parameter space of the
scattering amplitude. We have diplayed explicitly the saturation of
the impact parameter profiles for proton-proton and photon-proton
scattering at the black disc limit at high c.m. energies. The impact
parameter dependent gluon distribution of the proton is found to
saturate at small Bjorken $x$. We have compared the model results for
$pp$, $\pi p$, $Kp$, $\gamma^* p$ and $\gamma\gamma$ reactions with
experimental data and have shown saturation effects in experimental
observables.


The simplicity of minimal surfaces introduced in our model has allowed
us to show for the first time the QCD structure of the
non-perturbative dipole-dipole interaction in momentum space. This
contribution divides into two parts: The first part describes the
non-perturbative interaction between the quarks and antiquarks of the
two dipoles and exhibits the structure known from perturbative
two-gluon exchange~\cite{Low:1975sv+X,Gunion:iy}. The second part
describes the interaction between the strings of the two dipoles and
has a new structure originating from the geometry of the strings.

The contribution of the confining string to the total dipole-hadron
cross section $\sigma_{\!Dh}(x, |\vec{r}_{\!\mbox{\tiny\it D}}|)$ has
been studied. For small dipole sizes, $|\vec{r}_{\!\mbox{\tiny\it D}}|
\to 0$, the string contribution shows color transparency,
$\sigma_{\!Dh}(x,|\vec{r}_{\!\mbox{\tiny\it D}}|) \propto
r_{\!\mbox{\tiny\it D}}^2$, as known for the perturbative
contribution.  For large dipole sizes, $|\vec{r}_{\!\mbox{\tiny\it
    D}}|\,\gtsim\, 0.5\,\fm$, the non-perturbative contribution
increases linearly with increasing dipole size,
$\sigma_{\!Dh}(x,|\vec{r}_{\!\mbox{\tiny\it D}}|) \propto
|\vec{r}_{\!\mbox{\tiny\it D}}|$, in contrast to the perturbative
contribution which gives an $|\vec{r}_{\!\mbox{\tiny\it
    D}}|$\,-\,independent dipole-hadron cross section. This linear
increase is generated by the interaction of the string of the dipole
with the hadron: the longer the string, the larger the geometric cross
section with the hadron. String breaking, expected to stop the linear
increase for $|\vec{r}_{\!\mbox{\tiny\it D}}| \,\gtsim\, 1\, \fm$, is
excluded in our model that works in the quenched approximation. The
$|\vec{k}_{\!\perp}|$\,-\,factorization, known in perturbative
physics, has been found to be valid also for the non-perturbative
dipole-hadron interaction.

The most outstanding result of this work is the following feature of
the QCD string that confines the quark and antiquark in the dipole:
The QCD string of length $|\vec{r}_{\!\mbox{\tiny\it D}}|$ can be
exactly represented as an integral over stringless dipoles of sizes
$\xi|\vec{r}_{\!\mbox{\tiny\it D}}|$ with $0 \leq \xi \leq 1$ and
dipole number density $n(\xi) = 1/\xi^2$. A similar behavior has been
observed for the perturbatively computed wave function of a $q{\bar
  q}$ onium state in the large\,-\,$N_c$ limit where the numerous
emitted gluons inside the onium state represent
dipoles~\cite{Mueller:1994rr,Mueller:1994jq}. The decomposition of the
string into stringless dipoles has allowed us to rewrite the
string-hadron scattering process as an incoherent superposition of
dipole-hadron scattering processes and to extract the unintegrated
gluon distribution of hadrons and photons, respectively, from our
dipole-hadron and dipole-photon cross section via
$|\vec{k}_{\!\perp}|$\,-\,factorization.

We have shown explicitly the microscopic structure of the unintegrated
gluon distribution in hadrons and photons. For small momenta
$|\vec{k}_{\!\perp}|$, the unintegrated gluon distributions of hadrons
${\cal F}_h(x,k_{\!\perp}^2)$ are dominated by non-perturbative
physics and behave as $S_h^2/|\vec{k}_{\!\perp}|$ where $S_h$ denotes
the hadron extension.  The $1/|\vec{k}_{\!\perp}|$\,-\,behavior is a
string manifestation as it originates from the linear increase of the
total dipole-hadron cross section at large dipole sizes. For large
momenta, $|\vec{k}_{\!\perp}| \,\gtsim\, 1\,\GeV$, the unintegrated
gluon distributions of the hadrons are dominated by perturbative
physics and show the $1/k_{\!\perp}^2$\,-\,behavior induced by the
gluon propagator. In the perturbative region of large momenta the
valence constituents are resolved and the dependence of the
unintegrated gluon distribution on the hadron extension $S_h$
vanishes. In contrast, the unintegrated gluon distribution of photons
depends on the ``photon size'' for large $|\vec{k}_{\!\perp}|$ because
of the $Q^2$-dependent normalization of photon wave functions.

The $x$\,-\,dependence of the unintegrated gluon distribution of
hadrons and photons has been introduced phenomenologically into our
model. Motivated by experimental observations, we have given a strong
energy dependence to the perturbative contribution and a weak one to
the non-perturbative contribution. Consequently, with decreasing $x$
the perturbative contribution increases much stronger than the
non-perturbative contribution and extends into the
small\,-\,$|\vec{k}_{\!\perp}|$ region. A similar hard-to-soft
diffusion is seen also in the approach of Ivanov and
Nikolaev~\cite{Ivanov:2000cm} while a soft-to-hard diffusion is
obtained in the approach of the color glass
condensate~\cite{Iancu:2002xk}. Considering the integrated gluon
distribution of the proton $xG_p(x,Q^2)$, our non-perturbative
contribution dominates for $x \,\gtsim\, 10^{-3}$ while the
perturbative contribution becomes relevant for $x \,\ltsim\, 10^{-3}$
and generates the steep increase of $xG_p(x,Q^2)$ with decreasing $x$
at fixed $Q^2$. Also the rise of $xG_p(x,Q^2)$ with increasing $Q^2$
at fixed $x$ results from the strong energy dependence of the
perturbative contribution.

We have compared the unintegrated gluon distribution of the proton
${\cal F}_{p}(x,k_{\!\perp}^2)$ extracted from our loop-loop
correlation model (LLCM) with those obtained from the saturation model
of Golec-Biernat and W{\"u}sthoff (GBW)~\cite{Golec-Biernat:1999qd},
the derivative of the Gl{\"u}ck, Reya, and Vogt (GRV) parametrization
of $xG_p(x,Q^2)$~\cite{Gluck:1998xa}, and the approach of Ivanov and
Nikolaev (IN)~\cite{Ivanov:2000cm}. For $k_{\!\perp}^2 \to 0$, the
unintegrated gluon distribution of GBW decreases as $k_{\!\perp}^2$
and the one of IN as $k_{\!\perp}^4$ in contrast to the
$1/\sqrt{k_{\!\perp}^2}$\,-\,decrease found in our model.  In the
perturbative region, the LLCM, GRV, and IN unintegrated gluon
distributions become parallel for $x\,\ltsim\,10^{-2}$ and drop as
$1/k_{\!\perp}^2$ with increasing $k_{\!\perp}^2$. This perturbative
QCD behavior is not reproduced by the GBW unintegrated gluon
distribution which decreases exponentially with increasing
$k_{\!\perp}^2$. The $x$-dependence of the considered unintegrated
gluon distributions is weak in the non-perturbative region and becomes
stronger as $k_{\!\perp}^2$ increases.

We have studied saturation effects of the scattering amplitude in
impact parameter space.  
The impact parameter profiles have been found to respect the black
disc limit which is imposed by the $S$-matrix unitarity. We have
computed the impact parameter profiles for hadron-hadron and
longitudinal photon-proton scattering. They have shown the following
behavior for the evolution of the opacity and size of the interaction
particles with increasing energy: The particles become larger and
blacker as the energy increases. At energies beyond the black disc
limit, the opacity saturates while the expansion of the scattered
particles continues. The absolut value of the black disc limit and the
energy at which it is reached depend on the normalization of the wave
functions and the size of the interacting particles, respectively. For
asymptotic energies a universal result is obtained: the size of each
particle increases logarithmically with energy as required to
guarantee the Froissart bound~\cite{Froissart:1961ux+X}.

We have estimated the impact parameter dependent gluon distribution of
the proton $xG(x,Q^2,|\vec{b}_{\perp}|)$. We have shown a low-$x$
saturation of $xG(x,Q^2,|\vec{b}_{\perp}|)$ as a manifestation of the
$S$-matrix unitarity realized by multiple gluonic interactions. The
gluon density is found to decrease from the center towards the
periphery of the proton. The saturation value and the increase of
$xG(x,Q^2,|\vec{b}_{\perp}|=0)$ towards small $x$ depend on the photon
virtuality $Q^2$. In contrast, at fixed $Q^2$, the integrated gluon
distribution $xG(x,Q^2)$ does not saturate because of the growing
proton radius with decreasing $x$. In agreement with other
investigations of gluon
saturation~\cite{Mueller:1986wy,Mueller:1999wm,Iancu:2001md}, we have
found a slow down of $xG(x,Q^2)$ from a powerlike to a squared
logarithmic rise for Bjorken $x$ values beyond the black disc limit in
$xG(x,Q^2,|\vec{b}_{\perp}|)$.

The energies at which the saturation effects set in are model
dependent. In our approach the impact parameter profiles saturate the
black disc limit at zero impact parameter for $\sqrt{s} \gtsim
10^6\,\GeV$ in proton-proton scattering and for $\sqrt{s} \gtsim
10^7\,\GeV$ in longitudinal photon-proton scattering when $Q^2 \gtsim
1\,\GeV^2$. The saturation of $xG(x,Q^2,|\vec{b}_{\perp}|)$ occurs in
our approach for $Q^2\gtsim 1\,\GeV^2$ at values of $x \ltsim
10^{-10}$ which is far below the HERA and THERA range. Munier et
al.~\cite{Munier:2001gj} have also claimed that the black disc limit is not
reached in the energy range of HERA.

We have computed total cross sections $\sigma^{tot}$ for
proton-proton, pion-proton, kaon-proton, photon-proton, and
photon-photon reactions. A stronger energy rise of total cross
sections has been obtained going from large to small-size particles
involved in the interaction. This comes, of course, from the strong
energy dependence of the perturbative component that becomes
increasingly important with decreasing particle sizes. The growth of
the total cross sections becomes weaker for c.m.  energies $\sqrt{s}
\gtsim 10^6\,\GeV$ due to the onset of the black disc limit at
$|\vec{b}_{\perp}|=0$ in the profile functions. In fact, a transition
from a powerlike to an $\ln^2(s)$-increase of total cross sections of
hadrons and photons is found to set in at $\sqrt{s} \approx
10^6\,\GeV$. For asymptotic energies, the total hadronic cross
sections become universal, i.e., independent of the hadron species
considered, and increase in agreement with the Froissart
bound~\cite{Froissart:1961ux+X}. We predict for proton-proton
scattering at LHC ($\sqrt{s} = 14\,\TeV$) a total cross section of
$\sigma^{tot}_{pp} = 114.2\,\mb$ which is in good agreement with the
cosmic ray data.

For differential elastic cross sections $d\sigma^{el}/dt$ of
proton-proton, pion-proton and kaon-proton scattering, the diffraction
pattern and also the shrinkage of the diffraction peak with increasing
energy has been shown in good agreement with experimental data at
small momentum transfers $|t|$. Around the dip region, where a real
part is expected to be important, deviations from the data have
reflected that our $T$-matrix is purely imaginary. The smaller size of
the pion and kaon as compared to the proton has become visible in the
shift of the dip towards larger values of $|t|$ in the differential
elastic cross sections. A differential elastic cross section with a
negative curvature, $C<0$, and a dip at $|t| \approx 0.35\,\GeV^2$ is
predicted for LHC. The total elastic cross section, obtained by
integrating the differential elastic cross section over the momentum
transfer $|t|$, has shown similar features as the total hadronic cross
section discussed above. It takes a value of $\sigma^{el} \approx
30\,\mb$ at the LHC energy.

We have also studied the slope parameter $B(s)$ of the forward
differential elastic cross section for proton-proton scattering and
the ratios $\sigma^{el}/\sigma^{tot}$ and $\sigma^{tot}/B$ for
proton-proton, pion-proton and kaon-proton reactions. The slope
parameter and the ratios are a measure of the squared hadron extension
and hadron opacity, respectively. We have found an $\ln^2(s)$-increase
for the slope parameter at large c.m. energies. The ratios have shown
a universal behavior already for c.m. energies $\sqrt{s} \geq
10^6\,\GeV$. This behavior reflects most explicitly the saturation of
the impact parameter profiles at the black disc limit. The
slope parameter at LHC is predicted to be $B = 21.26\,\GeV^{-2}$.

The mentioned observables, namely, total cross sections, the structure
function of the proton, slope parameters, differential elastic cross
sections, elastic cross sections and the ratios
$\sigma^{el}/\sigma^{tot}$ and $\sigma^{tot}/B$ for proton-proton,
pion-proton, kaon-proton, photon-proton, and photon-photon reactions
are in good agreement with experimental data over a wide c.m. energy
range. We have provided a unified description of these reactions
using a universal mechanism for the interaction and energy
dependence. This mechanism seems to be appropriate and required
by experimental data.

In spite of the success of our combined perturbative and
non-perturbative QCD approach to high-energy scattering, there are
still several shortcomings and limitations of our model in its present
form that leave room for future improvements:
\begin{itemize}
\item Our approach does not generate energy dependent cross sections
  because of the missed gluon radiation.  Thus, for a fundamental
  understanding of the energy dependence of high-energy reactions one
  has to implement quantum evolution in our model analogous to
  complementary approaches~\cite{BFKL}. Attempts aiming at a
  description of high-energy scattering from the QCD
  Lagrangian have been recently reported: In~\cite{Hebecker:1999pb+X}
  structure functions of deep inelastic scattering at small Bjorken
  $x$ are related to an effective Euclidean field theory. Here one
  hopes that the limit $x\to 0$ corresponds to critical behavior in
  the effective theory. In another recent formalism, the energy
  dependence of the proton structure function has been related
  successfully to critical properties of an effective near light-cone
  Hamiltonian in a non-perturbative lattice
  approach~\cite{Pirner:2001pv+X}.
\item Our model works in the quenched approximation since dynamical
  fermion production is neglected. In addition also gluon vacuum
  polarization is missed. These shortcomings, of course, exclude
  string breaking in color-dipoles in the fundamental and adjoint
  representation, respectively. To be consistent with lattice QCD
  simulations~\cite{Laermann:1998gm,Bali:2000gf}, dynamical fermion
  and gluon production have to be introduced. Suggestions
  along these lines can be found in~\cite{Nachtmann:ed.kt}.
\item The Gaussian approximation of the functional integrals leads to
  a dependence of cross sections on the choice of the surface that is
  bounded the Wegner-Wislon loops.  It is a challenge to find the
  right surface that is favored by the Gaussian approximation.  There
  are several indications supporting the minimal surfaces used in this
  work.  Nevertheless, the minimal surfaces are not choosen by a variational
  principle and are not dynamical. Interpreting the surfaces as the
  worldsheets of the QCD strings, this means that our model can
  neither describe quantum fluctuations or excitations of the string
  nor string flips between two interacting color-dipoles.
  Accordingly, our model cannot reproduce the L\"uscher term
  (exitation of the string) in the static quark-antiquark potential
  recently confirmed on the lattice~\cite{Luscher:2002qv}. Interesting
  new developments towards a dynamical surface choice and a theory for
  the dynamics of the confining strings can be found
  in~\cite{Shevchenko:2002xi}.
\item Our ansatz for the gluon field strength correlator does not
  include explicitly the dependence on the path connecting the two
  gluon field strengths which has been recently measured on the
  lattice~\cite{DiGiacomo:2002mq}. In addition, the non-perturbative
  correlator, being appropriate to describe the
  scattering between large-size particles, overestimates the
  scattering between small-size particles as shown in deep
  inelastic scattering at high photon virtualities. Moreover, our
  decomposition of the correlator into a perturbative plus a
  non-perturbative component is not mandatory. Here it is not clear
  how to reconcile non-perturbative correlations with perturbative
  gluon exchange.  Methods complementary to our simple two component
  ansatz can be found
  in~\cite{Simonov:kt,Shevchenko:1998ej,Dosch:2000va,Shevchenko:2002xi}.
\item The static limit of our model becomes most explicit in the
  computation of the QCD van der Waals interaction between two static
  color-dipoles. A wrong result for the van der Waals potential is
  obtained from our model because of the energy degeneracy between the
  intermediate octet states and the intial (final) singlet states in
  the static limit. For a meaningful investigation of QCD van der
  Waals forces within our model, one has to go beyond the static limit
  and to describe the limited lifetime of the intermediate octet
  states appropriately.
\end{itemize}

Several other applications and extensions of the model can be
investigated in the future. The two most natural continuations of this
work would be
\begin{itemize}
\item to compute the vector meson production. Some work in this
  direction has already been done.
  
\item to generalize our model from two to many particle interactions,
  i.e. from the present particle-particle to particle-nucleus or
  nucleus-nucleus interactions. This would mean an extension of the
  model from the loop-loop correlation $\langle W_{r_1}[C_1] W_{r_2}[C_2]
  \rangle$ to multiple loop correlations $\langle W_{r_1}[C_1]
  \cdot\cdot\cdot W_{r_n}[C_n] \rangle$. There are already some first
  attempts in this direction.
\end{itemize}



%% file: Appendix.tex
%
\chapter{Conventions}
\label{Sec_Conventions}

For completeness the notational conventions used in this work are listed below.

\section{Units}

We work in natural units,
\be
        \hbar = c = 1
        \ .
\ee
The conversion constant reads $\hbar c = 197.32705\,\MeV\,\fm$.

\section{Lorentz Vectors}

Following the notations of the
textbooks~\cite{BJORKEN_1964_1965,Peskin:1995ev}, we write the
contravariant position four-vector $x^\mu$ in its instant form
\be
       x^\mu 
       = (x^0, x^1, x^2, x^3)
       = (t, x, y, z)
       = (x^0, \vec x_{\!\perp}, x^3)
       = (x^0, \vec x).
\label{Eq_contravariant_four-vector}
\ee
The covariant position four-vector 
\be
       x_\mu 
       = g_{\mu\nu} x^\nu        
       = (x_0, x_1, x_2, x_3)
       = (t, -x, -y, -z) 
\label{Eq_covariant_four-vector}
\ee
is obtained from the contravariant vector by the metric tensor
\be
        g_{\mu\nu}=g^{\mu\nu}
        =\pmatrix{
          +1&0&0&0\cr 
          0&-1&0&0\cr 
          0&0&-1&0\cr 
          0&0&0&-1\cr} 
        \ .
\label{Eq_metric_tensor}
\ee
Of course, implicit summation over repeated Lorentz indices is understood. The scalar product of $x^\mu$ with four-momentum $ p^\mu = (p^0, p^1, p^2, p^3) = (E, \vec p)$ is
\be
       x\cdot p = x^\mu p_\mu 
       = x^0 p_0 + x^1 p_1+ x^2 p_2 +x^3 p_3
       = tE - \vec x\cdot \vec p 
       \ .        
\label{Eq_scalar_product}
\ee
%

\section{Light-Cone Coordinates}

For the light-cone coordinates, we employ the convention that has been used by Lepage and Brodsky~\cite{Lepage:1980fj} and write the contravariant position four-vector as
\be
       x^\mu 
       = (x^+, x^-, x^1, x^2)
       = (x^+, x^-, \vec x_{\!\perp})
\label{Eq_contravariant_four-vector_light_cone}
\ee
with time-like (`light-cone time') and space-like (`light-cone position') components
\be
       x^+ = x^0 + x^3
       \quad{\rm and}\quad
       x^- = x^0 - x^3
       \ ,
\ee
respectively. The covariant position four-vector is again obtained by lowering the indices with the metric tensor, $x_\mu=g_{\mu\nu}x^\nu$, that has the following form in the Lepage-Brodsky light-cone coordinates
\be
        g^{\mu\nu}
        =\pmatrix{
          0&2&0&0 \cr 
          2&0&0&0 \cr 
          0&0&-1&0 \cr 
          0&0&0&-1 \cr}
        \qquad{\rm and}\qquad
        g_{\mu\nu}
        =\pmatrix{
          0&{1\over2}&0&0\cr 
          {1\over2}&0&0&0\cr  
          0&0&-1&0\cr 
          0&0&0&-1\cr}
        \ .
\ee
This form was required by the Lorentz-invariance of the scalar
product. The scalar product of the position four-vector with the momentum four-vector is written as
\be
       x\cdot p = x^\mu p_\mu 
       = x^+ p_+ + x^- p_- + x^1 p_1+ x^2 p_2 
       =  {1\over2}(x^+ p^- + x^- p^+) 
       - \vec x_{\!\perp} \vec p_{\!\perp}
       \ .
\label{Eq_scalar_product_light_cone}
\ee
The measure for a four-dimensional space-time integration, $d^4x$,
expressed in light-cone coordinates reads
\be
        d^4x = dx^0 dx^1 dx^2 dx^3 = \inv{2} dx^+ dx^- d^2x_{\!\perp}.
\label{Eq_d4x_light_cone}
\ee
%

\chapter{Wave Functions}
\label{Sec_Wave_Functions}

The light-cone wave functions $\psi_i(z_i,\vec{r}_i)$ provide the
distribution of transverse size and orientation ${\vec r}_{i}$ and
longitudinal quark momentum fraction $z_i$ to the light-like
Wegner-Wilson loops $W[C_i]$ that represent the scattering
color-dipoles. In this way, they specify the projectiles as mesons,
baryons described as quark-diquark systems, or photons that fluctuate
into a quark-antiquark pair before the interaction.

\section{Hadron Wave Function}

In this work mesons and baryons are assumed to have a quark-antiquark
and quark-diquark valence structure, respectively. As quark-diquark
systems are equivalent to quark-antiquark systems~\cite{Dosch:1989hu},
this allows us to model not only mesons but also baryons as
color-dipoles represented by Wegner-Wilson loops. We use for the
hadron wave function the phenomenological Gaussian Wirbel-Stech-Bauer
ansatz~\cite{Wirbel:1985ji}
\be
        \psi_h(z_i,\vec{r}_i) 
        = \sqrt{\frac{z_i(1-z_i)}{2 \pi S_h^2 N_h}}\, 
        e^{-(z_i-\inv{2})^2 / (4 \Delta z_h^2)}\,  
        e^{-|\vec{r}_i|^2 / (4 S_h^2)} 
        \ ,
\label{Eq_hadron_wave_function}
\ee
where the hadron wave function normalization to unity
\be
        \int \!\!dz_i d^2r_i \ |\psi_i(z_i,\vec{r}_i)|^2 = 1  
        \ ,
\label{Eq_hadron_wave_function_normalization}
\ee
requires the normalization constant
\be
        N_h = \int_0^1 dz_i \ z_i(1-z_i) \ e^{-(z_i-\inv{2})^2 / (2
        \Delta z_h^2)} 
        \ .
\label{Eq_N_h}
\ee
The different hadrons considered --- protons, pions, and kaons --- are
specified by $\Delta z_h$ and $S_h$ providing the width for the
distributions of the longitudinal momentum fraction carried by the
quark $z_i$ and transverse spatial extension $|\vec{r}_i|$,
respectively. In this work the extension parameter $S_h$ is a fit
parameter that should resemble approximately the electromagnetic
radius of the corresponding hadron~\cite{Dosch:2001jg}, while $\Delta z_h =
w/(\sqrt{2}\,m_h)$~\cite{Wirbel:1985ji} is fixed by the hadron mass
$m_h$ and the value $w = 0.35 - 0.5\,\GeV$ extracted from experimental
data. We find for (anti-)protons $\Delta z_p = 0.3$ and $S_p =
0.86\,\fm$, for pions $\Delta z_{\pi} = 2$ and $S_{\pi} = 0.607\,\fm$,
and for kaons $\Delta z_{K} = 0.57$ and $S_{K} = 0.55\,\fm$ which are
the values used in the main text. For convenience they are summarized
in Table~\ref{Tab_Hadron_Parameters}.
\begin{table}
\caption{\small Hadron Parameters} 
\vspace{0.3cm}
\centering      
\begin{tabular}{|l|l|l|}\hline
Hadron        & $\Delta z_h$    & $S_h\;[\fm]$  \\ [1ex] \hline\hline       
$p, \bar{p}$    & $0.3$           & $0.86$  \\ \hline
$\pi^{\pm}$     & $2$             & $0.607$ \\ \hline
$K^{\pm}$      & $0.57$          & $0.55$ \\ \hline
\end{tabular}
\label{Tab_Hadron_Parameters}
\end{table}

Concerning the quark-diquark structure of the baryons, the more
conventional three-quark structure of a baryon would complicate the
model significantly but would lead to similar predictions once the
model parameters are readjusted~\cite{Dosch:1994ym}. In fact, there
are also physical arguments that favor the quark-diquark structure of
the baryon such as the $\delta I = 1/2$ enhancement in
semi-leptonic decays of baryons~\cite{Dosch:1989hu} and the strong
attraction in the scalar diquark channel in the instanton
vacuum~\cite{Schafer:1994ra}.

\section{Photon Wave Function}   

The photon wave function $\psi_{\gamma}(z_i,\vec{r}_i,Q^2)$ describes
the fluctuation of a photon with virtuality $Q^2$ into a
quark-antiquark pair with longitudinal quark momentum fraction $z_i$
and spatial transverse size and orientation $\vec{r}_i$. The
computation of the corresponding transition amplitude $\langle
q\qbar(z_i,\vec{r}_i)|\gamma^*(Q^2)\rangle$ can be performed
conveniently in light-cone perturbation theory~\cite{Bjorken:1971ah+X}
and leads to the following squared wave functions for transverse $(T)$
and longitudinally $(L)$ polarized photons~\cite{Nikolaev:1991ja}
\bea
\!\!\!\!\!\!\!\!\!\!\!\!\!\!\!\!\!\!|\psi_{\gamma_T^*}(z_i,\vec{r}_i,Q^2)|^2\! 
        &\!\!\!=\!\!\!&\!\frac{3\,\alphaEM}{2\,\pi^2} \sum_f e_f^2
                \left\{\!\left[z_i^2\!+\!(1\!-\!z_i)^2\right]\!
                 \epsilon_f^2\,K_1^2(\epsilon_f\,|\vec{r}_i|) 
                  \!+\!m_f^2\,K_0^2(\epsilon_f\,|\vec{r}_i|) 
                \!\right\}
\label{Eq_photon_wave_function_T_squared} \\
\!\!\!\!\!\!\!\!\!\!\!\!\!\!\!\!\!\!|\psi_{\gamma_L^*}(z_i,\vec{r}_i,Q^2)|^2\! 
        &\!\!=\!\!&\!\frac{3\,\alphaEM}{2\,\pi^2} \sum_f e_f^2
                \left\{ 4\,Q^2\,z_i^2(1-z_i)^2\,K_0^2(\epsilon_f\,|\vec{r}_i|) \right\},
\label{Eq_photon_wave_function_L_squared}
\eea
where $\alphaEM$ is the fine-structure constant, $e_f$ is the electric
charge of the quark with flavor $f$, and $K_0$ and $K_1$ are the modified
Bessel functions (McDonald functions). In the above expressions,
\be
        \epsilon_f^2 = z_i(1-z_i)\,Q^2 + m_f^2
\label{Eq_photon_extension_parameter}
\ee
controlls the transverse size(-distribution) of the emerging dipole,
$|\vec{r}_i| \propto 1/ \epsilon_f$, that depends on the quark flavor
through the current quark mass $m_f$.

For small $Q^2$, the perturbatively derived wave functions,
(\ref{Eq_photon_wave_function_T_squared}) and
(\ref{Eq_photon_wave_function_L_squared}), are not appropriate since
the resulting large color-dipoles, i.e.\ $|\vec{r}_i| \propto 1/m_f
\gg 1\,\fm$, should encounter non-perturbative effects such as
confinement and chiral symmetry breaking. To take these effects into
account the vector meson dominance (VMD) model~\cite{Bauer:1978iq} is
usually used. However, the transition from the ``partonic'' behavior
at large $Q^2$ to the ``hadronic'' one at small $Q^2$ can be modelled
as well by introducing $Q^2$-dependent quark masses, $m_f = m_f(Q^2)$,
that interpolate between the current quarks at large $Q^2$ and the
constituent quarks at small $Q^2$~\cite{Dosch:1998nw}. Following this
approach, we use~(\ref{Eq_photon_wave_function_T_squared})
and~(\ref{Eq_photon_wave_function_L_squared}) also in the low-$Q^2$
region but with the running quark masses
\bea
        m_{u,d}(Q^2) 
        &=& 0.178\,\GeV\,\left(1-\frac{Q^2}{Q^2_{u,d}}\right)\,\Theta(Q^2_{u,d}-Q^2) 
        \ , 
        \label{Eq_m_ud_(Q^2)}\\
        m_s(Q^2) 
        &=& 0.121\,\GeV + 0.129\,\GeV\,\left(1-\frac{Q^2}{Q^2_s}\right)\,\Theta(Q^2_s-Q^2) 
        \label{Eq_m_s_(Q^2)}
        \ ,
\eea
and the fixed charm quark mass
\be
        m_c = 1.25\,\GeV
        \ ,
\ee
where the parameters $Q^2_{u,d} = 1.05\,\GeV^2$ and $Q^2_s =
1.6\,\GeV^2$ are taken directly from~\cite{Dosch:1998nw} while we
reduced the values for the constituent quark masses $m_f(Q^2 = 0)$
of~\cite{Dosch:1998nw} by about 20\%. The smaller constituent quark
masses are necessary in order to reproduce the total cross sections
for $\gamma^* p$ and $\gamma^* \gamma^*$ reactions at low $Q^2$.
Similar running quark masses are obtained in a QCD-motivated model of
the spontaneous chiral symmetry breaking in the instanton
vacuum~\cite{Petrov:1998kf} that improve the description of $\gamma^*
p$ scattering at low $Q^2$~\cite{Martin:1999bh+X}.

%

\chapter{Correlation Functions}
\label{Sec_Correlation_Functions}

In this appendix we describe explicitly the way from the simple
exponential correlation functions in Euclidean
space-time~(\ref{Eq_MSV_correlation_functions}) to their transverse
Fourier transforms in Minkowski space-time,
(\ref{Eq_F2[i_D_confining]})
and~(\ref{Eq_F2[i_D_non-confining_prime]}). The first step in this
procedure is the Fourier transformation of the exponential correlation
functions~(\ref{Eq_MSV_correlation_functions}) in four-dimensional
Euclidean space-time
\bea
        \tilde{D}^{E}(K^2) & \! = \! & \tilde{D}^{E}_{1}(K^2)
        = \int \! d^4Z \,D^{E}(Z^2/a^2)\,e^{iKZ} 
        \nonumber \\
        & \! = \! &   \int_0^\infty \!\!\! d|Z|\,|Z|^3 
                \int_0^\pi    \!\!\! d\phi_3\,\sin^2\!\phi_3 
                \int_0^\pi    \!\!\! d\phi_2\,\sin\!\phi_2 
                \int_0^{2\pi} \!\!\! d\phi_1 
                \,D^{E}(Z^2/a^2)\,e^{-i|K||Z|\cos\!\phi_3}
        \nonumber\\
        & \! = \! &   \frac{4\pi^2}{|K|} 
                \int_0^\infty \!\!\! d|Z|\,|Z|^2
                \,D^{E}(Z^2/a^2)\, J_1(|K||Z|) \, = \,\frac{12\pi^2}{a\,(K^2+a^{-2})^\frac{5}{2}}
        \ ,
\label{Eq_D^E(K^2)_for_exp_correlation}
\eea
where $J_1$ is the $1st$ order Bessel function of the first kind.
Here the Euclidean metric $-\delta_{\mu\nu}$ and four-dimensional
polar coordinates and the corresponding four-volume element $d^4Z =
d|Z|\,|Z|^3\,d\phi_3\,\sin^2\!\phi_3\,d\phi_2\,\sin\!\phi_2\,d\phi_1$
have been used. With~(\ref{Eq_D^E(K^2)_for_exp_correlation}), one
obtains 
\be
        \tilde{D}^{\prime\,E}_1(K^2) 
                := \frac{d}{dK^2}\,\tilde{D}^{E}_{1}(K^2) 
                 = -\,\frac{30\pi^2}{a\,(K^2+a^{-2})^\frac{7}{2}}
        \ .
\label{Eq_D_prime^E(K^2)_for_exp_correlation}
\ee
Now,~(\ref{Eq_D^E(K^2)_for_exp_correlation})
and~(\ref{Eq_D_prime^E(K^2)_for_exp_correlation}) are analytically
continued to Minkowski space-time, $K_4 \to i k^0$ or equivalently
$-\delta_{\mu\nu} \to g_{\mu\nu} = \mbox{diag}(1,-1,-1,-1)$,
\bea
        \tilde{D}(k^2) 
        =  -\,\frac{12\,\pi^2\,i}{a\,(-k^2+a^{-2})^\frac{5}{2}} \ , \quad\quad\quad
        \tilde{D}^{\prime}_1(k^2) 
        =  -\,\frac{30\,\pi^2\,i}{a(-k^2+a^{-2})^\frac{7}{2}} \ .
\label{Eq_D_(prime)(k^4)_for_exp_correlation}
\eea
Setting $k^0 = k^3 = 0$, which is enforced in the computation of
$\chi$ by $\delta$-functions, one finds $k^2=-{\vec k}^2_{\!\perp}$ and consequently
\be
        \hphantom{-}
        \tilde{D}^{(2)}({\vec k}^2_{\!\perp}) 
        =  -\,\frac{12\,\pi^2\,i}{a\,({\vec k}^2_{\!\perp} + a^{-2})^\frac{5}{2}} \ , \quad\quad\quad
        \hphantom{-}
        \tilde{D}^{\prime \,(2)}_1({\vec k}^2_{\!\perp}) 
        =  -\,\frac{30\,\pi^2\,i}{a({\vec k}^2_{\!\perp} + a^{-2})^\frac{7}{2}} \ .
\label{Eq_D_(prime)(k^2)_for_exp_correlation}
\ee
The transverse Fourier transformation~(\ref{Eq_transverse_Fourier_transform}) of these two expressions is the remaining step that leads directly to~(\ref{Eq_F2[i_D_confining]}) and~(\ref{Eq_F2[i_D_non-confining_prime]}).

\chapter{Non-Forward \boldmath$T$-Matrix Element}
\label{App_T_tneq0}

In this appendix we calculate the perturbative and non-confining
contribution to the non-forward ($t \neq 0$) $T$-matrix element. We
show explicitly that the non-forward $T$-matrix element depends on the
parameters which control the $z_i$\,-\,distribution of the wave
functions. These parameter are essential for a good description of
differential elastic cross sections and the slope parameter as shown
in Chapter~\ref{Sec_Comparison_Data}.

The confining contribution to the non-forward ($t \neq 0$) $T$-matrix
element is not presented. It is much more complicated since some of
the integrations cannot be performed analytically. Nevertheless, we
find numerically that it shows the same features concerning the
$z_i$\,-\,distributions as the perturbative and non-confining
contributions.

The perturbative contribution to the non-forward $T$-matrix element in
the small-$\chi$
limit~(\ref{Eq_model_purely_imaginary_T_amplitude_small_chi_limit}),
\be
        T^{\pert}(s_0,t) 
        =\frac{2is_0}{9} \!\int \!\!d^2b_{\!\perp} 
        e^{i {\vec q}_{\!\perp} {\vec b}_{\!\perp}}
        \!\int \!\!dz_1 d^2r_1 \!\int \!\!dz_2 d^2r_2\,\,
        |\psi_1(z_1,\vec{r}_1)|^2 \,\,
        |\psi_2(z_2,\vec{r}_2)|^2\,\left(\chi^{\pert}\right)^2 \ ,
\ee
reduces upon integration over the impact parameter $|{\vec
  b}_{\!\perp}|$ to
\bea
        &&\!\!\!\!\!\!\!\!\!
        T^{\pert}(s_0,t) =
        \frac{32is_0}{9}
        \int d^2k_{\!\perp}\,\, 
        \alphaS(k^2_{\!\perp})\,\,
        i\tilde{D}_{\pert}^{\prime \,(2)}(k^2_{\!\perp})\,\,
        \alphaS\!\left(\!(\vec{k}_{\!\perp}+
        \vec{q}_{\!\perp})^2\!\right)\,
        i\tilde{D}_{\pert}^{\prime\,(2)}\!
        \left(\!(\vec{k}_{\!\perp}+\vec{q}_{\!\perp})^2\right)
        \nonumber \\
        &&\!\!\!\!\!\!\!\!\!\times
        \int_0^1\!\! dz_1 \int_0^1\!\! dz_2
        \left[H_1\left(z_1^2q^2_{\!\perp}\right)-
        H_1\left((z_1\vec{q}_{\!\perp}+
        \vec{k}_{\!\perp})^2\right)\right]\!\! 
        \left [H_2\left(z_2^2q^2_{\!\perp}\right)-
        H_2\left((z_2\vec{q}_{\!\perp}+
        \vec{k}_{\!\perp})^2\right)\right] 
        \nonumber \\
\label{tneq0_p}
\eea
with 
\be
H_i\left((z_i\vec{q}_{\perp}+\vec{k}_{\!\perp})\right) := 
     \int \!d^2r_i\,|\psi_i(z_i,\vec{r}_i)|^2\,
     e^{i\,\vec{r}_i\,(z_i\vec{q}_{\perp}+\vec{k}_{\!\perp})} \ . 
\label{H_def}
\ee
and $|\psi_i(z_i,\vec{r}_i)|^2$ denoting hadron or photon wave
functions.

The non-confining contribution to the non-forward $T$-matrix
element in the small-$\chi$ limit~(\ref{Eq_model_purely_imaginary_T_amplitude_small_chi_limit}),
\be
        T^{\nprt}_{nc}(s_0,t) 
        =\frac{2is_0}{9} \!\int \!\!d^2b_{\!\perp} 
        e^{i {\vec q}_{\!\perp} {\vec b}_{\!\perp}}
        \!\int \!\!dz_1 d^2r_1 \!\int \!\!dz_2 d^2r_2\,\,
        |\psi_1(z_1,\vec{r}_1)|^2 \,\,
        |\psi_2(z_2,\vec{r}_2)|^2\,\left(\chi^{\nprt}_{nc}\right)^2 \ ,
\ee
becomes analogously 
\bea
        &&\!\!\!\!\!\!\!\!\!
        T^{\nprt}_{nc}(s_0,t) =
        \frac{8is_0}{9}\left(\frac{\pi^2 G_2\,(1-\kappa)}{24}\right)^{\!\!2} 
        \!\!\int \!\frac{d^2k_{\!\perp}}{(2\pi)^2}\,\,
        i\tilde{D}_{1}^{\prime \,(2)}(k^2_{\!\perp})\,\,
        i\tilde{D}_{1}^{\prime\,(2)}
        \left((\vec{k}_{\!\perp}+\vec{q}_{\!\perp})^2\right)
\label{tneq0_nc}\\
        &&\!\!\!\!\!\!\!\!\!\times
        \int_0^1\!\! dz_1 \int_0^1\!\! dz_2
        \left[H_1\left(z_1^2q^2_{\!\perp}\right)-
        H_1\left((z_1\vec{q}_{\!\perp}+
        \vec{k}_{\!\perp})^2\right)\right]\!\! 
        \left [H_2\left(z_2^2q^2_{\!\perp}\right)-
        H_2\left((z_2\vec{q}_{\!\perp}+
        \vec{k}_{\!\perp})^2\right)\right]
        \nonumber 
\eea
with $H_{1,2}$ defined in~(\ref{H_def}). 

For $t = -q^2_{\perp} \neq 0$, both contributions~(\ref{tneq0_p})
and~(\ref{tneq0_nc}) depend on the shape of the $z_i$-distribution of
the wave function, i.e., for the Gaussian hadron wave
function~(\ref{Eq_hadron_wave_function}), the contributions depend on
the width $\Delta z_h$. This $\Delta z_h$-dependence is transferred to
the differential elastic cross section $d\sigma^{el}/dt(s,t)$ given
in~(\ref{Eq_dsigma_el_dt}) and its local slope $B(s,t)$ given
in~(\ref{Eq_elastic_local_slope}).  At $t=0$, the dependence on the
shape of the $z_i$-distribution of the wave functions disappears
because of the normalization of the $z_i$-distribution as can be seen
immediately from~(\ref{H_def}).  Therefore, in our model the total
cross sections -- related via the optical theorem to the forward ($t =
0$) $T$-matrix element -- do not depend on the parameter that
characterize the $z_i$-distribution of the wave function.

%% file: Acknoledgements.tex
%
\centerline{\huge \bf Acknowledgments} 
\vspace*{1cm} 

First of all, I would like to express my gratitude to my adviser
Professor Hans-J{\"u}rgen Pirner for suggesting this work to me, for
numerous stimulating and insightful discussions, for constant
supervision and patient guidance, and for supporting me to give talks
at several conferences. His insightful advices have been essential for
the successful accomplishment of this research work. I appreciate very
much the friendly atmosphere in the Theory Group of Professor Pirner.

Next, I wish to thank Professor J{\"org} H{\"u}fner for his interest
in my work, for shairing his views, for willingness to referee this
thesis, and for taking part in the oral examination. Special thanks to
him for organizing the fridays ``tee meeting'', to his wife for
prepearing the delicious cakes, and to both for the tasty dinner and
the enjoyable evening at their home.

I want to thank Professor Hans-G{\"u}nter Dosch for numerous
discussions, his advice, and the fruitful collaboration.

I am grateful to Dr. habil. Hilmar Forkel for several insightful
discussions on physics and other topics.      

I want to express my gratitude to Dr. Yuri Ivanov for his continual
support in computational issues. Without Yuri's help most of the
formulas of this thesis could not have been turned into numbers.  I
was lucky to share the office with Yuri.

Now, I come to two colleagues and good friends of mine: Alberto
Polleri and Frank Steffen. Alberto earns special thanks for the
discussions about the physics, for the invitation to his extremely
good organized marriage, for the delicious dinner at his home, for the
nice time in New York, and particularly for the fun we had outside of
the physics in Heidelberg. The friendship with Frank started after
some time of close scientific collaboration. Special thanks to Frank
for this close and fruitful collaboration. I wish to thank him also
for the numerous discussions during lunch, for the donated photos made
during the holidays in Corsica, and for the really good time which we had
during the last three years.
 
Kai Schwenzer deserves a special thank since he took me to the
student's cafeteria on my first day in Heidelberg and offered to me
the first coffee in this city. I will never forget the fun we had in
Cargese.

I also thank J{\"o}rg Raufeisen, Alberto Accardi, and Yuan Feng for
interesting discussions, Carlo Ewerz and Otto Nachtmann for
suggestions which have helped me in finding the right perspective in
many cases, and Stephane Munier for discussions and proof-reading of the
abstract and introduction of the thesis. Sonja Bartsch, Carnelia
Merkel, and Melanie Steiert earn gratitude for the administrative
support, the careful reading of the manuscript, and the friendly
atmosphere which they create in the institute.

I want to thank my parents and brothers in a very special way for
their support, care, love, and advice. They are extremely important in
my life.

Finally, I am very thankful to my love, Hava, for all the wonderful
moments, for her support, and especially for her love. She is
of particular importance in my life.